\documentclass[fleqn,usenatbib]{mnras}
\usepackage{multirow}
\usepackage{newtxtext,newtxmath}
\usepackage{longtable}
\usepackage{indentfirst}
\usepackage{multicol}
\usepackage{graphicx}
\usepackage{subcaption}
\usepackage{multirow}
\usepackage{caption}
\usepackage{diagbox}
\usepackage[T1]{fontenc}
\usepackage{threeparttable}
\DeclareRobustCommand{\VAN}[3]{#2}
\let\VANthebibliography\thebibliography
\def\thebibliography{\DeclareRobustCommand{\VAN}[3]{##3}\VANthebibliography}
\usepackage{graphicx}
\usepackage{tablefootnote}
\usepackage{float}
\usepackage{placeins}
\usepackage{amsmath}	

\title[PN Morphologies \& CSPN from VLT, \emph{HST} \& Pan-STARRS]{Morphologies and Central Stars of 
Planetary Nebulae in the Galactic bulge from VLT, \emph{HST} and Pan-STARRS imaging}

\author[S. Tan et al.]{Shuyu Tan,$^{1,2}$\thanks{E-mail: shuyut@hku.hk} 
Quentin A. Parker$^{1,2}$\thanks{E-mail: quentinp@hku.hk}
Albert Zijlstra$^{3,2,1}$
Andreas Ritter$^{1,2}$
\\
$^{1}$Department of Physics, Chong Yuet Ming Physics Building, The University of Hong Kong, 
Pokfulam Road, Hong Kong\\
$^{2}$The Laboratory for Space Research, The University of Hong Kong, Cyberport 4, Hong Kong\\
$^{3}$Jodrell Bank Centre for Astrophysics, The University of Manchester, Oxford Road, M13 9PL,
Manchester, UK
}

\date{Accepted XXX. Received YYY; in original form ZZZ}

\pubyear{2015}

\begin{document}
\label{firstpage}
\pagerange{\pageref{firstpage}--\pageref{lastpage}}
\maketitle

\begin{abstract}
This is the first in a series of papers that present sets of different results for 136 
compact, known planetary nebulae within a 10~$\times$~10 degree 
region of the Galactic bulge. We use a large, previously unpublished sample 
of our own extensive ESO 8~m VLT deep imaging and spectroscopic data. 
This is combined with archival deep \emph{HST} imaging where 
available to provide a detailed morphological classification and study. 
The influence of angular resolution and sensitivity when assigning a morphology is discussed. 
A large fraction (68\%) of the sample are shown to be bipolars and the implications for 
this in the context of planetary nebulae progenitors are explored.
Four new planetary nebula central stars are also identified which are not in \emph{Gaia}.
This is based on both VLT and deep archival Pan-STARRS broad-band imagery. Some 11 putative 
central stars previously reported, based on \emph{Gaia} positions, are also not the true 
central star. In other cases the \emph{Gaia} central stars reported in the literature 
are actually based on the overall centroid position of a very compact planetary 
nebula rather than the actual 
central star within it. \emph{Gaia} parallax distances and kinematic ages for PNe in 
this sample are provided where possible based on fresh angular size measures from 
the new VLT imagery and \emph{Gaia} 
distances and literature expansion velocities when available. 
All these results are discussed within the context of the overall 
characteristics of the Galactic bulge and its planetary nebula population.
\end{abstract}



\section{Introduction}
\noindent
Planetary nebulae (PNe) are bright, gaseous emission envelopes ejected by low- 
to intermediate-mass stars ($M_{\star}\sim 0.8-8M_{\odot}$) during their 
post-AGB stage as the residual cores evolve to the white dwarf (WD) phase. 
Most PNe with hotter ($>$~45,000K) central stars (CSPN) are optically thin to Lyman 
continuum photons  \citep[e.g.][]{2007A&A...473..467S}. As such, outermost electrons 
of multiply-charged ions are ejected, giving PNe their predominate and strong emission lines. 
This makes PNe detectable even at large Galactocentric distances and in external 
galaxies like the Magellanic clouds \citep{2006MNRAS.373..521R} and other members of the 
local group \citep[e.g.][]{1980ApJS...42....1J,2006MNRAS.369..120M}. For an excellent, recent, 
general PNe review see \citet{2022PASP..134b2001K} and for a more specific PNe discovery and 
selection techniques review see \citet{2022FrASS...9.5287P}.

PNe exhibit a diversity of shapes (morphologies hereafter) that nevertheless can be 
allocated to a few basic underlying forms: elliptical (E), round (R), bipolar or butterfly 
(B), in rare cases irregular (I), asymmetric (A) and, when extremely compact, (S) for 
point-source like \citep[see, e.g.][]{1971A&A....10..161G, 1987ApJS...64..529C, 2000ASPC..199...17M, 2002ApJ...576..285S,  2003MNRAS.340..417C, 2006MNRAS.373...79P, 2011AJ....141..134S}.


Current morphological classifications assigned to all known Galactic PNe are contained 
in the "Gold Standard" HASH database\footnote{HASH: available online 
\url{http://www.hashpn.space}} \citet{parker2016hash}. HASH consolidates 
and federates available multi-wavelength imaging, spectroscopic and other data 
for both Galactic and Magellanic Cloud PNe. The assigned morphologies 
adopt the ERBIAS/sparm classification scheme introduced by 
\citet{2006MNRAS.373...79P} and used exclusively in HASH. The underlying physical causes 
for the observed morphological variety and their connection to their central stars 
(CSPN hereafter) have been extensively explored \citep[e.g.][]{2002ApJ...576..285S}, 
but mysteries concerning the definitive shaping mechanisms remain.

From an observational perspective, the interpretation of PN shapes has been largely 
dependent on the morphological characterisation of 2-D projections of largely optical
emission distributions from what are obviously 3-D nebula sources. The conversion from
2-D to a true 3-D representation of a PN depends on detailed kinematic sampling 
from high dispersion Integral Field Unit (IFU) spectroscopic observations
\citep[e.g.][]{2013MNRAS.434.1513D} and associated 3-D modelling with software 
such as \textsc{SHAPE} \citep{2006RMxAA..42...99S, 2015A&A...582A..60C}. 
While IFU observations are now becoming more common, 
e.g \citet{2016MNRAS.462.1393A, 2022MNRAS.516.2711C}, they remain, for now, the exception. 
This is at least partly due to the competition for access to limited, suitable IFU 
instrumentation (i.e. those with sufficient areal coverage, and angular and kinematic resolution). Consequently, morphological classifications and more fundamentally 
an understanding of them, is based on PNe 2-D imagery that will remain the default 
for the foreseeable future.

Modern, higher sensitivity and higher angular resolution observations, using dedicated 
ground-based and space-based telescopes and surveys, have further revealed a 
wealth of additional small-scale and large-scale features beyond the basic forms. These 
have further complicated this basic classification scheme. They include point-symmetric 
structures, jets, ansae, minor bipolar and multi-polar lobes, internal nebular striations, 
external, concentric rings and extensive "AGB" haloes and shells,  
\citep[e.g.][]{gorny1999atlas, sahai2011young, sabin2014first, akras2016low, 
stanghellini2016compact,sabin2021first}. These secondary morphological features 
are classified by the "sparm" system used by HASH (see later). 

Some background to the importance of PNe morphologies is given in Section~\ref{morph}.
The observations yielding the PN imaging data and details of the morphological and CSPN
analysis are described in Section~\ref{data} and Section~\ref{morph_n_cs} respectively. 
The various measurement results with the imaging data are listed and discussed in 
Section~\ref{res}. In Section~\ref{dis}, we briefly summarise current ideas about 
PN shaping mechanisms and compare the properties of PNe in our sample with 
theoretical predictions. Finally, Section~\ref{con} contains a summary of 
our main results with an emphasis on the prevalence of bipolar PNe uncovered.

\section{Why are PNe Morphologies Important?}
\label{morph}

PNe morphologies are a useful means to describe how gaseous ejected shells of dying 
stars are manifested. The most accurately determined morphologies from the best available 
(usually optical) imagery can permit detailed investigation  
as a function of different parameters. This is especially 
when combined with the statistical power from HASH classifications of large numbers 
of PNe.  Such parameters include Galactic location 
and environment, angular/physical size (so evolution), CSPN properties when available 
(including if in a binary system), nebular abundances and kinematic ages etc. 
Such studies can provide vital clues into the actual shaping mechanisms at 
play and their relative importance. 

Indeed, PNe morphologies, as a key component of multi-wavelength and spectroscopic observation and analysis, can deliver powerful insights into their chemical and 
physical conditions, central star and nebula evolutionary stages and other key properties. 
PNe morphologies can reveal effects from interactions with the local ISM, e.g. 
\citet{2007MNRAS.382.1233W}, the influence of possible external or internal magnetic 
fields, e.g. \citet{2007MNRAS.376..378S}, the results of fast winds interacting with 
previously ejected material at an earlier stage of evolution, e.g. 
\citet{2022ApJ...927..100K} and, as often invoked, the influence of CSPN binarity 
and common envelope evolution as perhaps the key driver of PNe shaping, e.g. 
\citet{2009PASP..121..316D}.

Being so diverse, PNe shapes are clearly telling us something fundamental about how and 
why PNe are formed and why most are not just a canonical, spherical gaseous envelope 
ejected from a single star. The observed diversity may result, at least in part,
from different types of progenitor stars where the basic 3 main bipolar, elliptical 
and round PNe morphologies are suggested to emerge from stars that have 
slightly different mass distributions \citep{gorny1997planetary} 
and where bipolar PNe in particular are thought produced by more massive 
progenitors \citep{corradi1995morphological, 1997A&A...321L..29M, 1998ApJS..117..341Z, 2001MNRAS.326.1041P}. 
This is especially given their preferred location at lower Galactic scale-heights 
in the mid-plane \citep{2006MNRAS.373...79P}. Here, any PNe found in such assumed 
younger environments should have derived from higher mass stars to be going 
through the PN phase now. Also see \citet{stanghellini1993correlations, 
amnuel1995asymmetrical, 1971A&A....10..161G, gorny1997planetary, 
stanghellini2002correlations} for further detailed comment and see \citet{2002ARA&A..40..439B}
for a review. Indeed, non-spherical features observed 
in the majority of PNe ($\sim$80\% - as determined from HASH) have long been attributed to 
stellar or sub-stellar companions. See \citet{2009PASP..121..316D} for a detailed case and 
review of the "binary hypotheses" of PNe formation and the supposition that most 
bipolar PNe arise from binaries \citep{1998ApJ...496..833S}. This counterpoint 
to the high mass, young progenitor bipolar origin scenario  was addressed by this latter work. 
Here, it is suggested that the observed correlation of more massive 
progenitors with bipolar PNe is due to larger ratios of the physical radii of 
red giant branch (RGB) stars compared to AGB stars that is exhibited by lower-mass 
stars compared to their higher mass equivalents. The larger radii of such stars 
on the RGB cause most stellar binary companions to interact with their lower-mass 
primaries that are already on the RGB. These could have formed bipolar PNe if the 
primary had been on the AGB.

\subsection{The importance of PNe in the Galactic bulge}

The Galactic bulge has been considered as an almost separate entity to the rest of the 
Galaxy consisting largely of older stars ($\sim$10~Gyrs) and low in gas and dust. 
For a good review see \cite{2016PASA...33...23N}. 
Previously, the bulge has been considered like an 
elliptical galaxy in overall colour and form. This simple picture has changed considerably 
over the last 20 years and the long-held view that the bulge is shaped like a tri-axial 
spheroid composed of old stars \citep{dwek1995morphology}, resembling an elliptical 
Galaxy shorn of its disk and spiral arms, is no longer valid. The bulge is now known 
to host an elongated central bar \citep{blitz1991direct, stanek1997modeling}. 
A detailed 3-D map using red clump stars as distance indicators (by assuming they have a 
constant absolute magnitude) also showed the bulge to be X-shaped \citep{saito2011mapping, 
saito2012vvv, wegg2013mapping}, i.e. a boxy peanut-shaped structure that becomes a bar in 
the inner bulge \citep[e.g.][]{athanassoula2005nature, zoccali2014giraffe} which has 
implications for feasible formation scenarios. Bulge dynamics
are different to the plane and in the inner region more chaotic. 

How well the currently available bulge PNe sample represents the underlying stellar population and its evolutionary history remains an open question. However, PNe 
represent a highly visible "touchstone" population from which to investigate certain bulge 
characteristics \citep[e.g.][]{peimbert1978chemical,peimbert1983type,corradi1995morphological}. Their range of inherent morphologies may play a significant role in 
aiding understanding late stage stellar evolution mass loss processes and, via spectroscopy
of the ejecta of different PNe types, large-scale chemical evolution and trends 
\citep{2013A&A...549A.147B,2017A&A...605A..89B}. 
This provides a key motivation for our work. The reliable 
morphological classification for compact Galactic bulge PNe, identification 
of their CSPN and estimation of their kinematic ages are the main focus of this first paper. 


\section{Sample Selection: New Observations and Archival Data}
\label{data}

In this work we present results from our new ESO VLT narrow-band imagery for
PNe located in the inner 10~$\times$~10 degree region of the Galactic bulge.
All the targets observed by the ESO VLT and in the \emph{HST} samples were 
selected to be highly likely physical members of the bulge.  This is 
based on the following criteria as detailed in \citet{rees2013alignment} 
and designed to reject contamination from bulge stars and foreground objects. 

\begin{enumerate}
    \item The PN lies within the inner 10 degrees of the Galactic Centre.
    \item The PN has a measured angular size of $>$~2~arcseconds and
    less than 35~arcseconds; e.g. see 
    \citet{acker2006400} that was further restricted to $\leq$~10~arcseconds (see below) 
    \item When a 5 GHz PN radio flux is available it must lie within the interval 
    (4.2 mJy, 59.1 mJy). \citep{siodmiak2001analysis, acker1992strasbourg}
\end{enumerate}

Whether a bulge PNe is compact or extended is irrelevant to their ages given the 
short overall PNe lifetimes c.f. that of their host stars. 
However, it is germane to their nebular
evolutionary state and ease of observation. The further restriction of the above 
selection criteria to angular extents $\leq$~10~arcseconds was applied 
to: i) yield young PNe that are usually of higher surface brightness; ii) be less 
affected by any seeing variations while iii) producing a sample size for the 
VLT that has realistic prospects of full completion. The restricted angular size also 
makes the sample more amenable to deep spectroscopy for detailed
abundance studies - a key aim of this overall project. Their compactness, though, 
does make their morphologies harder to discern from existing ground based
imagery. Here we re-examine both existing \emph{Hubble Space Telescope} (\emph{HST}) 
and our new VLT imagery to provide significantly improved classifications. 

The above final sample selection was based on vetting of all confirmed PNe from 
HASH in the bulge region that satisfied the above original criteria back in 2015. 
This produced 136 sources. However, HASH is an evolving database 
and in the interim many PNe entries have been updated, including some angular 
size measurements. Indeed, we provide fresh angular size measurements for all 
PNe in this sample (refer section~\ref{ang-size}) based largely on the new VLT narrow-band 
imagery. As a result many PNe would no longer satisfy the strict $\leq$~10~arcsecond 
diameter criterion. Many are now a little larger in these fresh estimates with an average 
value of 9.5~arcseconds with $\sigma$~=~4.2, but the integrity of the 
sample as compact PNe in the Galactic bulge remains.

All 136 original targets were spectroscopically observed with the FORS2 
instrument mounted on UT1 (Antu) of the European Southern Observatory (ESO) 
8.2 m Very Large Telescope (VLT) \citep{appenzeller1998successful}. 
The observations took place between 2015 and 2019  
under program IDs 095.D-0270(A), 097.D-0024(A), 099.D-0163(A),
and 0101.D-0192(A) for PIs Zijlstra and Parker. 
FORS2 narrow-band pre-imaging was obtained for all 136 sources. 
This imagery was supplemented by a further 40 objects previously observed using 
the Wide Field and Planetary Camera 2 Instrument (WFPC2) 
of the \emph{HST} in 2002, 2003 and 2009 through 
\emph{HST}-SNAP-8345 (PI: Sahai), \emph{HST}-SNAP-9356 (PI: Zijlstra) and 
\emph{HST}-GO-11185 (PI: Rubin). \emph{HST} imagery was used for morphological 
classification whenever available as it is obviously far superior. 


The main point of the VLT observations was to obtain deep, 
high quality PNe spectroscopy for abundance studies (the subject of 3 further 
papers in this series). The generally excellent narrow-band VLT imaging obtained for 
most was a mere by-product of the FORS2 spectroscopic observational process that often 
requires pre-imaging of the PNe in order to determine where best to place the 
spectroscopic slit.  The final observed VLT sample here comprises 136 PNe, 
representing about $\sim$20\% of the currently known bulge PNe population in HASH.

Our completed ESO VLT project was designated 
as a filler programme with no photometric quality requirement (and one of the reasons 
for choosing the target sample to be $\leq$~10~arcseconds across). These could proceed even 
under modest conditions, e.g. some clouds and/or high humidity and/or relatively 
poor seeing. This gave increased likelihood of our observations being obtained but
as a result care must be taken when attempting any reliable photometric assessment.
Fortunately, for the purposes of this particular paper, the adverse effects on 
discernable morphological detail and point source (CSPN) detection were modest 
over the range of observing conditions achieved in practice. 

Our VLT imagery and a careful assessment of the existing \emph{HST} imagery 
now enables us to better resolve both macro- and microscopic structures
for most of the objects not seen in previous studies and so present fresh, detailed  
morphological classification and analysis of this sample. These data are combined with the 
excellent resolution and deep Pan-STARRS broad band imagery \citep{2010SPIE.7733E..0EK}, 
when available. This includes the key blue 'g-band' and has allowed more 
accurate measurements and improvement to the position and information of CSPN where 
they are clearly identifiable, including 4 new CSPN discoveries not in \emph{Gaia}. 

\subsection{The ESO VLT instrumental configuration}

The FORS2 instrument is a multi-mode optical instrument mounted on the ESO VLT UT1 
telescope at the Cassegrain focus. It can perform imaging and long slit 
spectroscopy but also has a multi-object spectroscopy capability and even 
polarimetry \citep{1997SPIE.2871.1222N} that can cover a wavelength range 
from the atmospheric cut-off in the blue out to $\sim1\mu$m in the red.

FORS2 is equipped with a mosaic of two MIT CCD detectors, each a 4096~$\times$~2048
pixel array with an image scale of 0.126 arcsec/pixel (see FORS2 
manual\footnote{\url{http://www.eso.org/sci/facilities/paranal/instruments/fors/inst/pola.html}}). 
So called "pre-imaging" observations were performed under the direct imaging mode (IMG) 
using two narrow-band filters, H$\alpha$ (FWHM of 61\AA~at 6563\AA, transmits both H$\alpha$ and [N~II] $\lambda\lambda$ 6548, 6584 lines) and [O~III] (FWHM of 
57\AA~at 5001\AA). Among the 136 confirmed PNe, 83 PNe have images with both H$\alpha$ 
and [O~III] filters. A further 26 PNe have an image with the H$\alpha$ filter only and 
27 have an image with the [O~III] filter only. The integration times vary from 
2~seconds for the high surface brightness PNe to 240~seconds for those of lower surface 
brightness. Despite the filler program status the recorded seeing was typically 
better than 1.5~arcseconds, underlining the excellent nature of the ESO VLT site. 
A summary log of the VLT observations is provided in Table~\ref{tab:vlt_log} in 
Appendix~A. All images were bias-subtracted and flat-fielded, where the associated raw 
calibration files are available, using the standard ESO Reflex Data Reduction Pipeline 
\citep{freudling2013automated}. 

\subsection{The \emph{HST} imagery}

For the 40 \emph{HST} sub-sample of bulge PNe, the WFPC2/\emph{HST} has 
800~$\times$~800 pixel silicon CCD with a pixel size of 0.046~arcseconds. The 
narrow-band filter F656N (22\AA~wide at 6564\AA) was utilised for all observations. 
Exposure times were mostly 80 or 120~seconds but a few were of 0.5, 
40, 160, 200, 240, 320 and 400~seconds. The F502N filter (covering [O~III] 
5007\AA) was used for 31 objects in our sample. Here, the exposure 
time ranged from 60 to 600~seconds. The summary log of the \emph{HST} observations 
is given in Table~\ref{tab:HST_log} in Appendix~\ref{app:obs_logs}. Due to the higher 
spatial resolution of the \emph{HST}, these imaging data were used wherever possible 
for morphological classifications and for CSPN hunting and verification in this study. 

\subsection{Pan-STARRS imagery as a useful adjunct}
For some PNe the narrow-band VLT imagery 
was less sharp and indeed less deep for continuum sources
than the available broad-band Pan-STARRS imagery \citep{2010SPIE.7733E..0EK} due to occasional 
poorer seeing for our VLT filler program. Pan-STARRS is itself superior and better than 
the default main $\sim$2~arcsecond resolution HASH imagery from the SuperCOSMOS 
H$\alpha$ survey \citep{2005MNRAS.362..689P} that was originally used for 
morphological classification and that does not work well for the compact PNe sample here. 
Pan-STARRS is also particularly useful for CSPN 
verification as the broad multi-band photometry is quite deep. The typical 5$\sigma$ 
limiting magnitudes for point sources in the g, r, i, z and y bands are 
23.3, 23.2, 23.1, 22.3 and 21.4 respectively \citep{2016arXiv161205560C}. 
This is much deeper than \emph{Gaia}, especially for the g-band, 
and also allows CSPN candidate blue colours to be seen. 
This helps to confirm the CSPN identifications made here. 

\begin{figure}
    \centering
    \begin{subfigure}{\linewidth}
     \centering
    \includegraphics[height=0.44\linewidth]{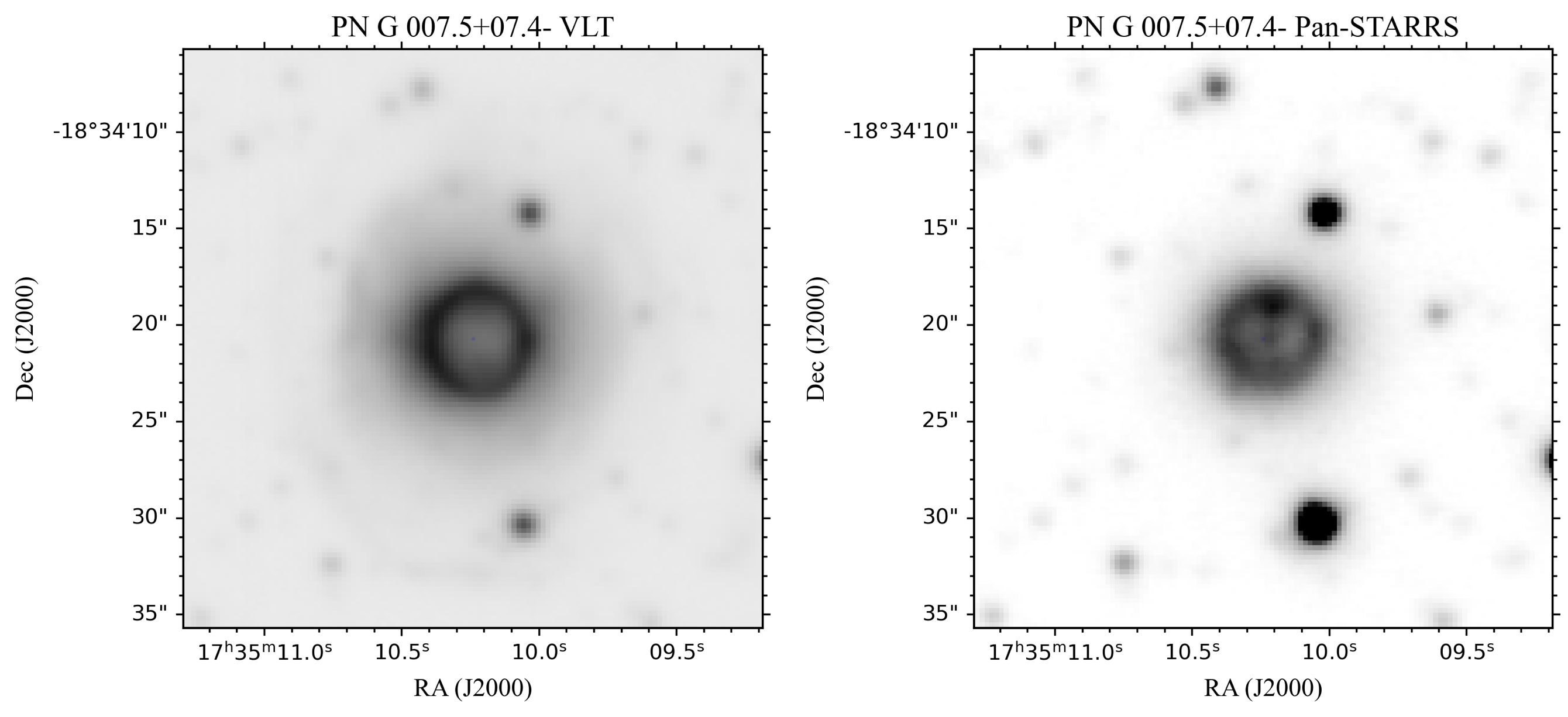}
    \caption{PNG~007.5+07.4}\vspace{0.3cm}
    \end{subfigure}
    \begin{subfigure}{\linewidth}
    \centering
    \includegraphics[height=0.44\linewidth]{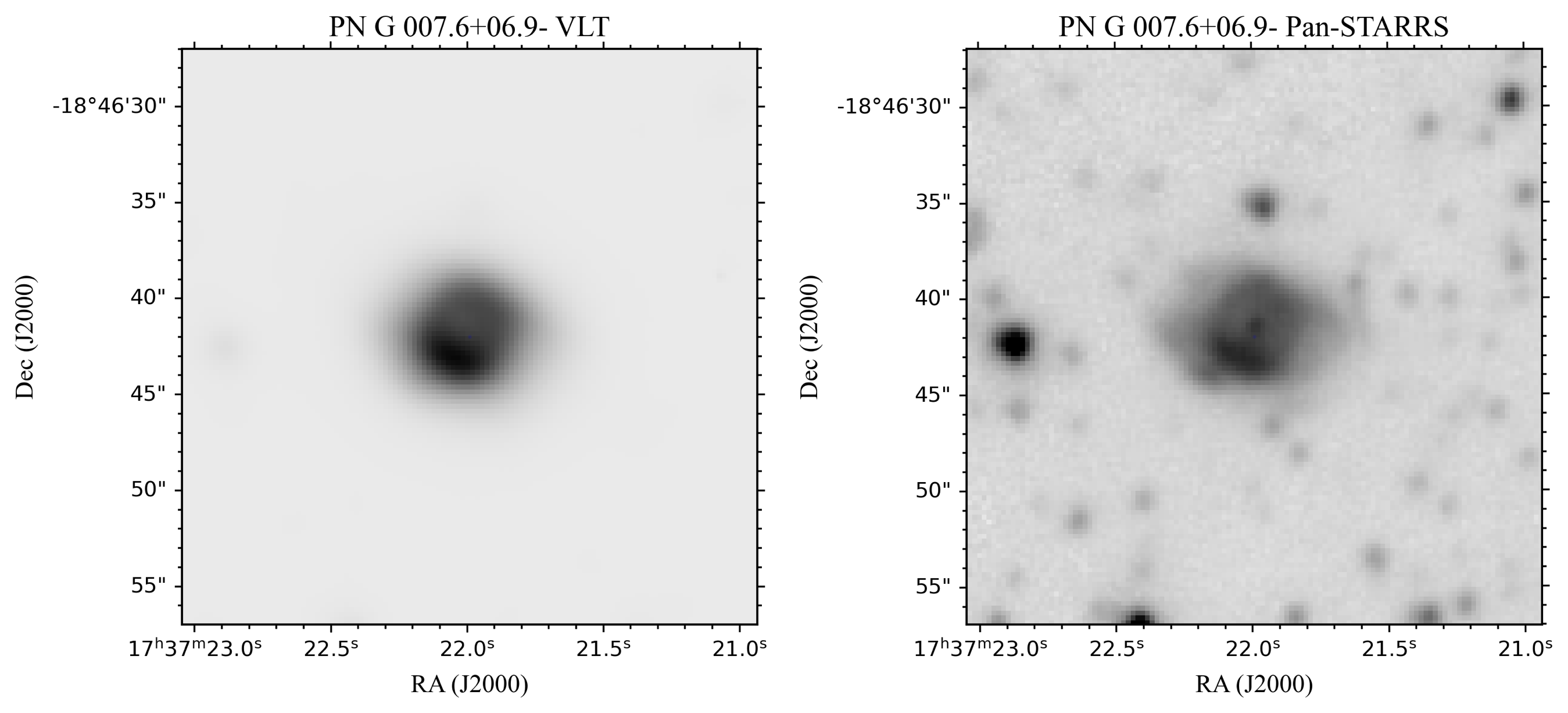}
    \caption{PNG 007.6+06.9}\vspace{0.1cm}
    \end{subfigure}
    \caption{Two comparison examples between the VLT (left) and Pan-STARRS (right) 
    imagery. The upper panel is of PNG~007.5+07.4. The CSPN 
    is only visible in the broad-band red Pan-STARRS image. Pan-STARRS goes deeper in 
    this example for continuum sources than the VLT equivalent. The lower panel 
    contrasts images of PNG~007.6+06.9. The Pan-STARRS image (right) in this case 
    is superior to the VLT both in terms of detail and depth. Again the CSPN is only 
    visible in the Pan-STARRS broad-band red image.}
    \label{fig:cs_comp}
\end{figure}

In Fig.~\ref{fig:cs_comp} we provide two examples showing a comparison of the VLT (left) and 
Pan-STARRS (right) imagery. The image quality of the narrow-band VLT imagery is usually 
superior to that of Pan-STARRS. The upper panel present the images of PNG~007.5+07.4. 
The VLT  image  clearly shows a well defined ring structure and a faint AGB halo but not the 
CSPN  which is only visible in the broad-band Pan-STARRS red image shown. It is also clear 
that Pan-STARRS goes deeper (in this example) for continuum sources than 
the VLT equivalent but with a slightly inferior stellar PSF. The lower panel contrasts 
images of PN~007.6+06.9 where the VLT image (left) exhibits a ring structure with an 
enhancement in brightness in the NW direction, as well as hints of an outer shell. 
The Pan-STARRS image (right) is superior to the VLT both in terms of detail and 
again the CSPN is only visible in the Pan-STARRS broad-band red image.

Although the two Pan-STARRS telescopes are only 1.8~m in diameter they benefit 
from generally excellent seeing conditions being situated at Haleakala Observatory, 
Hawaii. This yielded the high quality broad band filter images used here for both 
checking CSPN candidates and uncovering CSPN not evident in the VLT imaging. 
Often the VLT imaging does not go deep enough to detect faint CSPN in the narrow-band 
filters. Pan-STARRS data was only incorporated into HASH after original 
morphological classifications were assigned. 


\begin{figure}
    \centering
    \includegraphics[width=0.42\textwidth]{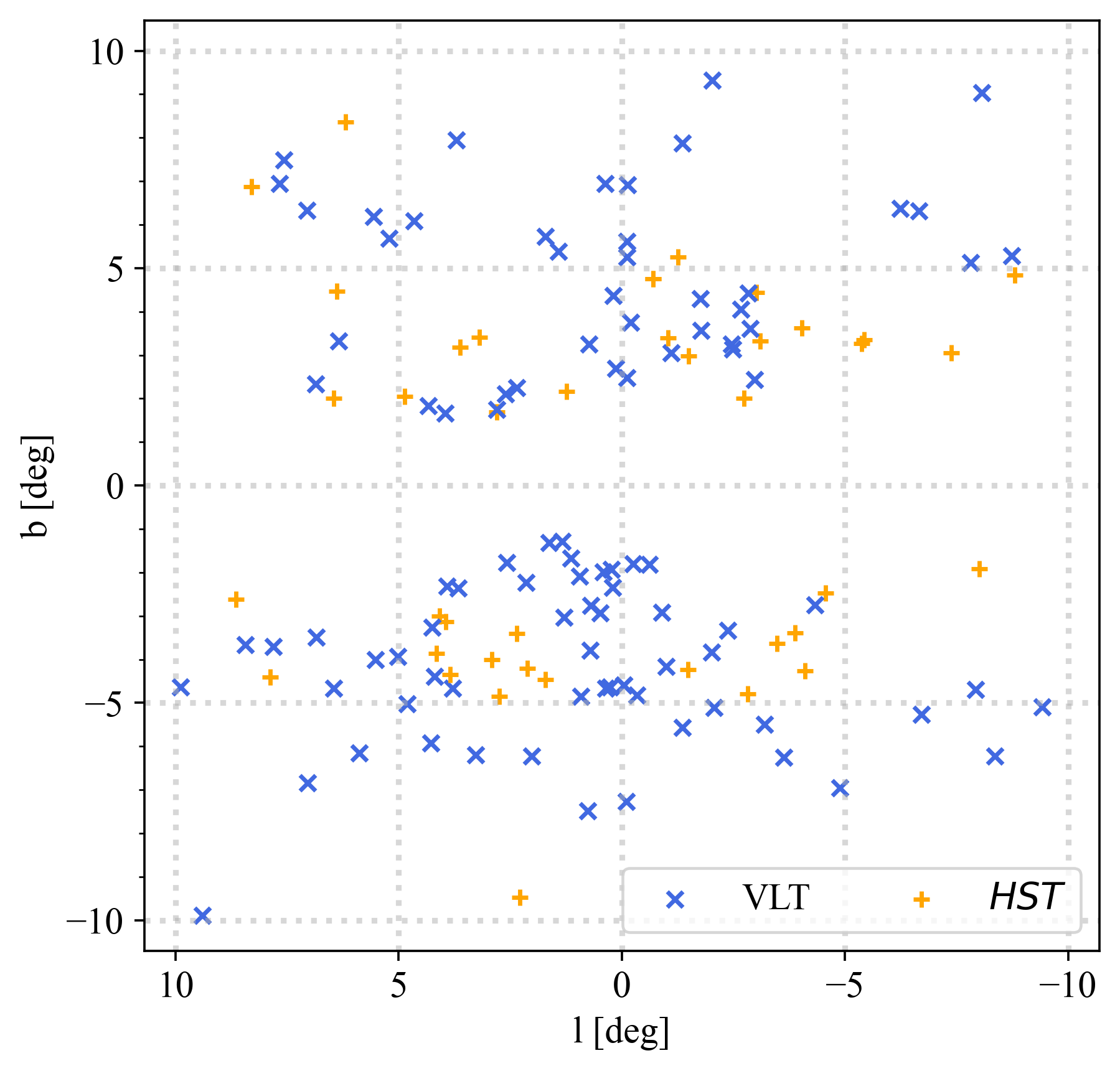}
    \caption{Distribution of the 136 bulge PNe sample observed with the VLT in 
    a Galactic latitude versus Galactic longitude plot. The blue crosses mark the 
    objects with no \emph{HST} data so the best images are usually from the VLT observations. 
    The orange crosses represent objects imaged with the \emph{HST} (even if they also have 
    VLT pre-imaging observations).}
    \label{fig:obs_dist}
\end{figure}

\section{Morphological Classification, CSPN and other Measurements}
\label{morph_n_cs}
\noindent Of the 136 PNe in our sample, imaging data of 96 PNe
are taken from the VLT observations while observations of 40 are 
from the \emph{HST} and used for morphological classification. 
The Galactic distribution of this sample is shown in 
Fig.~\ref{fig:obs_dist}. All spectroscopic data are used later in this series of 
papers for accurate abundance determinations and analysis (Tan et al., in preparation). 
The full set of the best available optical imagery for each of the 136 PNe 
is presented in Appendix~B.

Below we provide details of the updated morphological classifications obtained 
from the \emph{HST} and VLT imagery and the detection and positional 
measurements of any plausible CSPN seen in these and the supplementary Pan-STARRS images.

\subsection{Morphological classification}
\label{morph_class}

\begin{figure}
\centering
\subcaptionbox{ERBIAS\label{fig:main}}{\includegraphics[width=0.88\columnwidth]
{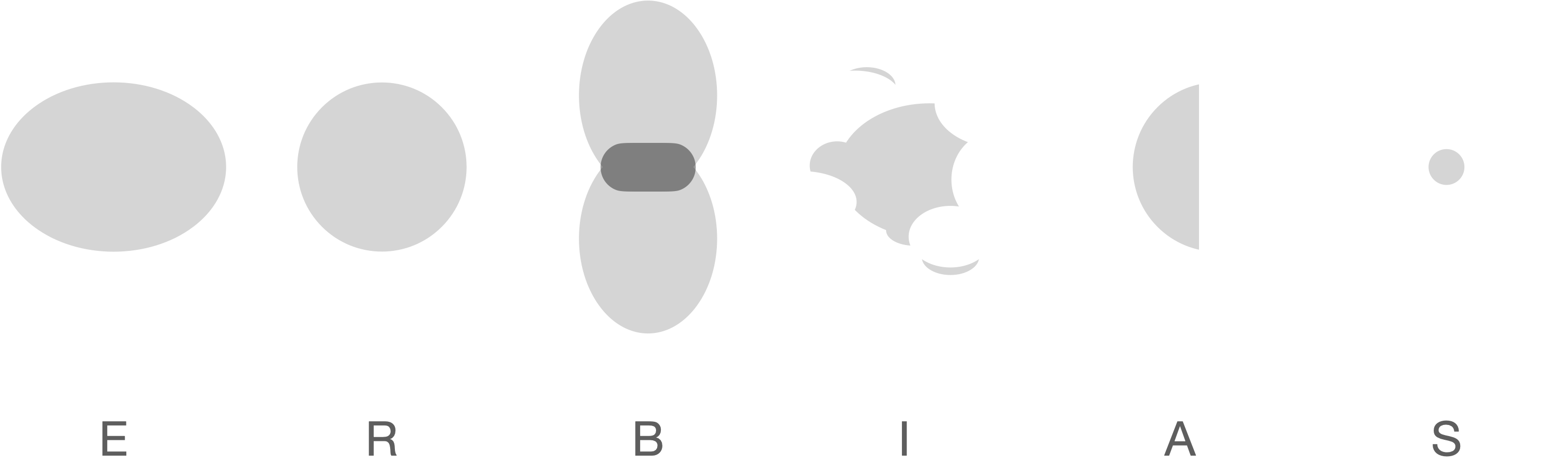}}\par
\vspace{0.3cm}
\subcaptionbox{sparm\label{fig:sub}}{\includegraphics[width=0.8\columnwidth]
{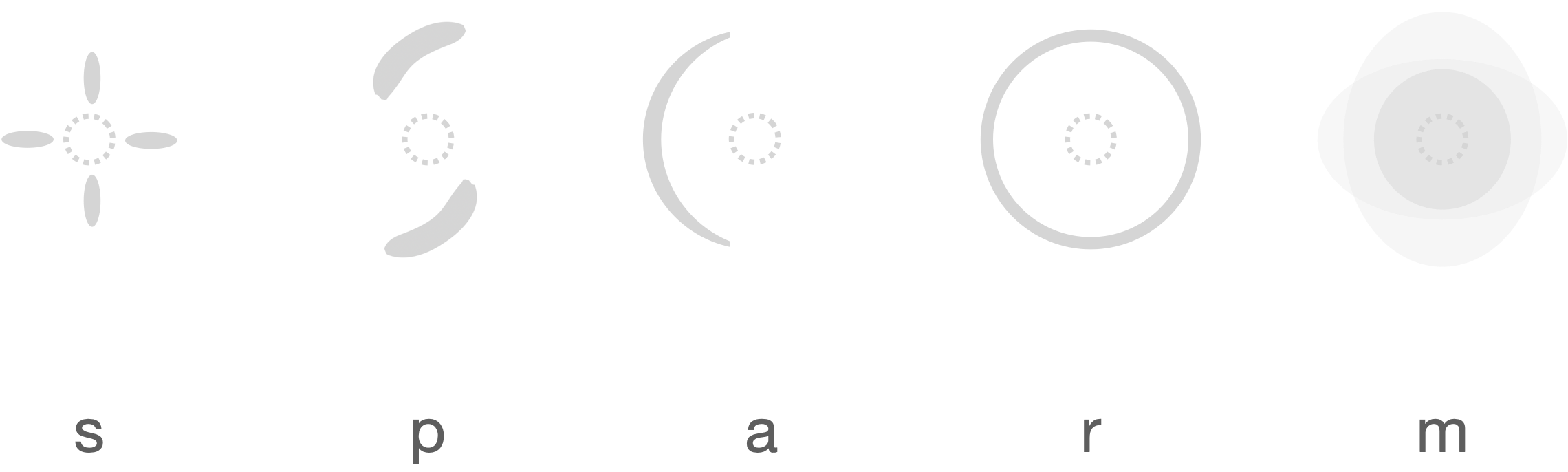}}\par 
\caption{A schematic of the main classifiers `ERBIAS' and the sub-classifiers `sparm' for
indicating the PN morphology taken from \citep{2006MNRAS.373...79P} and  
\citep{parker2016hash}. 
The dotted circles indicating the centre of the nebula are used 
as a reference to infer the secondary PN morphological features.}
\label{fig:classifier}
\end{figure}

\noindent As mentioned, the "ERBIAS sparm" morphological classification scheme, 
first described in \citet{2006MNRAS.373...79P}, was adopted. 
The upper case designator first indicates the basic class as Elliptical (E), 
Round (R), Bipolar (B), Irregular (I), Asymmetric 
(A) and the unresolved quasi-stellar PNe (S). To describe the secondary structural 
features evident in the nebulae, the lower-case `sparm' sub-classifiers were used immediately
following the main class. These are used to describe an object that also exhibits obvious 
internal structure with details such as filaments, knots, striations etc ‘s', possesses some 
point symmetry ‘p’, has an enhancement or one-sided brightness/asymmetry ‘a’, has an evident,
even if fractured, ring-like structure or annulus ‘r’ or shows multiple shells or external 
structure ‘m’. A scheme schematic of the main and sub-classifiers is presented in 
Fig.~\ref{fig:classifier}. PNe may have none, several or even all of these sub-classifiers.

This classification scheme was then applied to the best available multi-wavelength or
single, narrow-band image data for all PNe in our sample, whether it be from the \emph{HST}
or VLT. Of course this is inevitably a somewhat subjective judgement but the team is very 
experienced.

In Table~\ref{tb:sum_results} we present the updated morphological classifications 
for all 136 PNe in the selected bulge sample. Provided are the unique HASH ID number, 
the PNG designation, remeasured PNe centroid positions in J2000 
RA and DEC, our updated morphological classifications and new angular diameter 
estimates (in arcseconds) measured from the VLT,  \emph{HST} and, in three cases, Pan-STARRS
imagery. The telescope used is indicated in the last entry for each PN.
\renewcommand*{\arraystretch}{1.075}
\setlength{\tabcolsep}{3pt}
\begin{table*}
\centering
\caption{Listing of HASH \& PNG IDs, J2000 RA/DEC 
positions, morphological classifications, new, angular size 
measurements (arcseconds) for our PNe sample as measured from the 
VLT, \emph{HST} imagery and, in 3 cases, from Pan-STARRS. 
The telescope data used is indicated as the last entry for each PN.}
\label{tb:sum_results}
\begin{tabular}{l@{\hspace{0.2\tabcolsep}}cc@{\hspace{0.5\tabcolsep}}cclp{1.7cm}l@{\hspace{0.2\tabcolsep}}cc@{\hspace{0.5\tabcolsep}}ccll}
\hline
\multirow{2}{*}{\begin{tabular}[l]{@{}l@{}}HASH\\ ID \end{tabular}} & \multirow{2}{*}{PN
G} &  \multicolumn{2}{c}{PN Centroid Coords.} & \multirow{2}{*}{ Morph. } & \multirow{2}{*}{\begin{tabular}[l]{@{}l@{}}Ang.\\ Diam. \end{tabular}} & 
\multirow{2}{*}{Tele.} & \multirow{2}{*}{\begin{tabular}[l]{@{}l@{}}HASH\\ ID \end{tabular}}
& \multirow{2}{*}{PN
G} &  \multicolumn{2}{c}{PN Centroid Coords.} & \multirow{2}{*}{ Morph. } & \multirow{2}{*}{\begin{tabular}[l]{@{}l@{}}Ang.\\ Diam.\end{tabular}} & 
\multirow{2}{*}{Tele.} \\ \cline{3-4} \cline{10-11} &  & RA(J2000)  
& DEC(J2000)& &  & & &  & RA(J2000)   & DEC(J2000)    & 
& & \\
\hline
  12 &  000.1$+$02.6 &  17:35:35.44 &  -27:24:05.37 &     Ars &        12.6 &   VLT &      171 &  005.8$-$06.1 &  18:22:54.20 &  -26:49:17.00 &     Bms &         9.4 &   VLT \\
      16 &  000.1$+$04.3 &  17:29:23.41 &  -26:26:04.38 &       B &        12.0 &   VLT &     4108 &  006.1$+$08.3 &  17:28:57.61 &  -19:15:53.95 &     Ems &         2.4 &   \emph{HST} \\
      17 &  000.1$-$02.3 &  17:55:20.53 &  -29:57:36.78 &    Rars &        10.2 &   VLT &      179 &  006.3$+$03.3 &  17:47:33.95 &  -21:47:23.71 &    Bams &         9.0 &    PS \\
      19 &  000.2$-$01.9 &  17:53:45.64 &  -29:43:46.98 &     Brs &        19.4 &   VLT &      180 &  006.3$+$04.4 &  17:43:28.75 &  -21:09:51.77 &   Bamps &         5.4 &   \emph{HST} \\
      20 &  000.2$-$04.6 &  18:04:44.09 &  -31:02:48.88 &      Bs &         8.0 &   VLT &      182 &  006.4$+$02.0 &  17:52:41.44 &  -22:21:57.18 &    Bmps &         7.0 &   \emph{HST} \\
      24 &  000.3$+$06.9 &  17:20:22.04 &  -24:51:52.24 &     Brs &        11.0 &   VLT &      181 &  006.4$-$04.6 &  18:18:13.33 &  -25:38:07.95 &     Eas &         9.2 &   VLT \\
      22 &  000.3$-$04.6 &  18:05:02.68 &  -30:58:18.03 &     Bms &        11.8 &   VLT &      189 &  006.8$+$02.3 &  17:52:22.62 &  -21:51:13.51 &     Bms &         5.2 &   VLT \\
      26 &  000.4$-$01.9 &  17:54:25.37 &  -29:36:08.89 &     Brs &         8.4 &   VLT &      188 &  006.8$-$03.4 &  18:14:28.74 &  -24:43:38.12 &     Bms &         7.0 &   VLT \\
      27 &  000.4$-$02.9 &  17:58:19.40 &  -30:00:39.36 &   Bamrs &         8.8 &   VLT &      193 &  007.0$+$06.3 &  17:38:11.54 &  -19:37:35.83 &     Bps &         7.2 &   VLT \\
      40 &  000.7$+$03.2 &  17:34:54.72 &  -26:35:57.36 &    Bprs &        17.8 &   VLT &      192 &  007.0$-$06.8 &  18:27:59.62 &  -26:06:48.11 &     Bms &        12.4 &   VLT \\
      37 &  000.7$-$02.7 &  17:58:09.59 &  -29:44:19.98 &     Bps &         8.2 &   VLT &      197 &  007.5$+$07.4 &  17:35:10.24 &  -18:34:20.71 &   Eamrs &        18.2 &    PS \\
      38 &  000.7$-$03.7 &  18:02:19.26 &  -30:14:25.94 &    Bmrs &        11.4 &   VLT &      198 &  007.6$+$06.9 &  17:37:21.99 &  -18:46:42.00 &   Bamrs &        16.6 &    PS \\
      42 &  000.7$-$07.4 &  18:17:37.20 &  -31:56:46.68 &    Ears &        11.8 &   VLT &      200 &  007.8$-$03.7 &  18:17:16.01 &  -23:58:55.07 &    Bams &        17.0 &   VLT \\
      50 &  000.9$-$02.0 &  17:56:02.79 &  -29:11:16.69 &     Bps &         5.2 &   VLT &      201 &  007.8$-$04.4 &  18:20:08.84 &  -24:15:05.39 &   Eamrs &         7.2 &   \emph{HST} \\
      48 &  000.9$-$04.8 &  18:07:06.16 &  -30:34:17.39 &   Eamrs &        15.4 &   VLT &     4111 &  008.2$+$06.8 &  17:38:57.37 &  -18:17:35.92 &     Bms &         4.8 &   \emph{HST} \\
      54 &  001.1$-$01.6 &  17:54:52.07 &  -28:48:55.79 &     Eam &         7.8 &   VLT &      208 &  008.4$-$03.6 &  18:18:23.81 &  -23:24:57.50 &   Bmprs &        25.2 &   VLT \\
      58 &  001.2$+$02.1 &  17:40:12.83 &  -26:44:21.83 &   Bamrs &         5.2 &   \emph{HST} &      210 &  008.6$-$02.6 &  18:14:50.91 &  -22:43:55.52 &    Bams &         4.2 &   \emph{HST} \\
      60 &  001.2$-$03.0 &  18:00:37.62 &  -29:21:50.78 &       E &         6.6 &   VLT &      217 &  009.4$-$09.8 &  18:44:43.15 &  -25:21:34.06 &   Eamrs &         8.4 &   VLT \\
      62 &  001.3$-$01.2 &  17:53:47.16 &  -28:27:17.84 &    Bmrs &         6.2 &   VLT &      223 &  009.8$-$04.6 &  18:25:05.00 &  -22:34:52.64 &   Bamrs &        11.0 &   VLT \\
      64 &  001.4$+$05.3 &  17:28:37.63 &  -24:51:07.16 &    Baps &         8.6 &   VLT &     1147 &  350.5$-$05.0 &  17:42:54.08 &  -39:36:24.29 &    Emrs &         9.2 &   VLT \\
      70 &  001.6$-$01.3 &  17:54:34.89 &  -28:12:43.62 &     Brs &         5.4 &   VLT &     1151 &  351.1$+$04.8 &  17:03:46.84 &  -33:29:44.64 &    Emrs &         6.0 &   \emph{HST} \\
      73 &  001.7$+$05.7 &  17:28:01.74 &  -24:25:23.72 &   Eamrs &        12.2 &   VLT &     1152 &  351.2$+$05.2 &  17:02:19.10 &  -33:10:05.02 &      Bm &        14.4 &   VLT \\
      74 &  001.7$-$04.4 &  18:07:14.56 &  -29:41:24.67 &   Bamrs &         4.6 &   \emph{HST} &     1156 &  351.6$-$06.2 &  17:50:44.60 &  -39:17:25.98 &  Bamprs &        15.0 &   VLT \\
      81 &  002.0$-$06.2 &  18:15:06.50 &  -30:15:32.90 &    Eams &         6.4 &   VLT &     1159 &  351.9$+$09.0 &  16:50:17.08 &  -30:19:55.39 &   Bmprs &        15.4 &   VLT \\
      83 &  002.1$-$02.2 &  17:59:19.33 &  -28:13:48.31 &    Bmrs &         5.2 &   VLT &     1160 &  351.9$-$01.9 &  17:33:00.68 &  -36:43:52.94 &      Bs &         4.8 &   \emph{HST} \\
      82 &  002.1$-$04.2 &  18:07:07.28 &  -29:13:06.07 &    Bmps &         8.2 &   \emph{HST} &     1161 &  352.0$-$04.6 &  17:45:06.80 &  -38:08:49.49 &     Bmp &         7.4 &   VLT \\
      90 &  002.2$-$09.4 &  18:29:11.64 &  -31:29:59.24 &   Bmprs &        13.8 &   \emph{HST} &     1163 &  352.1$+$05.1 &  17:05:30.72 &  -32:32:08.22 &      Bs &        18.0 &   VLT \\
      91 &  002.3$+$02.2 &  17:42:30.02 &  -25:45:28.84 &   Eamrs &         6.6 &   VLT &     1164 &  352.6$+$03.0 &  17:14:42.93 &  -33:24:47.61 &  Bamprs &         5.2 &   \emph{HST} \\
      93 &  002.3$-$03.4 &  18:04:28.81 &  -28:37:38.29 &    Bmrs &         5.8 &   \emph{HST} &     1171 &  353.2$-$05.2 &  17:50:45.20 &  -37:23:53.09 &    Bmrs &        16.8 &   VLT \\
      98 &  002.5$-$01.7 &  17:58:31.20 &  -27:37:05.37 &    Ears &         5.6 &   VLT &     1172 &  353.3$+$06.3 &  17:04:18.33 &  -30:53:29.08 &      Bp &         8.2 &   VLT \\
    4326 &  002.6$+$02.1 &  17:43:39.48 &  -25:36:42.87 &      Bs &        18.4 &   VLT &     1178 &  353.7$+$06.3 &  17:05:13.91 &  -30:32:19.53 &   Bmprs &        10.0 &   VLT \\
     105 &  002.7$-$04.8 &  18:11:04.99 &  -28:58:59.22 &  Bamprs &        18.2 &   \emph{HST} &     1185 &  354.5$+$03.3 &  17:18:51.94 &  -31:39:06.52 &     Bas &         2.6 &   \emph{HST} \\
     106 &  002.8$+$01.7 &  17:45:39.80 &  -25:40:00.52 &   Bmars &         4.0 &   \emph{HST} &     1189 &  354.9$+$03.5 &  17:19:20.24 &  -31:12:40.95 &     Bas &         3.6 &   \emph{HST} \\
     107 &  002.8$+$01.8 &  17:45:28.34 &  -25:38:11.89 &    Bars &         9.4 &   VLT &     1191 &  355.1$-$06.9 &  18:02:32.32 &  -36:39:11.22 &      Em &         9.8 &   VLT \\
     109 &  002.9$-$03.9 &  18:08:05.76 &  -28:26:10.65 &   Bamrs &         6.6 &   \emph{HST} &     1199 &  355.4$-$02.4 &  17:44:20.61 &  -34:06:40.96 &     Bms &        10.6 &   \emph{HST} \\
     114 &  003.1$+$03.4 &  17:40:07.42 &  -24:25:42.77 &   Eamrs &         4.0 &   \emph{HST} &     1202 &  355.6$-$02.7 &  17:46:06.30 &  -34:03:45.54 &       E &         8.0 &   VLT \\
     116 &  003.2$-$06.2 &  18:17:41.44 &  -29:08:19.88 &     Bps &        11.0 &   VLT &     1209 &  355.9$+$03.6 &  17:21:31.91 &  -30:20:48.75 &     Eam &         3.8 &   \emph{HST} \\
     125 &  003.6$+$03.1 &  17:41:57.26 &  -24:11:16.33 &    Baps &         5.8 &   \emph{HST} &     1207 &  355.9$-$04.2 &  17:52:58.94 &  -34:38:22.99 &    Bmps &         8.2 &   \emph{HST} \\
     126 &  003.6$-$02.3 &  18:03:11.87 &  -26:58:32.76 &    Bars &        10.0 &   VLT &     1212 &  356.1$-$03.3 &  17:49:50.90 &  -34:00:31.31 &     Bas &         5.2 &   \emph{HST} \\
     128 &  003.7$+$07.9 &  17:24:45.28 &  -21:33:34.88 &    Bars &        11.2 &   VLT &     1217 &  356.3$-$06.2 &  18:02:32.00 &  -35:13:13.69 &     Ers &        11.6 &   VLT \\
     129 &  003.7$-$04.6 &  18:12:34.39 &  -27:58:10.70 &    Ears &         6.8 &   VLT &     1219 &  356.5$-$03.6 &  17:51:50.60 &  -33:47:35.59 &     Bps &        11.4 &   \emph{HST} \\
     131 &  003.8$-$04.3 &  18:11:29.26 &  -27:46:15.92 &    Baps &         5.4 &   \emph{HST} &     1230 &  356.8$+$03.3 &  17:25:06.10 &  -29:45:16.88 &      Bp &         3.6 &   \emph{HST} \\
     135 &  003.9$+$01.6 &  17:48:28.47 &  -24:41:23.79 &    Bars &         8.6 &   VLT &     1227 &  356.8$-$05.4 &  18:00:18.23 &  -34:27:40.73 &    Bmrs &        14.4 &   VLT \\
     133 &  003.9$-$02.3 &  18:03:39.30 &  -26:43:33.87 &     Bas &         7.6 &   VLT &     1232 &  356.9$+$04.4 &  17:21:04.46 &  -29:02:59.76 &    Bmps &         5.8 &   \emph{HST} \\
     134 &  003.9$-$03.1 &  18:06:50.03 &  -27:06:19.58 &    Ears &         7.8 &   \emph{HST} &     1234 &  357.0$+$02.4 &  17:28:50.35 &  -30:07:44.99 &    Ears &         8.6 &   VLT \\
     139 &  004.0$-$03.0 &  18:06:40.91 &  -26:54:56.28 &   Bamrs &         6.6 &   \emph{HST} &     1235 &  357.1$+$03.6 &  17:24:34.43 &  -29:24:19.59 &     Bas &         8.4 &   VLT \\
     140 &  004.1$-$03.8 &  18:10:12.28 &  -27:16:35.41 &     Bms &         3.4 &   \emph{HST} &     4139 &  357.1$+$04.4 &  17:21:37.98 &  -28:55:14.63 &     Ers &        10.8 &   VLT \\
     142 &  004.2$-$03.2 &  18:08:01.38 &  -26:54:02.29 &      Bs &         6.4 &   VLT &     1239 &  357.1$-$04.7 &  17:58:14.45 &  -33:47:37.45 &    Bprs &         3.6 &   \emph{HST} \\
     141 &  004.2$-$04.3 &  18:12:25.12 &  -27:29:13.19 &     Bps &         6.6 &   VLT &     1242 &  357.2$+$02.0 &  17:31:08.11 &  -30:10:28.15 &    Rmrs &         4.8 &   \emph{HST} \\
     143 &  004.2$-$05.9 &  18:18:38.34 &  -28:07:59.00 &      Bs &         9.0 &   VLT &     1246 &  357.3$+$04.0 &  17:23:24.94 &  -28:59:06.07 &      Ea &         8.6 &   VLT \\
    4315 &  004.3$+$01.8 &  17:48:36.54 &  -24:16:34.24 &      Bs &         8.8 &   VLT &     1252 &  357.5$+$03.1 &  17:27:24.36 &  -29:21:14.62 &      Ea &         6.0 &   VLT \\
     148 &  004.6$+$06.0 &  17:33:37.58 &  -21:46:24.80 &    Emrs &         8.0 &   VLT &     1253 &  357.5$+$03.2 &  17:26:59.83 &  -29:15:31.87 &     Bas &        10.20 &   VLT \\
     151 &  004.8$+$02.0 &  17:49:00.51 &  -23:42:54.90 &     Eas &         4.0 &   \emph{HST} &     1256 &  357.6$-$03.3 &  17:53:16.81 &  -32:40:38.58 &      Bs &        12.6 &   VLT \\
     150 &  004.8$-$05.0 &  18:16:11.43 &  -27:14:57.98 &    Bars &        11.4 &   VLT &     1259 &  357.9$-$03.8 &  17:56:13.93 &  -32:37:22.21 &    Bmrs &        13.0 &   VLT \\
     156 &  005.0$-$03.9 &  18:12:22.99 &  -26:32:54.50 &   Eamrs &        12.8 &   VLT &     4140 &  357.9$-$05.1 &  18:01:22.20 &  -33:17:43.08 &    Bmps &        26.4 &   VLT \\
     162 &  005.2$+$05.6 &  17:36:22.64 &  -21:31:12.37 &     Ers &         9.0 &   VLT &     1258 &  358.0$+$09.3 &  17:05:44.60 &  -25:25:01.49 &  Bamprs &        14.8 &   VLT \\
    4106 &  005.5$+$06.1 &  17:35:21.32 &  -20:57:20.52 &      Bs &         6.0 &   VLT &     1263 &  358.2$+$03.5 &  17:27:32.87 &  -28:31:06.94 &       E &         6.8 &   VLT \\
     165 &  005.5$-$04.0 &  18:13:40.62 &  -26:08:39.43 &    Ears &        12.4 &   VLT &     1266 &  358.2$+$04.2 &  17:24:52.10 &  -28:05:54.60 &      Em &        14.0 &   VLT \\
\hline
\end{tabular}
\end{table*}

\begin{table*}
\centering
\contcaption{}
\begin{tabular}{l@{\hspace{0.2\tabcolsep}}cc@{\hspace{0.5\tabcolsep}}cclp{1.7cm}l@{\hspace{0.2\tabcolsep}}cc@{\hspace{0.5\tabcolsep}}ccll}
\hline
\multirow{2}{*}{\begin{tabular}[l]{@{}l@{}}HASH\\ ID \end{tabular}} & \multirow{2}{*}{PN
G} &  \multicolumn{2}{c}{PN Centroid Coords.} & \multirow{2}{*}{ Morph. } & \multirow{2}{*}{\begin{tabular}[l]{@{}l@{}}Ang.\\ Diam. \end{tabular}} & 
\multirow{2}{*}{Tele.} & \multirow{2}{*}{\begin{tabular}[l]{@{}l@{}}HASH\\ ID \end{tabular}}
& \multirow{2}{*}{PN
G} &  \multicolumn{2}{c}{PN Centroid Coords.} & \multirow{2}{*}{ Morph. } & \multirow{2}{*}{\begin{tabular}[l]{@{}l@{}}Ang.\\ Diam.\end{tabular}} & 
\multirow{2}{*}{Tele.} \\ \cline{3-4} \cline{10-11} &  & RA(J2000)  
& DEC(J2000)& &  & & &  & RA(J2000)   & DEC(J2000)    & 
& & \\
\hline
1275 &  358.5$+$02.9 &  17:30:30.43 &  -28:35:54.90 &    Bas &         3.8 &   \emph{HST} &     1312 &  359.3$-$01.8 &  17:51:18.93 &  -30:23:53.23 &     Bp &         6.8 &   VLT \\
    1276 &  358.5$-$04.2 &  17:59:02.51 &  -32:21:43.63 &   Bmps &         7.8 &   \emph{HST} &     1319 &  359.6$-$04.8 &  18:04:07.75 &  -31:39:10.74 &    Brs &        16.6 &   VLT \\
    1280 &  358.6$+$07.8 &  17:12:39.17 &  -25:43:37.50 &     Bs &        13.4 &   VLT &     1322 &  359.7$-$01.8 &  17:52:05.97 &  -30:05:14.12 &  Ramrs &         8.0 &   VLT \\
    1281 &  358.6$-$05.5 &  18:04:56.21 &  -32:54:01.25 &  Bamps &        16.8 &   VLT &     1327 &  359.8$+$02.4 &  17:35:48.12 &  -27:43:20.40 &     Em &         5.2 &   VLT \\
    1286 &  358.7$+$05.2 &  17:22:28.29 &  -27:08:42.51 &  Eamrs &         3.6 &   \emph{HST} &     1324 &  359.8$+$03.7 &  17:30:46.74 &  -27:05:59.76 &     Bp &         7.2 &   VLT \\
    1293 &  358.8$+$03.0 &  17:31:09.28 &  -28:14:50.17 &   Ears &         8.8 &   VLT &     1328 &  359.8$+$05.2 &  17:25:23.63 &  -26:11:53.00 &   Bprs &        21.4 &   VLT \\
    1295 &  358.9$+$03.4 &  17:30:02.55 &  -27:59:18.19 &  Bamps &         3.2 &   \emph{HST} &     1329 &  359.8$+$05.6 &  17:24:01.48 &  -25:59:23.39 &      B &         7.0 &   VLT \\
    1298 &  359.0$-$04.1 &  17:59:56.74 &  -31:54:27.90 &    Bms &         9.2 &   VLT &     1326 &  359.8$+$06.9 &  17:19:13.39 &  -25:17:17.32 &   Bmps &        31.8 &   VLT \\
    1302 &  359.1$-$02.9 &  17:55:05.70 &  -31:12:16.85 &     Bs &        10.0 &   VLT &     1330 &  359.8$-$07.2 &  18:14:50.60 &  -32:36:55.30 &     Em &        11.6 &   VLT \\
    1308 &  359.2$+$04.7 &  17:25:44.08 &  -26:57:48.04 &   Ears &         2.6 &   \emph{HST} &     1334 &  359.9$-$04.5 &  18:03:52.62 &  -31:17:46.79 &      E &         9.0 &   VLT \\
\hline
\end{tabular}
\end{table*}

\subsection{Astrometric integrity of the \emph{HST}, Pan-STARRS and VLT imagery}
Prior to detection and measurement of any CSPN evident in our data and comparison 
with \emph{Gaia} it is necessary to establish the astrometric integrity across 
the various image data sets used. The applied astrometric solution for the \emph{HST} 
\citep{2000PASP..112.1360A} and Pan-STARRS data \citep{2020ApJS..251....6M} are 
reported as highly accurate. However, for the VLT imagery the astrometric fits 
provided from the ESO pipelines are not sufficiently accurate. It was necessary 
to boot-strap the VLT data to the equivalent \emph{Gaia} 
data co-ordinate system before CSPN position comparisons. Small-scale translation and
rotations were needed with typical offsets found to be 1 to a few VLT image pixels. 
This task was accomplished using \texttt{Python} scripts after identifying 6-10 
stars in common widely spread across the images and \emph{Gaia} data. 

We also noticed a small systematic $\sim$1~pixel offset to the North and the West 
between the \emph{Gaia} and Pan-STARRS frames and so a similar exercise was done 
here. Once completed we could then confidently assess the CSPN detected 
in the VLT, \emph{HST} and Pan-STARRS imagery with the co-ordinates
for the CSPN reported from \emph{Gaia} data, e.g. \citet{chornay2021one} and 
\citet{gonzalez2021planetary} (see later). 

\subsection{Identification of CSPN}
\label{Central stars}

CSPN exhibit a wide range of heterogeneous properties and 
range from weak emission line stars (wels) to PG1159 stars, Wolf-Rayet type stars of 
the [WO], [WC] and even [WN] sequence (where the square bracket demarcates them 
from their population I high mass equivalents), different kinds of 
white dwarfs (DA, DAO, DO) and early and late O(H) and Of(H) stars etc, e.g. see
\citet{2018A&A...614A.135W}. How these stars have influenced the characteristics 
of the surrounding PNe is a key question and it depends on the timescales 
for the CSPN evolution and the visibility lifetime of the PN itself as it 
expands to ultimate dissolution in the ambient ISM e.g. see \citet{2022ApJ...935L..35F}.

In Table~\ref{tab:cs_gaia} we present the results of our careful investigation 
into the CSPN that we have independently identified in the \emph{HST}, and 
astrometrically corrected VLT and Pan-STARRS imagery for all 136 PNe in our 
sample. 

A total of 78 PNe have high confidence CSPN detected in the combined 
imaging data for a 57.4\% detection rate. 
In most cases (58/78) the \emph{HST} and VLT imagery provided the best data to 
identify the CSPN. However, in 18  cases the Pan-STARRS data was the best 
(though in 2 cases the CSPN was seen only in the 
Vista Variables in the Via Lactea (VVV) Near Infrared survey data \citep{2012A&A...537A.107S} 
due to dust extinction. This is because not only is the Pan-STARRS data 
deeper in many cases than that from the VLT, but for some sources the resolution 
is also better (due to the filler program nature of the VLT observations). Also, because 
Pan-STARRS used broad-band filters, the CSPN (and their colours) are easier to see.

These candidates are selected as being almost exclusively at the exact geometric centres of
their respective host, compact PNe. Of these, 22 are already listed in the putative CSPN
catalogue of \citet{weidmann2020catalogue} that reports 
not the actual accurate CSPN co-ordinates in many cases, but rather the centroid 
positions of these high surface brightness, compact PNe. These are quoted only to be accurate to 0.1~seconds of time for Right Ascension and 0.1~seconds of arc for Declination. 
This is insufficiently accurate for reporting 
the positions of the true CSPN visible in the \emph{HST} and VLT imagery. 
The offsets of 0.20 to 0.5~arcseconds typical in both co-ordinates in  
\citet{weidmann2020catalogue} can miss the CSPN completely in the high resolution 
imagery. The spectral types of 22 CSPN reported in 
\citet{weidmann2020catalogue} are also shown in the last column.
They largely comprise O-type stars, \emph{wels} and [WR] stars.

Of our 78 CSPN, 74 also have \emph{Gaia} EDR3 counterparts included in 
Table~\ref{tab:cs_gaia}. 
We also provide \emph{Gaia} CSPN G~magnitudes, proper motions and distance estimates 
where available. An assessment of the reported \emph{Gaia} CSPN identifications 
is given below.

\setlength{\tabcolsep}{3.5pt}
\renewcommand{\arraystretch}{1.17}
\begin{table*}
\centering
\caption{A total of 78 PNe with CSPN detected in the combined imaging data 
of our sample. Cols. 1-2 list HASH IDs and PN G numbers of each nebula, cols 3-4 refer to the CSPN coordinates measured in this work, telescopes used for the CSPN detection are given in col. 5, cols. 6-7 are coordinates of \emph{Gaia} CSPN identifications. The 63 we consider to have been correctly identified in 
\emph{Gaia} EDR3 are listed with "Y" in the "True?" column (col. 8). The angular separations between true CSPNe and \emph{Gaia} identifications are listed in col. 9. Cols. 10-14 are the apparent magnitudes in \emph{Gaia} G band, parallaxes and uncertainties in parallaxes respectively. The spectral types of 22 CSPN that were reported in \citet{weidmann2020catalogue} are also shown in the last column. There are 3 CSPN spectra types that are indicated as being in binary systems by '+?' where the nature of the binary companion is unknown.}
\label{tab:cs_gaia}
\begin{tabular}{l@{\hspace{0.2\tabcolsep}}cc@{\hspace{0.8\tabcolsep}}ccc@{\hspace{0.8\tabcolsep}}ccccccc@{\hspace{0.2\tabcolsep}}c}
\hline
HASH & \multirow{2}{*}{PN G} & \multicolumn{2}{c}{This Work}          &  \multirow{2}{*}{Tele.} &  
\multicolumn{2}{c}{Gaia EDR3}         & \multirow{2}{*}{True?} & ${\Delta r_{{\star}}}$ & \emph{G}  & 
$\omega$   & $\sigma_{\omega}$      & $D$      &    \multirow{2}{*}{Spec. Type}        \\ \cline{3-4} 
\cline{6-7} ID & & RAJ2000 & \multicolumn{1}{c}{DECJ2000} &     &  RAJ2000     & 
\multicolumn{1}{c}{DECJ2000} &   &  (arcsec) & (mag)   & (mas)                & (mas)          & (kpc)
&       \\
\hline
      12 &  000.1+02.6 &  17:35:35.44 &  -27:24:05.37 &   VLT &  17:35:35.45 &  -27:24:05.40 &        Y &  0.07 &   17.9 &        &        &                                &                        \\
      17 &  000.1-02.3 &  17:55:20.55 &  -29:57:36.40 &   VLT &              &               &      New &       &        &        &        &                                &                        \\
      19 &  000.2-01.9 &  17:53:45.64 &  -29:43:46.98 &   VLT &  17:53:45.65 &  -29:43:46.94 &        Y &  0.13 &   16.9 & -0.164 &  0.112 &   6.49$_{-1.16}^{+1.27}$ &      O(H)6-8 III-V+? \\
      27 &  000.4-02.9 &  17:58:19.40 &  -30:00:39.36 &   VLT &  17:58:19.39 &  -30:00:39.36 &        Y &   0.10 &   18.3 &  0.367 &  0.189 &    4.63$_{-1.38}^{+1.5}$ &                        \\
      40 &  000.7+03.2 &  17:34:54.73 &  -26:35:57.28 &    PS &  17:34:54.72 &  -26:35:57.33 &        Y &  0.14 &   20.9 &        &        &                                &                        \\
      38$^{\bullet}$ &  000.7-03.7 &  18:02:19.26 &  -30:14:25.94 &   VLT &  18:02:19.26 &  -30:14:25.89 &        Y? &  0.05 &   18.4 &  1.391 &  0.299 &   {\bf 0.98$_{-0.28}^{+2.18}$} &                        \\
      42$^{\diamond}$ &  000.7-07.4 &  18:17:37.20 &  -31:56:46.87 &   VVV &  18:17:37.20 &  -31:56:46.99 &        N &  0.12 &        &        &        &                                &                        \\
      58 &  001.2+02.1 &  17:40:12.83 &  -26:44:21.83 &   \emph{HST} &  17:40:12.84 &  -26:44:21.75 &        Y &  0.11 &   18.7 & -0.116 &  0.404 &   5.92$_{-1.71}^{+1.78}$ &                        \\
      60 &  001.2-03.0 &  18:00:37.62 &  -29:21:50.67 &    PS &  18:00:37.62 &  -29:21:50.73 &        Y &  0.08 &   15.7 &  0.091 &  0.052 &   8.92$_{-1.82}^{+2.11}$ &               [WC 11]? \\
      62 &  001.3-01.2 &  17:53:47.17 &  -28:27:17.99 &   VLT &  17:53:47.17 &  -28:27:18.09 &        Y &  0.12 &   18.0 &  0.751 &  0.297 &   4.84$_{-2.84}^{+3.77}$ &                        \\
      73 &  001.7+05.7 &  17:28:01.74 &  -24:25:23.72 &   VLT &  17:28:01.75 &  -24:25:23.43 &        N &  0.32 &        &        &        &                                &                        \\
      74$^{*}$ &  001.7-04.4 &  18:07:14.56 &  -29:41:24.67 &   \emph{HST} &  18:07:14.56 &  -29:41:24.61 &        N &  0.07 &    &    &     &     &                [WC 11] \\
      90 &  002.2-09.4 &  18:29:11.64 &  -31:29:59.24 &   \emph{HST} &  18:29:11.65 &  -31:29:59.19 &        Y &  0.09 &   15.2 &  0.276 &  0.086 &   4.57$_{-1.03}^{+1.24}$ &              [WO 4]pec \\
      91 &  002.3+02.2 &  17:42:30.02 &  -25:45:28.84 &   VLT &  17:42:29.90 &     -25:45:27 &        N &  2.42 &        &        &        &                                &                        \\
      93 &  002.3-03.4 &  18:04:28.81 &  -28:37:38.29 &   \emph{HST} &  18:04:28.81 &  -28:37:38.29 &        Y &  0.05 &   18.9 &        &        &                                &                        \\
      98 &  002.5-01.7 &  17:58:31.20 &  -27:37:05.37 &   VLT &  17:58:31.22 &  -27:37:04.60 &        N &  0.82 &        &        &        &                                &                        \\
    4326 &  002.6+02.1 &  17:43:39.48 &  -25:36:42.87 &   VLT &  17:43:39.55 &  -25:36:42.84 &        N &  0.97 &        &        &        &                                &                        \\
     105 &  002.7-04.8 &  18:11:04.99 &  -28:58:59.22 &   \emph{HST} &  18:11:04.99 &  -28:58:59.09 &        Y &  0.13 &   18.3 & -1.359 &  0.524 &    4.31$_{-1.03}^{+1.1}$ &  cont.                 \\
     106 &  002.8+01.7 &  17:45:39.80 &  -25:40:00.52 &   \emph{HST} &  17:45:39.80 &  -25:40:00.50 &        Y &  0.02 &   17.8 & -0.148 &   0.16 &  11.85$_{-3.31}^{+3.59}$ &                        \\
     109 &  002.9-03.9 &  18:08:05.76 &  -28:26:10.32 &   \emph{HST} &  18:08:05.77 &  -28:26:10.81 &        N &  0.51 &        &        &        &                                &                        \\
     114 &  003.1+03.4 &  17:40:07.42 &  -24:25:42.77 &   \emph{HST} &  17:40:07.42 &  -24:25:42.67 &        Y &  0.11 &   16.5 &  0.083 &  0.062 &   7.88$_{-1.53}^{+1.73}$ &                        \\
     125 &  003.6+03.1 &  17:41:57.26 &  -24:11:16.33 &   \emph{HST} &  17:41:57.26 &  -24:11:16.24 &        Y &   0.10 &   17.6 &  -0.45 &  0.581 &   8.43$_{-2.31}^{+2.35}$ &  \emph{wels}                 \\
     129 &  003.7-04.6 &  18:12:34.39 &  -27:58:10.70 &    PS &  18:12:34.39 &  -27:58:10.65 &        Y &  0.06 &   16.9 &        &        &                                &  \emph{wels}                 \\
     131 &  003.8-04.3 &  18:11:29.26 &  -27:46:15.92 &   \emph{HST} &  18:11:29.31 &  -27:46:16.16 &        N &  0.69 &        &        &        &                                &                        \\
     134 &  003.9-03.1 &  18:06:50.03 &  -27:06:19.58 &   \emph{HST} &  18:06:50.03 &  -27:06:19.51 &        Y &  0.08 &   19.4 &        &        &                                &                        \\
     139 &  004.0-03.0 &  18:06:40.91 &  -26:54:56.28 &   \emph{HST} &  18:06:40.91 &  -26:54:56.38 &        Y &  0.11 &   14.4 &  0.053 &  0.038 &   9.69$_{-1.55}^{+1.74}$ &  O(H)f+?             \\
     142 &  004.2-03.2 &  18:08:01.39 &  -26:54:02.14 &    PS &  18:08:01.39 &  -26:54:02.18 &        Y &  0.04 &   17.2 & -0.104 &  0.109 &                                &                        \\
     141 &  004.2-04.3 &  18:12:25.11 &  -27:29:13.18 &    PS &  18:12:25.11 &  -27:29:13.20 &        Y &  0.04 &   16.4 &  0.283 &  0.082 &    6.10$_{-2.11}^{+3.21}$ &  \emph{wels}                 \\
     143 &  004.2-05.9 &  18:18:38.35 &  -28:07:58.53 &   VLT &  18:18:38.35 &  -28:07:58.60 &        Y &  0.07 &   17.3 &  0.286 &  0.151 &                                &                        \\
    4315 &  004.3+01.8 &  17:48:36.54 &  -24:16:34.24 &   VLT &  17:48:36.54 &  -24:16:34.23 &        Y &  0.05 &   16.2 & -0.024 &  0.142 &   5.73$_{-1.18}^{+1.28}$ &                        \\
     148 &  004.6+06.0 &  17:33:37.58 &  -21:46:24.80 &   VLT &  17:33:37.58 &  -21:46:24.88 &        Y &  0.08 &   17.6 &  0.048 &   0.15 &   6.41$_{-1.52}^{+1.65}$ &  \emph{wels}                 \\
     151 &  004.8+02.0 &  17:49:00.51 &  -23:42:54.90 &   \emph{HST} &  17:49:00.51 &  -23:42:54.84 &        Y &  0.06 &   16.1 &  0.169 &  0.052 &   8.86$_{-2.64}^{+3.73}$ &                        \\
     150 &  004.8-05.0 &  18:16:11.44 &  -27:14:57.83 &    PS &  18:16:11.44 &  -27:14:57.86 &        Y &  0.04 &   17.8 & -0.194 &  0.151 &   7.55$_{-1.65}^{+1.79}$ &                        \\
     156 &  005.0-03.9 &  18:12:22.99 &  -26:32:54.50 &   VLT &  18:12:22.98 &  -26:32:54.52 &        Y &  0.09 &   19.0 &        &        &                                &                        \\
     162 &  005.2+05.6 &  17:36:22.65 &  -21:31:12.60 &   VLT &  17:36:22.64 &  -21:31:12.39 &        N &  0.25 &        &        &        &                                &                        \\
    4106 &  005.5+06.1 &  17:35:21.43 &  -20:57:23.47 &    PS &  17:35:21.43 &  -20:57:23.45 &        Y &  0.02 &   16.6 &  0.014 &  0.095 &                                &                        \\
     165 &  005.5-04.0 &  18:13:40.60 &  -26:08:39.53 &    PS &  18:13:40.60 &  -26:08:39.60 &        Y &  0.07 &   19.3 &  0.608 &  0.429 &                                &                        \\
     179 &  006.3+03.3 &  17:47:33.95 &  -21:47:23.29 &   VLT &  17:47:33.94 &  -21:47:23.29 &        Y &  0.11 &   18.5 & -0.022 &  0.246 &                                &                        \\
     180 &  006.3+04.4 &  17:43:28.75 &  -21:09:51.77 &   \emph{HST} &  17:43:28.76 &  -21:09:51.68 &        Y &  0.12 &   18.1 &  0.018 &  0.177 &                                &  Of?                   \\
     182 &  006.4+02.0 &  17:52:41.44 &  -22:21:57.18 &   \emph{HST} &  17:52:41.45 &  -22:21:57.14 &        Y &   0.10 &   17.6 &        &        &                                &  \emph{wels}                 \\
     193 &  007.0+06.3 &  17:38:11.59 &  -19:37:37.62 &    PS &  17:38:11.60 &  -19:37:37.75 &        Y &  0.21 &   15.7 &  0.177 &  0.041 &                                &                        \\
     197 &  007.5+07.4 &  17:35:10.22 &  -18:34:20.76 &    PS &  17:35:10.22 &  -18:34:20.70 &        Y &  0.06 &   19.3 &  0.886 &  0.335 &                                &                        \\
     198 &  007.6+06.9 &  17:37:21.99 &  -18:46:41.76 &    PS &  17:37:21.99 &  -18:46:41.75 &        Y &  0.03 &   18.6 & -0.119 &  0.257 &                                &                        \\
     200 &  007.8-03.7 &  18:17:16.03 &  -23:58:54.89 &    PS &  18:17:16.03 &  -23:58:54.84 &        Y &  0.06 &   17.1 &        &        &                                &  [WC]                  \\
     201 &  007.8-04.4 &  18:20:08.84 &  -24:15:05.39 &   \emph{HST} &  18:20:08.84 &  -24:15:05.36 &        Y &  0.05 &   14.5 &  0.033 &  0.032 &   9.93$_{-1.32}^{+1.46}$ &              O(H)8-9 I \\
    4111 &  008.2+06.8 &  17:38:57.37 &  -18:17:35.92 &   \emph{HST} &  17:38:57.37 &  -18:17:35.80 &        Y &  0.12 &   13.6 &   0.05 &  0.029 &                                &                O(H)7-8 \\
     208 &  008.4-03.6 &  18:18:23.83 &  -23:24:57.50 &    PS &  18:18:23.83 &  -23:24:57.57 &        Y &  0.07 &   17.7 &  0.231 &  0.114 &                                &                        \\
     210 &  008.6-02.6 &  18:14:50.91 &  -22:43:55.52 &   \emph{HST} &  18:14:50.91 &  -22:43:55.51 &        Y &  0.06 &   17.9 &  0.089 &  0.188 &                                &                        \\
     217 &  009.4-09.8 &  18:44:43.16 &  -25:21:33.80 &    PS &  18:44:43.15 &  -25:21:33.89 &        Y &  0.14 &   17.5 & -0.089 &  0.146 &    8.10$_{-1.79}^{+1.91}$ &                        \\
     223$^{\bullet}$ &  009.8-04.6 &  18:25:04.97 &  -22:34:52.48 &    PS &  18:25:04.96 &  -22:34:52.48 &        Y? &  0.09 &   18.7 &  3.371 &  0.551 &   {\bf 0.32$_{-0.05}^{+0.08}$} &                 [WO 2] \\
\hline
\end{tabular}
\end{table*}
\begin{table*}
\centering
\contcaption{}
\begin{tabular}{l@{\hspace{0.2\tabcolsep}}cc@{\hspace{0.8\tabcolsep}}ccc@{\hspace{0.8\tabcolsep}}ccccccc@{\hspace{0.2\tabcolsep}}c}
\hline
HASH & \multirow{2}{*}{PN G} & \multicolumn{2}{c}{This Work}          &  \multirow{2}{*}{Tele.} &  \multicolumn{2}{c}{Gaia EDR3}         & \multirow{2}{*}{True?} & ${\Delta r_{{\star}}}$ & \emph{G}  & $\omega$   & $\sigma_{\omega}$      & $D$      &      \multirow{2}{*}{Spec. Type}    \\ \cline{3-4} \cline{6-7} ID & & RAJ2000 & \multicolumn{1}{c}{DECJ2000} &     &  RAJ2000     & \multicolumn{1}{c}{DECJ2000} &   &  (arcsec) & (mag)   & (mas)                & (mas)          & (kpc) &        \\
\hline
    1151 &  351.1+04.8 &  17:03:46.84 &  -33:29:44.64 &   \emph{HST} &  17:03:46.84 &  -33:29:44.52 &        Y &  0.12 &   16.1 &  0.102 &  0.071 &    7.08$_{-1.4}^{+1.57}$ &  wels?                 \\
  1156$^{\diamond}$ &  351.6-06.2 &  17:50:44.62 &  -39:17:26.80 &   VVV &  17:50:44.60 &  -39:17:26.64 &        N &  0.32 &        &        &        &                               &                        \\
      1159 &  351.9+09.0 &  16:50:17.08 &  -30:19:55.39 &   VLT &  16:50:17.08 &  -30:19:55.28 &        Y &  0.12 &   16.6 &  0.122 &  0.082 &  7.49$_{-1.66}^{+1.84}$ &                        \\
    1160 &  351.9-01.9 &  17:33:00.68 &  -36:43:52.94 &   \emph{HST} &  17:33:00.68 &  -36:43:52.90 &        Y &  0.05 &   18.5 &        &        &                               &                        \\
    1164 &  352.6+03.0 &  17:14:42.93 &  -33:24:47.71 &   \emph{HST} &              &               &      New &       &        &        &        &                               &                        \\
    1178 &  353.7+06.3 &  17:05:13.91 &  -30:32:19.53 &   VLT &  17:05:13.91 &  -30:32:19.60 &        Y &  0.08 &   17.2 &        &        &                               &  O                     \\
    1189 &  354.9+03.5 &  17:19:20.24 &  -31:12:40.95 &   \emph{HST} &  17:19:20.24 &  -31:12:40.96 &        Y &  0.02 &   19.4 &  0.438 &  0.447 &                               &                        \\
    1199$^{\#}$ &  355.4-02.4 &  17:44:20.61 &  -34:06:40.96 &   \emph{HST} &  17:44:20.64 &  -34:06:40.80 &        N &  0.34 &   &        &        &                               &                        \\
    1207 &  355.9-04.2 &  17:52:58.94 &  -34:38:22.99 &   \emph{HST} &  17:52:58.94 &  -34:38:22.85 &        Y &  0.14 &   16.3 &  0.197 &   0.08 &  6.28$_{-1.53}^{+1.81}$ &  wels                  \\
    1235 &  357.1+03.6 &  17:24:34.45 &  -29:24:19.80 &    PS &  17:24:34.45 &  -29:24:19.79 &        Y &  0.02 &   16.2 &  0.172 &  0.054 &  5.67$_{-0.99}^{+1.17}$ &  wels                  \\
    1242 &  357.2+02.0 &  17:31:08.11 &  -30:10:28.08 &   \emph{HST} &              &               &      New &       &        &        &        &                               &                        \\
    1246 &  357.3+04.0 &  17:23:24.93 &  -28:59:05.91 &    PS &  17:23:24.93 &  -28:59:06.03 &        Y &  0.13 &   18.7 &  0.007 &  0.254 &  7.49$_{-2.13}^{+2.25}$ &                        \\
    1252 &  357.5+03.1 &  17:27:24.36 &  -29:21:14.63 &    PS &  17:27:24.36 &  -29:21:14.68 &        Y &  0.06 &   15.2 &  0.093 &  0.037 &                               &                        \\
    1256 &  357.6-03.3 &  17:53:16.82 &  -32:40:38.50 &   VLT &  17:53:16.82 &  -32:40:38.58 &        Y &  0.08 &   19.1 &  0.953 &  0.447 &  5.37$_{-2.92}^{+2.99}$ &  O(H)+?              \\
    1259 &  357.9-03.8 &  17:56:13.93 &  -32:37:22.21 &   VLT &  17:56:13.93 &  -32:37:22.20 &        Y &  0.05 &   18.7 & -2.383 &  0.341 &  8.31$_{-1.66}^{+1.75}$ &                        \\
    1258 &  358.0+09.3 &  17:05:44.60 &  -25:25:01.49 &   VLT &  17:05:44.60 &  -25:25:01.52 &        Y &  0.07 &   17.2 &  0.086 &  0.098 &                               &                        \\
    1263 &  358.2+03.5 &  17:27:32.87 &  -28:31:06.86 &    PS &  17:27:32.86 &  -28:31:06.93 &        Y &  0.11 &   19.1 & -1.589 &  0.486 &  8.98$_{-2.29}^{+2.39}$ &                        \\
    1275 &  358.5+02.9 &  17:30:30.43 &  -28:35:54.90 &   \emph{HST} &  17:30:30.44 &  -28:35:54.92 &        Y &  0.13 &   19.0 &  1.294 &  0.397 &  10.8$_{-7.25}^{+5.02}$ &                        \\
    1276 &  358.5-04.2 &  17:59:02.51 &  -32:21:43.63 &   \emph{HST} &  17:59:02.51 &  -32:21:43.50 &        Y &  0.13 &   15.1 & -0.447 &  0.242 &   7.90$_{-1.81}^{+1.91}$ &                        \\
    1280 &  358.6+07.8 &  17:12:39.17 &  -25:43:37.50 &   VLT &  17:12:39.16 &  -25:43:37.56 &        Y &  0.11 &   16.8 & -0.032 &  0.073 &                               &                        \\
    1281 &  358.6-05.5 &  18:04:56.22 &  -32:54:01.05 &   VLT &  18:04:56.23 &  -32:54:01.08 &        Y &  0.11 &   18.6 &   0.36 &  0.206 &  5.33$_{-1.79}^{+1.95}$ &                        \\
    1286 &  358.7+05.2 &  17:22:28.29 &  -27:08:42.51 &   \emph{HST} &  17:22:28.29 &  -27:08:42.46 &        Y &  0.05 &   16.9 &  0.152 &  0.084 &                               &                        \\
    1293 &  358.8+03.0 &  17:31:09.28 &  -28:14:50.17 &   VLT &              &               &      New &       &        &        &        &                               &                        \\
    1308 &  359.2+04.7 &  17:25:44.08 &  -26:57:48.04 &   \emph{HST} &  17:25:44.08 &  -26:57:48.01 &        Y &  0.04 &   15.3 &  0.064 &  0.037 &                               &                        \\
    1312 &  359.3-01.8 &  17:51:18.93 &  -30:23:53.23 &   VLT &  17:51:18.92 &  -30:23:53.18 &        Y &   0.10 &   16.0 &  0.083 &  0.062 &   6.92$_{-1.3}^{+1.51}$ &                [WC 11] \\
    1322 &  359.7-01.8 &  17:52:05.97 &  -30:05:14.12 &   VLT &  17:52:05.96 &  -30:05:14.15 &        Y &   0.10 &   17.7 &  0.135 &  0.126 &  5.16$_{-1.25}^{+1.46}$ &                        \\
    1324 &  359.8+03.7 &  17:30:46.73 &  -27:05:59.91 &    PS &  17:30:46.73 &  -27:05:59.86 &        Y &  0.05 &   18.1 &  0.172 &   0.22 &  8.09$_{-2.42}^{+2.51}$ &                        \\
    1328 &  359.8+05.2 &  17:25:23.63 &  -26:11:53.00 &   VLT &  17:25:23.62 &  -26:11:53.00 &        Y &  0.09 &   19.0 & -1.284 &  0.353 &                               &                        \\
\hline
\multicolumn{14}{l}{\small $^{\bullet}$ These two PNe gave suspiciously low \emph{Gaia} distances which give unfeasibly short kinematic lifetimes. See main text for caveats.}\\
\multicolumn{14}{l}{\small $^{*}$ \emph{HST} shows two stars within the PN envelope 
$\sim$0.2~arcseconds apart. \emph{Gaia} cannot resolve these. The reported astrometric solution is of low quality}\\
\multicolumn{14}{l}{\small $^{\ }$ (Goodness of fit statistic $>$ 100).}\\
\multicolumn{14}{l}{\small $^{\#}$ \emph{Gaia} has selected a compact nebula knot 
$\sim$0.3~arcseconds from the actual CSPN evident in the short exposure broad-band 457M filter \emph{HST} 
imagery}\\
\multicolumn{14}{l}{\small $^{\diamond}$ Two examples where the VVV NIR imagery 
has revealed a faint CSPN within an inner oval NIR nebula ring}
\end{tabular}
\end{table*}

\section{Results}
\label{res}
\noindent Here we present results from our work on morphological 
classifications, CSPN identification and measurement and kinematic age estimates. 
PNe centroid positions, morphological classifications, 
updated angular size estimates, CSPN coordinates have been determined (where available) 
from the combined \emph{HST}, VLT, Pan-STARRS and in two cases, near 
infrared J, H and K$_{s}$ data from the  VVV \citep{2012A&A...537A.107S}.
These measurements are summarised in Table~\ref{tb:sum_results} and 
Table~\ref{tab:cs_gaia}. All centroid coordinates of PNe in this sample were 
remeasured with this unique set of combined high angular resolution and 
high sensitivity imaging data. Currently none of the VLT 
images have yet been ingested into HASH. 

\subsection{Updated angular size measurements}
\label{ang-size}
The new VLT imagery provides an opportunity to improve the angular 
size estimates currently in HASH as previously estimated from the SHS 
imagery. The generally better seeing in the VLT imagery c.f. the 2~arcseconds 
typical for the SHS, and the better VLT resolution allows improved estimation 
of PNe angular size. This is an important parameter when estimating kinematic age. 
Angular size measures are taken from the main body of the PNe and not from any very
faint outer halos that likely reflect earlier mass loss off the AGB. 
All new angular size values are provided in Table~\ref{tb:sum_results} and 
were consistently taken from the 98\% 
intensity plots from the VLT and \emph{HST} 
fits images with histogram equalized scaling. We believe this choice better reflects 
the true extent of the ejected material that constitutes these PNe.

Estimated errors in angular size measurements are 4 image pixels which is 0.5 
and 0.2~arcseconds for the VLT and \emph{HST} images respectively. 
Comparison with current HASH values show a mean difference of 3.4~arcseconds. 
The average angular size of this re-measured $\leq$~10~arcsecond PNe bulge 
sample is 9.5 with $\sigma$~=~5.0. Interestingly, 47 PNe now  exceed (mostly marginally) 
the original $\leq$~10~arcsecond major diameter HASH based angular diameter 
selection criterion. 

\subsection{Morphological classification}
\label{morph_class_res}
Below we present the results from our re-evaluated morphological classifications for all 
136 PNe in the bulge sample. We examine the effect of image quality, show the 
new distribution of morphological classes, reveal the surprising prevalence of 
bipolars and the incidence of point-symmetric and non-symmetric structures.
This discussion supplements the PN by PN listing given in Table~\ref{tb:sum_results}.  
Improved image resolution for many compact bulge PNe from the VLT 
(and in some cases from Pan-STARRS) together with a fresh 
assessment of the \emph{HST} imagery has revealed additional morphological 
structures for many PNe in this sample. All PNe \emph{HST} 
imagery is already in HASH and was usually used for the original HASH 
morphological classifications. In some cases even the true basic PNe form, 
not apparent from the lower resolution imagery previously used, 
has been revealed for the first time.

\subsubsection{The effect of image quality}
\label{image_quality_PA}

\begin{figure*}
    \centering
    \includegraphics[height=0.232\linewidth]{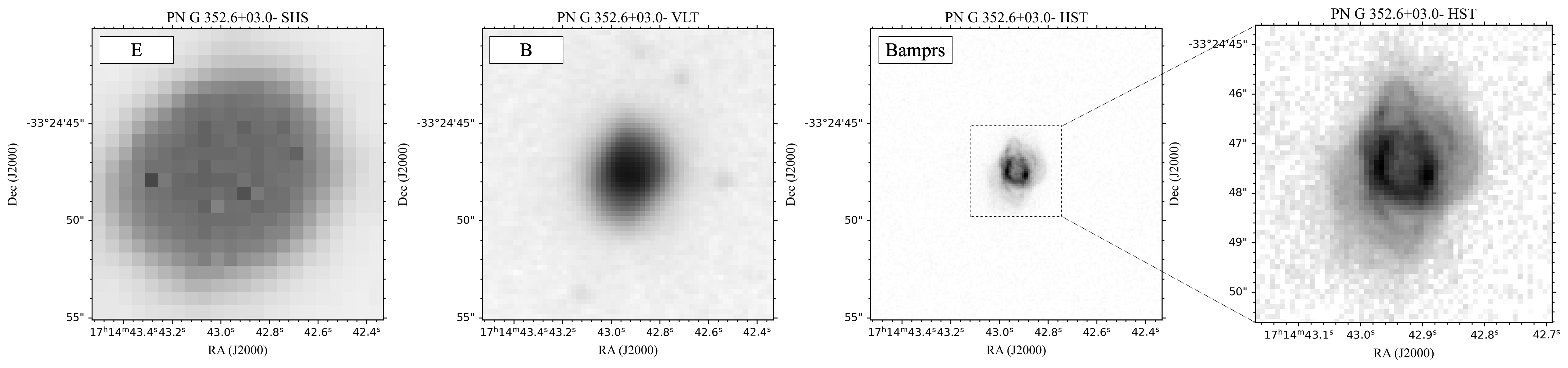}
    \includegraphics[height=0.23\linewidth]{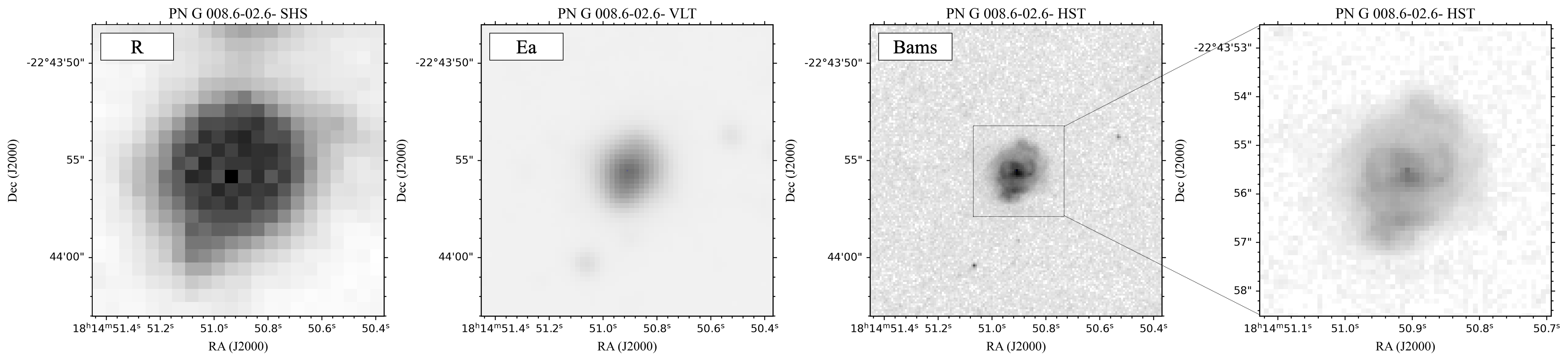}
    \caption{Two cases that contrast the difference in PN morphological classification 
    with varying image quality. PNG~352.6+03.0 (upper panel) shows a change in 
    morphological class assignment from `E' $\rightarrow$ `B' when going from the 
    SHS to the VLT image. The final \emph{HST} image reveals more structural details, 
    including the bipolar lobes and central ring structure allowing for 
    additional `sparm' classifiers to be added - better seen in the expanded, final image. 
    The lower panel show images of PNG~008.6-02.6. The main morphological class 
    has been changed from `R' in the SHS to `E' in the VLT while the \emph{HST} 
    image shows the PN is actually a bipolar with outer 
    structures and internal over densities. In both cases the CSPN are also now 
    visible thanks to \emph{HST}.}
    \label{fig:morph-comp}
\end{figure*}

It is clear from this work that assigned PNe morphology, both in terms of basic 
classes and also sub-classifications is, unsurprisingly, a strong function of the
resolution, depth and sensitivity of the available imagery especially when the
PN is compact. In Fig.~\ref{fig:morph-comp} we show two examples 
that demonstrate the image quality comparison between the original SHS 
imagery and the VLT and \emph{HST} imagery selected from the 40 PNe where 
both observational data sets are available. This demonstrates the effect 
in going from lower to intermediate to high image resolution and how this can 
alter assigned morphological classifications. 
This is only really an issue for compact PNe in HASH with diameters 
$<$~10~arcseconds, as for our bulge sample.

In the top panel PNG~352.6+03.0 is presented. It shows a main morphology shift from class 
from `E' to `B' when going from the SHS to the VLT image. The final \emph{HST} images at right 
reveal more structural details that allow for additional `sparm' classifiers. The lower 
panel show images of PNG~008.6-02.6. The main morphological class has been changed from `R' 
in the SHS to `E' in the VLT while the \emph{HST} image shows the PN is actually a bipolar 
with outer structures and internal over densities. The CSPN is now also visible. 
Both examples demonstrate the importance of image quality and resolution in PNe 
morphological classification. We note 32 of the 40 objects 
classified from the VLT imagery have morphological results somewhat inconsistent with 
their \emph{HST} counterparts which are used in preference. This demonstrates that
using images where small-scale internal and outer PN features 
are not well-resolved or seen can lead to basic classification errors and 
differences in the sub-class assignments. 

\setlength{\tabcolsep}{8pt}
\begin{table}
\centering
\begin{tabular}{lccccc}
\hline
\multirow{2}{*}{Morphology} & \multirow{2}{*}{Change} & \multicolumn{3}{c}{Telescope} & \multirow{2}{*}{Total} \\ \cline{3-5}
&   & VLT       & \emph{HST} &  PS             &\\ \hline
Main Class             & E $\rightarrow$ B       & 28            & 2   &  -  & 30   \\
                        & R $\rightarrow$ B       & 4             &-     &  -  & 4    \\
                        & S $\rightarrow$ B       & 4             &-     &   -   & 4    \\
                        & R $\rightarrow$ E       & 6             &-     & 1   & 7    \\
                        & S $\rightarrow$ E       & 7             & 1   &  -   & 8    \\
                        & E $\rightarrow$ A       & 1             &-     &  -  & 1    \\\hline
Total &   &               &     &          & 54   \\ \hline
Sub-class    & s & 61            & 3    &  2  & 66   \\          
            & p & 16            &  2  &  - & 18   \\
            & a & 25            & 5   & 1 & 31   \\
            & r & 27            & 3   &  1 & 31   \\
            & m & 34            &  6  &  2   & 42   \\
 \hline
Total &   &   &        &      & 188  \\ \hline
\end{tabular}
\caption{Improvement in PN morphological classifications possible mainly 
from the combined \emph{HST} and VLT imagery. The integers show the numbers 
of PNe that have undergone a classification change in this study. 
A total of 56 PNe have had their major class changed with the bulk (38) being 
reclassified to bipolar for a total change of 54/136 or 70\%. The bottom part of the table illustrates detection of the newly resolved sub-structures. The combined imagery allows to add 188 sub-classifiers for the whole sample (1.4 new `sparm' sub-classifiers per nebula on average).}
\label{tab:morph_improve}
\end{table}

\subsubsection{An enrichment in bipolar morphology revealed}
\label{sec:more_bipolar}
\noindent The revised morphological classification of the bulge PNe were 
compared with those recorded in the HASH database. Changes in the main and 
sub-classifiers are summarised in Table~\ref{tab:morph_improve}. PNe of the 
morphological class I and A are rare: only one was found in this sample. 
In HASH overall they represent a mere 0.7\% and 0.4\% of all Galactic PNe.

Among the 96 of 136 PNe observed with the VLT and where the pre-imaging 
provides the best available data, 5 now reveal an elliptical nebulosity with no 
resolved internal structure, resulting in a new `E' class designation. 
They were previously classified as `S' for point-like objects in 
HASH from the previous lower resolution optical imagery. For the remaining 91 PNe, 
the main "ERBIAS" morphology of 45 PNe have changed in this new assessment 
from their current HASH values. Some 36 PNe classified as E, R or S actually exhibit clear 
bipolar features in the VLT observations. For 76 PNe, we have also made changes 
to the secondary "sparm" classification. 

All \emph{HST} observations are accessible in HASH via a clickable icon 
on the main database page for each PNe.
Based on assessment of the \emph{HST} imagery we now re-classify 
PNG~355.9+03.6 from `S' to `E', PNG~001.2+02.1 and PNG~004.0-03.0 from `E' to `B' 
(the old HASH morphologies for these two were out of date). A further 11 PNe with 
\emph{HST} imagery have had their `sparm' sub-classifiers upgraded after careful examination.

\begin{figure}
    \centering
    \includegraphics[width=0.49\textwidth]{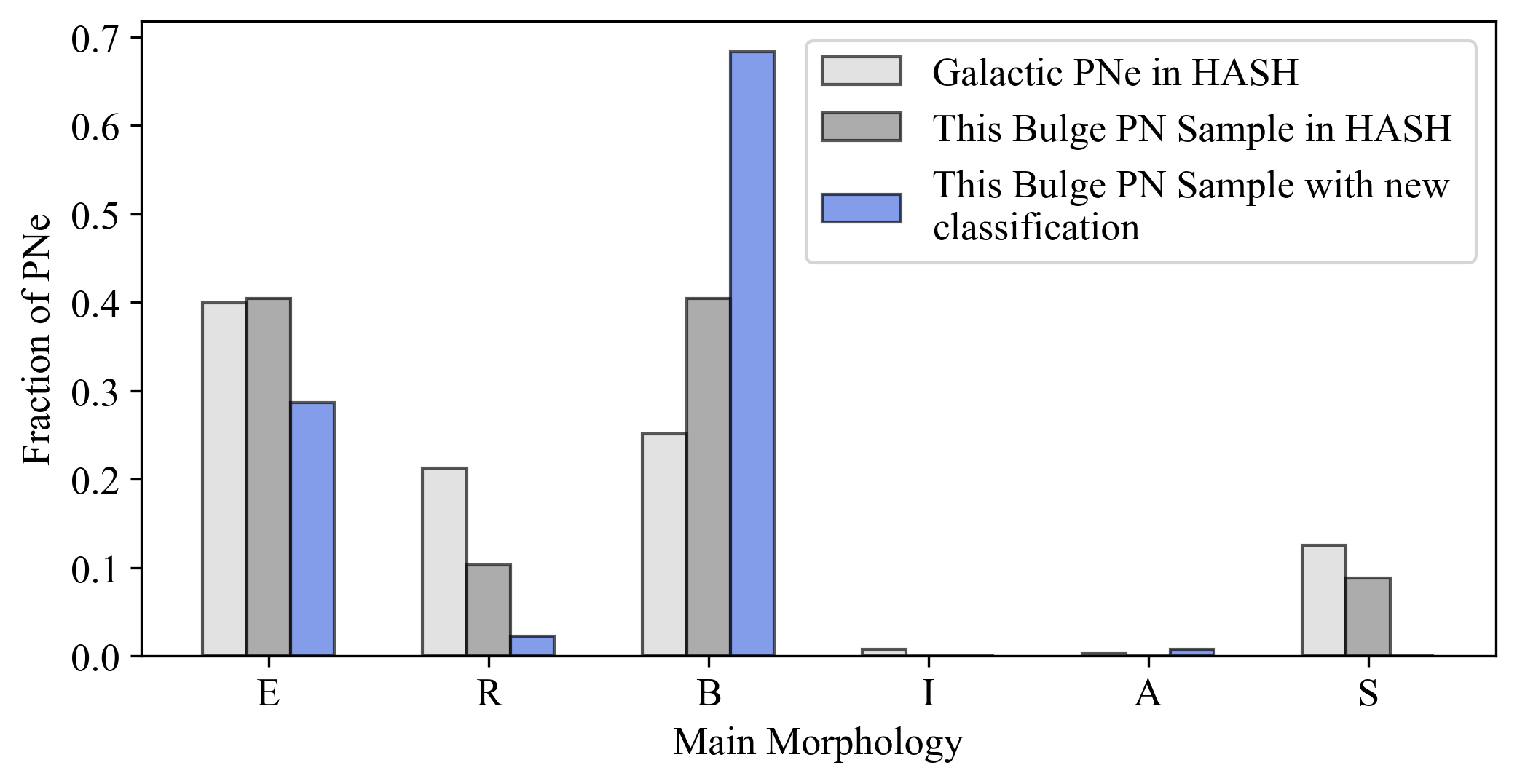}
    \caption{The fractional populations of different, basic PNe morphological classes. 
    The blue histogram shows the sample of 136 PNe in the combined VLT and \emph{HST} 
    observations. The dark grey histogram indicates the previous fraction of types 
    of this sample in the HASH database. The light grey histogram presents the 
    fractional morphological results for all 2676 'True" Galactic PNe with a 
    morphological classification in HASH.}
    \label{fig:morph_frac}
\end{figure}

We present the fraction of objects in the different main morphological classes
in the histogram in Fig.~\ref{fig:morph_frac}. We include the re-assessed morphologies 
of this PNe sample as well as for all Galactic "True" PNe in the HASH 
database plotted for comparison. From Table~\ref{tab:morph_improve}, 70\% of the changes
made to main HASH classifications are for newly-resolved bipolar PNe. As a result, evident
bipolar PN now represent the majority (68\%) of bulge PNe in our sample. 
Interestingly, even with the old HASH classifications, the fraction of bipolar PNe in the
Galactic bulge exceeds that for general Galactic disk PNe population and accounts for the 
largest proportion of this population. This reveals an underestimation of the actual
fraction of bipolar PNe compared to what lower resolution and poorer sensitivity 
imaging had indicated. For example, when bipolar lobes are not seen (too faint) or 
resolved (too compact), the morphology can appear to belong to class E or even S.

This 68\% fraction must be a lower limit as inclination of the bipolar to the line 
of sight can also hide the bipolar nature of some PNe \citep{1997MNRAS.284...32P}. 
Indeed, the thick nebular annular ring of PNG~357.2+02.0 shown in the upper right middle panel in Fig.~\ref{fig:new_cspn} is probably one such case. It is a very close analogue to that modelled for a bipolar 
inclined only 10~degrees to the line of sight as in Fig.2 of \citet{1997MNRAS.284...32P}. 
It is also reminiscent of the example for PN Sp~1 (PNG~329.0+01.9) presented by \citet{mitchell2006structures} where additional kinematic evidence indicate 
we are seeing a bipolar PN pole on.

\subsubsection{Resolved point-symmetric features in bipolar PNe}\noindent

The fraction of PNe with a specific secondary "sparm" morphology among each main class 
is summarised in Table~\ref{tab:sub_fraction}. With the combined VLT, \emph{HST} and 
occasional Pan-STARSS observations, the number of PNe with internal structure resolved 
(sub-class s) doubled. The fraction of bipolar PNe in the sub-class `s' is larger than 
that of the elliptical PNe, which supports the idea that determination of bipolar 
morphology requires better-resolved imaging.  Three class `E' PNe, PNG~003.6-02.3, 
PNG~008.4-03.6 and PNG~353.7+06.3, and one class `R' PN, PNG~002.5-01.7, were observed 
with point-symmetric features in HASH. However, all PNe in sub-class `p' in this work 
have bipolar morphology. There is strong observational evidence that PNe with such 
point-symmetric features house short-period binaries \citep{2009A&A...505..249M}. 
In our sample such features were also seen in PNG~002.5-01.7, PNG~003.6-02.3 and 
PNG~008.4-03.6 in their H$\alpha$ outer shells in catalogued images in HASH. High-resolution images of PNG~002.5-01.7 and PNG~003.6-02.3 in this combined data show no point-symmetric structure while our VLT image of PNG~008.4-03.6 exhibits clearly lobes and the equatorial
concentration of matter expected for bipolar PNe, although VLT H$\alpha$ 
images of PNG~002.5-01.7 and PNG~003.6-02.3 are not available.
\setlength{\tabcolsep}{9pt}
\begin{table*}
\centering
\begin{tabular}{crccccccc}
\hline
\multicolumn{1}{c}{Class} &      &  -      &  \multicolumn{1}{c}{s} & \multicolumn{1}{c}{p} & \multicolumn{1}{c}{r} & \multicolumn{1}{c}{a} & \multicolumn{1}{c}{m} &  Total \\ \hline
\multirow{2}{*}{E}  & HASH & 19 [0.35] & 13 [0.24] & $\ $3 [0.05] & 15 [0.27] & 22 [0.40]  & 14 [0.25]   & 55    \\
                      & This Work & $\ $4 [0.10]  & 27 [0.69] & -  & 24 [0.62]  & 23 [0.59] & 20 [0.51]   & 39    \\ \hline
\multirow{2}{*}{R}  & HASH &$\ $8 [0.57]& $\ $3 [0.21] & $\ $1 [0.07]     & $\ $5 [0.36]     & $\ $2 [0.14]       & $\ $1 [0.07]   & 14    \\ &                        This Work & - & $\ $3 [1.00] &  - & $\ $2 [0.67] & $\ $ 3 [1.00] & $\ $ 2 [0.67]  & 3     \\ \hline
\multirow{2}{*}{B}    & HASH & 14 [0.25]       & 34 [0.62] & 16 [0.29] & 17 [0.31]   & $\ $8 [0.15]  & 19 [0.35]      & 55  \\  
                    & This Work & $\ $2 [0.02]   & 85 [0.91]        & 35 [0.38]    & 33 [0.35]       & 36 [0.39]    & 47 [0.51]      & 93    \\ \hline
\end{tabular}
\caption{Number of objects in a given main and sub-class for the sample of bulge PN in 
this work. The fraction of the total PNe of each main class is indicated in the parentheses.
The upper and the lower rows compared the results in HASH and in this work.}
\label{tab:sub_fraction}
\end{table*}

\subsubsection{The incidence of non-symmetric structures}
\label{unsymm} 
\noindent All sub-classifiers, except for sparm sub-classifier `a' which describes a one-side enhancement, refer to either axisymmetrical or spherical or point-symmetrical features about the geometric centroid centre of the nebulae. The majority of PNe in this sample show morphological structures with one or more of these symmetries. However, 5 PNe, PNG~002.1-02.2, PNG~008.6-02.6, PNG~354.9+03.5,
PNG~356.1-03.3 and PNG~357.5+03.2, show certain structural features that are neither axisymmetrical nor 
point-symmetrical. Unlike a one-side protrusion or asymmetry (PNe with sub-class a) or unequal lobes that could be explained by binary interactions, such features that show a departure from axisymmetry and point-symmetry simultaneously may have resulted from the presence of a triple stellar system \citep[e.g.][]{jones2016central}. We do not explore this issue further in this paper.



\subsection{Summary of findings for CSPN}
Below we provide summary details of the main findings from our search for and 
identification of CSPN in our complete bulge sample.

\subsubsection{Existing catalogues of CSPN and \emph{Gaia}}

Searching for CSPN in the overall Galactic PNe population from \emph{Gaia} DR2 data was
first conducted by \citet{gonzalez2019properties}, \citet{stanghellini2020population} 
through a point-based matching approach, and in \citet{chornay2020searching} through an 
automated matching process that considered both \emph{Gaia} relative positions and 
colour information. The results were updated with the latest Gaia EDR3 data in 
\citet{chornay2021one} and \citet{gonzalez2021planetary}, hereafter CW21 and GSM21 
respectively. A preliminary analysis by \citet{2022Galax..10...32P}, developed 
further at the WD~2022 conference in Tubingen in August 2022 by 
co-author Parker, indicates that the \citet{chornay2021one} catalogue is the more 
reliable, but that both suffer problems of correct CSPN identification (see later). 
This is at least in part because many true CSPN are actually fainter than the 
\emph{Gaia} limits and the bluest, most central remaining candidate gets selected 
instead. We compare our detected CSPN sample with those reported in CW21 and GSM21. 
The full CW21 sample used the entire HASH Galactic PN catalogue as the input list.  


\subsubsection{Comparison with existing \emph{Gaia} CSPN listings}
\begin{figure*}
    \centering
    \begin{subfigure}{\linewidth}
    \centering
 \includegraphics[height=0.232\linewidth]{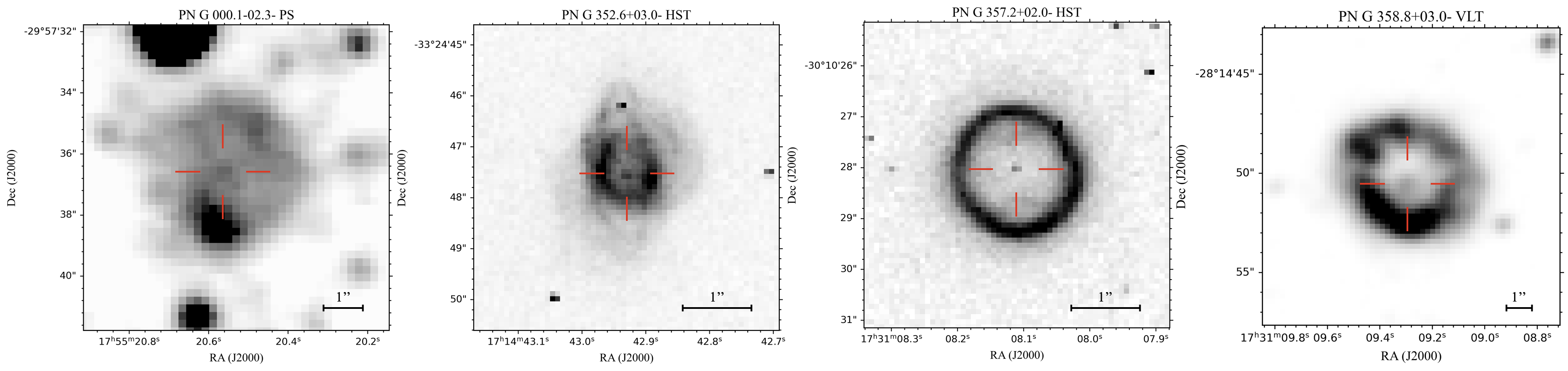}   
 \caption{Images of the 4 PNe with newly discovered CSPN: PNG~000.1-02.3, PNG~352.6+03.0, 
 PNG~357.2+02.0, and PNG~358.8+03.0. The CSPN are indicated as the centres 
 with open red crosses. The origin of the discovery image as VLT, \emph{HST} or Pan-STARRS is given 
 at the top with the PNG designation}
  \end{subfigure}\par\medskip
\begin{subfigure}{\linewidth}
    \centering
 \includegraphics[height=0.23\linewidth]{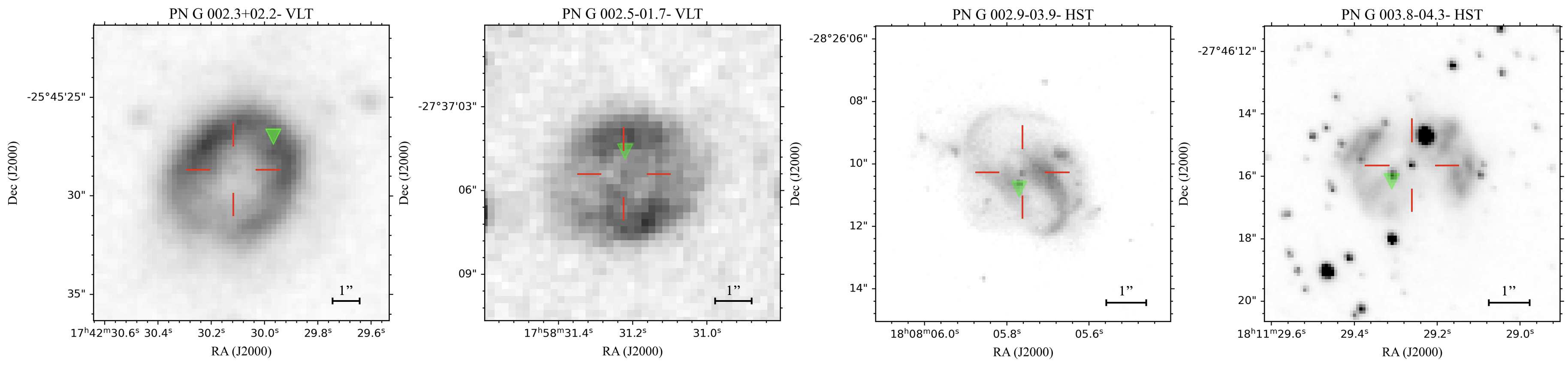}    
 \caption{Examples of false \emph{Gaia} CSPN identifications from CW21/CSM21. The open 
 red crosses are the new CSPN positions while green triangles show the corresponding 
 \emph{Gaia} putative CSPN detections reported in CW21/CSM21. PNG~002.3-02.2 (left panel) 
 and PNG~002.5-01.7 (left middle panel) are examples from VLT imagery where the 
 reported \emph{Gaia} CSPN is just part of a bright nebula region at the rim. 
 PNG~002.9-03.9 (right middle panel) and PNG~003.8-04.3 (right panel) are examples (which is too faint for \emph{Gaia}) was selected. The true CSPN are fainter and bluer 
 and both located at the nebula's geometric centroid.}
  \end{subfigure}\par\medskip
    \caption{Upper panel: Identifications of 4 new CSPN discovered in this work. 
    Lower panel: Examples of 4 PNe showing that the previously reported \emph{Gaia} 
    CSPN detections (green triangles) are false with the actual CSPN indicated via open red 
    crosses.}
    \label{fig:new_cspn}
\end{figure*}

A quick comparison revealed that 70 of 78 CSPN recovered from our work have a nominal 
identification in CW21 while 71 are identified in GSM21 and 67 CSPNe appear in results 
of both studies for a combined total of 74 \emph{Gaia} CSPN. Four CSPN observed, 
those of PNG~000.1-02.3, PNG 352.6+03.0, PNG 357.2+02.0,and PNG~358.8+03.0, 
were not in either of these CSPN catalogues, nor have they a \emph{Gaia} 
identification, so they are newly discovered CSPN in this work. 
These are shown in the upper four panels of Fig.~\ref{fig:new_cspn}.

A further 11 CSPN we have identified have relatively large offsets from the stars 
listed as the \emph{Gaia} CSPN in CW21/GSM21 or have other issues where we believe 
\emph{Gaia} has not actually seen the true CSPN (e.g. PNG~355.4-02.4). 
These are false \emph{Gaia} CSPN identifications. They are identified with the 
letter 'N' in Table~\ref{tab:cs_gaia} under 
the column heading "True?" and comprise 14.9\% of the 
reported \emph{Gaia} CSPN in this sample. Four examples are shown in the lower 
panels of Fig.~\ref{fig:new_cspn}. In the figure the centre of the open red crosses 
represents the new CSPN positions while green triangles show the 
corresponding \emph{Gaia} CSPN reported in CW21/CSM21. 
PNG~002.3+02.2 (left panel) and PNG~002.5-01.7 (left middle panel) 
are examples from VLT imagery where the reported \emph{Gaia} CSPN is just part of a bright nebula region at the rim. PNG~002.9-03.9 (right middle panel) and 
PNG~003.8-04.3 (right panel) are examples from \emph{HST} imagery where a brighter 
\emph{Gaia} star other than the actual 
CSPN (which is too faint for \emph{Gaia}) was selected. The true CSPN are fainter 
and bluer and both are located at the nebula's geometric centroid. 
There are finally two examples of CSPN identified from the VVV survey 
\citep{2012A&A...537A.107S} for PNG~000.7-07.4 and PNG~351.6-06.2. 
There is no evidence in their \emph{HST} imagery of any visible CSPN but instead an 
obscured zone. We believe the reported \emph{Gaia} CSPN merely reflects
fits to the overall compact PNe and not to any true CSPN which we consider 
obscured by dust but visible in the NIR. The VVV imagery of both is similar 
- fractured oval rings surrounding a clear, faint CSPN at the centre. 
The optical nebula spectra of both PNe are high excitation with He~II detected so there
is clearly a hot star present, even if obscured. 
There are a further two PNe, PNG~000.7-03.7 and PNG~009.8-04.6 whose \emph{Gaia} 
distances are suspiciously low. They have been assigned a Y? entry in 
Table~\ref{tab:cs_gaia} and are discussed in Section~\ref{sec:cs_dist} in this paper.
 
\subsubsection{Issues with \emph{Gaia} as a resource for CSPN}

The \emph{Gaia} satellite telescopes have relatively small rectangular 
mirrors of 1.45~$\times$~0.5~m yielding a 
typical angular resolution never smaller than $\sim$0.18~arcseconds 
\citet{2021A&A...649A...2L}. In practice very few \emph{Gaia} stars are found with 
separations $<$~0.6~arcseconds due to other limitations. Furthermore, as can 
be seen via the typical ground based imagery for these Bulge PNe in HASH (see 
Fig.~\ref{fig:classifier}), that without the benefit of high resolution \emph{HST} and VLT and even Pan-STARRS image data, these largely high-surface brightness PNe appear 
very compact. Consequently, it is very difficult to both clearly assess the PNe morphology 
or actually see the CSPN, as \citet{weidmann2020catalogue} notes. Hence, for some of 
the putative CSPN identified in \emph{Gaia} for very compact PNe (e.g. for PNG~001.7-04.4), 
they actually likely reflect the fitted position of the compact PNe itself. 
This is rather than the actual resolved position and identification of the 
true CSPN, which in some cases are below the \emph{Gaia} magnitude limits 
\citep{2022Galax..10...32P}. This is even if the reported positions are sufficiently 
accurate as they are based on the centroid estimates of a very compact PNe.

In CW21 CSPN are attributed with a reliability factor (certainty of the CSPN
identification) of between 0 and 1, and a quality label A, B or C in GSM21. 
These are not listed in Table~\ref{tab:cs_gaia}. The reliability factor depends 
on the angular separation between the match and the position of the PN relative 
to the size of the nebulae and the colour. For 61 PNe with \emph{Gaia} 
identifications consistent with the observed CSPN, 33 have a reliability 
factor $>$ 0.85 and 26 are concluded as less-reliable with a reliability 
$<$ 0.85 in CW21. When considering the reliability groups in GSM21, 11 of 15 
objects with a quality label A in GSM21 show a decent match between 
the \emph{Gaia} identification and with what we consider to be the actual CSPN.

The angular separations between our 
independently measured positions from our high resolution VLT and archival \emph{HST} 
images and the reported \emph{Gaia} CSPN coordinates in CW21 
are listed in Table~\ref{tab:cs_gaia} as parameter ${\Delta r_{{\star}}}$. The values of 
${\Delta r_{{\star}}}$ are generally less than 10\% of the major radii of the nebulae in 
HASH. This is followed, when available, by the astrometric measurements of parallaxes 
and the distance measurement in CW21. 


\subsubsection{CSPN parallax distances from \emph{Gaia}}
\label{sec:cs_dist}
The parallax distances of 33 of the \emph{Gaia} sources that match our CSPN 
observations are available. For most we consider the CSPN well determined. 

For PNG~000.7-03.7 and PNG~009.8-04.6 their distances are $0.98^{+1.27}_{-0.28}$~kpc 
and $0.32^{+0.08}_{-0.05}$~kpc respectively. If these CSPN are correctly identified 
then these two PNe are in the near foreground to the bulge. However, we believe 
these distances are anomalous. For PNG~000.7-03.7 (HASH ID 38) the identified CSPN has a reasonable "Goodness of Fit" (GoF) statistic 
value \citep[][]{lindegren2021gaia} of $\sim4$ but is 
0.25~arcsec from the PN's geometric centre which is unexpected for such 
a compact PN (angular diameter is $\sim$13~arcseconds). The PN is high 
excitation with He~II~$\lambda$~4686 $\sim$ H$\beta$. In this case neither size nor spectrum 
are compatible with the CSPNs resultant kinematic age of 800$_{-300}^{+1800}$ years. 
PNG~009.8-04.6 (HASH ID 223) has a \emph{Gaia} GoF value of 18. Values greater than 
3 are considered poor fits \citep[][]{lindegren2021gaia}. 
The putative CSPN is also not at the PN's 
geometric centre. It could even be just a dense nebula knot as it appears 
somewhat indistinct in the VLT imagery. The PN itself is well resolved 
in Pan-STARRS and has a high excitation spectrum, neither of which fits with a very
low kinematic age of only 200$\pm$10 years based on the identified CSPN.

If we adopt the canonical bulge distance of 8~kpc for these two PNe then the 
kinematic ages are 5200 and 4300 years for PNG~000.7-03.7 and PNG~009.8-04.6 
respectively. This fits with expectations and our other results.

For five further PNe, PNG~002.7-04.8, PNG~002.2-09.4, 
PNG~000.4-02.9, PNG~357.1+03.6 and PNG~359.7-01.8, the \emph{Gaia} 
sources are at $4.31^{+1.1}_{-1.03}$, $4.57^{+1.24}_{-1.03}$, $4.63^{+1.5}_{-1.38}$, 
$5.67^{+1.17}_{-0.99}$ and $5.16^{+1.46}_{-1.25}$~kpc, these distances are 
3.4$\sigma$, 2.8$\sigma$, 2.2$\sigma$ and 1.9$\sigma$ smaller than 8 kpc respectively, 
so these five sources could possibly be PNe in the middle foreground to the bulge. 
Finally PNG~358.5+02.9 and PNG~002.8+01.7 have parallax distances of $10.8^{+5.02}_{-7.25}$ 
and $11.85^{+3.59}_{-3.31}$~kpc respectively, but still in the bulge to within the large 
errors. All other distances for the sample are compatible with being in the bulge.



We provide new, accurate CSPN positions for many compact bulge CSPN in this work.
Our new results show that $\sim$14.9\% (11 out of 74 with a further two doubtful) 
CSPN identifications based on \emph{Gaia} EDR3 data presented in CW21 (and GSM21) 
do not identify the correct CSPN nor reflect the actual CSPN positions properly. 
As reported by \citet{2022Galax..10...32P} this is a broader issue 
where for many PNe the true CSPN can be extremely faint and may fall 
below the Gaia limits. This may lead to the bluest star closest to 
the PNe geometric centroid being selected. As we are working in
the dense bulge region there is usually such an alternative star choice available.
Our preliminary, independent assessment of Gaia CSPN lists from GSM21 and CW21 
for 430 checked HASH CSPN across the general Galaxy (Parker et al., in preparation), 
show 28\% and 12\% mis-identifications respectively. Correct astronomical 
and photometric measurements of true CSPN is vital to help understand the 
resultant PNe and to help ascertain their actual bulge membership when \emph{Gaia} 
distances are available so care should be taken when using these compilations. 

\begin{figure}
    \centering
    \includegraphics[width=8.5cm]{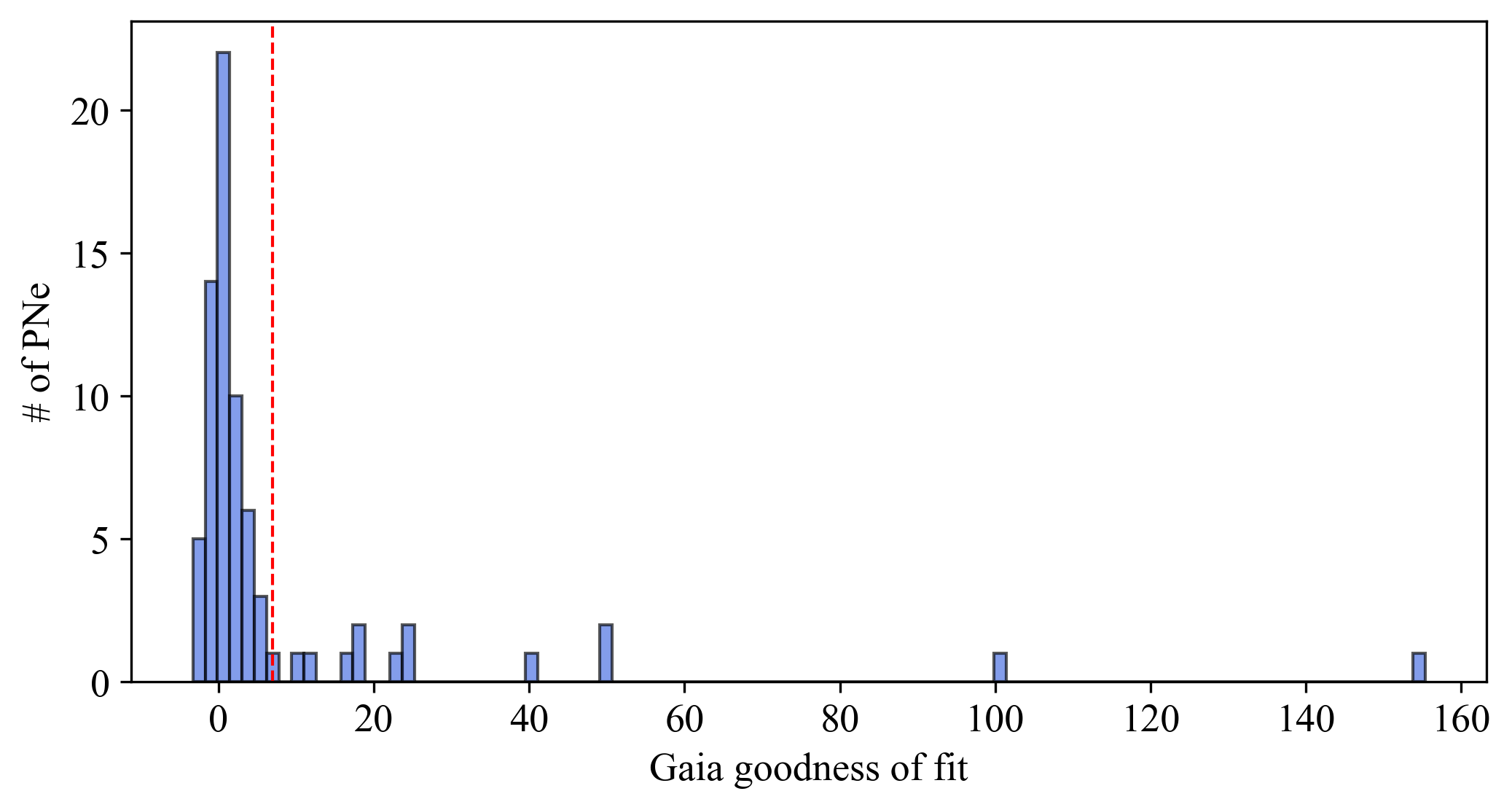}
    \caption{Distribution of available \emph{Gaia} "Goodness of Fit" (GoF) 
    parameter values for PNe in our sample. The red dotted vertical line is 
    at GoF~=~7, a value which includes the apparent Gaussian distribution of 
    all but 14 of the 74 PNe with available GoF measures.}
    \label{fig:gaia_gof}
\end{figure}

\subsection{Kinematic Age Calculations}

In Table~\ref{tab:kinematic-age} we present results for kinematic ages for all 
PNe in the sample using the best available data. This includes
\emph{Gaia} parallax distances and literature expansion velocity estimates.
For 24 PNe kinematic ages are derived from the updated angular size 
measurements in arcseconds (from the HST, VLT and 
Pan-STARRS imagery) converted to a physical size in parsecs from the distances 
from \emph{Gaia} data (34 cases) and literature expansion velocities in 
km s$^{-1}$ (77 cases) where both are available. These are assigned quality class A. 
When only one of either the \emph{Gaia} distance or literature expansion velocity 
are available these are assigned quality class B where a canonical bulge distance of 8~kpc is 
assumed if the distance is missing or an average expansion velocity of 20~km s$^{-1}$ $\sigma$~=~8 is adopted if no literature value is available. This accounts 
for 64 PNe in Table~\ref{tab:kinematic-age} part (b). Quality class C is where the canonical bulge distance  and average expansion velocity (as calculated from all the literature expansion velocity values available in class A and B) is adopted and 
this covers 48 PNe. According to \citet{gesicki2014accelerated}, a correction 
factor of 1.4 was applied to the mass-averaged expansion velocities in the 
literature to account for the mean propagation speed. The 
errors on the kinematic ages were calculated using Monte Carlo simulations with 
10,000 realizations. Lower and upper uncertainties take the 16 and 84 percentile 
values of the simulation estimates. The average kinematic age obtained here is 
6400~$\pm~$3800~years. The results for HASH PNe 38 and 223 are excluded from 
the calculations as there are serious doubts about their very low \emph{Gaia} 
distances (see earlier discussion). This work represents the most accurate 
estimations ever provided for the kinematic ages of PNe in the Galactic bulge.

\begin{figure}
    \centering
    \includegraphics[width=8.5cm]{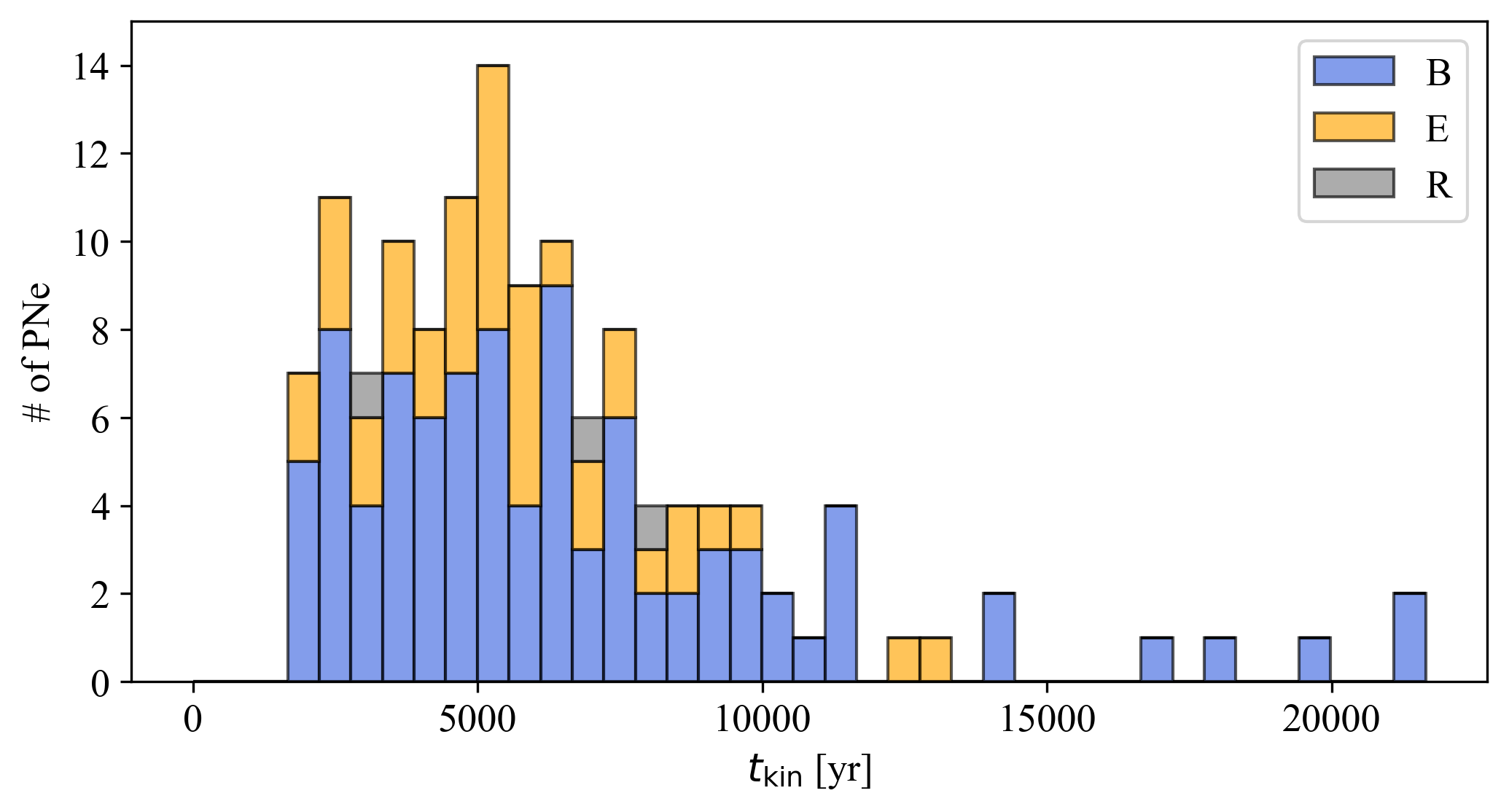}
    \caption{Distribution of kinematic ages determined for this bulge PN sample. 
    Histograms are color-coded according to the main morphological classes.}
    \label{fig:tk_dist}
\end{figure}

The histogram of kinematic ages as a function of morphological type is shown in 
Fig.~\ref{fig:tk_dist}. The histogram bin size is about the same as 
the typical calculated error on the individual kinematic ages. The 7 ostensibly 
oldest PNe are all bipolars and they completely dominate after 10000~years 
with 14/16 cases. Over the range 1500 to 10000~years the fraction of ellipticals 
to bipolars remains approximately constant.

\setlength{\tabcolsep}{4pt}
\renewcommand*{\arraystretch}{1.13}
\begin{table*}
\centering
\caption{Kinematic age estimates for all PNe in the sample. For 23 PNe these are 
derived from the updated angular size measurements in arcseconds converted to a 
physical size in parsecs from the distances from \emph{Gaia} data and literature 
expansion velocities in km s$^{-1}$. These are assigned quality class (Q) of "A". 
When only one of either \emph{Gaia} distance or literature expansion velocity 
are available these are Q of "B" (65 PNe). A Q of "C" is where a canonical bulge 
distance of 8~kpc is assumed in lieu if a \emph{Gaia} distance and an average 
expansion velocity of 20~km s$^{-1}$ assumed as calculated from all the literature 
values available in Q of A and B (48 PNe).}
\label{tab:kinematic-age}
\begin{subtable}{\textwidth}
\centering
\caption*{(a)}
\begin{tabular}{l@{\hspace{0.6\tabcolsep}}cccccp{2.45cm}l@{\hspace{0.6\tabcolsep}}cccccl}
\hline
\multirow{2}{*}{\begin{tabular}[c]{@{}l@{}}HASH \\ID\end{tabular}} & \multirow{2}{*}{PNG} & \multirow{2}{*}{Q} & \multirow{2}{*}{\begin{tabular}[c]{@{}c@{}}$D$\\(kpc)\end{tabular}} & \multirow{2}{*}{\begin{tabular}[c]{@{}c@{}}$R_{\mathrm{out}}$\\ (pc)\end{tabular}} & \multirow{2}{*}{\begin{tabular}[c]{@{}c@{}}$V_{\mathrm{exp}}$\\ (km s$^{-1}$)\end{tabular}} & \multirow{2}{*}{\begin{tabular}[c]{@{}c@{}}$t_{\mathrm{kin}}$\\ (yr)\end{tabular}} & \multirow{2}{*}{\begin{tabular}[c]{@{}l@{}}HASH \\ ID\end{tabular}} & \multirow{2}{*}{PNG} & \multirow{2}{*}{Q} & \multirow{2}{*}{\begin{tabular}[c]{@{}l@{}}$D$\\ (kpc)\end{tabular}} & \multirow{2}{*}{\begin{tabular}[c]{@{}l@{}}$R_{\mathrm{out}}$\\(pc)\end{tabular}} & \multirow{2}{*}{\begin{tabular}[c]{@{}l@{}}$V_{\mathrm{exp}}$\\ (km s$^{-1}$)\end{tabular}} & \multirow{2}{*}{\begin{tabular}[c]{@{}l@{}}$t_{\mathrm{kin}}$\\ (yr)\end{tabular}} \\ &    &    &      &          &       &          &          & &   &         &             &      &          \\\hline
 19 &  000.2-01.9 &     A &   6.49$_{-1.16}^{+1.27}$ &      0.31 &  10$^{[2]}$ &  21600$_{-4400}^{+4700}$ &   217 &  009.4-09.8 &     A &   8.10$_{-1.79}^{+1.91}$ &      0.16 &  23$^{[3]}$ &  4800$_{-1100}^{+1300}$ \\
   38 &  000.7-03.7 &     A &   0.98$_{-0.28}^{+2.18}$ &      0.03 &  26$^{[2]}$ &     800$_{-200}^{+1600}$ &   223 &  009.8-04.6 &     A &  0.32$_{-0.05}^{+0.08}$ &      0.01 &  30$^{[3]}$ &                 200$\pm$000 \\
   90 &  002.2-09.4 &     A &   4.57$_{-1.03}^{+1.24}$ &      0.15 &  28$^{[3]}$ &    3700$_{-800}^{+1000}$ &  1151 &  351.1+04.8 &     A &   7.08$_{-1.4}^{+1.57}$ &       0.10 &  23$^{[1]}$ &    3000$_{-600}^{+700}$ \\
  105 &  002.7-04.8 &     A &    4.31$_{-1.03}^{+1.1}$ &      0.19 &  19$^{[3]}$ &   6900$_{-1700}^{+1800}$ &  1159 &  351.9+09.0 &     A &  7.49$_{-1.66}^{+1.84}$ &      0.28 &  29$^{[5]}$ &  6700$_{-1500}^{+1600}$ \\
  106 &  002.8+01.7 &     A &  11.85$_{-3.31}^{+3.59}$ &      0.11 &  29$^{[1]}$ &     2600$_{-800}^{+700}$ &  1207 &  355.9-04.2 &     A &  6.28$_{-1.53}^{+1.81}$ &      0.13 &  15$^{[2]}$ &  6000$_{-1600}^{+1400}$ \\
  114 &  003.1+03.4 &     A &   7.88$_{-1.53}^{+1.73}$ &      0.07 &  24$^{[1]}$ &     2000$_{-400}^{+500}$ &  1235 &  357.1+03.6 &     A &  5.67$_{-0.99}^{+1.17}$ &      0.11 &  23$^{[4]}$ &    3300$_{-600}^{+700}$ \\
  125 &  003.6+03.1 &     A &   8.43$_{-2.31}^{+2.35}$ &      0.12 &  16$^{[1]}$ &                5200$\pm$1400 &  1256 &  357.6-03.3 &     A &  5.37$_{-2.92}^{+2.99}$ &      0.16 &  15$^{[2]}$ &  7400$_{-3900}^{+4300}$ \\
  139 &  004.0-03.0 &     A &   9.69$_{-1.55}^{+1.74}$ &      0.16 &  12$^{[3]}$ &   9300$_{-1600}^{+1900}$ &  1263 &  358.2+03.5 &     A &  8.98$_{-2.29}^{+2.39}$ &      0.15 &  22$^{[3]}$ &               4700$\pm$1200 \\
  141 &  004.2-04.3 &     A &    6.10$_{-2.11}^{+3.21}$ &       0.10 &  18$^{[3]}$ &   3800$_{-1400}^{+1900}$ &  1275 &  358.5+02.9 &     A &  10.8$_{-7.25}^{+5.02}$ &       0.10 &  28$^{[1]}$ &  2400$_{-1600}^{+1000}$ \\
  148 &  004.6+06.0 &     A &   6.41$_{-1.52}^{+1.65}$ &      0.12 &  14$^{[2]}$ &   5900$_{-1300}^{+1900}$ &  1276 &  358.5-04.2 &     A &   7.90$_{-1.81}^{+1.91}$ &      0.15 &  14$^{[1]}$ &  7400$_{-1600}^{+1900}$ \\
  151 &  004.8+02.0 &     A &   8.86$_{-2.64}^{+3.73}$ &      0.08 &  18$^{[1]}$ &    3100$_{-900}^{+1300}$ &  1322 &  359.7-01.8 &     A &  5.16$_{-1.25}^{+1.46}$ &       0.10 &  23$^{[3]}$ &    3000$_{-800}^{+900}$ \\
  150 &  004.8-05.0 &     A &   7.55$_{-1.65}^{+1.79}$ &      0.21 &  32$^{[3]}$ &                4500$\pm$1000 &       &           &     &                              &         &                         &                              \\ \hline
\end{tabular}
\end{subtable}
\newline
\vspace*{0.45cm}
\newline
\begin{subtable}{\textwidth}
\centering
\caption*{(b)}
\begin{tabular}{l@{\hspace{0.6\tabcolsep}}cccccp{2.45cm}l@{\hspace{0.6\tabcolsep}}cccccl}
\hline
\multirow{2}{*}{\begin{tabular}[c]{@{}l@{}}HASH \\ ID\end{tabular}} & \multirow{2}{*}{PNG} & \multirow{2}{*}{\begin{tabular}[c]{@{}c@{}}Q\\ \end{tabular}} & \multirow{2}{*}{\begin{tabular}[c]{@{}c@{}}$D$\\ (kpc)\end{tabular}} & \multirow{2}{*}{\begin{tabular}[c]{@{}c@{}}$R_{\mathrm{out}}$\\ (pc)\end{tabular}} & \multirow{2}{*}{\begin{tabular}[c]{@{}c@{}}$V_{\mathrm{exp}}$\\ (km s$^{-1}$)\end{tabular}} & \multirow{2}{*}{\begin{tabular}[c]{@{}c@{}}$t_{\mathrm{kin}}$\\ (yr)\end{tabular}} & \multirow{2}{*}{\begin{tabular}[c]{@{}l@{}}HASH \\ ID\end{tabular}} & \multirow{2}{*}{PNG} & \multirow{2}{*}{\begin{tabular}[c]{@{}l@{}}Q\\ \end{tabular}} & \multirow{2}{*}{\begin{tabular}[c]{@{}l@{}}$D$\\ (kpc)\end{tabular}} & \multirow{2}{*}{\begin{tabular}[c]{@{}l@{}}$R_{\mathrm{out}}$\\(pc)\end{tabular}} & \multirow{2}{*}{\begin{tabular}[c]{@{}l@{}}$V_{\mathrm{exp}}$\\ (km s$^{-1}$)\end{tabular}} & \multirow{2}{*}{\begin{tabular}[c]{@{}l@{}}$t_{\mathrm{kin}}$\\ (yr)\end{tabular}} \\
         &    &    &         &             &               &          &          &&               &         &             &      &          \\\hline    16 &  000.1+04.3 &     B &                 8.0$\pm$1.0  &      0.23 &  20$^{[3]}$ &   8000$_{-1000}^{+1200}$ &   210 &  008.6-02.6 &     B &                 8.0$\pm$1.0  &      0.08 &  28$^{[1]}$ &     1900$_{-200}^{+300}$ \\
   26 &  000.4-01.9 &     B &                 8.0$\pm$1.0  &      0.16 &  24$^{[3]}$ &     4600$_{-600}^{+800}$ &  1160 &  351.9-01.9 &     B &                 8.0$\pm$1.0  &      0.09 &  13$^{[1]}$ &     4800$_{-600}^{+900}$ \\
   27 &  000.4-02.9 &     B &   4.63$_{-1.38}^{+1.5}$ &       0.10 &    20$^{\mathrm{a}}$ &   3400$_{-1300}^{+2500}$ &  1161 &  352.0-04.6 &     B &                 8.0$\pm$1.0  &      0.14 &  17$^{[3]}$ &    5700$_{-800}^{+1100}$ \\
   40 &  000.7+03.2 &     B &                 8.0$\pm$1.0  &      0.35 &  23$^{[3]}$ &  10600$_{-1300}^{+1400}$ &  1163 &  352.1+05.1 &     B &                 8.0$\pm$1.0  &      0.35 &  17$^{[3]}$ &               14300$\pm$2000 \\
   37 &  000.7-02.7 &     B &                 8.0$\pm$1.0  &      0.16 &  27$^{[3]}$ &                 4100$\pm$600 &  1164 &  352.6+03.0 &     B &                 8.0$\pm$1.0  &       0.10 &  23$^{[1]}$ &                 3000$\pm$400 \\
   50 &  000.9-02.0 &     B &                 8.0$\pm$1.0  &       0.10 &  16$^{[3]}$ &                 4300$\pm$800 &  1172 &  353.3+06.3 &     B &                 8.0$\pm$1.0  &      0.16 &  22$^{[5]}$ &     5000$_{-700}^{+800}$ \\
   48 &  000.9-04.8 &     B &                 8.0$\pm$1.0  &       0.30 &  30$^{[2]}$ &                 6900$\pm$900 &  1178 &  353.7+06.3 &     B &                 8.0$\pm$1.0  &      0.19 &  14$^{[2]}$ &   9400$_{-1300}^{+1600}$ \\
   58 &  001.2+02.1 &     B &  5.92$_{-1.71}^{+1.78}$ &      0.08 &    20$^{\mathrm{a}}$ &   2700$_{-1000}^{+2000}$ &  1185 &  354.5+03.3 &     B &                 8.0$\pm$1.0  &      0.05 &  19$^{[1]}$ &     1800$_{-200}^{+300}$ \\
   60 &  001.2-03.0 &     B &  8.92$_{-1.82}^{+2.11}$ &      0.14 &    20$^{\mathrm{a}}$ &   4800$_{-1700}^{+3400}$ &  1189 &  354.9+03.5 &     B &                 8.0$\pm$1.0  &      0.07 &  16$^{[1]}$ &                 3000$\pm$400 \\
   62 &  001.3-01.2 &     B &  4.84$_{-2.84}^{+3.77}$ &      0.07 &    20$^{\mathrm{a}}$ &   2400$_{-1600}^{+2900}$ &  1191 &  355.1-06.9 &     B &                 8.0$\pm$1.0  &      0.19 &  15$^{[3]}$ &                8800$\pm$1300 \\
   73 &  001.7+05.7 &     B &                 8.0$\pm$1.0  &      0.24 &  39$^{[3]}$ &     4200$_{-500}^{+600}$ &  1199 &  355.4-02.4 &     B &                 8.0$\pm$1.0  &      0.21 &  23$^{[3]}$ &     6300$_{-800}^{+600}$ \\
   74 &  001.7-04.4 &     B &                 8.0$\pm$1.0  &      0.09 &  26$^{[1]}$ &     2400$_{-300}^{+200}$ &  1202 &  355.6-02.7 &     B &                 8.0$\pm$1.0  &      0.15 &  11$^{[3]}$ &   9500$_{-1600}^{+1800}$ \\
   81 &  002.0-06.2 &     B &                 8.0$\pm$1.0  &      0.13 &   7$^{[3]}$ &  12900$_{-2800}^{+2500}$ &  1209 &  355.9+03.6 &     B &                 8.0$\pm$1.0  &      0.07 &  21$^{[1]}$ &                 2300$\pm$300 \\
   83 &  002.1-02.2 &     B &                 8.0$\pm$1.0  &       0.10 &  22$^{[3]}$ &     3100$_{-600}^{+400}$ &  1212 &  356.1-03.3 &     B &                 8.0$\pm$1.0  &       0.10 &  20$^{[1]}$ &     3400$_{-400}^{+500}$ \\
   82 &  002.1-04.2 &     B &                 8.0$\pm$1.0  &      0.16 &  15$^{[3]}$ &    7400$_{-900}^{+1000}$ &  1219 &  356.5-03.6 &     B &                 8.0$\pm$1.0  &      0.22 &  25$^{[1]}$ &                 6100$\pm$800 \\
   91 &  002.3+02.2 &     B &                 8.0$\pm$1.0  &      0.13 &  17$^{[3]}$ &     5300$_{-800}^{+900}$ &  1230 &  356.8+03.3 &     B &                 8.0$\pm$1.0  &      0.07 &   9$^{[1]}$ &     5400$_{-700}^{+800}$ \\
   93 &  002.3-03.4 &     B &                 8.0$\pm$1.0  &      0.11 &  29$^{[1]}$ &                 2600$\pm$300 &  1232 &  356.9+04.4 &     B &                 8.0$\pm$1.0  &      0.11 &   9$^{[1]}$ &   8500$_{-1100}^{+1700}$ \\
 4326 &  002.6+02.1 &     B &                 8.0$\pm$1.0  &      0.36 &  41$^{[3]}$ &     6100$_{-700}^{+800}$ &  1239 &  357.1-04.7 &     B &                 8.0$\pm$1.0  &      0.07 &  27$^{[1]}$ &                 1800$\pm$200 \\
  109 &  002.9-03.9 &     B &                 8.0$\pm$1.0  &      0.13 &  39$^{[1]}$ &     2300$_{-200}^{+300}$ &  1242 &  357.2+02.0 &     B &                 8.0$\pm$1.0  &      0.09 &   8$^{[1]}$ &    7800$_{-800}^{+1700}$ \\
  116 &  003.2-06.2 &     B &                 8.0$\pm$1.0  &      0.21 &  22$^{[5]}$ &    6600$_{-900}^{+1100}$ &  1246 &  357.3+04.0 &     B &  7.49$_{-2.13}^{+2.25}$ &      0.15 &    20$^{\mathrm{a}}$ &   5200$_{-2000}^{+3900}$ \\
  126 &  003.6-02.3 &     B &                 8.0$\pm$1.0  &       0.20 &  28$^{[3]}$ &     4900$_{-600}^{+700}$ &  1253 &  357.5+03.2 &     B &                 8.0$\pm$1.0  &       0.20 &  27$^{[3]}$ &                 5100$\pm$700 \\
  129 &  003.7-04.6 &     B &                 8.0$\pm$1.0  &      0.13 &  24$^{[3]}$ &                 3700$\pm$500 &  1259 &  357.9-03.8 &     B &  8.31$_{-1.66}^{+1.75}$ &      0.26 &    20$^{\mathrm{a}}$ &   9000$_{-3000}^{+6400}$ \\
  131 &  003.8-04.3 &     B &                 8.0$\pm$1.0  &       0.10 &  29$^{[3]}$ &                 2400$\pm$300 &  1281 &  358.6-05.5 &     B &  5.33$_{-1.79}^{+1.95}$ &      0.22 &    20$^{\mathrm{a}}$ &   7600$_{-3100}^{+5900}$ \\
  133 &  003.9-02.3 &     B &                 8.0$\pm$1.0  &      0.15 &  20$^{[3]}$ &     5200$_{-800}^{+600}$ &  1286 &  358.7+05.2 &     B &                 8.0$\pm$1.0  &      0.07 &  22$^{[1]}$ &                 2200$\pm$300 \\
  140 &  004.1-03.8 &     B &                 8.0$\pm$1.0  &      0.06 &  22$^{[1]}$ &     1900$_{-200}^{+300}$ &  1293 &  358.8+03.0 &     B &                 8.0$\pm$1.0  &      0.17 &  20$^{[2]}$ &    5900$_{-700}^{+1000}$ \\
 4315 &  004.3+01.8 &     B &  5.73$_{-1.18}^{+1.28}$ &      0.12 &    20$^{\mathrm{a}}$ &   4100$_{-1400}^{+2700}$ &  1295 &  358.9+03.4 &     B &                 8.0$\pm$1.0  &      0.06 &  17$^{[1]}$ &                 2400$\pm$400 \\
  165 &  005.5-04.0 &     B &                 8.0$\pm$1.0  &      0.24 &  32$^{[2]}$ &     5200$_{-600}^{+800}$ &  1308 &  359.2+04.7 &     B &                 8.0$\pm$1.0  &      0.05 &  15$^{[1]}$ &                 2300$\pm$400 \\
 4108 &  006.1+08.3 &     B &                 8.0$\pm$1.0  &      0.05 &  16$^{[1]}$ &     2100$_{-200}^{+300}$ &  1312 &  359.3-01.8 &     B &   6.92$_{-1.3}^{+1.51}$ &      0.12 &    20$^{\mathrm{a}}$ &   4100$_{-1300}^{+2900}$ \\
  180 &  006.3+04.4 &     B &                 8.0$\pm$1.0  &      0.11 &  46$^{[1]}$ &     1600$_{-200}^{+100}$ &  1324 &  359.8+03.7 &     B &  8.09$_{-2.42}^{+2.51}$ &      0.14 &    20$^{\mathrm{a}}$ &   4800$_{-1900}^{+3600}$ \\
  182 &  006.4+02 &     B &                 8.0$\pm$1.0  &      0.14 &  19$^{[1]}$ &     5100$_{-600}^{+700}$ &  1329 &  359.8+05.6 &     B &                 8.0$\pm$1.0  &      0.14 &   5$^{[2]}$ &  19500$_{-4000}^{+5200}$ \\
  181 &  006.4-04.6 &     B &                 8.0$\pm$1.0  &      0.18 &  26$^{[2]}$ &     4800$_{-700}^{+600}$ &  1330 &  359.8-07.2 &     B &                 8.0$\pm$1.0  &      0.22 &  29$^{[5]}$ &                 5200$\pm$600 \\
  201 &  007.8-04.4 &     B &  9.93$_{-1.32}^{+1.46}$ &      0.17 &    20$^{\mathrm{a}}$ &   5900$_{-1800}^{+4100}$ &  1334 &  359.9-04.5 &     B &                 8.0$\pm$1.0  &      0.17 &  17$^{[3]}$ &    6900$_{-800}^{+1300}$ \\
 4111 &  008.2+06.8 &     B &                 8.0$\pm$1.0  &      0.09 &  12$^{[1]}$ &    5200$_{-700}^{+1000}$ &       &           &     &                              &         &                         &                               \\
 \hline
\end{tabular}
\end{subtable}
\end{table*}
\begin{table*}
\centering
\contcaption{}
\begin{subtable}{\textwidth}
\centering
\caption*{(c)}
\vspace*{2mm}
\begin{tabular}{l@{\hspace{0.6\tabcolsep}}cccccp{2.45cm}l@{\hspace{0.6\tabcolsep}}cccccl}
\hline
\multirow{2}{*}{\begin{tabular}[c]{@{}l@{}}HASH \\ ID\end{tabular}} & \multirow{2}{*}{PNG} & \multirow{2}{*}{\begin{tabular}[c]{@{}c@{}}Q \\ \end{tabular}} & \multirow{2}{*}{\begin{tabular}[c]{@{}c@{}}$D$\\ (kpc)\end{tabular}} & \multirow{2}{*}{\begin{tabular}[c]{@{}c@{}}$R_{\mathrm{out}}$\\ (pc)\end{tabular}} & \multirow{2}{*}{\begin{tabular}[c]{@{}c@{}}$V_{\mathrm{exp}}$\\ (km s$^{-1}$)\end{tabular}} & \multirow{2}{*}{\begin{tabular}[c]{@{}c@{}}$t_{\mathrm{kin}}$\\ (yr)\end{tabular}} & \multirow{2}{*}{\begin{tabular}[c]{@{}l@{}}HASH \\ ID\end{tabular}} & \multirow{2}{*}{PNG} & \multirow{2}{*}{\begin{tabular}[c]{@{}c@{}}Q \\ \end{tabular}} & \multirow{2}{*}{\begin{tabular}[c]{@{}l@{}}$D$\\ (kpc)\end{tabular}} & \multirow{2}{*}{\begin{tabular}[c]{@{}l@{}}$R_{\mathrm{out}}$\\(pc)\end{tabular}} & \multirow{2}{*}{\begin{tabular}[c]{@{}l@{}}$V_{exp}$\\ (km s$^{-1}$)\end{tabular}} & \multirow{2}{*}{\begin{tabular}[c]{@{}l@{}}$t_{\mathrm{kin}}$\\ (yr)\end{tabular}} \\
         &    &    &         &             &               &          &          &&               &         &             &      &          \\\hline   12 &  000.1+02.6 &     C &  8.0$\pm$1.0 &       0.24 &  20$^{\mathrm{a}}$ &  8300$_{-2600}^{+5500}$ &      192 &  007.0-06.8 &     C &  8.0$\pm$1.0 &       0.24 &  20$^{\mathrm{a}}$ &    8300$_{-2500}^{+5500}$ \\
      17 &  000.1-02.3 &     C &  8.0$\pm$1.0 &       0.20 &  20$^{\mathrm{a}}$ &  6900$_{-2200}^{+4300}$ &      197 &  007.5+07.4 &     C &  8.0$\pm$1.0 &       0.35 &  20$^{\mathrm{a}}$ &   12200$_{-3700}^{+8000}$ \\
      20 &  000.2-04.6 &     C &  8.0$\pm$1.0 &       0.15 &  20$^{\mathrm{a}}$ &  5200$_{-1600}^{+3500}$ &      198 &  007.6+06.9 &     C &  8.0$\pm$1.0 &       0.32 &  20$^{\mathrm{a}}$ &   11100$_{-3400}^{+7500}$ \\
      24 &  000.3+06.9 &     C &  8.0$\pm$1.0 &       0.21 &  20$^{\mathrm{a}}$ &  7300$_{-2200}^{+5000}$ &      200 &  007.8-03.7 &     C &  8.0$\pm$1.0 &       0.33 &  20$^{\mathrm{a}}$ &   11500$_{-3400}^{+7600}$ \\
      22 &  000.3-04.6 &     C &  8.0$\pm$1.0 &       0.23 &  20$^{\mathrm{a}}$ &  8000$_{-2500}^{+5400}$ &      208 &  008.4-03.6 &     C &  8.0$\pm$1.0 &       0.49 &  20$^{\mathrm{a}}$ &  17100$_{-5200}^{+10900}$ \\
      42 &  000.7-07.4 &     C &  8.0$\pm$1.0 &       0.23 &  20$^{\mathrm{a}}$ &  8000$_{-2400}^{+5100}$ &     1147 &  350.5-05.0 &     C &  8.0$\pm$1.0 &       0.18 &  20$^{\mathrm{a}}$ &    6200$_{-1900}^{+3900}$ \\
      54 &  001.1-01.6 &     C &  8.0$\pm$1.0 &       0.15 &  20$^{\mathrm{a}}$ &  5200$_{-1600}^{+3400}$ &     1152 &  351.2+05.2 &     C &  8.0$\pm$1.0 &       0.28 &  20$^{\mathrm{a}}$ &    9700$_{-3000}^{+6500}$ \\
      64 &  001.4+05.3 &     C &  8.0$\pm$1.0 &       0.17 &  20$^{\mathrm{a}}$ &  5900$_{-1800}^{+3700}$ &     1156 &  351.6-06.2 &     C &  8.0$\pm$1.0 &       0.29 &  20$^{\mathrm{a}}$ &   10100$_{-3100}^{+6700}$ \\
      70 &  001.6-01.3 &     C &  8.0$\pm$1.0 &       0.11 &  20$^{\mathrm{a}}$ &  3800$_{-1100}^{+2600}$ &     1171 &  353.2-05.2 &     C &  8.0$\pm$1.0 &       0.33 &  20$^{\mathrm{a}}$ &   11500$_{-3400}^{+7500}$ \\
      98 &  002.5-01.7 &     C &  8.0$\pm$1.0 &       0.11 &  20$^{\mathrm{a}}$ &  3800$_{-1100}^{+2600}$ &     1217 &  356.3-06.2 &     C &  8.0$\pm$1.0 &       0.22 &  20$^{\mathrm{a}}$ &    7600$_{-2300}^{+5100}$ \\
     107 &  002.8+01.8 &     C &  8.0$\pm$1.0 &       0.18 &  20$^{\mathrm{a}}$ &  6200$_{-1900}^{+4200}$ &     1227 &  356.8-05.4 &     C &  8.0$\pm$1.0 &       0.28 &  20$^{\mathrm{a}}$ &    9700$_{-2900}^{+6400}$ \\
     128 &  003.7+07.9 &     C &  8.0$\pm$1.0 &       0.22 &  20$^{\mathrm{a}}$ &  7600$_{-2200}^{+4900}$ &     1234 &  357.0+02.4 &     C &  8.0$\pm$1.0 &       0.17 &  20$^{\mathrm{a}}$ &    5900$_{-1700}^{+4000}$ \\
     135 &  003.9+01.6 &     C &  8.0$\pm$1.0 &       0.17 &  20$^{\mathrm{a}}$ &  5900$_{-1700}^{+3900}$ &     4139 &  357.1+04.4 &     C &  8.0$\pm$1.0 &       0.21 &  20$^{\mathrm{a}}$ &    7300$_{-2200}^{+4700}$ \\
     134 &  003.9-03.1 &     C &  8.0$\pm$1.0 &       0.15 &  20$^{\mathrm{a}}$ &  5200$_{-1600}^{+3600}$ &     1252 &  357.5+03.1 &     C &  8.0$\pm$1.0 &       0.12 &  20$^{\mathrm{a}}$ &    4100$_{-1200}^{+2900}$ \\
     142 &  004.2-03.2 &     C &  8.0$\pm$1.0 &       0.12 &  20$^{\mathrm{a}}$ &  4100$_{-1300}^{+2700}$ &     4140 &  357.9-05.1 &     C &  8.0$\pm$1.0 &       0.51 &  20$^{\mathrm{a}}$ &  17800$_{-5200}^{+12000}$ \\
     143 &  004.2-05.9 &     C &  8.0$\pm$1.0 &       0.18 &  20$^{\mathrm{a}}$ &  6200$_{-1800}^{+4000}$ &     1258 &  358.0+09.3 &     C &  8.0$\pm$1.0 &       0.29 &  20$^{\mathrm{a}}$ &   10100$_{-2900}^{+6400}$ \\
     156 &  005.0-03.9 &     C &  8.0$\pm$1.0 &       0.25 &  20$^{\mathrm{a}}$ &  8700$_{-2600}^{+5800}$ &     1266 &  358.2+04.2 &     C &  8.0$\pm$1.0 &       0.27 &  20$^{\mathrm{a}}$ &    9400$_{-2900}^{+6200}$ \\
     162 &  005.2+05.6 &     C &  8.0$\pm$1.0 &       0.17 &  20$^{\mathrm{a}}$ &  5900$_{-1900}^{+4100}$ &     1280 &  358.6+07.8 &     C &  8.0$\pm$1.0 &       0.26 &  20$^{\mathrm{a}}$ &    9000$_{-2800}^{+6100}$ \\
     171 &  005.8-06.1 &     C &  8.0$\pm$1.0 &       0.18 &  20$^{\mathrm{a}}$ &  6200$_{-1800}^{+4400}$ &     1302 &  359.1-02.9 &     C &  8.0$\pm$1.0 &       0.19 &  20$^{\mathrm{a}}$ &    6600$_{-2000}^{+4600}$ \\
     179 &  006.3+03.3 &     C &  8.0$\pm$1.0 &       0.18 &  20$^{\mathrm{a}}$ &  6200$_{-1900}^{+4000}$ &     1319 &  359.6-04.8 &     C &  8.0$\pm$1.0 &       0.32 &  20$^{\mathrm{a}}$ &   11100$_{-3500}^{+7600}$ \\
     189 &  006.8+02.3 &     C &  8.0$\pm$1.0 &       0.10 &  20$^{\mathrm{a}}$ &  3400$_{-1100}^{+2300}$ &     1327 &  359.8+02.4 &     C &  8.0$\pm$1.0 &       0.10 &  20$^{\mathrm{a}}$ &    3400$_{-1000}^{+2300}$ \\
     188 &  006.8-03.4 &     C &  8.0$\pm$1.0 &       0.14 &  20$^{\mathrm{a}}$ &  4800$_{-1300}^{+3100}$ &     1328 &  359.8+05.2 &     C &  8.0$\pm$1.0 &       0.41 &  20$^{\mathrm{a}}$ &   14300$_{-4300}^{+9500}$ \\
     193 &  007.0+06.3 &     C &  8.0$\pm$1.0 &       0.14 &  20$^{\mathrm{a}}$ &  4800$_{-1500}^{+3300}$ &     1326 &  359.8+06.9 &     C &  8.0$\pm$1.0 &       0.62 &  20$^{\mathrm{a}}$ &  21600$_{-6300}^{+14000}$ \\
\hline \vspace{-0.28cm}\\
\multicolumn{14}{l}{References: \small $^{[1]}$ \citet{gesicki2014accelerated}; \small $^{[2]}$ \citet{richer2010evolution}; \small $^{[3]}$ \citet{richer2008acceleration}; \small $^{[4]}$\citet{gesicki2006planetary}; \small $^{[5]}$ \citet{gesicki2000expansion}.}\\
\multicolumn{14}{l}{\small $^{\mathrm{a}}$ Mean expansion velocity: 20 $\pm$ 8 km s$^{-1}$ used.}\\
\end{tabular}
\end{subtable}
\end{table*}

\section{Discussion}
\label{dis}
\subsection{PN Morphologies and the Shaping mechanism of PNe}
\label{mechanism}

\noindent Round (spherical) PNe occupy a small fraction ($\sim$10\%) in our 
bulge sample while the overall fraction in HASH is 21\%, with many being at 
larger Galactic scale heights. With higher-resolution imaging more aspherical 
features of PNe previously classed as `R' are 
found, by either revealing an elliptical outer envelope or even faint bipolar lopes. 
In one case, PNG~357.2+02.0, the round annular ring likely represents a bipolar 
PN seen pole on (refer earlier discussion in Section~\ref{sec:more_bipolar}). Such "adjustments" are generally for 
compact PNe where significantly improved imaging resolution and sensitivity 
make the difference. 


The generalised interacting stellar wind (GISW) model \citep{balick1987evolution}, 
has been the cornerstone paradigm
in explaining the observed aspherical features of elliptical and bipolar morphologies 
of PNe. Here the slow wind is enhanced in an equatorial plane expulsion of material from 
the progenitor post-AGB star. Under such a generalised scenario, the 
fast wind from the CSPN would be preferentially around an axis orthogonal to the 
equatorial density enhancement. A follow-up theoretical study 
\citet{kahn1985shapes} conjectured that by varying the pole-to-equator density contrast 
in the AGB slow wind, a nebula could have a full-range of morphologies from mildly 
elliptical to strongly bipolar. This was substantiated by analytical models in 
\citet{icke1988blowing}, \citet{balick1987evolution} and \citet{icke1989evolution} and 
later numerical simulations in \citet{soker1989interacting}, 
\citet{mellema1991hydrodynamical} and \citet{icke1992hydrodynamics}.

There are also strong morphological clues as to whether a bipolar PN is likely to 
host a binary, e.g. \citet{2009A&A...505..249M}. However, whether the CSPN is in a wide or close binary system cannot generally be inferred just from the PN shape \citep[e.g.][]{soker2001collimated, boffin2019importance}. The formation of bipolar lobes could be due to aspherical winds or jets and 
may fade on a timescale an order of magnitude smaller than the typical visibility 
time of a PN \citep[][]{jacob2013evolution, gesicki20163d} but see \citet{parker2022planetary}. 

According to best current understanding, observed PNe bipolar morphologies are largely due 
to central star binarity \citep[e.g.][]{soker1994disks, boffin2012interacting, 
de2015identifying, garcia2018common, akashi2021shaping, chamandy2018accretion}, while 
point-symmetric structures are explained by the precession of the accretion disk around the 
secondary \citep[][]{livio1996formation, boffin2012interacting}.

Therefore, for cases of PNe where the shape of a bipolar waist is seen while lobes are 
faint or unseen, they should fit the bipolar class. Further, whether the lobes or thin 
rings formed in common envelope evolution (CEE) appear or not, depends 
largely on the kinematic age of the PN, the viewing angle and whether the internal 
structures are resolved. Our classification for the bipolar PNe should be more 
representative to this population compared with \citet{rees2013alignment}. 

\subsection{Implications for the large fraction of bipolar PNe found in the Galactic bulge}

The large fraction of bipolar PNe in our sample highlights not only 
the likely importance of binary interaction in PN shaping but raises 
issues of the bulge stellar population. 
It is also interesting to confirm that the point-symmetric features, 
including occurrences of multi-lobes (PNe with sub-class m and p), spiral 
structures or over-densities in the outer lobes are only seen in bipolar PNe and 
these have been strongly argued to be specific morphological traits of PNe with 
close binary cores \citep{2009A&A...505..249M}. 

However, the question of why an overall ostensibly very old Galactic bulge stellar 
population should have a PN population highly dominated by bipolars, i.e. 68\% 
of all compact bulge PNe known, is important. As also known there is compelling 
evidence that bipolars, from Galactic disk scale height arguments alone, may also 
at least partially emerge from higher-mass and 
therefore younger progenitors, e.g. \citet{corradi1995morphological, 2006MNRAS.373...79P} 
which may or may not host binaries. Another key point is the previous best estimates for the close binary fraction for PNe in the Galactic bulge is 
only 15-20\%, where up to 60\% have bipolar morphologies if the inclination effects are considered \citep{miszalski2009binary}.  This compares to the lower limit bipolar fraction of 68\% for all compact bulge PNe and 
indeed a high bipolar fraction of all bulge PNe. So do all bipolars host both more 
massive progenitors and binaries while close binaries in particular 
have additional morphological features that give them away? If so why are so 
many bulge PNe bipolars? If they are of higher surface brightness and so are easier 
to detect this does not seem to be a feature of the general Galactic disk PNe 
population where the apparent bipolar fraction is $\sim$20\% from HASH.
We have shown that many compact elliptical PNe are actually bipolars in higher resolution
imaging so the current HASH bipolar fraction is an underestimate but even so this cannot 
easily explain what we see in the Galactic bulge. 

Our results for PN morphologies, therefore, suggest an underestimation of the 
fraction of younger and more massive stars and/or binary systems in the Galactic bulge. 
With a bin size of 1000 years, the number of observed bipolar and elliptical PNe 
start to decrease rapidly when they are kinematically older than 5000 and 6000 yr 
respectively in our compact sample. Taking those as estimations of 
typical PN life times and combining with the total PN number of 2000 in the Galactic bulge 
predicted in \citet{gesicki2014accelerated} and our results of the fractions of PN in these two 
morphological classes (29\% for elliptical and 68\% for bipolar), the birth rates of elliptical 
and bipolar PNs are $\sim$0.12~$\mathrm{yr}^{-1}$ and $\sim$0.23~$\mathrm{yr}^{-1}$ respectively. 
This indicates a higher, nearly doubled, death rate of progenitors of observed PNe showing bipolar
morphologies. 

Models show that stellar death rate decreases after the star burst and plateaus around 7~Gyr (see 
Fig. 10 in \citet{gesicki2014accelerated}. Thus, if progenitors of both the elliptical and bipolar
PNe are older than 7~Gyr, our PN sample should trace both populations. The higher birth rate of 
bipolar PNe might suggest the existence of a younger population $<$~7 Gyr assuming our selection 
criteria on the PN size leads to no morphological bias. Alternatively, the true fraction of binary
systems may be much higher than expected. This is suggested by the search of binary CSPNe 
through the infrared (IR) excess method in \citep[][]{de2013binary, de2015identifying}. Also
more bipolar PNe may have been formed from long-period wide orbit binaries but that so far
has limited observational evidence, or PNe might just be
preferentially formed in binary systems and so are over-representing the underlying binary 
fraction compared with the general main-sequence population that their progenitors belong to. This
is the idea of the PN \textit{Binary Hypothesis} in \citet{de2009origin}. However, there are as 
yet no clear answers to this conundrum. 


\section{Conclusions}
\label{con}
We have undertaken a detailed study of a carefully selected sample of compact PNe 
in the Galactic bulge. We have used our own VLT narrow-band imagery and archival 
narrow and broad band \emph{HST} and occasionally Pan-STARRS broad-band imagery 
to provide significantly improved morphological classifications and angular size 
measurements for a well defined sample of compact bulge PNe. 
Many PNe have been re-classified as bipolars at a level 
that is likely to be replicated in the general HASH Galactic disk sample 
for PNe $\leq$~10~arcseconds across, assuming deeper 
and higher resolution imaging eventually becomes available. 
The increase in the fraction of bipolar PNe comprising our bulge sample is very 
significant for a population ostensibly dominated by old stars. 
It is clear the high visibility of compact PNe allows us to 
find and study such a key touchstone population of late-stage stellar 
evolution in the Galactic bulge. If bipolars can derive 
from higher mass progenitors, as inferred by their preponderance at lower Galactic
scale heights in the Galactic disk, as well as being in binaries, then there has to be a 
significant, young stellar population in the bulge feeding the observed population 
today. They dominate all PNe emerging from the general, several Giga year old bulge 
stellar population.

We have also used these data to provide a more complete 
and accurate listing of bulge CSPN including 4 new discoveries. 
We also show that 11 previous \emph{Gaia} CSPN identified are incorrect. Many actual 
CSPN are beyond \emph{Gaia} limits but visible in the deep Pan-STARRS and VLT data 
allowing their proper identification. For others \emph{Gaia} has selected the wrong 
star or even a compact emission region as the CSPN. For some the PNe are so compact 
that we believe \emph{Gaia} has simply taken the centroid of this as the CSPN 
given there are no other \emph{Gaia} stars identified in the immediate region. We find 
for this sample that 15\% of \emph{Gaia} CSPN listed in CW21 are incorrect. As such we 
emphasise that great care must be taken when using the \emph{Gaia} based CSPN 
compilations of CW21 and CSM21. We strongly recommend HASH \citep{parker2016hash} 
for such studies as every CSPN listed has been holistically vetted. 

Finally, using the best available data, we provide the most complete 
and accurate kinematic age estimates ever compiled for all known 
compact PNe in the bulge. This includes \emph{Gaia} parallax distances for 33, 
accurate expansion velocities for 77, improved angular size determinations for all 
(from the HST, VLT and Pan-STARRS imagery) and average distance and expansion 
velocities for the rest. The expected young age of the PNe sample is confirmed 
with an average kinematic age of 6400~$\pm$~3800 years. Kinematically older 
PNe are dominated by bipolars with 14/16 PNe with estimated ages $>$~10000 years 
being bipolars. Even in this compact sample these older PNe are more extended and it
is easier to discern bipolar lobes. Hence a selection effect is likely at play. 
It is also likely that some of the very compact PNe currently classified 
as elliptical from the VLT imagery may turn out to be bipolars in 
higher resolution imagery. This would make the bipolar fraction even larger. 
The best current work also puts the short-period binary fraction of 
bulge PNe at 15-20\% - maximum 60\% in bipolars \citep{miszalski2009binary}, giving 9-12\% of PNe in binaries are bipolars. This is a factor of up to 7 lower than the bipolar fraction we find. A younger stellar 
population in the bulge, wider period binaries and the relationship between PN 
formation and the CSPN binarity may pay a role here but the issue of the progenitor 
masses of the bipolar family remains. Finally, all updated angular sizes and 
morphological classifications presented here have also now been ingested into 
HASH. It is clear more work is needed to understand the observed bulge 
PNe morphological distribution and the associated
progenitor stellar populations.

\section*{Acknowledgements}
QAP thanks the Hong Kong Research Grants Council for GRF research support
under grants 17326116 and 17300417. ST thanks HKU and QAP for provision of an MPhil scholarship. AR thanks HKU and QAP for the provision of a 
postdoctoral fellowship. AAZ acknowledges support from STFC under grant ST/T000414/1. This work has made use of data from the European
Space Agency (ESA) mission
{\it Gaia~} (\url{https://www.cosmos.esa.int/gaia}), processed by the {\it Gaia}
Data Processing and Analysis Consortium (DPAC,
\url{https://www.cosmos.esa.int/web/gaia/dpac/consortium}. Funding for 
the DPAC has been provided by national institutions, in particular the 
institutions participating in the {\it Gaia} Multilateral Agreement.

This research made use of Astropy, a community-developed core \texttt{Python} package for Astronomy \citep[][]{collaboration2013astronomy} and APLpy, an open-source plotting package for \texttt{Python} hosted at \url{http://aplpy.github.io}. This work used observations made with the NASA/ESA \emph{Hubble Space Telescope}. 
Some data were obtained from the Hubble Legacy Archive, which is a collaboration 
between the Space Telescope Science Institute (STScI/NASA), the Space Telescope 
European Coordinating Facility (ST-ECF/ESA), and the Canadian Astronomy 
Data Centre (CADC/NRC/CSA).

\section*{Data Availability}

The data underlying this article are available in the article itself and 
in its associated online material freely accessible from the HASH database found here: \url{http://hashpn.space} by simply entering the unique HASH ID number for 
each source as provided.



\bibliographystyle{mnras}
\bibliography{example} 

\begin{thebibliography}{}
\makeatletter
\relax
\def\mn@urlcharsother{\let\do\@makeother \do\$\do\&\do\#\do\^\do\_\do\%\do\~}
\def\mn@doi{\begingroup\mn@urlcharsother \@ifnextchar [ {\mn@doi@}
  {\mn@doi@[]}}
\def\mn@doi@[#1]#2{\def\@tempa{#1}\ifx\@tempa\@empty \href
  {http://dx.doi.org/#2} {doi:#2}\else \href {http://dx.doi.org/#2} {#1}\fi
  \endgroup}
\def\mn@eprint#1#2{\mn@eprint@#1:#2::\@nil}
\def\mn@eprint@arXiv#1{\href {http://arxiv.org/abs/#1} {{\tt arXiv:#1}}}
\def\mn@eprint@dblp#1{\href {http://dblp.uni-trier.de/rec/bibtex/#1.xml}
  {dblp:#1}}
\def\mn@eprint@#1:#2:#3:#4\@nil{\def\@tempa {#1}\def\@tempb {#2}\def\@tempc
  {#3}\ifx \@tempc \@empty \let \@tempc \@tempb \let \@tempb \@tempa \fi \ifx
  \@tempb \@empty \def\@tempb {arXiv}\fi \@ifundefined
  {mn@eprint@\@tempb}{\@tempb:\@tempc}{\expandafter \expandafter \csname
  mn@eprint@\@tempb\endcsname \expandafter{\@tempc}}}

\bibitem[\protect\citeauthoryear{Acker, Marcout, Ochsenbein, Stenholm, Tylenda
  \& Schohn}{Acker et~al.}{1992}]{acker1992strasbourg}
Acker A.,  Marcout J.,  Ochsenbein F.,  Stenholm B.,  Tylenda R.,   Schohn C.,
  1992, The Strasbourg-ESO Catalogue of Galactic Planetary Nebulae. Parts I,
  II.

\bibitem[\protect\citeauthoryear{Acker, Peyaud  \& Parker}{Acker
  et~al.}{2006}]{acker2006400}
Acker A.,  Peyaud A.~E.,   Parker Q.,  2006, Proceedings of the International
  Astronomical Union, 2, 355

\bibitem[\protect\citeauthoryear{Akashi \& Soker}{Akashi \&
  Soker}{2021}]{akashi2021shaping}
Akashi M.,  Soker N.,  2021, The Astrophysical Journal, 913, 91

\bibitem[\protect\citeauthoryear{Akras \& Gon{\c{c}}alves}{Akras \&
  Gon{\c{c}}alves}{2016}]{akras2016low}
Akras S.,  Gon{\c{c}}alves D.~R.,  2016, Monthly Notices of the Royal
  Astronomical Society, 455, 930

\bibitem[\protect\citeauthoryear{{Ali}, {Dopita}, {Basurah}, {Amer}, {Alsulami}
   \& {Alruhaili}}{{Ali} et~al.}{2016}]{2016MNRAS.462.1393A}
{Ali} A.,  {Dopita} M.~A.,  {Basurah} H.~M.,  {Amer} M.~A.,  {Alsulami} R.,
  {Alruhaili} A.,  2016, \mn@doi [\mnras] {10.1093/mnras/stw1744}, \href
  {https://ui.adsabs.harvard.edu/abs/2016MNRAS.462.1393A} {462, 1393}

\bibitem[\protect\citeauthoryear{Amnuel}{Amnuel}{1995}]{amnuel1995asymmetrical}
Amnuel P.,  1995, Astrophysics and Space Science, 225, 275

\bibitem[\protect\citeauthoryear{{Anderson} \& {King}}{{Anderson} \&
  {King}}{2000}]{2000PASP..112.1360A}
{Anderson} J.,  {King} I.~R.,  2000, \mn@doi [\pasp] {10.1086/316632}, \href
  {https://ui.adsabs.harvard.edu/abs/2000PASP..112.1360A} {112, 1360}

\bibitem[\protect\citeauthoryear{Appenzeller et~al.,}{Appenzeller
  et~al.}{1998}]{appenzeller1998successful}
Appenzeller I.,  et~al., 1998, The messenger, 94

\bibitem[\protect\citeauthoryear{Athanassoula}{Athanassoula}{2005}]{athanassoula2005nature}
Athanassoula E.,  2005, Monthly Notices of the Royal Astronomical Society, 358,
  1477

\bibitem[\protect\citeauthoryear{Balick}{Balick}{1987}]{balick1987evolution}
Balick B.,  1987, The Astronomical Journal, 94, 671

\bibitem[\protect\citeauthoryear{{Balick} \& {Frank}}{{Balick} \&
  {Frank}}{2002}]{2002ARA&A..40..439B}
{Balick} B.,  {Frank} A.,  2002, \mn@doi [\araa]
  {10.1146/annurev.astro.40.060401.093849}, \href
  {https://ui.adsabs.harvard.edu/abs/2002ARA&A..40..439B} {40, 439}

\bibitem[\protect\citeauthoryear{{Bensby} et~al.,}{{Bensby}
  et~al.}{2013}]{2013A&A...549A.147B}
{Bensby} T.,  et~al., 2013, \mn@doi [\aap] {10.1051/0004-6361/201220678}, \href
  {https://ui.adsabs.harvard.edu/abs/2013A&A...549A.147B} {549, A147}

\bibitem[\protect\citeauthoryear{{Bensby} et~al.,}{{Bensby}
  et~al.}{2017}]{2017A&A...605A..89B}
{Bensby} T.,  et~al., 2017, \mn@doi [\aap] {10.1051/0004-6361/201730560}, \href
  {https://ui.adsabs.harvard.edu/abs/2017A&A...605A..89B} {605, A89}

\bibitem[\protect\citeauthoryear{Blitz \& Spergel}{Blitz \&
  Spergel}{1991}]{blitz1991direct}
Blitz L.,  Spergel D.~N.,  1991, The Astrophysical Journal, 379, 631

\bibitem[\protect\citeauthoryear{Boffin \& Jones}{Boffin \&
  Jones}{2019}]{boffin2019importance}
Boffin H.~M.,  Jones D.,  2019, The Importance of Binaries in the Formation and
  Evolution of Planetary Nebulae.
Springer

\bibitem[\protect\citeauthoryear{Boffin, Miszalski, Rauch, Jones, Corradi,
  Napiwotzki, Day-Jones  \& K{\"o}ppen}{Boffin
  et~al.}{2012}]{boffin2012interacting}
Boffin H.~M.,  Miszalski B.,  Rauch T.,  Jones D.,  Corradi R.~L.,  Napiwotzki
  R.,  Day-Jones A.~C.,   K{\"o}ppen J.,  2012, Science, 338, 773

\bibitem[\protect\citeauthoryear{Chamandy et~al.,}{Chamandy
  et~al.}{2018}]{chamandy2018accretion}
Chamandy L.,  et~al., 2018, Monthly Notices of the Royal Astronomical Society,
  480, 1898

\bibitem[\protect\citeauthoryear{{Chambers} et~al.,}{{Chambers}
  et~al.}{2016}]{2016arXiv161205560C}
{Chambers} K.~C.,  et~al., 2016, arXiv e-prints, \href
  {https://ui.adsabs.harvard.edu/abs/2016arXiv161205560C} {p. arXiv:1612.05560}

\bibitem[\protect\citeauthoryear{Chornay \& Walton}{Chornay \&
  Walton}{2020}]{chornay2020searching}
Chornay N.,  Walton N.,  2020, Astronomy \& Astrophysics, 638, A103

\bibitem[\protect\citeauthoryear{Chornay \& Walton}{Chornay \&
  Walton}{2021}]{chornay2021one}
Chornay N.,  Walton N.,  2021, Astronomy \& Astrophysics, 656, A110

\bibitem[\protect\citeauthoryear{{Chu}, {Jacoby}  \& {Arendt}}{{Chu}
  et~al.}{1987}]{1987ApJS...64..529C}
{Chu} Y.-H.,  {Jacoby} G.~H.,   {Arendt} R.,  1987, \mn@doi [\apjs]
  {10.1086/191207}, \href
  {https://ui.adsabs.harvard.edu/abs/1987ApJS...64..529C} {64, 529}

\bibitem[\protect\citeauthoryear{{Clairmont}, {Steffen}  \&
  {Koning}}{{Clairmont} et~al.}{2022}]{2022MNRAS.516.2711C}
{Clairmont} R.,  {Steffen} W.,   {Koning} N.,  2022, \mn@doi [\mnras]
  {10.1093/mnras/stac2375}, \href
  {https://ui.adsabs.harvard.edu/abs/2022MNRAS.516.2711C} {516, 2711}

\bibitem[\protect\citeauthoryear{{Clyne}, {Akras}, {Steffen}, {Redman},
  {Gon{\c{c}}alves}  \& {Harvey}}{{Clyne} et~al.}{2015}]{2015A&A...582A..60C}
{Clyne} N.,  {Akras} S.,  {Steffen} W.,  {Redman} M.~P.,  {Gon{\c{c}}alves}
  D.~R.,   {Harvey} E.,  2015, \mn@doi [\aap] {10.1051/0004-6361/201526585},
  \href {https://ui.adsabs.harvard.edu/abs/2015A&A...582A..60C} {582, A60}

\bibitem[\protect\citeauthoryear{Collaboration, Robitaille, Tollerud
  et~al.}{Collaboration et~al.}{2013}]{collaboration2013astronomy}
Collaboration A.,  Robitaille T.,  Tollerud E.,   et~al., 2013, A33

\bibitem[\protect\citeauthoryear{Corradi \& Schwarz}{Corradi \&
  Schwarz}{1995}]{corradi1995morphological}
Corradi R.,  Schwarz H.,  1995, Astronomy and Astrophysics, 293, 871

\bibitem[\protect\citeauthoryear{{Corradi}, {Sch{\"o}nberner}, {Steffen}  \&
  {Perinotto}}{{Corradi} et~al.}{2003}]{2003MNRAS.340..417C}
{Corradi} R.~L.~M.,  {Sch{\"o}nberner} D.,  {Steffen} M.,   {Perinotto} M.,
  2003, \mn@doi [\mnras] {10.1046/j.1365-8711.2003.06294.x}, \href
  {https://ui.adsabs.harvard.edu/abs/2003MNRAS.340..417C} {340, 417}

\bibitem[\protect\citeauthoryear{{Danehkar}, {Parker}  \&
  {Ercolano}}{{Danehkar} et~al.}{2013}]{2013MNRAS.434.1513D}
{Danehkar} A.,  {Parker} Q.~A.,   {Ercolano} B.,  2013, \mn@doi [\mnras]
  {10.1093/mnras/stt1116}, \href
  {https://ui.adsabs.harvard.edu/abs/2013MNRAS.434.1513D} {434, 1513}

\bibitem[\protect\citeauthoryear{De~Marco}{De~Marco}{2009b}]{de2009origin}
De~Marco O.,  2009b, Publications of the Astronomical Society of the Pacific,
  121, 316

\bibitem[\protect\citeauthoryear{{De Marco}}{{De
  Marco}}{2009a}]{2009PASP..121..316D}
{De Marco} O.,  2009a, \mn@doi [\pasp] {10.1086/597765}, \href
  {https://ui.adsabs.harvard.edu/abs/2009PASP..121..316D} {121, 316}

\bibitem[\protect\citeauthoryear{De~Marco, Passy, Frew, Moe  \&
  Jacoby}{De~Marco et~al.}{2013}]{de2013binary}
De~Marco O.,  Passy J.-C.,  Frew D.~J.,  Moe M.,   Jacoby G.~H.,  2013, Monthly
  Notices of the Royal Astronomical Society, 428, 2118

\bibitem[\protect\citeauthoryear{De~Marco, Long, Jacoby, Hillwig, Kronberger,
  Howell, Reindl  \& Margheim}{De~Marco et~al.}{2015}]{de2015identifying}
De~Marco O.,  Long J.,  Jacoby G.~H.,  Hillwig T.,  Kronberger M.,  Howell
  S.~B.,  Reindl N.,   Margheim S.,  2015, Monthly Notices of the Royal
  Astronomical Society, 448, 3587

\bibitem[\protect\citeauthoryear{Dwek et~al.,}{Dwek
  et~al.}{1995}]{dwek1995morphology}
Dwek E.,  et~al., 1995, The Astrophysical Journal, 445, 716

\bibitem[\protect\citeauthoryear{{Fragkou}, {Parker}, {Zijlstra},
  {V{\'a}zquez}, {Sabin}  \& {Rechy-Garcia}}{{Fragkou}
  et~al.}{2022}]{2022ApJ...935L..35F}
{Fragkou} V.,  {Parker} Q.~A.,  {Zijlstra} A.~A.,  {V{\'a}zquez} R.,  {Sabin}
  L.,   {Rechy-Garcia} J.~S.,  2022, \mn@doi [\apjl]
  {10.3847/2041-8213/ac88c1}, \href
  {https://ui.adsabs.harvard.edu/abs/2022ApJ...935L..35F} {935, L35}

\bibitem[\protect\citeauthoryear{Freudling, Romaniello, Bramich, Ballester,
  Forchi, Garc{\'\i}a-Dabl{\'o}, Moehler  \& Neeser}{Freudling
  et~al.}{2013}]{freudling2013automated}
Freudling W.,  Romaniello M.,  Bramich D.,  Ballester P.,  Forchi V.,
  Garc{\'\i}a-Dabl{\'o} C.,  Moehler S.,   Neeser M.,  2013, Astronomy \&
  Astrophysics, 559, A96

\bibitem[\protect\citeauthoryear{Garc{\'\i}a-Segura, Ricker  \&
  Taam}{Garc{\'\i}a-Segura et~al.}{2018}]{garcia2018common}
Garc{\'\i}a-Segura G.,  Ricker P.~M.,   Taam R.~E.,  2018, The Astrophysical
  Journal, 860, 19

\bibitem[\protect\citeauthoryear{Gesicki \& Zijlstra}{Gesicki \&
  Zijlstra}{2000}]{gesicki2000expansion}
Gesicki K.,  Zijlstra A.~A.,  2000, Astronomy and Astrophysics, 358, 1058

\bibitem[\protect\citeauthoryear{Gesicki, Zijlstra, Acker, G{\'o}rny,
  Gozdziewski  \& Walsh}{Gesicki et~al.}{2006}]{gesicki2006planetary}
Gesicki K.,  Zijlstra A.,  Acker A.,  G{\'o}rny S.,  Gozdziewski K.,   Walsh
  J.,  2006, Astronomy \& Astrophysics, 451, 925

\bibitem[\protect\citeauthoryear{Gesicki, Zijlstra, Hajduk  \& Szyszka}{Gesicki
  et~al.}{2014}]{gesicki2014accelerated}
Gesicki K.,  Zijlstra A.,  Hajduk M.,   Szyszka C.,  2014, Astronomy \&
  Astrophysics, 566, A48

\bibitem[\protect\citeauthoryear{Gesicki, Zijlstra  \& Morisset}{Gesicki
  et~al.}{2016}]{gesicki20163d}
Gesicki K.,  Zijlstra A.,   Morisset C.,  2016, Astronomy \& Astrophysics, 585,
  A69

\bibitem[\protect\citeauthoryear{Gonz{\'a}lez-Santamar{\'\i}a, Manteiga,
  Manchado, Ulla  \& Dafonte}{Gonz{\'a}lez-Santamar{\'\i}a
  et~al.}{2019}]{gonzalez2019properties}
Gonz{\'a}lez-Santamar{\'\i}a I.,  Manteiga M.,  Manchado A.,  Ulla A.,
  Dafonte C.,  2019, Astronomy \& Astrophysics, 630, A150

\bibitem[\protect\citeauthoryear{Gonz{\'a}lez-Santamar{\'\i}a, Manteiga,
  Manchado, Ulla, Dafonte  \& L{\'o}pez~Varela}{Gonz{\'a}lez-Santamar{\'\i}a
  et~al.}{2021}]{gonzalez2021planetary}
Gonz{\'a}lez-Santamar{\'\i}a I.,  Manteiga M.,  Manchado A.,  Ulla A.,  Dafonte
  C.,   L{\'o}pez~Varela P.,  2021, Astronomy \& Astrophysics, 656, A51

\bibitem[\protect\citeauthoryear{G{\'o}rny, Stasinska  \& Tylenda}{G{\'o}rny
  et~al.}{1997}]{gorny1997planetary}
G{\'o}rny S.,  Stasinska G.,   Tylenda R.,  1997, Astronomy and Astrophysics,
  318, 256

\bibitem[\protect\citeauthoryear{G{\'o}rny, Schwarz, Corradi  \&
  Van~Winckel}{G{\'o}rny et~al.}{1999}]{gorny1999atlas}
G{\'o}rny S.,  Schwarz H.,  Corradi R.,   Van~Winckel H.~V.,  1999, Astronomy
  and Astrophysics Supplement Series, 136, 145

\bibitem[\protect\citeauthoryear{{Greig}}{{Greig}}{1971}]{1971A&A....10..161G}
{Greig} W.~E.,  1971, \aap, \href
  {https://ui.adsabs.harvard.edu/abs/1971A&A....10..161G} {10, 161}

\bibitem[\protect\citeauthoryear{Icke}{Icke}{1988}]{icke1988blowing}
Icke V.,  1988, Astronomy and Astrophysics, 202, 177

\bibitem[\protect\citeauthoryear{Icke, Preston  \& Balick}{Icke
  et~al.}{1989}]{icke1989evolution}
Icke V.,  Preston H.~L.,   Balick B.,  1989, Astronomical Journal, 97, 462

\bibitem[\protect\citeauthoryear{Icke, Balick  \& Frank}{Icke
  et~al.}{1992}]{icke1992hydrodynamics}
Icke V.,  Balick B.,   Frank A.,  1992, Astronomy and Astrophysics, 253, 224

\bibitem[\protect\citeauthoryear{Jacob, Sch{\"o}nberner  \& Steffen}{Jacob
  et~al.}{2013}]{jacob2013evolution}
Jacob R.,  Sch{\"o}nberner D.,   Steffen M.,  2013, Astronomy \& Astrophysics,
  558, A78

\bibitem[\protect\citeauthoryear{{Jacoby}}{{Jacoby}}{1980}]{1980ApJS...42....1J}
{Jacoby} G.~H.,  1980, \mn@doi [\apjs] {10.1086/190642}, \href
  {https://ui.adsabs.harvard.edu/abs/1980ApJS...42....1J} {42, 1}

\bibitem[\protect\citeauthoryear{Jones}{Jones}{2016}]{jones2016central}
Jones D.,  2016, Proceedings of the International Astronomical Union, 12, 169

\bibitem[\protect\citeauthoryear{Kahn \& West}{Kahn \&
  West}{1985}]{kahn1985shapes}
Kahn F.,  West K.~A.,  1985, Monthly Notices of the Royal Astronomical Society,
  212, 837

\bibitem[\protect\citeauthoryear{{Kaiser} et~al.,}{{Kaiser}
  et~al.}{2010}]{2010SPIE.7733E..0EK}
{Kaiser} N.,  et~al., 2010, in {Stepp} L.~M.,  {Gilmozzi} R.,   {Hall} H.~J.,
  eds,  Society of Photo-Optical Instrumentation Engineers (SPIE) Conference
  Series Vol. 7733, Ground-based and Airborne Telescopes III. p. 77330E,
  \mn@doi{10.1117/12.859188}

\bibitem[\protect\citeauthoryear{{Kastner}, {Moraga Baez}, {Balick}, {Bublitz},
  {Montez}, {Frank}  \& {Blackman}}{{Kastner}
  et~al.}{2022}]{2022ApJ...927..100K}
{Kastner} J.~H.,  {Moraga Baez} P.,  {Balick} B.,  {Bublitz} J.,  {Montez} R.,
  {Frank} A.,   {Blackman} E.,  2022, \mn@doi [\apj]
  {10.3847/1538-4357/ac51cd}, \href
  {https://ui.adsabs.harvard.edu/abs/2022ApJ...927..100K} {927, 100}

\bibitem[\protect\citeauthoryear{{Kwitter} \& {Henry}}{{Kwitter} \&
  {Henry}}{2022}]{2022PASP..134b2001K}
{Kwitter} K.~B.,  {Henry} R.~B.~C.,  2022, \mn@doi [\pasp]
  {10.1088/1538-3873/ac32b1}, \href
  {https://ui.adsabs.harvard.edu/abs/2022PASP..134b2001K} {134, 022001}

\bibitem[\protect\citeauthoryear{{Lindegren} et~al.,}{{Lindegren}
  et~al.}{2021a}]{2021A&A...649A...2L}
{Lindegren} L.,  et~al., 2021a, \mn@doi [\aap] {10.1051/0004-6361/202039709},
  \href {https://ui.adsabs.harvard.edu/abs/2021A&A...649A...2L} {649, A2}

\bibitem[\protect\citeauthoryear{Lindegren et~al.,}{Lindegren
  et~al.}{2021b}]{lindegren2021gaia}
Lindegren L.,  et~al., 2021b, Astronomy \& Astrophysics, 649, A2

\bibitem[\protect\citeauthoryear{Livio \& Pringle}{Livio \&
  Pringle}{1996}]{livio1996formation}
Livio M.,  Pringle J.,  1996, The Astrophysical Journal, 465, L55

\bibitem[\protect\citeauthoryear{{Magnier} et~al.,}{{Magnier}
  et~al.}{2020}]{2020ApJS..251....6M}
{Magnier} E.~A.,  et~al., 2020, \mn@doi [\apjs] {10.3847/1538-4365/abb82a},
  \href {https://ui.adsabs.harvard.edu/abs/2020ApJS..251....6M} {251, 6}

\bibitem[\protect\citeauthoryear{{Manchado}, {Villaver}, {Stanghellini}  \&
  {Guerrero}}{{Manchado} et~al.}{2000}]{2000ASPC..199...17M}
{Manchado} A.,  {Villaver} E.,  {Stanghellini} L.,   {Guerrero} M.~A.,  2000,
  in {Kastner} J.~H.,  {Soker} N.,   {Rappaport} S.,  eds,  Astronomical
  Society of the Pacific Conference Series Vol. 199, Asymmetrical Planetary
  Nebulae II: From Origins to Microstructures. p.~17 (\mn@eprint {arXiv}
  {astro-ph/0002073})

\bibitem[\protect\citeauthoryear{{Mellema}}{{Mellema}}{1997}]{1997A&A...321L..29M}
{Mellema} G.,  1997, \aap, \href
  {https://ui.adsabs.harvard.edu/abs/1997A&A...321L..29M} {321, L29}

\bibitem[\protect\citeauthoryear{Mellema, Eulderink  \& Icke}{Mellema
  et~al.}{1991}]{mellema1991hydrodynamical}
Mellema G.,  Eulderink F.,   Icke V.,  1991, Astronomy and Astrophysics, 252,
  718

\bibitem[\protect\citeauthoryear{{Merrett} et~al.,}{{Merrett}
  et~al.}{2006}]{2006MNRAS.369..120M}
{Merrett} H.~R.,  et~al., 2006, \mn@doi [\mnras]
  {10.1111/j.1365-2966.2006.10268.x}, \href
  {https://ui.adsabs.harvard.edu/abs/2006MNRAS.369..120M} {369, 120}

\bibitem[\protect\citeauthoryear{Miszalski, Acker, Moffat, Parker  \&
  Udalski}{Miszalski et~al.}{2009a}]{miszalski2009binary}
Miszalski B.,  Acker A.,  Moffat A.~F.,  Parker Q.~A.,   Udalski A.,  2009a,
  Astronomy \& Astrophysics, 496, 813

\bibitem[\protect\citeauthoryear{{Miszalski}, {Acker}, {Parker}  \&
  {Moffat}}{{Miszalski} et~al.}{2009b}]{2009A&A...505..249M}
{Miszalski} B.,  {Acker} A.,  {Parker} Q.~A.,   {Moffat} A.~F.~J.,  2009b,
  \mn@doi [\aap] {10.1051/0004-6361/200912176}, \href
  {https://ui.adsabs.harvard.edu/abs/2009A&A...505..249M} {505, 249}

\bibitem[\protect\citeauthoryear{Mitchell, Pollacco, O'Brien, Bryce, L{\'o}pez
  \& Meaburn}{Mitchell et~al.}{2006}]{mitchell2006structures}
Mitchell D.~L.,  Pollacco D.,  O'Brien T.,  Bryce M.,  L{\'o}pez J.,   Meaburn
  J.,  2006, Proceedings of the International Astronomical Union, 2, 139

\bibitem[\protect\citeauthoryear{{Nataf}}{{Nataf}}{2016}]{2016PASA...33...23N}
{Nataf} D.~M.,  2016, \mn@doi [\pasa] {10.1017/pasa.2015.38}, \href
  {https://ui.adsabs.harvard.edu/abs/2016PASA...33...23N} {33, e023}

\bibitem[\protect\citeauthoryear{{Nicklas}, {Seifert}, {Boehnhardt},
  {Kiesewetter-Koebinger}  \& {Rupprecht}}{{Nicklas}
  et~al.}{1997}]{1997SPIE.2871.1222N}
{Nicklas} H.,  {Seifert} W.,  {Boehnhardt} H.,  {Kiesewetter-Koebinger} S.,
  {Rupprecht} G.,  1997, in {Ardeberg} A.~L.,  ed.,  Society of Photo-Optical
  Instrumentation Engineers (SPIE) Conference Series Vol. 2871, Optical
  Telescopes of Today and Tomorrow. pp 1222--1230, \mn@doi{10.1117/12.269011}

\bibitem[\protect\citeauthoryear{{Parker}}{{Parker}}{2022a}]{2022FrASS...9.5287P}
{Parker} Q.~A.,  2022a, \mn@doi [Frontiers in Astronomy and Space Sciences]
  {10.3389/fspas.2022.895287}, \href
  {https://ui.adsabs.harvard.edu/abs/2022FrASS...9.5287P} {9, 895287}

\bibitem[\protect\citeauthoryear{Parker}{Parker}{2022b}]{parker2022planetary}
Parker Q.~A.,  2022b, Frontiers in Astronomy and Space Sciences, 9, 895287

\bibitem[\protect\citeauthoryear{{Parker} et~al.,}{{Parker}
  et~al.}{2005}]{2005MNRAS.362..689P}
{Parker} Q.~A.,  et~al., 2005, \mn@doi [\mnras]
  {10.1111/j.1365-2966.2005.09350.x}, \href
  {https://ui.adsabs.harvard.edu/abs/2005MNRAS.362..689P} {362, 689}

\bibitem[\protect\citeauthoryear{{Parker} et~al.,}{{Parker}
  et~al.}{2006}]{2006MNRAS.373...79P}
{Parker} Q.~A.,  et~al., 2006, \mn@doi [\mnras]
  {10.1111/j.1365-2966.2006.10950.x}, \href
  {https://ui.adsabs.harvard.edu/abs/2006MNRAS.373...79P} {373, 79}

\bibitem[\protect\citeauthoryear{Parker, Boji{\v{c}}i{\'c}  \& Frew}{Parker
  et~al.}{2016}]{parker2016hash}
Parker Q.~A.,  Boji{\v{c}}i{\'c} I.~S.,   Frew D.~J.,  2016, in Journal of
  Physics: Conference Series. p. 032008

\bibitem[\protect\citeauthoryear{{Parker}, {Xiang}  \& {Ritter}}{{Parker}
  et~al.}{2022}]{2022Galax..10...32P}
{Parker} Q.~A.,  {Xiang} Z.,   {Ritter} A.,  2022, \mn@doi [Galaxies]
  {10.3390/galaxies10010032}, \href
  {https://ui.adsabs.harvard.edu/abs/2022Galax..10...32P} {10, 32}

\bibitem[\protect\citeauthoryear{Peimbert}{Peimbert}{1978}]{peimbert1978chemical}
Peimbert M.,  1978, in Symposium-International Astronomical Union. pp 215--224

\bibitem[\protect\citeauthoryear{Peimbert \& Torres-Peimbert}{Peimbert \&
  Torres-Peimbert}{1983}]{peimbert1983type}
Peimbert M.,  Torres-Peimbert S.,  1983, in Symposium-International
  Astronomical Union. pp 233--242

\bibitem[\protect\citeauthoryear{{Phillips}}{{Phillips}}{2001}]{2001MNRAS.326.1041P}
{Phillips} J.~P.,  2001, \mn@doi [\mnras] {10.1046/j.1365-8711.2001.04715.x},
  \href {https://ui.adsabs.harvard.edu/abs/2001MNRAS.326.1041P} {326, 1041}

\bibitem[\protect\citeauthoryear{{Pollacco} \& {Bell}}{{Pollacco} \&
  {Bell}}{1997}]{1997MNRAS.284...32P}
{Pollacco} D.~L.,  {Bell} S.~A.,  1997, \mn@doi [\mnras]
  {10.1093/mnras/284.1.32}, \href
  {https://ui.adsabs.harvard.edu/abs/1997MNRAS.284...32P} {284, 32}

\bibitem[\protect\citeauthoryear{Rees \& Zijlstra}{Rees \&
  Zijlstra}{2013}]{rees2013alignment}
Rees B.,  Zijlstra A.,  2013, Monthly Notices of the Royal Astronomical
  Society, 435, 975

\bibitem[\protect\citeauthoryear{{Reid} \& {Parker}}{{Reid} \&
  {Parker}}{2006}]{2006MNRAS.373..521R}
{Reid} W.~A.,  {Parker} Q.~A.,  2006, \mn@doi [\mnras]
  {10.1111/j.1365-2966.2006.11087.x}, \href
  {https://ui.adsabs.harvard.edu/abs/2006MNRAS.373..521R} {373, 521}

\bibitem[\protect\citeauthoryear{Richer, L{\'o}pez, Pereyra, Riesgo,
  Garc{\'\i}a-D{\'\i}az  \& B{\'a}ez}{Richer
  et~al.}{2008}]{richer2008acceleration}
Richer M.~G.,  L{\'o}pez J.~A.,  Pereyra M.,  Riesgo H.,  Garc{\'\i}a-D{\'\i}az
  M.~T.,   B{\'a}ez S.-H.,  2008, The Astrophysical Journal, 689, 203

\bibitem[\protect\citeauthoryear{Richer, L{\'o}pez, Garc{\'\i}a-D{\'\i}az,
  Clark, Pereyra  \& D{\'\i}az-M{\'e}ndez}{Richer
  et~al.}{2010}]{richer2010evolution}
Richer M.~G.,  L{\'o}pez J.~A.,  Garc{\'\i}a-D{\'\i}az M.~T.,  Clark D.~M.,
  Pereyra M.,   D{\'\i}az-M{\'e}ndez E.,  2010, The Astrophysical Journal, 716,
  857

\bibitem[\protect\citeauthoryear{{Sabin}, {Zijlstra}  \& {Greaves}}{{Sabin}
  et~al.}{2007}]{2007MNRAS.376..378S}
{Sabin} L.,  {Zijlstra} A.~A.,   {Greaves} J.~S.,  2007, \mn@doi [\mnras]
  {10.1111/j.1365-2966.2007.11445.x}, \href
  {https://ui.adsabs.harvard.edu/abs/2007MNRAS.376..378S} {376, 378}

\bibitem[\protect\citeauthoryear{Sabin et~al.,}{Sabin
  et~al.}{2014}]{sabin2014first}
Sabin L.,  et~al., 2014, Monthly Notices of the Royal Astronomical Society,
  443, 3388

\bibitem[\protect\citeauthoryear{Sabin et~al.,}{Sabin
  et~al.}{2021}]{sabin2021first}
Sabin L.,  et~al., 2021, Monthly Notices of the Royal Astronomical Society,
  508, 1599

\bibitem[\protect\citeauthoryear{{Sahai}, {Morris}  \& {Villar}}{{Sahai}
  et~al.}{2011a}]{2011AJ....141..134S}
{Sahai} R.,  {Morris} M.~R.,   {Villar} G.~G.,  2011a, \mn@doi [\aj]
  {10.1088/0004-6256/141/4/134}, \href
  {https://ui.adsabs.harvard.edu/abs/2011AJ....141..134S} {141, 134}

\bibitem[\protect\citeauthoryear{Sahai, Morris  \& Villar}{Sahai
  et~al.}{2011b}]{sahai2011young}
Sahai R.,  Morris M.~R.,   Villar G.~G.,  2011b, The Astronomical Journal, 141,
  134

\bibitem[\protect\citeauthoryear{Saito, Zoccali, McWilliam, Minniti, Gonzalez
  \& Hill}{Saito et~al.}{2011}]{saito2011mapping}
Saito R.~K.,  Zoccali M.,  McWilliam A.,  Minniti D.,  Gonzalez O.~A.,   Hill
  V.,  2011, The Astronomical Journal, 142, 76

\bibitem[\protect\citeauthoryear{Saito et~al.,}{Saito
  et~al.}{2012a}]{saito2012vvv}
Saito R.~K.,  et~al., 2012a, Astronomy \& Astrophysics, 537, A107

\bibitem[\protect\citeauthoryear{{Saito} et~al.,}{{Saito}
  et~al.}{2012b}]{2012A&A...537A.107S}
{Saito} R.~K.,  et~al., 2012b, \mn@doi [\aap] {10.1051/0004-6361/201118407},
  \href {https://ui.adsabs.harvard.edu/abs/2012A&A...537A.107S} {537, A107}

\bibitem[\protect\citeauthoryear{{Sch{\"o}nberner}, {Jacob}, {Steffen}  \&
  {Sandin}}{{Sch{\"o}nberner} et~al.}{2007}]{2007A&A...473..467S}
{Sch{\"o}nberner} D.,  {Jacob} R.,  {Steffen} M.,   {Sandin} C.,  2007, \mn@doi
  [\aap] {10.1051/0004-6361:20077437}, \href
  {https://ui.adsabs.harvard.edu/abs/2007A&A...473..467S} {473, 467}

\bibitem[\protect\citeauthoryear{Si{\'o}dmiak \& Tylenda}{Si{\'o}dmiak \&
  Tylenda}{2001}]{siodmiak2001analysis}
Si{\'o}dmiak N.,  Tylenda R.,  2001, Astronomy \& Astrophysics, 373, 1032

\bibitem[\protect\citeauthoryear{{Soker}}{{Soker}}{1998}]{1998ApJ...496..833S}
{Soker} N.,  1998, \mn@doi [\apj] {10.1086/305407}, \href
  {https://ui.adsabs.harvard.edu/abs/1998ApJ...496..833S} {496, 833}

\bibitem[\protect\citeauthoryear{Soker}{Soker}{2001}]{soker2001collimated}
Soker N.,  2001, The Astrophysical Journal, 558, 157

\bibitem[\protect\citeauthoryear{Soker \& Livio}{Soker \&
  Livio}{1989}]{soker1989interacting}
Soker N.,  Livio M.,  1989, The Astrophysical Journal, 339, 268

\bibitem[\protect\citeauthoryear{Soker \& Livio}{Soker \&
  Livio}{1994}]{soker1994disks}
Soker N.,  Livio M.,  1994, The Astrophysical Journal, 421, 219

\bibitem[\protect\citeauthoryear{Stanek, Udalski, Szyma{\'n}ski,
  Ka{\l}u{\.z}ny, Kubiak, Mateo  \& Krzemi{\'n}ski}{Stanek
  et~al.}{1997}]{stanek1997modeling}
Stanek K.,  Udalski A.,  Szyma{\'n}ski M.,  Ka{\l}u{\.z}ny J.,  Kubiak Z.~M.,
  Mateo M.,   Krzemi{\'n}ski W.,  1997, The Astrophysical Journal, 477, 163

\bibitem[\protect\citeauthoryear{Stanghellini, Corradi  \&
  Schwarz}{Stanghellini et~al.}{1993}]{stanghellini1993correlations}
Stanghellini L.,  Corradi R.~L.,   Schwarz H.~E.,  1993, Astronomy and
  astrophysics, 279, 521

\bibitem[\protect\citeauthoryear{Stanghellini, Villaver, Manchado  \&
  Guerrero}{Stanghellini et~al.}{2002b}]{stanghellini2002correlations}
Stanghellini L.,  Villaver E.,  Manchado A.,   Guerrero M.~A.,  2002b, The
  Astrophysical Journal, 576, 285

\bibitem[\protect\citeauthoryear{{Stanghellini}, {Villaver}, {Manchado}  \&
  {Guerrero}}{{Stanghellini} et~al.}{2002a}]{2002ApJ...576..285S}
{Stanghellini} L.,  {Villaver} E.,  {Manchado} A.,   {Guerrero} M.~A.,  2002a,
  \mn@doi [\apj] {10.1086/341340}, \href
  {https://ui.adsabs.harvard.edu/abs/2002ApJ...576..285S} {576, 285}

\bibitem[\protect\citeauthoryear{Stanghellini, Shaw  \& Villaver}{Stanghellini
  et~al.}{2016}]{stanghellini2016compact}
Stanghellini L.,  Shaw R.~A.,   Villaver E.,  2016, The Astrophysical Journal,
  830, 33

\bibitem[\protect\citeauthoryear{Stanghellini, Bucciarelli, Lattanzi  \&
  Morbidelli}{Stanghellini et~al.}{2020}]{stanghellini2020population}
Stanghellini L.,  Bucciarelli B.,  Lattanzi M.~G.,   Morbidelli R.,  2020, The
  Astrophysical Journal, 889, 21

\bibitem[\protect\citeauthoryear{{Steffen} \& {L{\'o}pez}}{{Steffen} \&
  {L{\'o}pez}}{2006}]{2006RMxAA..42...99S}
{Steffen} W.,  {L{\'o}pez} J.~A.,  2006, \rmxaa, \href
  {https://ui.adsabs.harvard.edu/abs/2006RMxAA..42...99S} {42, 99}

\bibitem[\protect\citeauthoryear{{Wareing}, {Zijlstra}  \& {O'Brien}}{{Wareing}
  et~al.}{2007}]{2007MNRAS.382.1233W}
{Wareing} C.~J.,  {Zijlstra} A.~A.,   {O'Brien} T.~J.,  2007, \mn@doi [\mnras]
  {10.1111/j.1365-2966.2007.12459.x}, \href
  {https://ui.adsabs.harvard.edu/abs/2007MNRAS.382.1233W} {382, 1233}

\bibitem[\protect\citeauthoryear{Wegg \& Gerhard}{Wegg \&
  Gerhard}{2013}]{wegg2013mapping}
Wegg C.,  Gerhard O.,  2013, Monthly Notices of the Royal Astronomical Society,
  435, 1874

\bibitem[\protect\citeauthoryear{{Weidmann} et~al.,}{{Weidmann}
  et~al.}{2018}]{2018A&A...614A.135W}
{Weidmann} W.,  et~al., 2018, \mn@doi [\aap] {10.1051/0004-6361/201731805},
  \href {https://ui.adsabs.harvard.edu/abs/2018A&A...614A.135W} {614, A135}

\bibitem[\protect\citeauthoryear{Weidmann et~al.,}{Weidmann
  et~al.}{2020}]{weidmann2020catalogue}
Weidmann W.~A.,  et~al., 2020, Astronomy \& Astrophysics, 640, A10

\bibitem[\protect\citeauthoryear{{Zhang} \& {Kwok}}{{Zhang} \&
  {Kwok}}{1998}]{1998ApJS..117..341Z}
{Zhang} C.~Y.,  {Kwok} S.,  1998, \mn@doi [\apjs] {10.1086/313118}, \href
  {https://ui.adsabs.harvard.edu/abs/1998ApJS..117..341Z} {117, 341}

\bibitem[\protect\citeauthoryear{Zoccali et~al.,}{Zoccali
  et~al.}{2014}]{zoccali2014giraffe}
Zoccali M.,  et~al., 2014, Astronomy \& Astrophysics, 562, A66

\makeatother
\end{thebibliography}



\clearpage
\onecolumn
\appendix
\section{Observing logs for the VLT (Table A1) and \emph{HST} (Table A2) imagery used}
\label{app:obs_logs}
\setlength{\tabcolsep}{15pt}
\renewcommand*{\arraystretch}{0.93}
\begin{table*}
\centering
\caption{Observing log of the VLT PNe imagery. Provided are PNG ID, narrow-band filter used, exposure time in seconds date of observation, official ESO observing program ID and the reported seeing in arecseconds (complete table is available online).}
\label{tab:vlt_log}
\begin{tabular}{lcrccc}
\hline
\hline
PN G &   Filter &  Exposure &        Date &        Prog. ID &  Seeing \\
\hline
PNG 000.1+02.6 &     [O III] &       6.0 &  2015-04-17 &   095.D-0270(A) &    0.62 \\
 PNG 000.1+04.3 &     [O III] &      10.0 &  2015-04-18 &   095.D-0270(A) &    1.23 \\
 PNG 000.1-02.3 &     [O III] &       5.0 &  2015-04-17 &   095.D-0270(A) &    1.16 \\
 PNG 000.2-01.9 &  H$\alpha$ &       5.0 &  2015-04-17 &   095.D-0270(A) &    0.63 \\
 PNG 000.2-04.6 &     [O III] &       5.0 &  2015-04-17 &   095.D-0270(A) &    0.76 \\
 PNG 000.3+06.9 &     [O III] &       6.0 &  2015-04-19 &   095.D-0270(A) &    1.06 \\
 PNG 000.3-04.6 &     [O III] &       5.0 &  2015-04-18 &   095.D-0270(A) &    1.32 \\
 PNG 000.4-01.9 &  H$\alpha$ &       5.0 &  2015-04-18 &   095.D-0270(A) &    1.07 \\
 PNG 000.4-02.9 &  H$\alpha$ &       4.0 &  2015-04-19 &   095.D-0270(A) &    0.83 \\
 PNG 000.7+03.2 &  H$\alpha$ &       5.0 &  2015-04-19 &   095.D-0270(A) &    0.83 \\
 .. & .. & .. & .. & .. & .. 
 \\
 .. & .. & .. & .. & .. & .. \\ \vspace{-0.15cm} \\
\hline
\end{tabular}
\end{table*}
\setlength{\tabcolsep}{15pt}
\begin{table*}
\centering
\caption{Observing logs of the \emph{HST} imagery used.Provided are the PN usual names or PNG IDs, the \emph{HST} narrow band filters used to make our measurements, the exposure time in seconds, date of observation, program ID and data set (complete table is available online).}
\label{tab:HST_log}
\begin{tabular}{lcrccc}
\hline
\hline
\\
Objects &   Filter &  Exposure &  Date &        Prog. ID &  Data Set \\
\hline
  M1-42 &  F502N &    0.5 &  2009-04-22 &   11185 &  ua2o010dm \\
  M1-42 &  F656N &    0.5 &  2009-04-22 &   11185 &  ua2o010bm \\
  M1-42 &  F502N &  600.0 &  2009-04-22 &   11185 &  ua2o010am \\
  M1-42 &  F656N &  300.0 &  2009-04-22 &   11185 &  ua2o0108m \\
  M1-42-REPEAT &  F502N &  600.0 &  2009-06-09 &   11185 &  ua2o1008m \\
  M1-42-REPEAT &  F656N &  300.0 &  2009-06-09 &   11185 &  ua2o1006m \\
    PK002-04D1 &  F656N &   20.0 &  2000-02-17 &    8345 &  u5hh4101r \\
    PK002-04D1 &  F656N &  140.0 &  2000-02-17 &    8345 &  u5hh4102r \\
    PK002-04D1 &  F656N &  400.0 &  2000-02-17 &    8345 &  u5hh4103r \\
    PK002-09D1 &  F656N &  230.0 &  2000-08-26 &    8345 &  u5hh5602r \\
    .. & .. & .. & .. & .. & .. 
 \\
 .. & .. & .. & .. & .. & .. \\ \vspace{-0.1cm} \\
 \hline
 \end{tabular}
\end{table*}
\section{The best optical PN images.}
\begin{figure*}
\caption{The best optical PN images from the VLT or \emph{HST} (narrow-band) and in a few cases, Pan-STARRS (broad-band) used in this study. The data origin is given at the top of each image along with the PNG ID. Images are generally that obtained with H$\alpha$ filters or otherwise indicated. Pan-STARRS images are in g-band, where used. Complete set of PN images is available online.}
\begin{subfigure}{\linewidth} 
\caption{PNe with angular sizes greater than 20 arcsec. Each image has a field of view of 30" $\times$ 30".}
  \includegraphics[width=.32\linewidth]{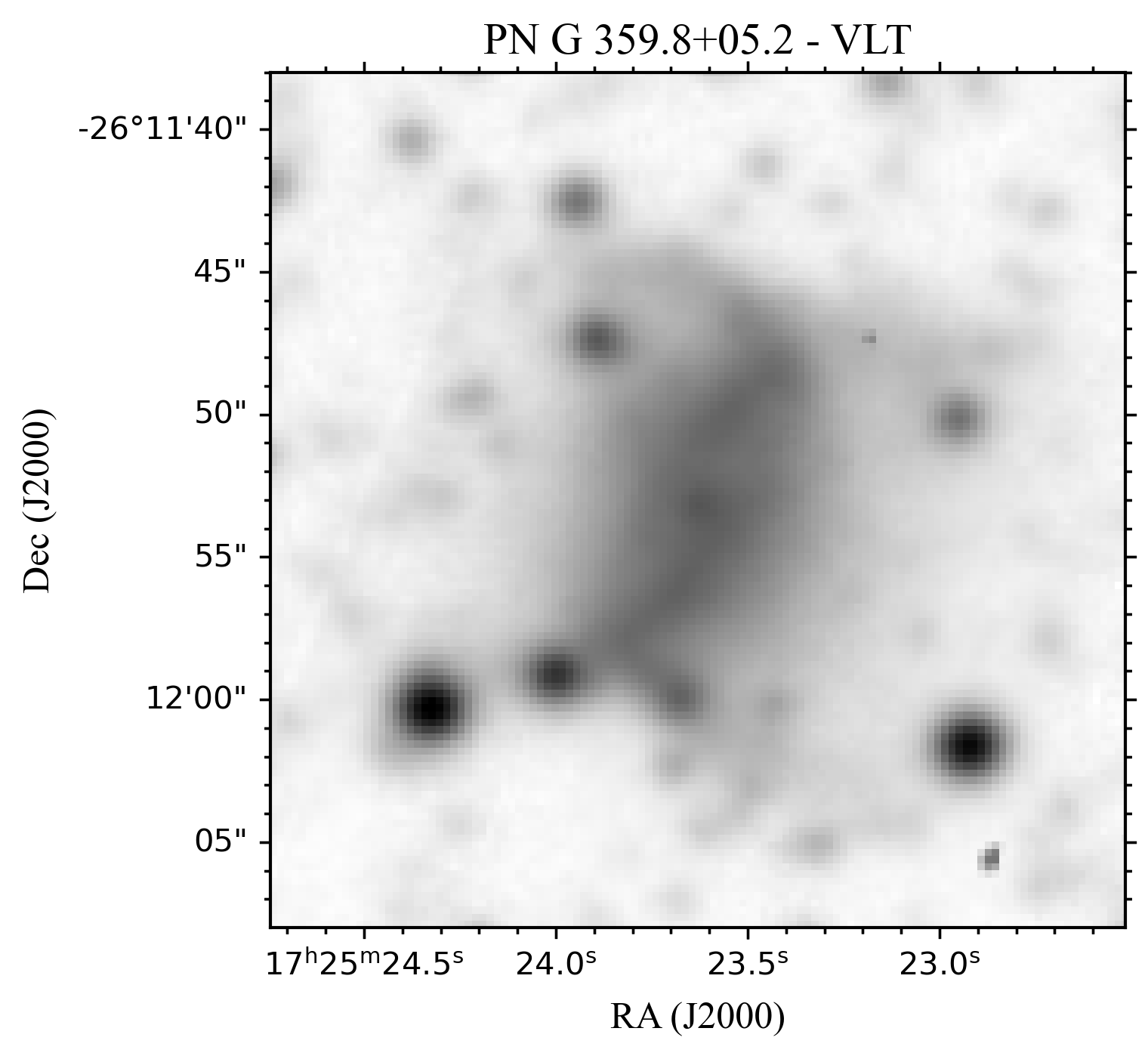}\hfill 
  \includegraphics[width=.32\linewidth]{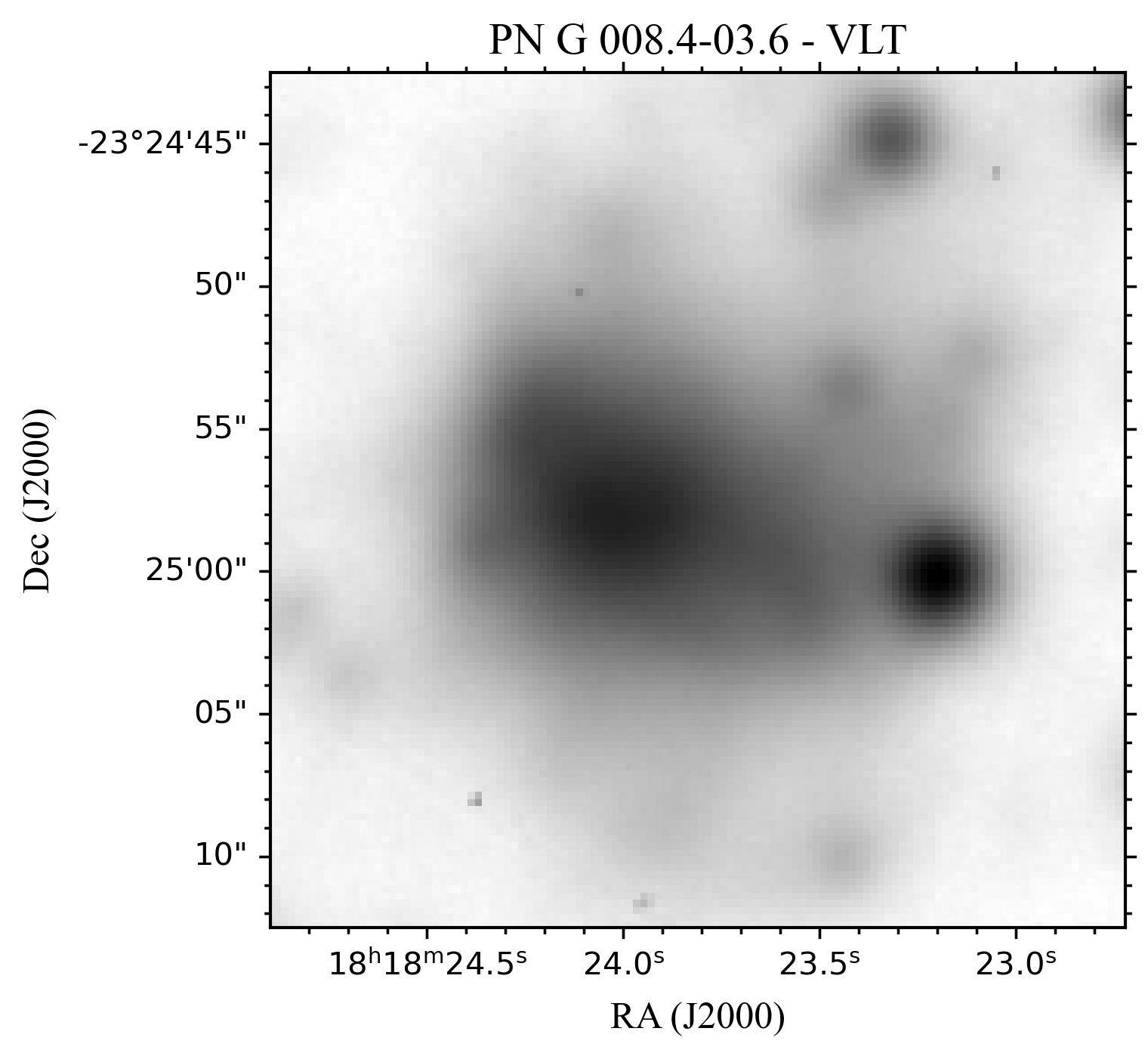}\hfill 
  \includegraphics[width=.32\linewidth]{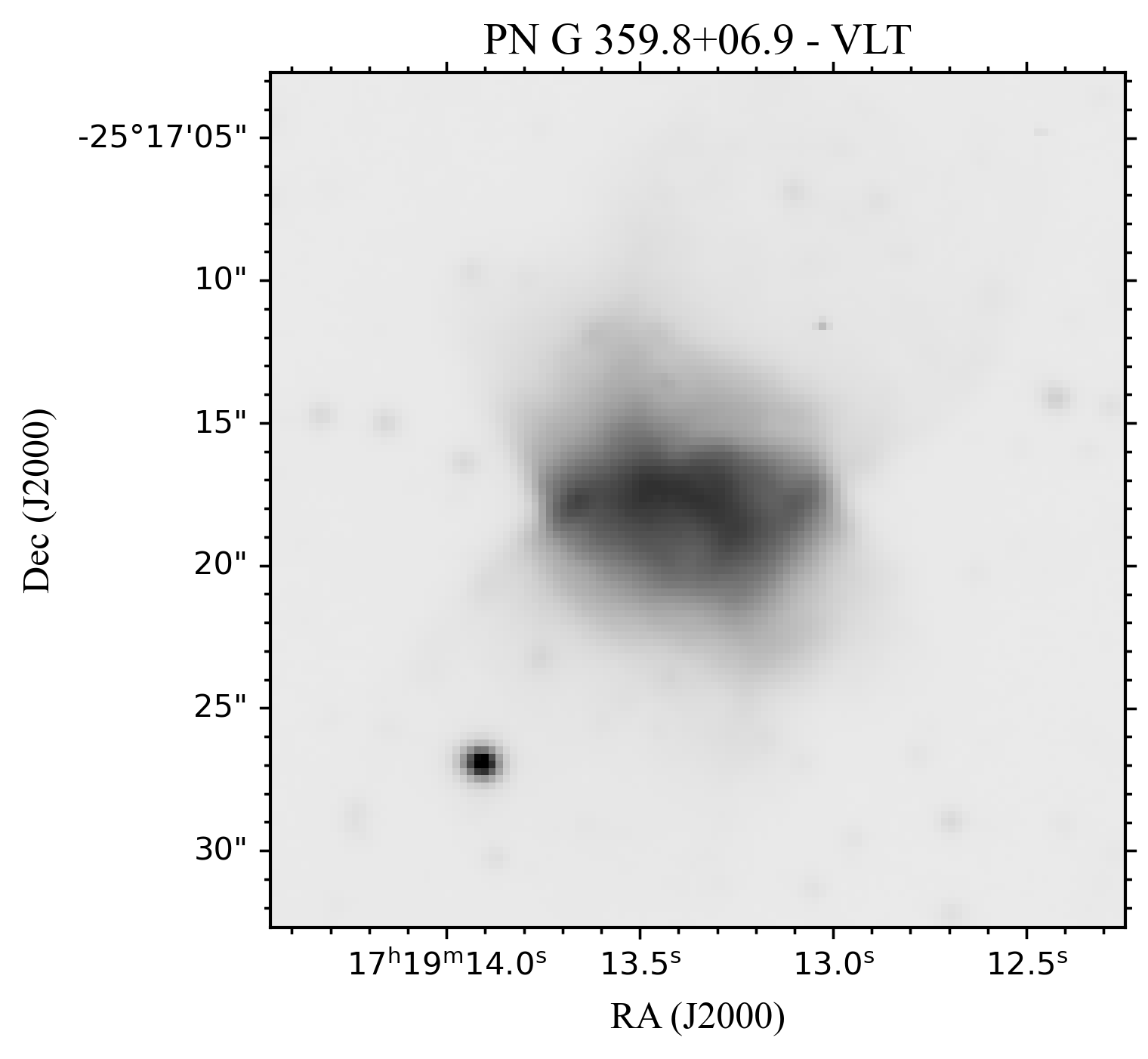}\hfill 
 \end{subfigure}
 .. ..
 \end{figure*}
\clearpage
\bsp
\label{lastpage}
\end{document}


\appendix
\onecolumn
\section{Observing logs for the VLT (Table A1) and \emph{HST} (Table A2) imagery used}
\label{app:obs_logs}
\setlength{\tabcolsep}{15pt}
\renewcommand*{\arraystretch}{0.93}
\begin{longtable}{lcrccc}
\caption{Observing log of the VLT PNe imagery. Provided are PNG ID, narrow-band filter used, exposure time in seconds date of observation, official ESO observing program ID and the reported seeing in arecseconds.}
\label{tab:vlt_log}\\
\hline
\hline
PN G &   Filter &  Exposure &        Date &        Prog. ID &  Seeing \\
\hline
\endfirsthead
\caption{Continued:}\\
\hline
\hline
PN G &   Filter &  Exposure &        Date &        Prog. ID &  Seeing \\
\hline
\endhead
\hline
\endfoot
PNG 000.1+02.6 &     [O III] &       6.0 &  2015-04-17 &   095.D-0270(A) &    0.62 \\
 PNG 000.1+04.3 &     [O III] &      10.0 &  2015-04-18 &   095.D-0270(A) &    1.23 \\
 PNG 000.1-02.3 &     [O III] &       5.0 &  2015-04-17 &   095.D-0270(A) &    1.16 \\
 PNG 000.2-01.9 &  H$\alpha$ &       5.0 &  2015-04-17 &   095.D-0270(A) &    0.63 \\
 PNG 000.2-04.6 &     [O III] &       5.0 &  2015-04-17 &   095.D-0270(A) &    0.76 \\
 PNG 000.3+06.9 &     [O III] &       6.0 &  2015-04-19 &   095.D-0270(A) &    1.06 \\
 PNG 000.3-04.6 &     [O III] &       5.0 &  2015-04-18 &   095.D-0270(A) &    1.32 \\
 PNG 000.4-01.9 &  H$\alpha$ &       5.0 &  2015-04-18 &   095.D-0270(A) &    1.07 \\
 PNG 000.4-02.9 &  H$\alpha$ &       4.0 &  2015-04-19 &   095.D-0270(A) &    0.83 \\
 PNG 000.7+03.2 &  H$\alpha$ &       5.0 &  2015-04-19 &   095.D-0270(A) &    0.83 \\
 PNG 000.7-02.7 &     [O III] &       5.0 &  2015-04-19 &   095.D-0270(A) &    0.67 \\
 PNG 000.7-03.7 &     [O III] &       5.0 &  2015-05-19 &   095.D-0270(A) &    1.05 \\
 PNG 000.7-07.4 &     [O III] &       7.0 &  2015-05-19 &   095.D-0270(A) &    1.26 \\
 PNG 000.9-02.0 &  H$\alpha$ &       4.0 &  2015-05-27 &   095.D-0270(A) &    0.74 \\
 PNG 000.9-04.8 &     [O III] &       5.0 &  2015-06-22 &   095.D-0270(A) &    0.57 \\
 PNG 001.1-01.6 &  H$\alpha$ &       5.0 &  2015-06-18 &   095.D-0270(A) &    1.47 \\
 PNG 001.2+02.1 &     [O III] &       7.0 &  2015-05-27 &   095.D-0270(A) &    0.80 \\
 PNG 001.2-03.0 &  H$\alpha$ &       5.0 &  2015-06-18 &   095.D-0270(A) &    1.24 \\
 PNG 001.3-01.2 &  H$\alpha$ &       4.0 &  2015-06-10 &   095.D-0270(A) &    0.76 \\
 PNG 001.3-01.2 &  H$\alpha$ &       4.0 &  2015-06-10 &   095.D-0270(A) &    0.86 \\
 PNG 001.3-01.2 &  H$\alpha$ &       4.0 &  2015-07-12 &   095.D-0270(A) &    0.84 \\
 PNG 001.3-01.2 &     [O III] &       4.0 &  2015-07-12 &   095.D-0270(A) &    0.70 \\
 PNG 001.4+05.3 &  H$\alpha$ &       5.0 &  2015-05-23 &   095.D-0270(A) &    1.01 \\
 PNG 001.6-01.3 &  H$\alpha$ &       4.0 &  2015-06-20 &   095.D-0270(A) &    0.87 \\
 PNG 001.7+05.7 &     [O III] &      10.0 &  2015-05-27 &   095.D-0270(A) &    0.73 \\
 PNG 001.7-04.4 &  H$\alpha$ &       4.0 &  2015-06-21 &   095.D-0270(A) &    0.76 \\
 PNG 002.0-06.2 &  H$\alpha$ &       5.0 &  2015-07-12 &   095.D-0270(A) &    0.67 \\
 PNG 002.1-02.2 &     [O III] &       5.0 &  2015-06-22 &   095.D-0270(A) &    0.79 \\
 PNG 002.1-04.2 &  H$\alpha$ &       5.0 &  2015-06-22 &   095.D-0270(A) &    0.64 \\
 PNG 002.2-09.4 &  H$\alpha$ &       5.0 &  2015-06-18 &   095.D-0270(A) &    0.99 \\
 PNG 002.2-09.4 &     [O III] &       5.0 &  2015-06-18 &   095.D-0270(A) &    1.18 \\
 PNG 002.3+02.2 &  H$\alpha$ &       4.0 &  2015-07-12 &   095.D-0270(A) &    1.02 \\
 PNG 002.3+02.2 &     [O III] &       4.0 &  2015-07-12 &   095.D-0270(A) &    0.76 \\
 PNG 002.3-03.4 &  H$\alpha$ &       5.0 &  2015-07-12 &   095.D-0270(A) &    1.36 \\
 PNG 002.5-01.7 &  H$\alpha$ &       4.0 &  2015-07-12 &   095.D-0270(A) &    0.64 \\
 PNG 002.5-01.7 &     [O III] &       4.0 &  2015-07-12 &   095.D-0270(A) &    0.63 \\
 PNG 002.6+02.1 &  H$\alpha$ &       7.0 &  2015-07-13 &   095.D-0270(A) &    1.09 \\
 PNG 002.7-04.8 &  H$\alpha$ &       5.0 &  2015-07-12 &   095.D-0270(A) &    1.02 \\
 PNG 002.8+01.7 &     [O III] &       4.0 &  2015-07-13 &   095.D-0270(A) &    0.91 \\
 PNG 002.8+01.8 &     [O III] &       4.0 &  2015-07-13 &   095.D-0270(A) &    0.84 \\
 PNG 002.9-03.9 &  H$\alpha$ &       5.0 &  2015-07-13 &   095.D-0270(A) &    0.71 \\
 PNG 003.1+03.4 &     [O III] &       4.0 &  2015-07-13 &   095.D-0270(A) &    0.97 \\
 PNG 003.2-06.2 &  H$\alpha$ &       5.0 &  2015-07-13 &   095.D-0270(A) &    0.83 \\
 PNG 003.6+03.1 &     [O III] &       4.0 &  2015-07-13 &   095.D-0270(A) &    0.95 \\
 PNG 003.6-02.3 &     [O III] &       4.0 &  2015-07-13 &   095.D-0270(A) &    0.73 \\
 PNG 003.7+07.9 &     [O III] &      10.0 &  2015-05-24 &   095.D-0270(A) &    0.94 \\
 PNG 003.7-04.6 &  H$\alpha$ &      55.0 &  2016-08-10 &   097.D-0024(A) &    0.75 \\
 PNG 003.7-04.6 &  H$\alpha$ &      55.0 &  2016-08-10 &   097.D-0024(A) &    0.75 \\
 PNG 003.7-04.6 &     [O III] &       1.0 &  2016-08-10 &   097.D-0024(A) &    0.92 \\
 PNG 003.7-04.6 &     [O III] &       1.0 &  2016-08-10 &   097.D-0024(A) &    0.87 \\
 PNG 003.7-04.6 &     [O III] &      28.0 &  2016-08-10 &   097.D-0024(A) &    0.80 \\
 PNG 003.7-04.6 &     [O III] &      28.0 &  2016-08-10 &   097.D-0024(A) &    0.80 \\
 PNG 003.8-04.3 &  H$\alpha$ &       5.0 &  2015-07-24 &   095.D-0270(A) &    0.97 \\
 PNG 003.9+01.6 &  H$\alpha$ &       6.0 &  2015-07-25 &   095.D-0270(A) &    0.81 \\
 PNG 003.9-02.3 &     [O III] &       4.0 &  2015-07-26 &   095.D-0270(A) &    0.94 \\
 PNG 003.9-03.1 &  H$\alpha$ &       5.0 &  2015-08-11 &   095.D-0270(A) &    0.97 \\
 PNG 004.0-03.0 &  H$\alpha$ &       5.0 &  2015-08-08 &   095.D-0270(A) &    1.32 \\
 PNG 004.1-03.8 &  H$\alpha$ &       5.0 &  2015-08-08 &   095.D-0270(A) &    0.94 \\
 PNG 004.2-03.2 &  H$\alpha$ &       6.0 &  2015-08-16 &   095.D-0270(A) &    0.82 \\
 PNG 004.2-04.3 &  H$\alpha$ &       5.0 &  2015-08-08 &   095.D-0270(A) &    1.39 \\
 PNG 004.2-04.3 &     [O III] &       5.0 &  2015-08-08 &   095.D-0270(A) &    1.05 \\
 PNG 004.2-04.3 &     [O III] &       5.0 &  2015-08-08 &   095.D-0270(A) &    0.98 \\
 PNG 004.2-05.9 &     [O III] &       4.0 &  2015-07-13 &   095.D-0270(A) &    1.09 \\
 PNG 004.3+01.8 &     [O III] &       4.0 &  2015-07-20 &   095.D-0270(A) &    0.60 \\
 PNG 004.6+06.0 &  H$\alpha$ &       5.0 &  2015-07-21 &   095.D-0270(A) &    0.99 \\
 PNG 004.6+06.0 &  H$\alpha$ &       5.0 &  2015-07-21 &   095.D-0270(A) &    1.00 \\
 PNG 004.6+06.0 &     [O III] &       5.0 &  2015-07-21 &   095.D-0270(A) &    1.04 \\
 PNG 004.8+02.0 &     [O III] &       4.0 &  2015-07-21 &   095.D-0270(A) &    1.19 \\
 PNG 004.8-05.0 &  H$\alpha$ &       5.0 &  2015-07-21 &   095.D-0270(A) &    0.85 \\
 PNG 005.0-03.9 &  H$\alpha$ &     150.0 &  2018-07-11 &  0101.D-0192(A) &    0.73 \\
 PNG 005.0-03.9 &     [O III] &     150.0 &  2018-07-11 &  0101.D-0192(A) &    0.71 \\
 PNG 005.2+05.6 &  H$\alpha$ &       5.0 &  2015-07-21 &   095.D-0270(A) &    0.72 \\
 PNG 005.2+05.6 &     [O III] &       5.0 &  2015-07-21 &   095.D-0270(A) &    0.82 \\
 PNG 005.2+05.6 &     [O III] &       5.0 &  2015-07-21 &   095.D-0270(A) &    0.76 \\
 PNG 005.5+06.1 &     [O III] &       5.0 &  2015-07-25 &   095.D-0270(A) &    1.11 \\
 PNG 005.5-04.0 &  H$\alpha$ &     150.0 &  2018-07-04 &  0101.D-0192(A) &    1.26 \\
 PNG 005.5-04.0 &     [O III] &     150.0 &  2018-07-04 &  0101.D-0192(A) &    1.21 \\
 PNG 005.8-06.1 &  H$\alpha$ &       4.0 &  2016-07-24 &   097.D-0024(A) &    0.58 \\
 PNG 005.8-06.1 &     [O III] &      60.0 &  2017-06-23 &   099.D-0163(A) &    1.10 \\
 PNG 006.1+08.3 &     [O III] &       4.0 &  2015-07-25 &   095.D-0270(A) &    0.91 \\
 PNG 006.3+03.3 &  H$\alpha$ &       5.0 &  2015-07-25 &   095.D-0270(A) &    1.06 \\
 PNG 006.3+03.3 &     [O III] &       5.0 &  2015-07-25 &   095.D-0270(A) &    0.83 \\
 PNG 006.3+04.4 &  H$\alpha$ &       7.0 &  2015-08-11 &   095.D-0270(A) &    1.04 \\
 PNG 006.4+02.0 &     [O III] &       4.0 &  2015-08-16 &   095.D-0270(A) &    0.75 \\
 PNG 006.4-04.6 &  H$\alpha$ &     150.0 &  2018-07-04 &  0101.D-0192(A) &    0.82 \\
 PNG 006.4-04.6 &     [O III] &     150.0 &  2018-07-04 &  0101.D-0192(A) &    1.33 \\
 PNG 006.8+02.3 &  H$\alpha$ &       7.0 &  2015-08-16 &   095.D-0270(A) &    0.86 \\
 PNG 006.8-03.4 &  H$\alpha$ &     150.0 &  2018-07-11 &  0101.D-0192(A) &    0.64 \\
 PNG 006.8-03.4 &     [O III] &     150.0 &  2018-07-11 &  0101.D-0192(A) &    0.65 \\
 PNG 007.0+06.3 &     [O III] &       4.0 &  2015-08-18 &   095.D-0270(A) &    1.01 \\
 PNG 007.0-06.8 &  H$\alpha$ &       5.0 &  2018-07-10 &  0101.D-0192(A) &    0.68 \\
 PNG 007.0-06.8 &     [O III] &      25.0 &  2018-07-10 &  0101.D-0192(A) &    0.75 \\
 PNG 007.5+07.4 &  H$\alpha$ &     180.0 &  2016-06-12 &   097.D-0024(A) &    0.97 \\
 PNG 007.5+07.4 &     [O III] &     180.0 &  2016-06-12 &   097.D-0024(A) &    0.99 \\
 PNG 007.6+06.9 &  H$\alpha$ &     150.0 &  2017-07-18 &   099.D-0163(A) &    1.42 \\
 PNG 007.6+06.9 &     [O III] &     150.0 &  2017-07-18 &   099.D-0163(A) &    1.43 \\
 PNG 007.8-03.7 &  H$\alpha$ &     180.0 &  2017-07-18 &   099.D-0163(A) &    1.36 \\
 PNG 007.8-03.7 &     [O III] &     180.0 &  2017-07-18 &   099.D-0163(A) &    1.42 \\
 PNG 007.8-04.4 &  H$\alpha$ &       6.0 &  2016-07-07 &   097.D-0024(A) &    0.59 \\
 PNG 007.8-04.4 &     [O III] &      10.0 &  2016-07-07 &   097.D-0024(A) &    0.71 \\
 PNG 008.2+06.8 &  H$\alpha$ &       6.0 &  2017-07-29 &   099.D-0163(A) &    0.89 \\
 PNG 008.2+06.8 &     [O III] &      90.0 &  2017-07-29 &   099.D-0163(A) &    1.01 \\
 PNG 008.4-03.6 &  H$\alpha$ &     180.0 &  2017-07-18 &   099.D-0163(A) &    1.29 \\
 PNG 008.4-03.6 &     [O III] &     180.0 &  2017-07-18 &   099.D-0163(A) &    1.62 \\
 PNG 008.6-02.6 &  H$\alpha$ &      60.0 &  2016-05-14 &   097.D-0024(A) &    0.96 \\
 PNG 008.6-02.6 &     [O III] &      50.0 &  2016-05-14 &   097.D-0024(A) &    0.85 \\
 PNG 009.4-09.8 &  H$\alpha$ &      30.0 &  2016-06-11 &   097.D-0024(A) &    0.61 \\
 PNG 009.4-09.8 &     [O III] &      30.0 &  2016-06-11 &   097.D-0024(A) &    0.63 \\
 PNG 009.8-04.6 &  H$\alpha$ &      90.0 &  2016-06-11 &   097.D-0024(A) &    0.78 \\
 PNG 009.8-04.6 &     [O III] &      60.0 &  2016-06-11 &   097.D-0024(A) &    0.78 \\
 PNG 350.5-05.0 &  H$\alpha$ &      40.0 &  2016-05-12 &   097.D-0024(A) &    1.12 \\
 PNG 350.5-05.0 &     [O III] &      40.0 &  2016-05-12 &   097.D-0024(A) &    0.99 \\
 PNG 351.1+04.8 &  H$\alpha$ &      14.0 &  2016-05-12 &   097.D-0024(A) &    1.60 \\
 PNG 351.1+04.8 &     [O III] &      42.0 &  2016-05-12 &   097.D-0024(A) &    1.54 \\
 PNG 351.2+05.2 &  H$\alpha$ &      13.0 &  2016-06-09 &   097.D-0024(A) &    1.13 \\
 PNG 351.2+05.2 &     [O III] &      46.0 &  2016-06-09 &   097.D-0024(A) &    1.72 \\
 PNG 351.6-06.2 &  H$\alpha$ &      40.0 &  2016-06-04 &   097.D-0024(A) &    0.89 \\
 PNG 351.6-06.2 &     [O III] &      40.0 &  2016-06-04 &   097.D-0024(A) &    1.00 \\
 PNG 351.9+09.0 &  H$\alpha$ &     150.0 &  2016-06-11 &   097.D-0024(A) &    0.95 \\
 PNG 351.9+09.0 &     [O III] &      55.0 &  2016-06-11 &   097.D-0024(A) &    0.94 \\
 PNG 351.9-01.9 &  H$\alpha$ &      26.0 &  2016-07-06 &   097.D-0024(A) &    1.06 \\
 PNG 351.9-01.9 &     [O III] &      91.0 &  2016-07-06 &   097.D-0024(A) &    1.03 \\
 PNG 352.0-04.6 &  H$\alpha$ &      17.0 &  2016-07-02 &   097.D-0024(A) &    0.58 \\
 PNG 352.0-04.6 &     [O III] &      60.0 &  2016-07-02 &   097.D-0024(A) &    0.59 \\
 PNG 352.1+05.1 &  H$\alpha$ &      16.0 &  2016-07-06 &   097.D-0024(A) &    1.69 \\
 PNG 352.1+05.1 &     [O III] &      25.0 &  2016-07-06 &   097.D-0024(A) &    1.69 \\
 PNG 352.6+03.0 &  H$\alpha$ &      15.0 &  2016-07-07 &   097.D-0024(A) &    0.81 \\
 PNG 352.6+03.0 &     [O III] &      45.0 &  2016-07-07 &   097.D-0024(A) &    0.85 \\
 PNG 353.2-05.2 &  H$\alpha$ &      45.0 &  2016-07-09 &   097.D-0024(A) &    0.99 \\
 PNG 353.2-05.2 &     [O III] &      66.0 &  2016-07-09 &   097.D-0024(A) &    0.88 \\
 PNG 353.3+06.3 &  H$\alpha$ &      19.0 &  2017-07-23 &   099.D-0163(A) &    0.82 \\
 PNG 353.3+06.3 &     [O III] &      20.0 &  2017-07-23 &   099.D-0163(A) &    0.85 \\
 PNG 353.7+06.3 &  H$\alpha$ &      25.0 &  2016-07-11 &   097.D-0024(A) &    0.81 \\
 PNG 353.7+06.3 &     [O III] &      30.0 &  2016-07-11 &   097.D-0024(A) &    0.77 \\
 PNG 354.5+03.3 &  H$\alpha$ &      21.0 &  2016-07-14 &   097.D-0024(A) &    0.74 \\
 PNG 354.5+03.3 &     [O III] &      20.0 &  2016-07-14 &   097.D-0024(A) &    0.75 \\
 PNG 354.9+03.5 &  H$\alpha$ &      22.0 &  2016-08-08 &   097.D-0024(A) &    0.78 \\
 PNG 354.9+03.5 &     [O III] &      90.0 &  2016-08-08 &   097.D-0024(A) &    0.78 \\
 PNG 355.1-06.9 &  H$\alpha$ &       6.0 &  2016-07-14 &   097.D-0024(A) &    0.85 \\
 PNG 355.1-06.9 &     [O III] &       7.0 &  2016-07-14 &   097.D-0024(A) &    0.88 \\
 PNG 355.4-02.4 &  H$\alpha$ &      14.0 &  2016-09-12 &   097.D-0024(A) &    0.96 \\
 PNG 355.4-02.4 &     [O III] &      45.0 &  2016-09-12 &   097.D-0024(A) &    0.82 \\
 PNG 355.6-02.7 &  H$\alpha$ &      30.0 &  2017-07-24 &   099.D-0163(A) &    1.66 \\
 PNG 355.6-02.7 &     [O III] &      25.0 &  2017-07-24 &   099.D-0163(A) &    1.34 \\
 PNG 355.9+03.6 &  H$\alpha$ &       2.0 &  2017-07-24 &   099.D-0163(A) &    1.21 \\
 PNG 355.9+03.6 &     [O III] &       9.0 &  2017-07-24 &   099.D-0163(A) &    1.21 \\
 PNG 355.9-04.2 &  H$\alpha$ &      38.0 &  2018-06-15 &  0101.D-0192(A) &    0.99 \\
 PNG 355.9-04.2 &     [O III] &      45.0 &  2018-06-15 &  0101.D-0192(A) &    0.91 \\
 PNG 356.1-03.3 &  H$\alpha$ &     150.0 &  2017-07-29 &   099.D-0163(A) &    0.78 \\
 PNG 356.1-03.3 &     [O III] &      90.0 &  2017-07-29 &   099.D-0163(A) &    0.84 \\
 PNG 356.3-06.2 &  H$\alpha$ &      70.0 &  2017-07-31 &   099.D-0163(A) &    1.44 \\
 PNG 356.3-06.2 &     [O III] &      70.0 &  2017-07-31 &   099.D-0163(A) &    1.23 \\
 PNG 356.5-03.6 &  H$\alpha$ &     180.0 &  2017-07-31 &   099.D-0163(A) &    1.36 \\
 PNG 356.5-03.6 &     [O III] &     180.0 &  2017-07-31 &   099.D-0163(A) &    1.18 \\
 PNG 356.8+03.3 &  H$\alpha$ &      16.0 &  2018-06-16 &  0101.D-0192(A) &    0.79 \\
 PNG 356.8+03.3 &     [O III] &     120.0 &  2018-06-16 &  0101.D-0192(A) &    0.76 \\
 PNG 356.8-05.4 &  H$\alpha$ &     120.0 &  2017-07-31 &   099.D-0163(A) &    1.03 \\
 PNG 356.8-05.4 &     [O III] &     120.0 &  2017-07-31 &   099.D-0163(A) &    1.02 \\
 PNG 356.9+04.4 &  H$\alpha$ &      10.0 &  2018-06-18 &  0101.D-0192(A) &    1.81 \\
 PNG 356.9+04.4 &     [O III] &       9.0 &  2018-06-18 &  0101.D-0192(A) &    1.81 \\
 PNG 357.0+02.4 &  H$\alpha$ &     120.0 &  2017-08-01 &   099.D-0163(A) &    0.96 \\
 PNG 357.0+02.4 &     [O III] &      85.0 &  2017-08-01 &   099.D-0163(A) &    1.08 \\
 PNG 357.1+03.6 &  H$\alpha$ &       4.0 &  2018-05-09 &  0101.D-0192(A) &    1.25 \\
 PNG 357.1+03.6 &  H$\alpha$ &      25.0 &  2018-05-09 &  0101.D-0192(A) &    1.32 \\
 PNG 357.1+03.6 &  H$\alpha$ &       4.0 &  2018-05-09 &  0101.D-0192(A) &    1.27 \\
 PNG 357.1+03.6 &     [O III] &      60.0 &  2018-05-09 &  0101.D-0192(A) &    1.24 \\
 PNG 357.1+04.4 &  H$\alpha$ &     150.0 &  2018-05-09 &  0101.D-0192(A) &    1.05 \\
 PNG 357.1+04.4 &     [O III] &     150.0 &  2018-05-09 &  0101.D-0192(A) &    1.11 \\
 PNG 357.1-04.7 &  H$\alpha$ &      12.0 &  2017-09-11 &   099.D-0163(A) &    1.24 \\
 PNG 357.1-04.7 &     [O III] &     120.0 &  2017-09-11 &   099.D-0163(A) &    1.23 \\
 PNG 357.2+02.0 &  H$\alpha$ &      57.0 &  2016-06-28 &   097.D-0024(A) &    1.02 \\
 PNG 357.2+02.0 &     [O III] &      60.0 &  2016-06-28 &   097.D-0024(A) &    1.08 \\
 PNG 357.3+04.0 &  H$\alpha$ &      30.0 &  2016-05-14 &   097.D-0024(A) &    0.78 \\
 PNG 357.3+04.0 &     [O III] &      65.0 &  2016-05-14 &   097.D-0024(A) &    0.78 \\
 PNG 357.5+03.1 &  H$\alpha$ &      20.0 &  2016-06-06 &   097.D-0024(A) &    1.23 \\
 PNG 357.5+03.1 &  H$\alpha$ &      20.0 &  2016-06-06 &   097.D-0024(A) &    1.10 \\
 PNG 357.5+03.1 &  H$\alpha$ &      40.0 &  2016-06-06 &   097.D-0024(A) &    1.27 \\
 PNG 357.5+03.1 &  H$\alpha$ &      30.0 &  2016-06-06 &   097.D-0024(A) &    2.53 \\
 PNG 357.5+03.1 &     [O III] &      10.0 &  2016-06-06 &   097.D-0024(A) &    2.22 \\
 PNG 357.5+03.1 &     [O III] &      60.0 &  2016-06-06 &   097.D-0024(A) &    1.80 \\
 PNG 357.5+03.2 &  H$\alpha$ &     240.0 &  2016-06-11 &   097.D-0024(A) &    0.74 \\
 PNG 357.5+03.2 &     [O III] &      60.0 &  2016-06-11 &   097.D-0024(A) &    0.77 \\
 PNG 357.6-03.3 &  H$\alpha$ &     130.0 &  2016-06-11 &   097.D-0024(A) &    0.97 \\
 PNG 357.6-03.3 &     [O III] &     120.0 &  2016-06-11 &   097.D-0024(A) &    1.03 \\
 PNG 357.9-03.8 &  H$\alpha$ &      50.0 &  2016-06-28 &   097.D-0024(A) &    1.13 \\
 PNG 357.9-03.8 &     [O III] &      70.0 &  2016-06-28 &   097.D-0024(A) &    1.15 \\
 PNG 357.9-05.1 &  H$\alpha$ &      28.0 &  2016-06-15 &   097.D-0024(A) &    0.85 \\
 PNG 357.9-05.1 &     [O III] &      30.0 &  2016-06-15 &   097.D-0024(A) &    0.85 \\
 PNG 358.0+09.3 &  H$\alpha$ &      16.0 &  2016-07-02 &   097.D-0024(A) &    0.52 \\
 PNG 358.0+09.3 &     [O III] &      60.0 &  2016-07-02 &   097.D-0024(A) &    0.51 \\
 PNG 358.2+03.5 &  H$\alpha$ &      30.0 &  2016-07-06 &   097.D-0024(A) &    1.02 \\
 PNG 358.2+03.5 &     [O III] &      35.0 &  2016-07-06 &   097.D-0024(A) &    1.09 \\
 PNG 358.2+04.2 &  H$\alpha$ &      22.0 &  2016-07-06 &   097.D-0024(A) &    1.30 \\
 PNG 358.2+04.2 &     [O III] &      35.0 &  2016-07-06 &   097.D-0024(A) &    1.57 \\
 PNG 358.5+02.9 &  H$\alpha$ &      45.0 &  2016-07-07 &   097.D-0024(A) &    0.89 \\
 PNG 358.5+02.9 &     [O III] &      70.0 &  2016-07-07 &   097.D-0024(A) &    0.79 \\
 PNG 358.5-04.2 &  H$\alpha$ &       5.0 &  2016-07-07 &   097.D-0024(A) &    0.75 \\
 PNG 358.5-04.2 &     [O III] &       7.0 &  2016-07-07 &   097.D-0024(A) &    0.75 \\
 PNG 358.6+07.8 &  H$\alpha$ &      35.0 &  2016-07-09 &   097.D-0024(A) &    0.79 \\
 PNG 358.6+07.8 &     [O III] &      35.0 &  2016-07-09 &   097.D-0024(A) &    0.76 \\
 PNG 358.6-05.5 &  H$\alpha$ &      62.0 &  2016-07-09 &   097.D-0024(A) &    0.70 \\
 PNG 358.6-05.5 &     [O III] &      90.0 &  2016-07-09 &   097.D-0024(A) &    0.71 \\
 PNG 358.7+05.2 &  H$\alpha$ &      35.0 &  2016-07-11 &   097.D-0024(A) &    0.94 \\
 PNG 358.7+05.2 &     [O III] &      60.0 &  2016-07-11 &   097.D-0024(A) &    1.15 \\
 PNG 358.8+03.0 &  H$\alpha$ &      11.0 &  2016-07-28 &   097.D-0024(A) &    0.51 \\
 PNG 358.8+03.0 &     [O III] &      30.0 &  2016-07-28 &   097.D-0024(A) &    0.51 \\
 PNG 358.9+03.4 &  H$\alpha$ &      35.0 &  2018-06-10 &  0101.D-0192(A) &    1.17 \\
 PNG 358.9+03.4 &     [O III] &      45.0 &  2018-06-10 &  0101.D-0192(A) &    1.11 \\
 PNG 359.0-04.1 &  H$\alpha$ &      50.0 &  2016-08-08 &   097.D-0024(A) &    0.80 \\
 PNG 359.0-04.1 &     [O III] &      40.0 &  2016-08-08 &   097.D-0024(A) &    0.93 \\
 PNG 359.1-02.9 &  H$\alpha$ &      40.0 &  2016-09-27 &   097.D-0024(A) &    0.76 \\
 PNG 359.1-02.9 &     [O III] &      98.0 &  2016-09-27 &   097.D-0024(A) &    0.72 \\
 PNG 359.2+04.7 &  H$\alpha$ &      91.0 &  2016-09-12 &   097.D-0024(A) &    0.72 \\
 PNG 359.2+04.7 &     [O III] &      60.0 &  2016-09-12 &   097.D-0024(A) &    0.72 \\
 PNG 359.3-01.8 &  H$\alpha$ &      90.0 &  2018-06-10 &  0101.D-0192(A) &    0.92 \\
 PNG 359.3-01.8 &     [O III] &     120.0 &  2018-06-10 &  0101.D-0192(A) &    0.95 \\
 PNG 359.6-04.8 &  H$\alpha$ &     150.0 &  2018-05-09 &  0101.D-0192(A) &    1.29 \\
 PNG 359.6-04.8 &     [O III] &     150.0 &  2018-05-09 &  0101.D-0192(A) &    1.31 \\
 PNG 359.7-01.8 &  H$\alpha$ &     150.0 &  2018-06-10 &  0101.D-0192(A) &    0.80 \\
 PNG 359.7-01.8 &     [O III] &     150.0 &  2018-06-10 &  0101.D-0192(A) &    0.83 \\
 PNG 359.8+02.4 &  H$\alpha$ &      32.0 &  2018-06-12 &  0101.D-0192(A) &    0.97 \\
 PNG 359.8+02.4 &     [O III] &     120.0 &  2018-06-12 &  0101.D-0192(A) &    0.98 \\
 PNG 359.8+03.7 &  H$\alpha$ &     100.0 &  2018-06-16 &  0101.D-0192(A) &    0.78 \\
 PNG 359.8+03.7 &     [O III] &     150.0 &  2018-06-16 &  0101.D-0192(A) &    0.72 \\
 PNG 359.8+05.2 &  H$\alpha$ &     150.0 &  2018-06-17 &  0101.D-0192(A) &    1.04 \\
 PNG 359.8+05.2 &     [O III] &     150.0 &  2018-06-17 &  0101.D-0192(A) &    1.16 \\
 PNG 359.8+05.6 &  H$\alpha$ &       6.0 &  2018-07-11 &  0101.D-0192(A) &    0.75 \\
 PNG 359.8+05.6 &     [O III] &      10.0 &  2018-07-11 &  0101.D-0192(A) &    0.71 \\
 PNG 359.8+06.9 &  H$\alpha$ &     150.0 &  2018-07-10 &  0101.D-0192(A) &    0.80 \\
 PNG 359.8+06.9 &     [O III] &     150.0 &  2018-07-10 &  0101.D-0192(A) &    0.81 \\
 PNG 359.8-07.2 &  H$\alpha$ &      26.0 &  2018-05-09 &  0101.D-0192(A) &    1.48 \\
 PNG 359.8-07.2 &     [O III] &      12.0 &  2018-05-09 &  0101.D-0192(A) &    1.33 \\
 PNG 359.9-04.5 &  H$\alpha$ &      24.0 &  2018-06-10 &  0101.D-0192(A) &    1.11 \\
 PNG 359.9-04.5 &     [O III] &      36.0 &  2018-06-10 &  0101.D-0192(A) &    1.10 \\
\hline
\end{longtable}
\setlength{\tabcolsep}{15pt}
\begin{longtable}{lcrccc}
\caption{Observing logs of the \emph{HST} imagery used.Provided are the PN usual names or PNG IDs, the \emph{HST} narrow band filters used to make our measurements, the exposure time in seconds, date of observation, program ID and Data set.}
\label{tab:HST_log}\\
\hline
\hline
Objects &   Filter &  Exposure &        Date &        Prog. ID &  Data Set \\
\hline
\endfirsthead
\caption{Continued:}\\
\hline
\hline
Objects &   Filter &  Exposure &        Date &        Prog. ID &  Data Set \\
\hline
\endhead
\hline
\endfoot
  M1-42 &  F502N &    0.5 &  2009-04-22 &   11185 &  ua2o010dm \\
         M1-42 &  F656N &    0.5 &  2009-04-22 &   11185 &  ua2o010bm \\
         M1-42 &  F502N &  600.0 &  2009-04-22 &   11185 &  ua2o010am \\
         M1-42 &  F656N &  300.0 &  2009-04-22 &   11185 &  ua2o0108m \\
  M1-42-REPEAT &  F502N &  600.0 &  2009-06-09 &   11185 &  ua2o1008m \\
  M1-42-REPEAT &  F656N &  300.0 &  2009-06-09 &   11185 &  ua2o1006m \\
    PK002-04D1 &  F656N &   20.0 &  2000-02-17 &    8345 &  u5hh4101r \\
    PK002-04D1 &  F656N &  140.0 &  2000-02-17 &    8345 &  u5hh4102r \\
    PK002-04D1 &  F656N &  400.0 &  2000-02-17 &    8345 &  u5hh4103r \\
    PK002-09D1 &  F656N &  230.0 &  2000-08-26 &    8345 &  u5hh5602r \\
    PK002-09D1 &  F656N &   20.0 &  2000-08-26 &    8345 &  u5hh5601r \\
    PK006+02D5 &  F656N &  300.0 &  2000-03-26 &    8345 &  u5hh6902r \\
    PK006+02D5 &  F656N &   20.0 &  2000-03-26 &    8345 &  u5hh6901r \\
    PK007-04D1 &  F656N &   20.0 &  2000-03-25 &    8345 &  u5hh0701r \\
    PK007-04D1 &  F656N &  300.0 &  2000-03-25 &    8345 &  u5hh0702m \\
    PK355-04D2 &  F656N &  300.0 &  2000-08-22 &    8345 &  u5hh3702r \\
    PK355-04D2 &  F656N &   20.0 &  2000-08-22 &    8345 &  u5hh3701r \\
 PNG001.2+02.1 &  F656N &  140.0 &  2003-06-20 &    9356 &  u6mg1001m \\
 PNG001.2+02.1 &  F502N &  140.0 &  2003-06-20 &    9356 &  u6mg1004m \\
 PNG001.7-04.4 &  F656N &  100.0 &  2002-08-13 &    9356 &  u6mg1301m \\
 PNG001.7-04.4 &  F502N &  140.0 &  2002-08-13 &    9356 &  u6mg1304m \\
 PNG002.3-03.4 &  F502N &  140.0 &  2003-04-26 &    9356 &  u6mg0504m \\
 PNG002.3-03.4 &  F656N &  140.0 &  2003-04-26 &    9356 &  u6mg0501m \\
 PNG002.8+01.7 &  F656N &  140.0 &  2003-05-02 &    9356 &  u6mg0401m \\
 PNG002.9-03.9 &  F656N &  140.0 &  2003-04-28 &    9356 &  u6mg0601m \\
 PNG002.9-03.9 &  F502N &  140.0 &  2003-04-28 &    9356 &  u6mg0604m \\
 PNG003.1+03.4 &  F656N &  140.0 &  2002-07-02 &    9356 &  u6mg1401m \\
 PNG003.6+03.1 &  F656N &  140.0 &  2003-04-27 &    9356 &  u6mg1501m \\
 PNG003.6+03.1 &  F502N &  140.0 &  2003-04-27 &    9356 &  u6mg1504m \\
 PNG003.9-03.1 &  F502N &  140.0 &  2003-04-27 &    9356 &  u6mg1704m \\
 PNG003.9-03.1 &  F656N &  140.0 &  2003-04-27 &    9356 &  u6mg1701m \\
 PNG004.0-03.0 &  F502N &   80.0 &  2003-05-21 &    9356 &  u6mg5904m \\
 PNG004.0-03.0 &  F656N &  100.0 &  2003-05-21 &    9356 &  u6mg5901m \\
 PNG004.1-03.8 &  F502N &  140.0 &  2002-09-11 &    9356 &  u6mg1804m \\
 PNG004.1-03.8 &  F656N &  140.0 &  2002-09-11 &    9356 &  u6mg1801n \\
 PNG004.8+02.0 &  F656N &  200.0 &  2002-08-31 &    9356 &  u6mg1901m \\
 PNG004.8+02.0 &  F502N &  200.0 &  2002-08-31 &    9356 &  u6mg1904m \\
 PNG006.1+08.3 &  F502N &   80.0 &  2002-07-07 &    9356 &  u6mg2204m \\
 PNG006.1+08.3 &  F656N &  100.0 &  2002-07-07 &    9356 &  u6mg2201m \\
 PNG006.3+04.4 &  F656N &  140.0 &  2003-04-03 &    9356 &  u6mg6001m \\
 PNG006.3+04.4 &  F502N &  140.0 &  2003-04-03 &    9356 &  u6mg6004m \\
 PNG006.4+02.0 &  F656N &   80.0 &  2003-03-19 &    9356 &  u6mg2301m \\
 PNG006.4+02.0 &  F502N &   80.0 &  2003-03-19 &    9356 &  u6mg2304m \\
 PNG008.2+06.8 &  F502N &  230.0 &  2003-03-18 &    9356 &  u6mg2604m \\
 PNG008.2+06.8 &  F656N &  100.0 &  2003-03-18 &    9356 &  u6mg2601m \\
 PNG008.6-02.6 &  F502N &  140.0 &  2002-09-15 &    9356 &  u6mg4704m \\
 PNG008.6-02.6 &  F656N &  140.0 &  2002-09-15 &    9356 &  u6mg4701m \\
 PNG351.1+04.8 &  F656N &   80.0 &  2003-06-06 &    9356 &  u6mg2901m \\
 PNG351.1+04.8 &  F502N &   80.0 &  2003-06-06 &    9356 &  u6mg2904m \\
 PNG351.9-01.9 &  F656N &  100.0 &  2003-05-27 &    9356 &  u6mg4801m \\
 PNG351.9-01.9 &  F502N &  140.0 &  2003-05-27 &    9356 &  u6mg4804m \\
 PNG352.6+03.0 &  F656N &  100.0 &  2002-08-12 &    9356 &  u6mg3001m \\
 PNG352.6+03.0 &  F502N &  140.0 &  2002-08-12 &    9356 &  u6mg3004m \\
 PNG354.5+03.3 &  F502N &  140.0 &  2003-05-05 &    9356 &  u6mg5004m \\
 PNG354.5+03.3 &  F656N &  140.0 &  2003-05-05 &    9356 &  u6mg5001m \\
 PNG354.9+03.5 &  F656N &  140.0 &  2002-10-15 &    9356 &  u6mg4901m \\
 PNG354.9+03.5 &  F502N &  200.0 &  2002-10-15 &    9356 &  u6mg4904m \\
 PNG355.4-02.4 &  F656N &  100.0 &  2002-08-11 &    9356 &  u6mg3101m \\
 PNG355.4-02.4 &  F502N &   80.0 &  2002-08-11 &    9356 &  u6mg3104m \\
 PNG355.9+03.6 &  F656N &  140.0 &  2003-06-21 &    9356 &  u6mg3301m \\
 PNG355.9+03.6 &  F502N &  140.0 &  2003-06-21 &    9356 &  u6mg3304m \\
 PNG356.1-03.3 &  F656N &  140.0 &  2003-03-27 &    9356 &  u6mg3401m \\
 PNG356.1-03.3 &  F502N &  140.0 &  2003-03-27 &    9356 &  u6mg3404m \\
 PNG356.5-03.6 &  F656N &  180.0 &  2002-09-19 &    9356 &  u6mg4101m \\
 PNG356.5-03.6 &  F502N &  200.0 &  2002-09-19 &    9356 &  u6mg4104m \\
 PNG356.8+03.3 &  F656N &  140.0 &  2002-07-16 &    9356 &  u6mg3501m \\
 PNG356.9+04.4 &  F656N &  140.0 &  2003-04-18 &    9356 &  u6mg5601m \\
 PNG356.9+04.4 &  F502N &  140.0 &  2003-04-18 &    9356 &  u6mg5604m \\
 PNG357.1-04.7 &  F656N &  100.0 &  2003-06-20 &    9356 &  u6mg3601m \\
 PNG357.1-04.7 &  F502N &  140.0 &  2003-06-20 &    9356 &  u6mg3604m \\
 PNG357.2+02.0 &  F502N &  140.0 &  2002-08-14 &    9356 &  u6mg4204m \\
 PNG357.2+02.0 &  F656N &  140.0 &  2002-08-14 &    9356 &  u6mg4201m \\
 PNG358.5+02.9 &  F656N &  140.0 &  2003-05-02 &    9356 &  u6mg5101m \\
 PNG358.5+02.9 &  F502N &  140.0 &  2003-05-02 &    9356 &  u6mg5104m \\
 PNG358.5-04.2 &  F656N &   80.0 &  2003-06-20 &    9356 &  u6mg3801m \\
 PNG358.5-04.2 &  F502N &   80.0 &  2003-06-20 &    9356 &  u6mg3804m \\
 PNG358.7+05.2 &  F656N &  140.0 &  2002-09-19 &    9356 &  u6mg3901m \\
 PNG358.9+03.4 &  F656N &  100.0 &  2003-06-20 &    9356 &  u6mg5301m \\
 PNG358.9+03.4 &  F502N &  140.0 &  2003-06-20 &    9356 &  u6mg5304m \\
 PNG359.2+04.7 &  F656N &  140.0 &  2003-04-01 &    9356 &  u6mg5201m \\
 PNG359.2+04.7 &  F502N &  200.0 &  2003-03-31 &    9356 &  u6mg5204m \\
\end{longtable}


\appendix
\vspace{-1cm}
\setcounter{section}{1}
\onecolumn
\section{The best optical PN images.}
\begin{figure}
\caption{The best optical PN images from the VLT or \emph{HST} (narrow-band) and in a few cases, Pan-STARRS (broad-band) used in this study. The data origin is given at the top of each image along with the PNG ID. Images are generally that obtained with H$\alpha$ filters or otherwise indicated. Pan-STARRS images are in g-band, where used.}\vspace{-0.1cm}
\begin{subfigure}{\linewidth} 
\caption{PNe with angular sizes greater than 20 arcsec. Each image has a field of view of 30" $\times$ 30".}\vspace{-0.1cm}
 \includegraphics[width=.32\linewidth]{PN_images/PNG008.4-03.6_2d_.png}\hfill
  \includegraphics[width=.32\linewidth]{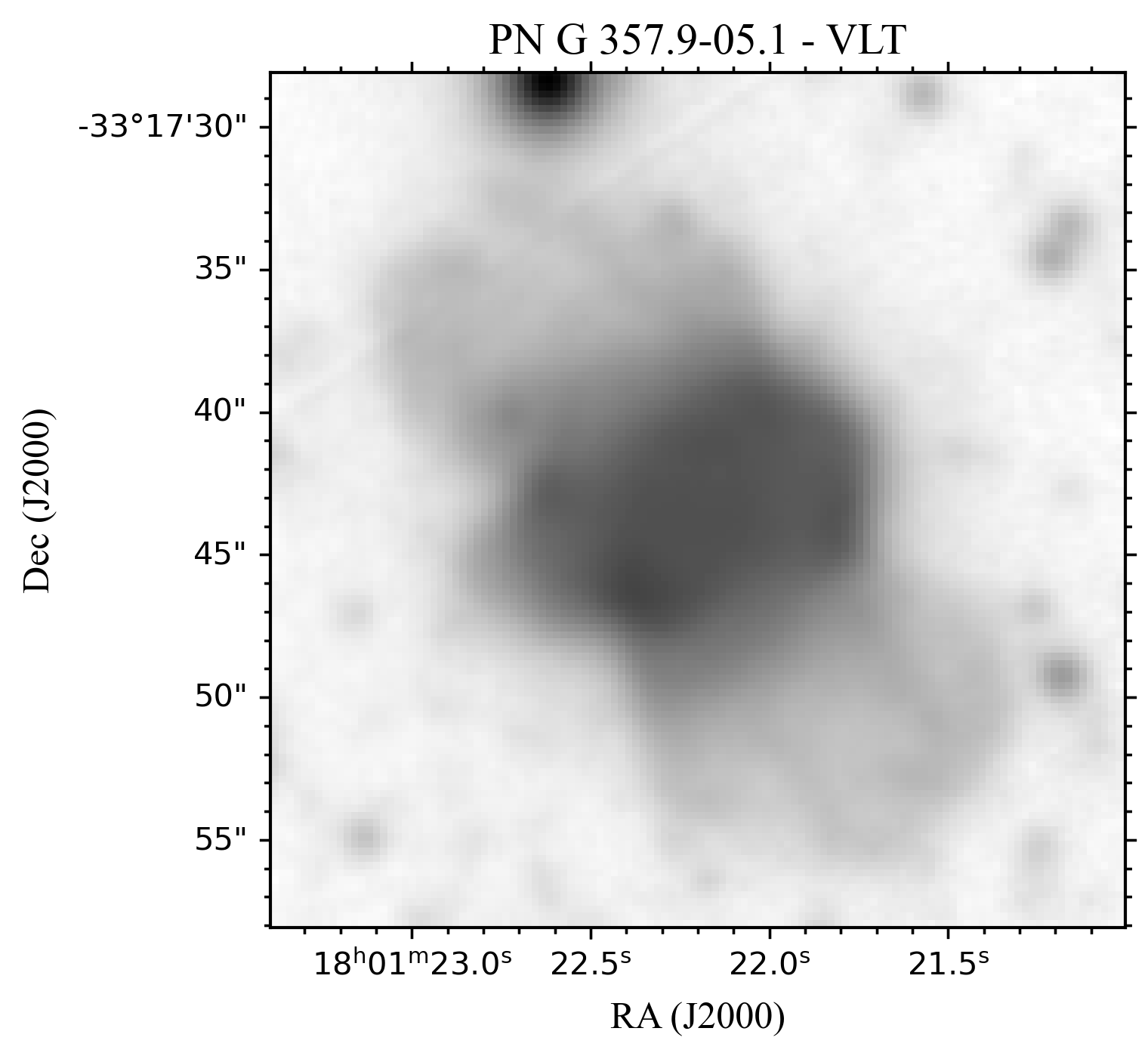}\hfill
  \includegraphics[width=.32\linewidth]{PN_images/PNG359.8+05.2_2d.png}\hfill
 \end{subfigure}
 \begin{subfigure}{\linewidth} 
  \includegraphics[width=.32\linewidth]{PN_images/PNG359.8+06.9_2d.png}\hfill
 \end{subfigure}\par\medskip 
 \begin{subfigure}{\linewidth} 
 \caption{PNe with angular sizes between 10 and 20 arcsec. Each image has a field of view of 20" $\times$ 20".}\vspace{-0.1cm}
  \includegraphics[width=.32\linewidth]{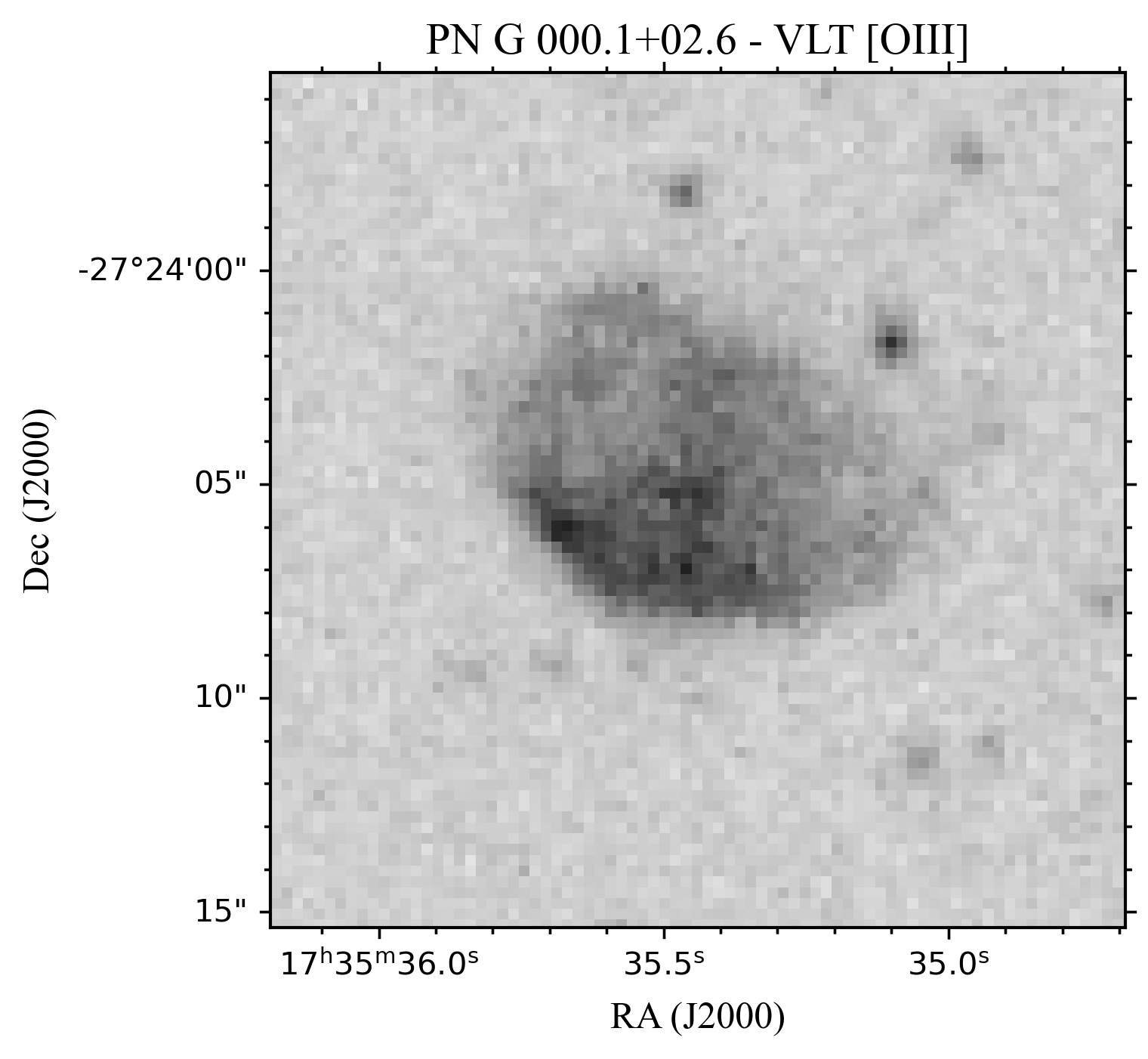}\hfill 
  \includegraphics[width=.32\linewidth]{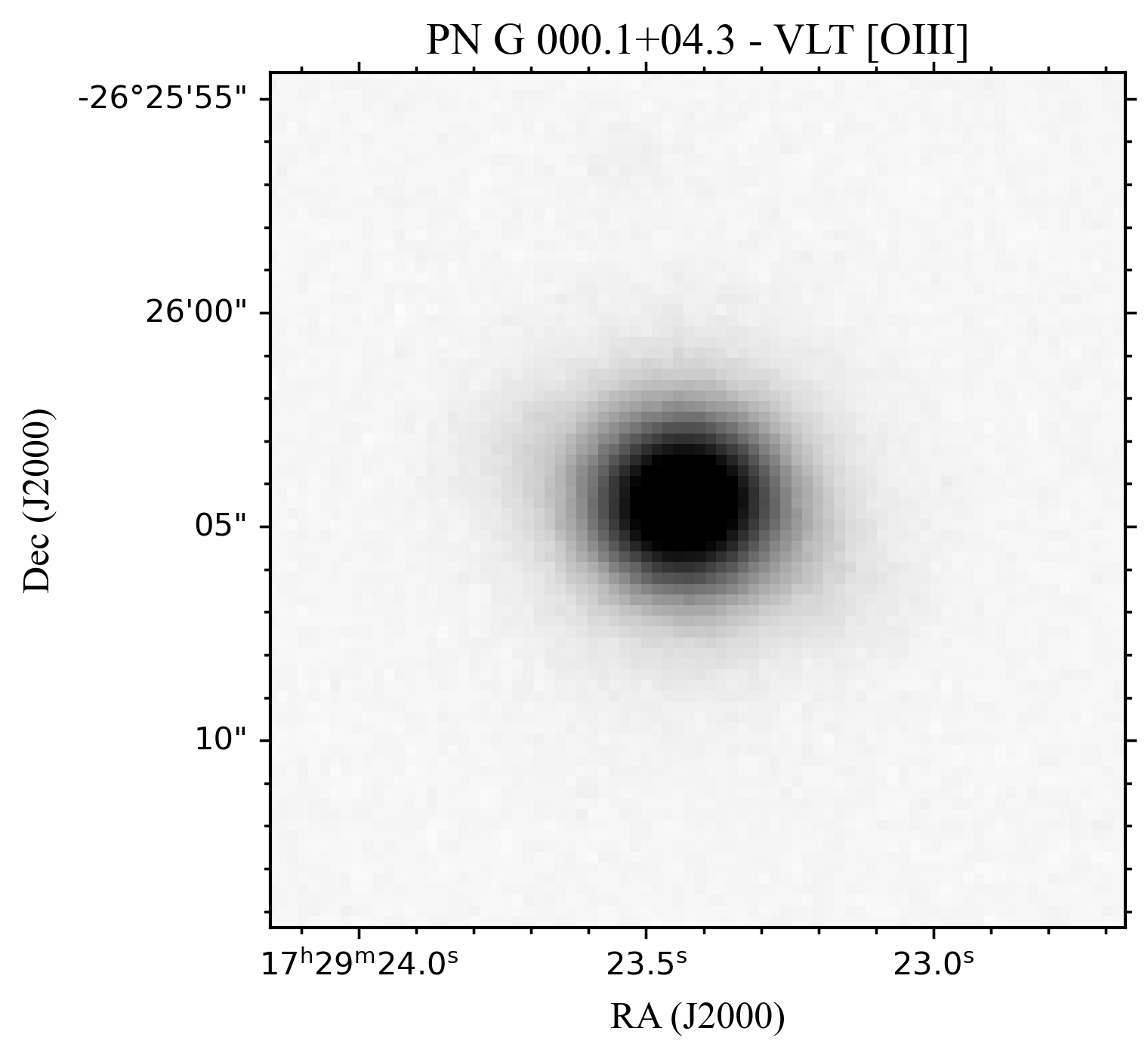}\hfill 
  \includegraphics[width=.32\linewidth]{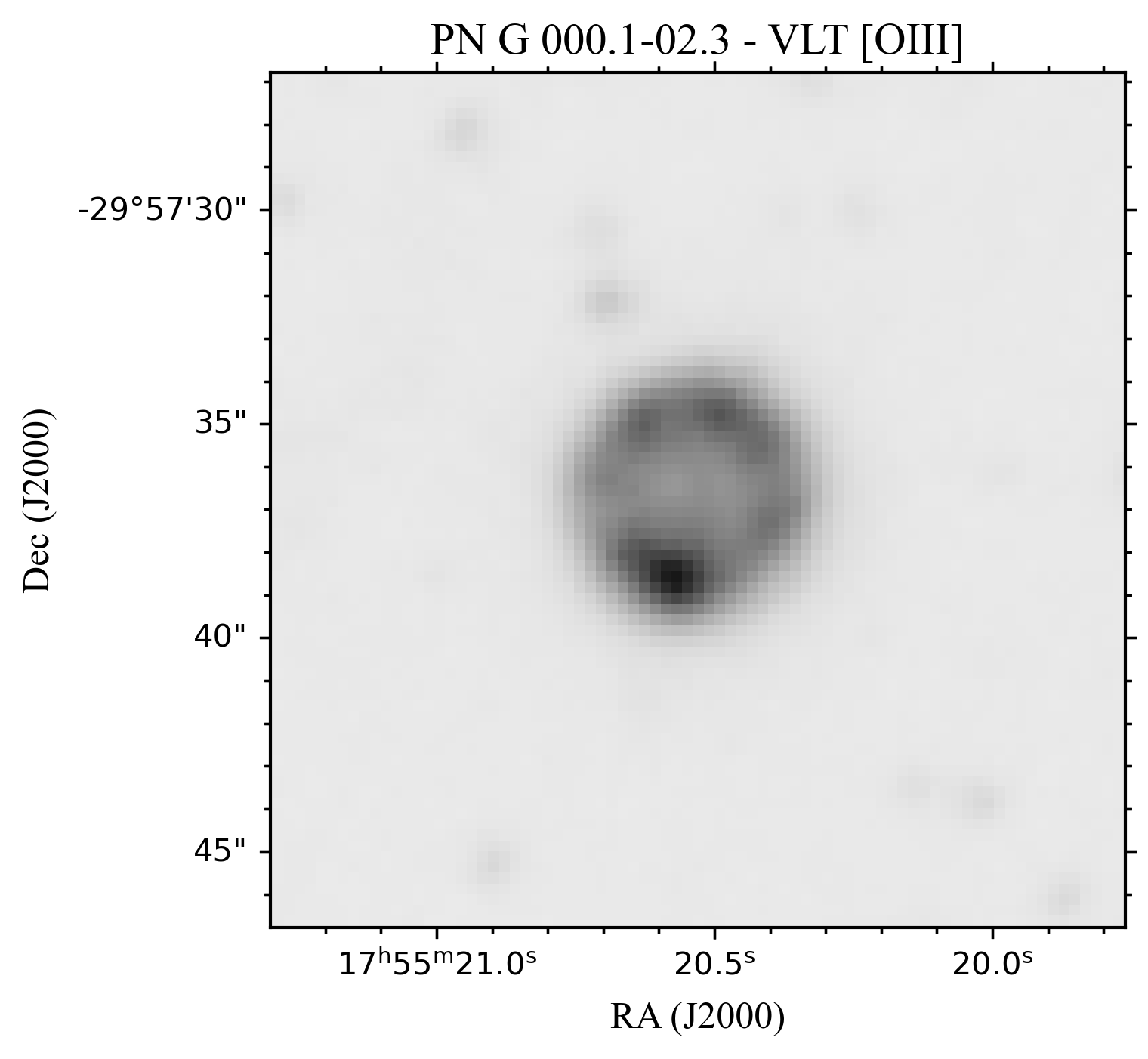}\hfill 
 \end{subfigure}\par\medskip 
  \end{figure} 
 \begin{figure} 
 \ContinuedFloat 
 \caption[]{continued:} 
\begin{subfigure}{\linewidth} 
  \includegraphics[width=.32\linewidth]{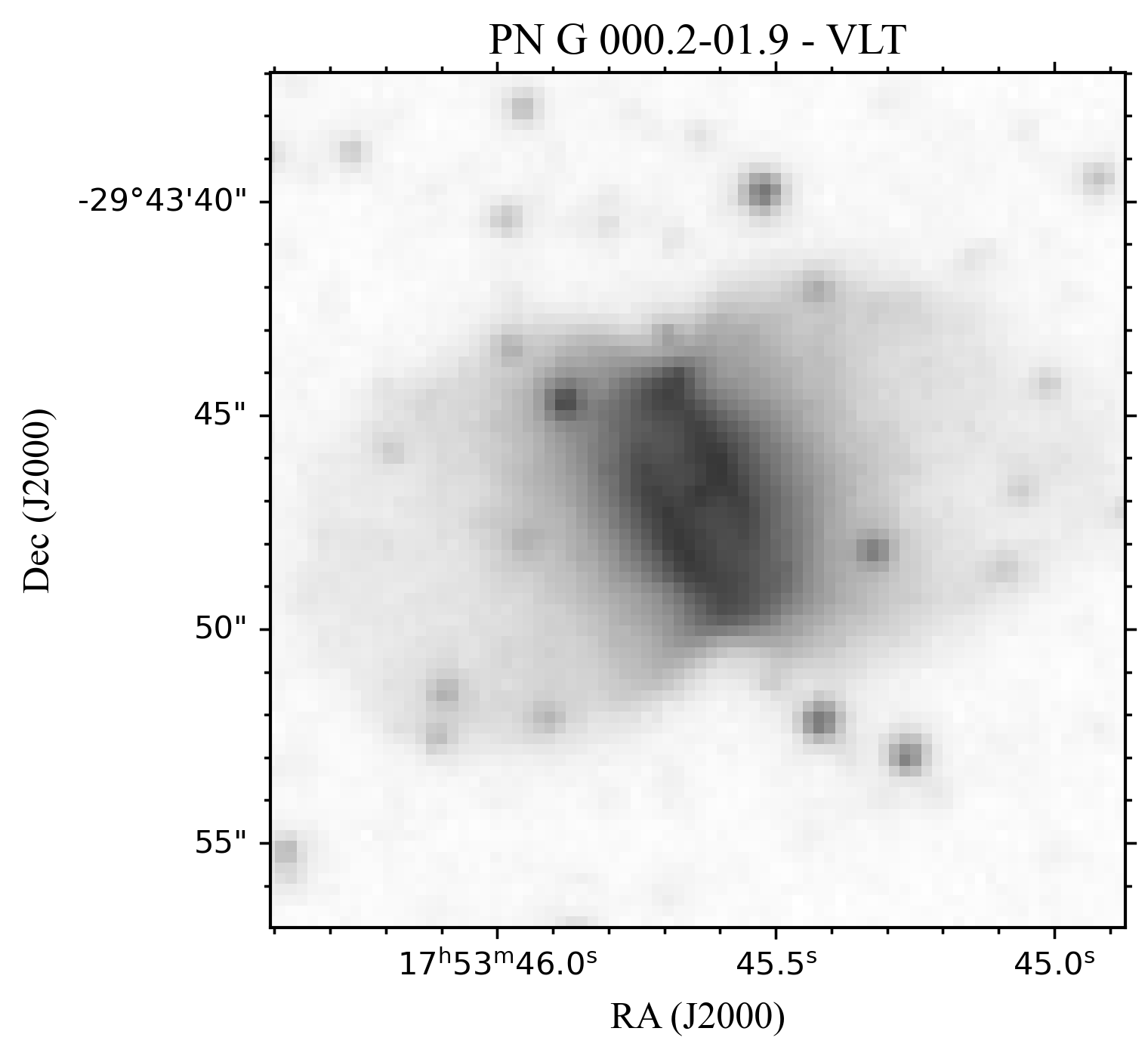}\hfill 
  \includegraphics[width=.32\linewidth]{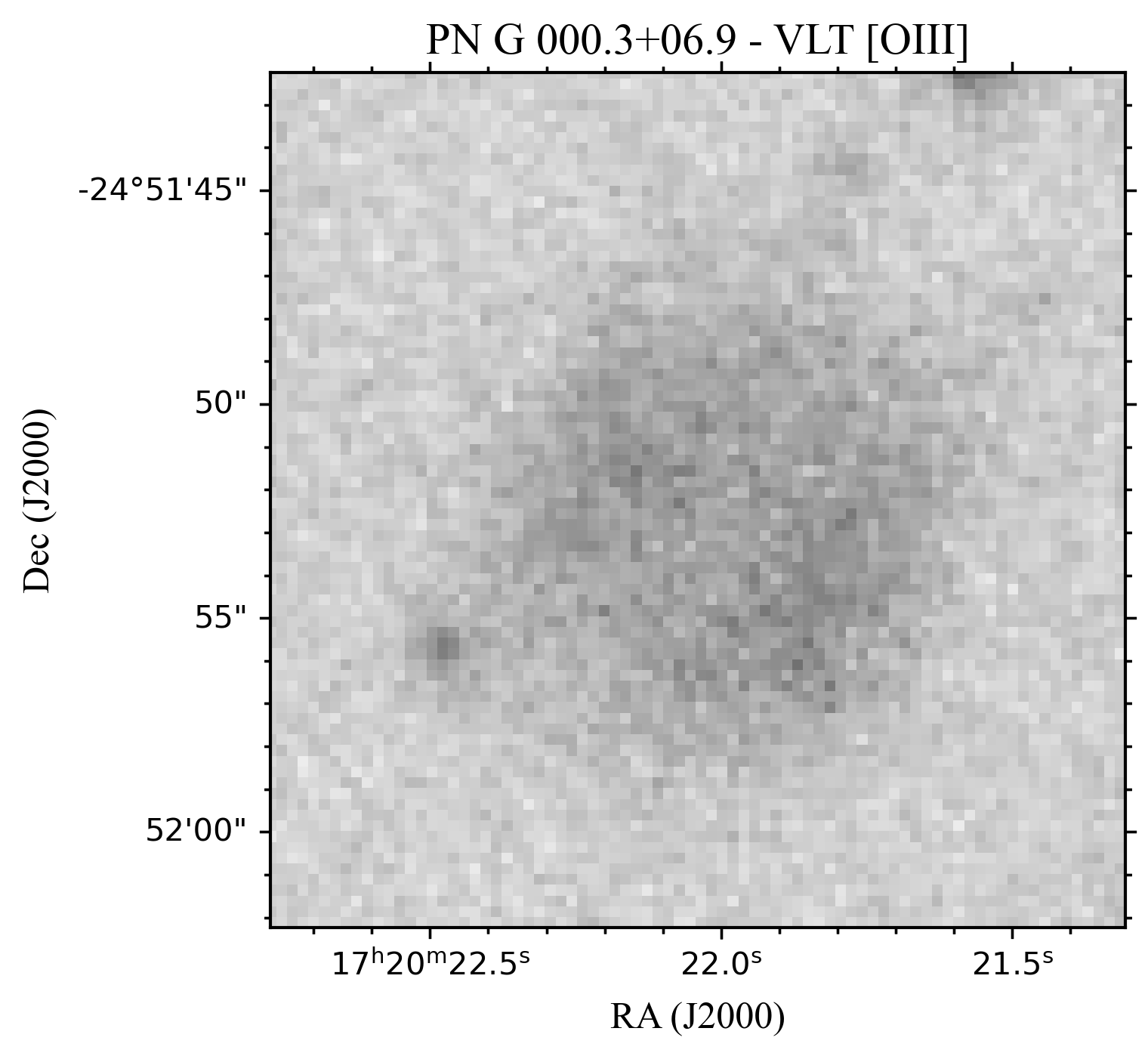}\hfill 
  \includegraphics[width=.32\linewidth]{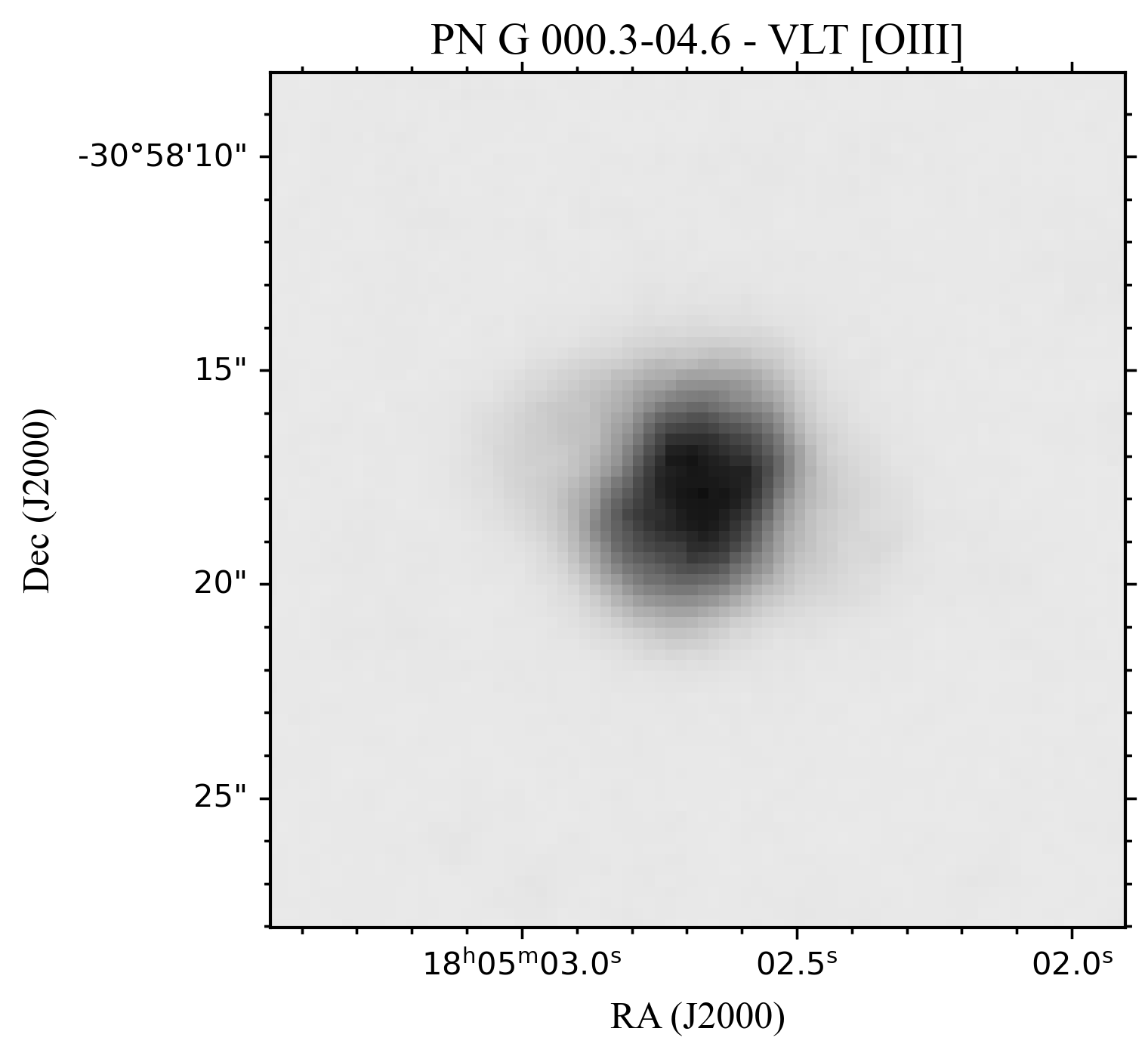}\hfill 
 \end{subfigure}\par\medskip 
\begin{subfigure}{\linewidth} 
  \includegraphics[width=.32\linewidth]{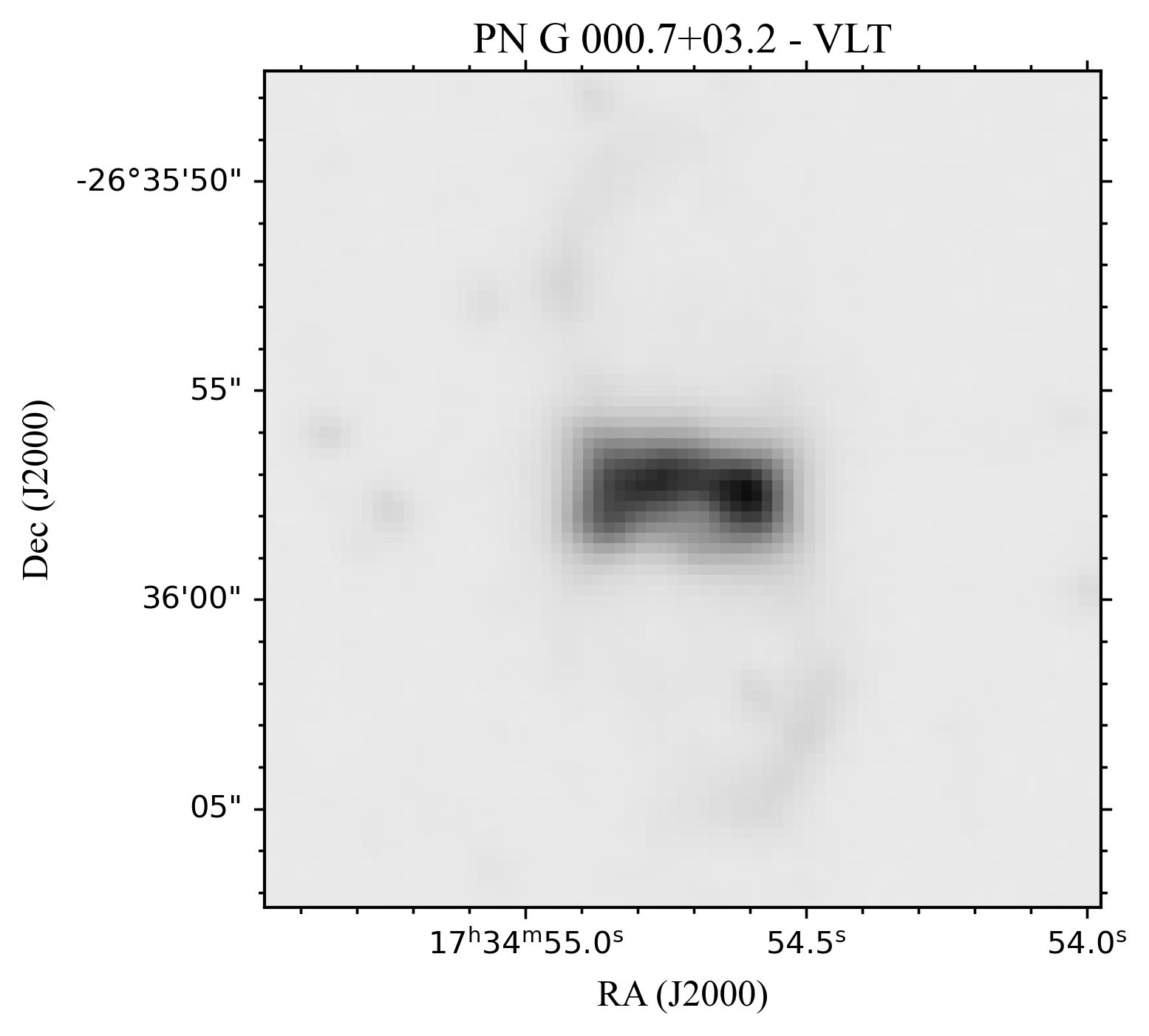}\hfill 
  \includegraphics[width=.32\linewidth]{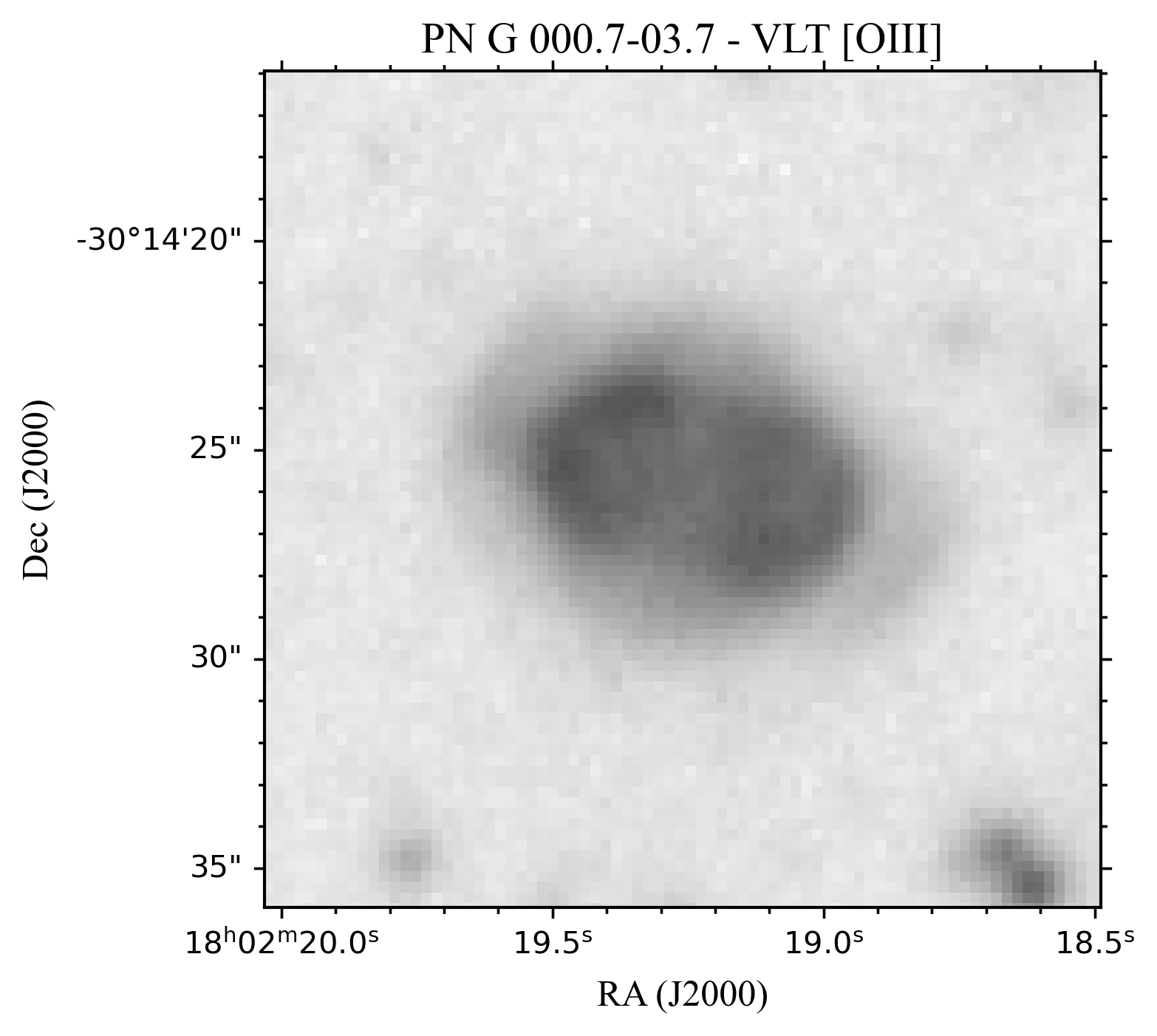}\hfill 
  \includegraphics[width=.32\linewidth]{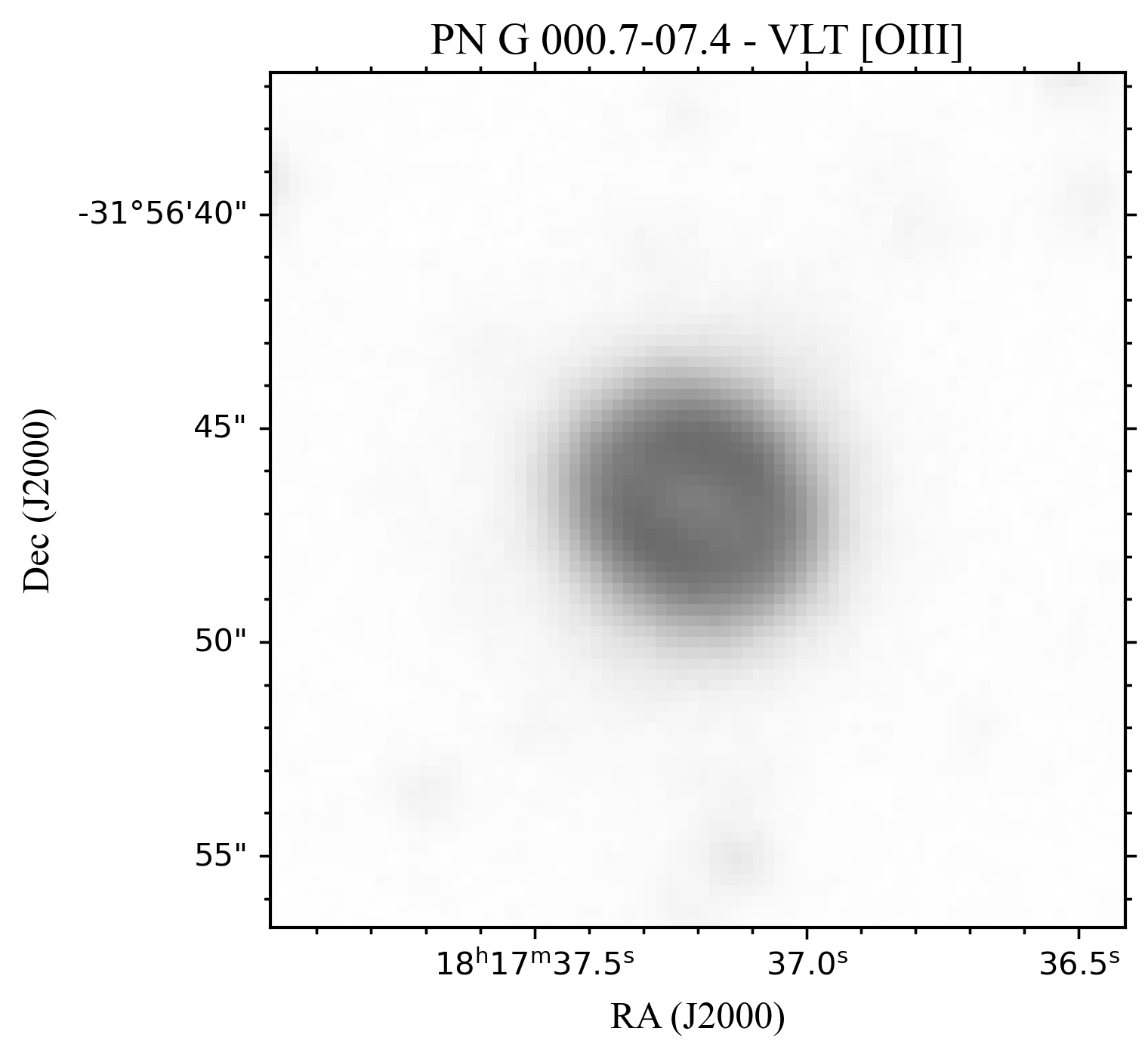}\hfill 
 \end{subfigure}\par\medskip 
\begin{subfigure}{\linewidth} 
  \includegraphics[width=.32\linewidth]{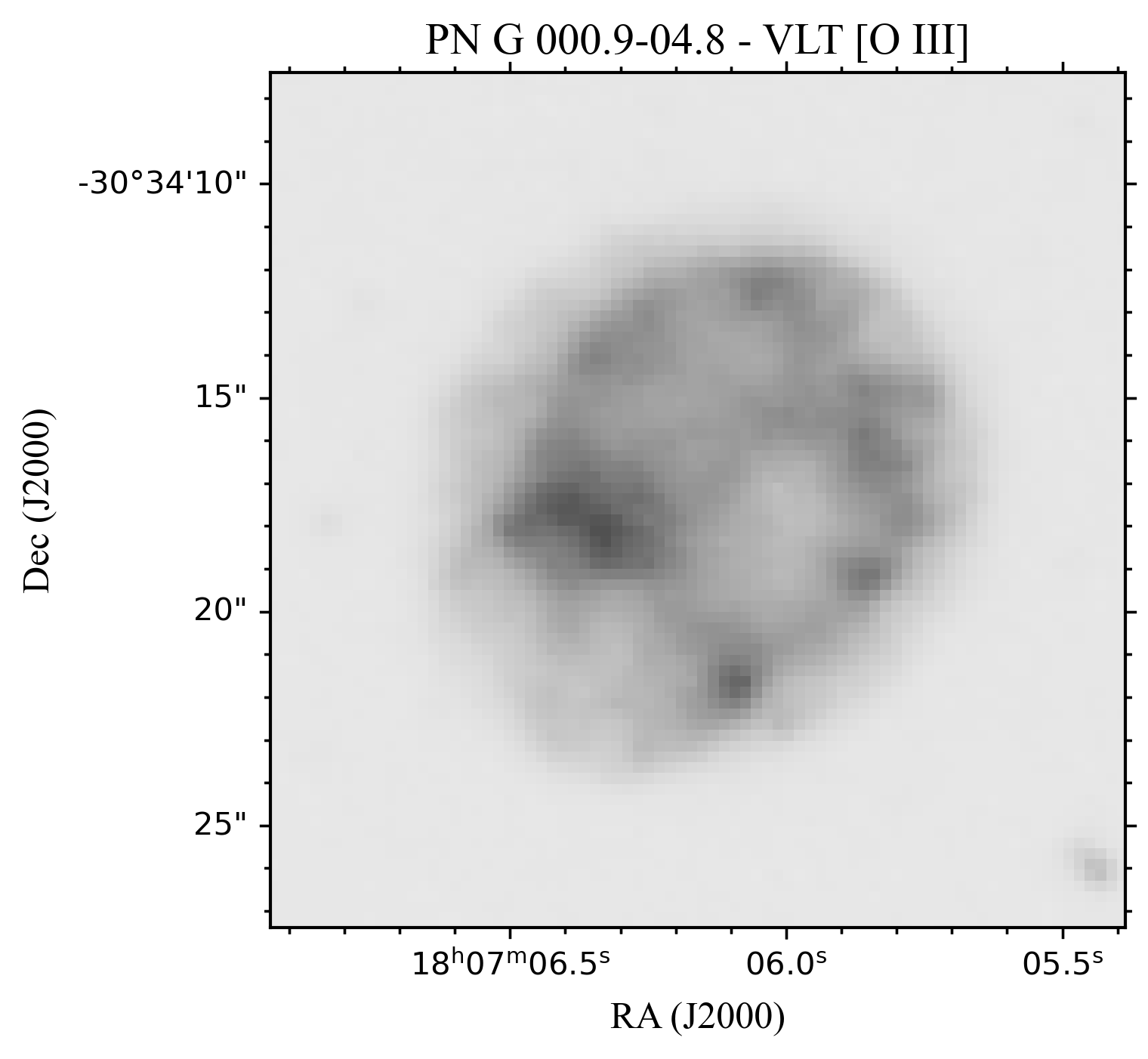}\hfill 
  \includegraphics[width=.32\linewidth]{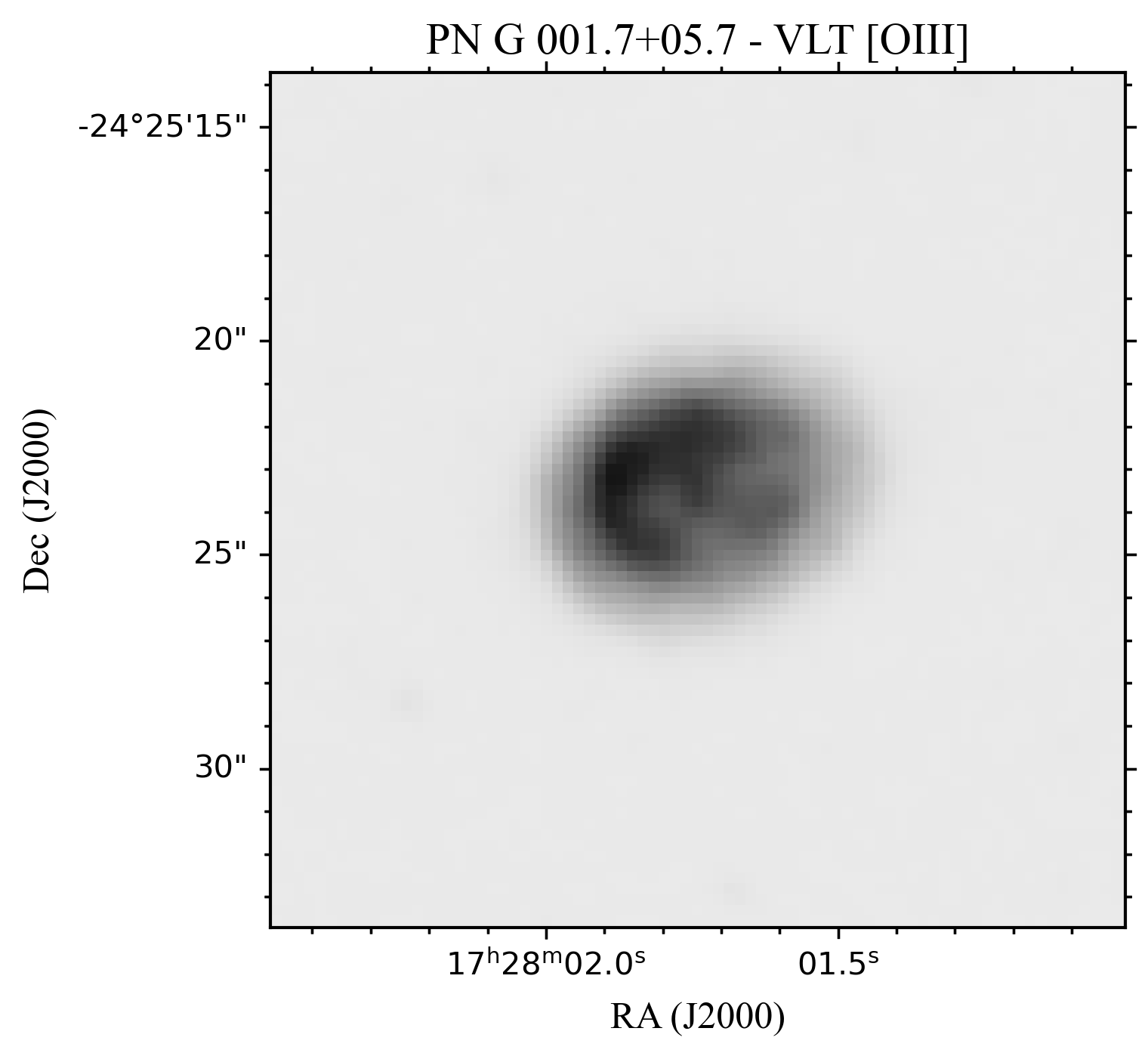}\hfill 
  \includegraphics[width=.32\linewidth]{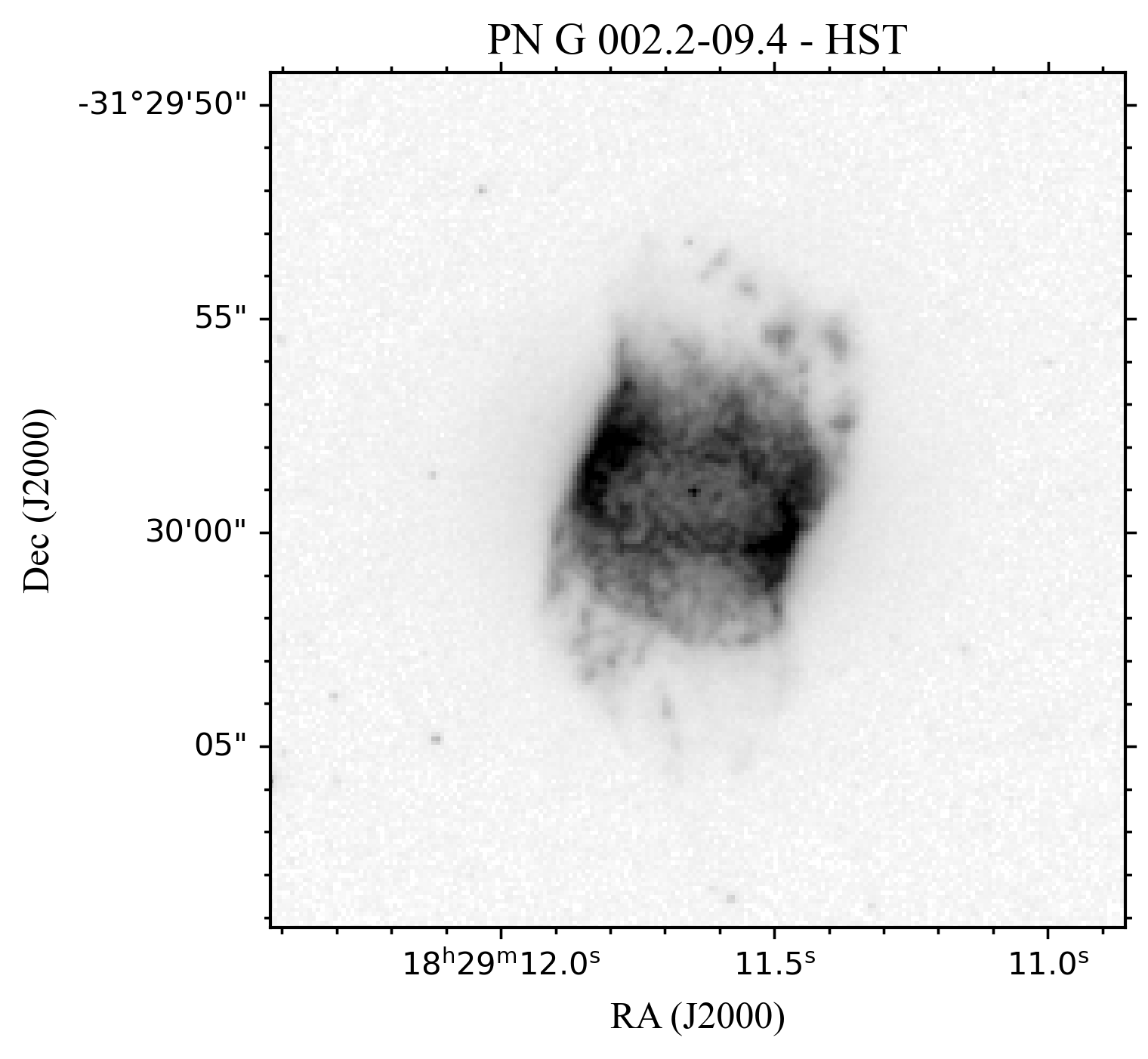}\hfill 
 \end{subfigure}\par\medskip 
\begin{subfigure}{\linewidth} 
  \includegraphics[width=.32\linewidth]{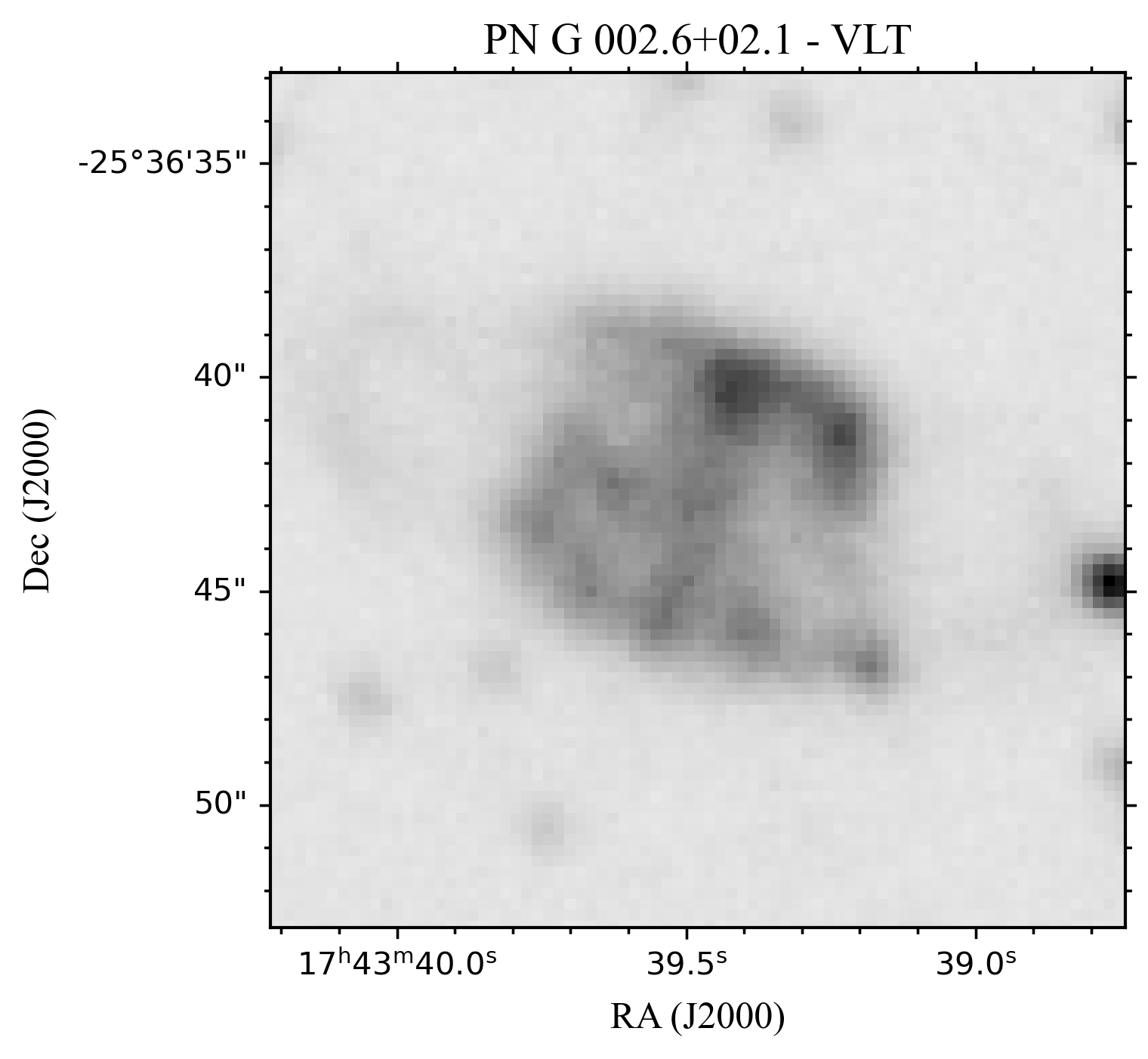}\hfill 
  \includegraphics[width=.32\linewidth]{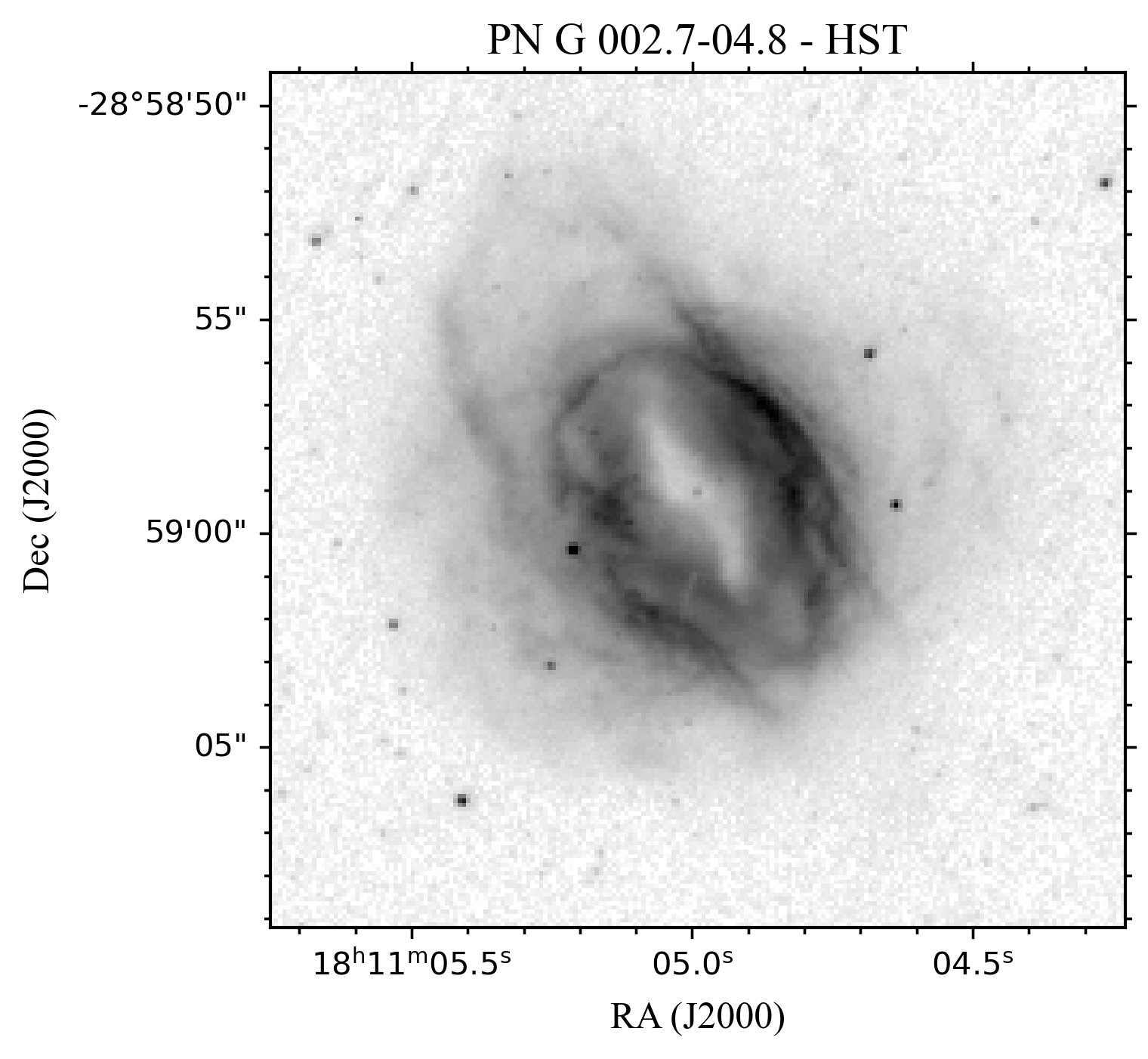}\hfill 
  \includegraphics[width=.32\linewidth]{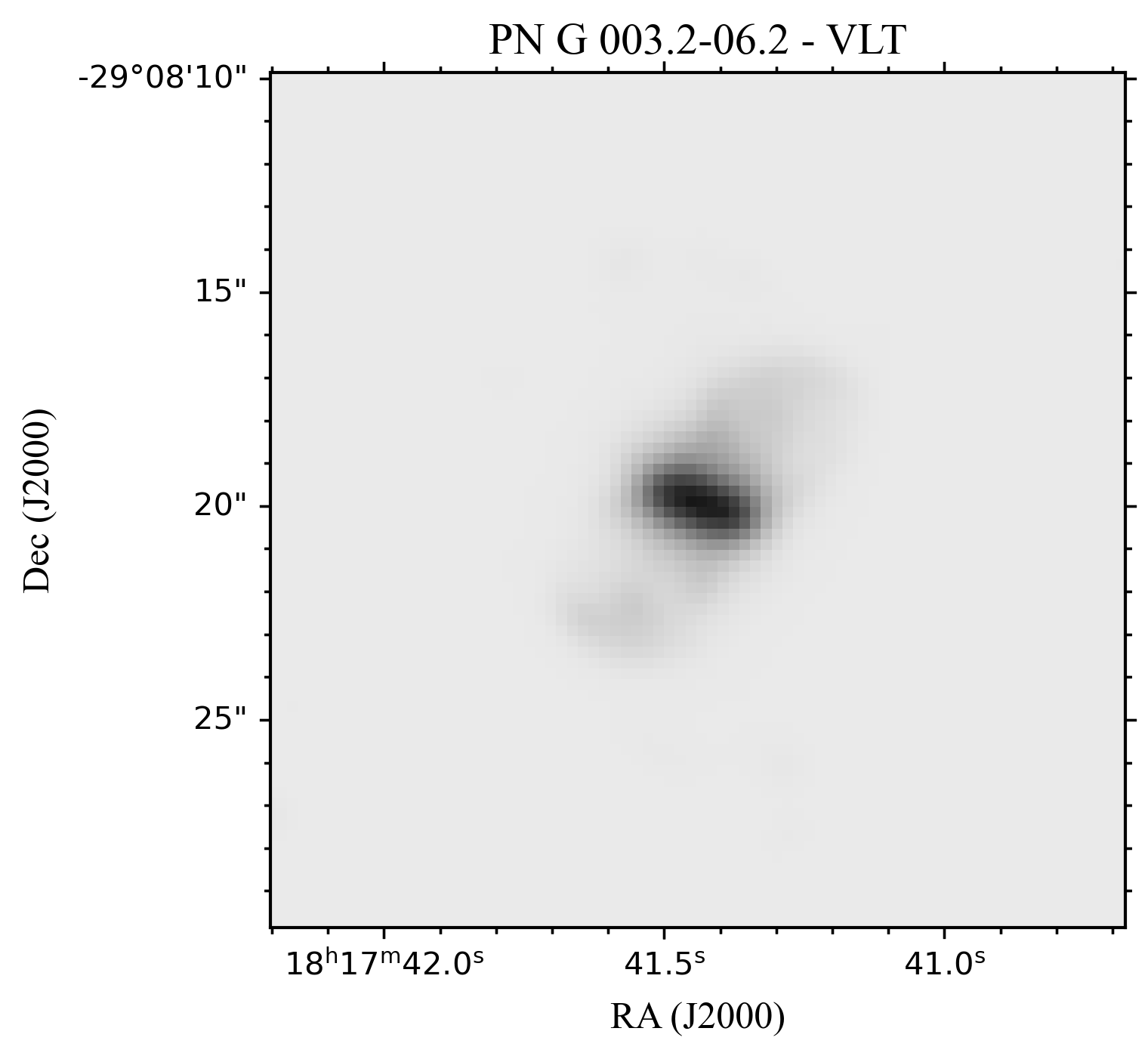}\hfill 
 \end{subfigure}\par\medskip 
  \end{figure} 
 \begin{figure} 
 \ContinuedFloat 
 \caption[]{continued:} 
\begin{subfigure}{\linewidth} 
  \includegraphics[width=.32\linewidth]{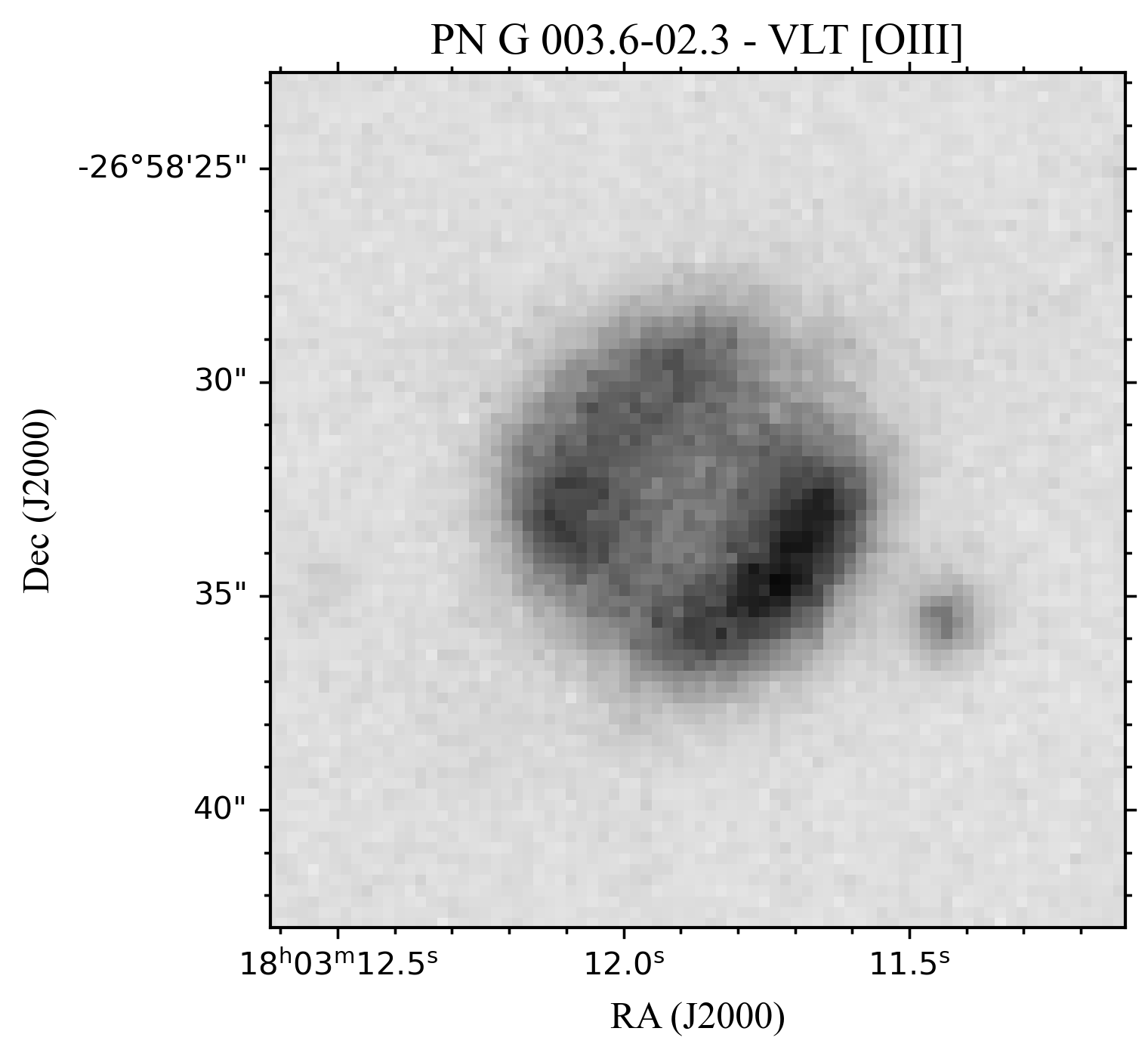}\hfill 
  \includegraphics[width=.32\linewidth]{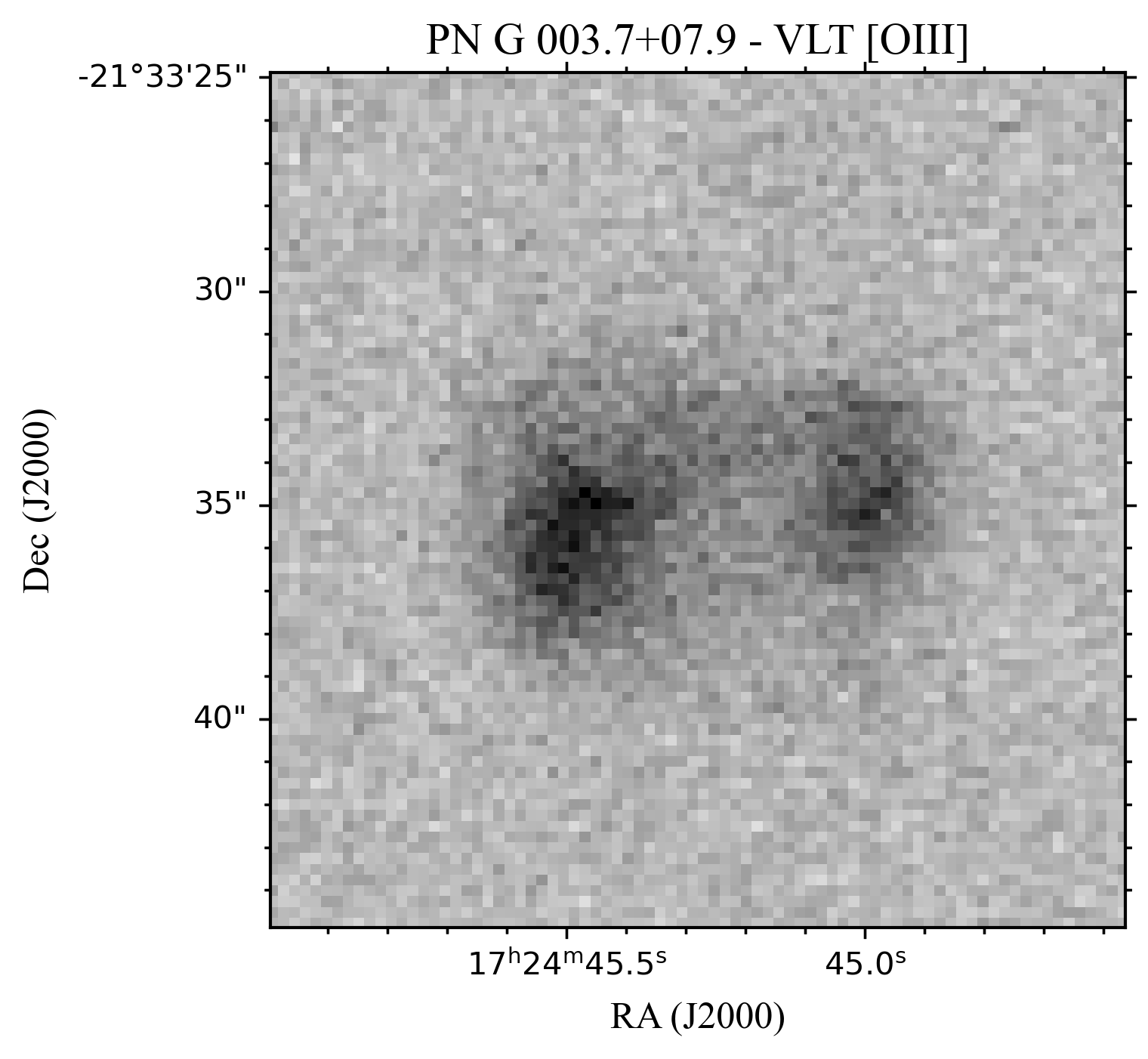}\hfill 
  \includegraphics[width=.32\linewidth]{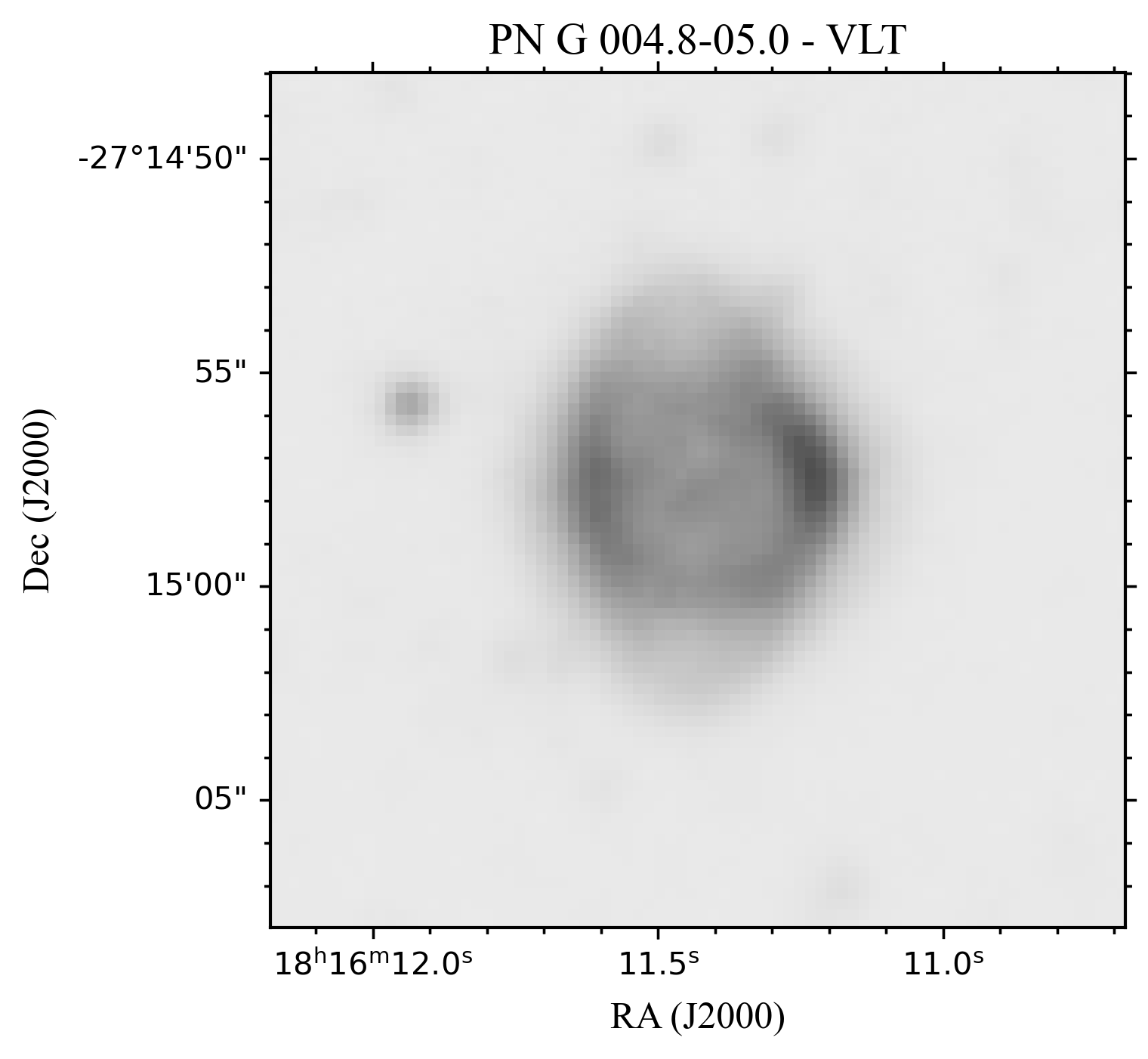}\hfill 
 \end{subfigure}\par\medskip 
\begin{subfigure}{\linewidth} 
  \includegraphics[width=.32\linewidth]{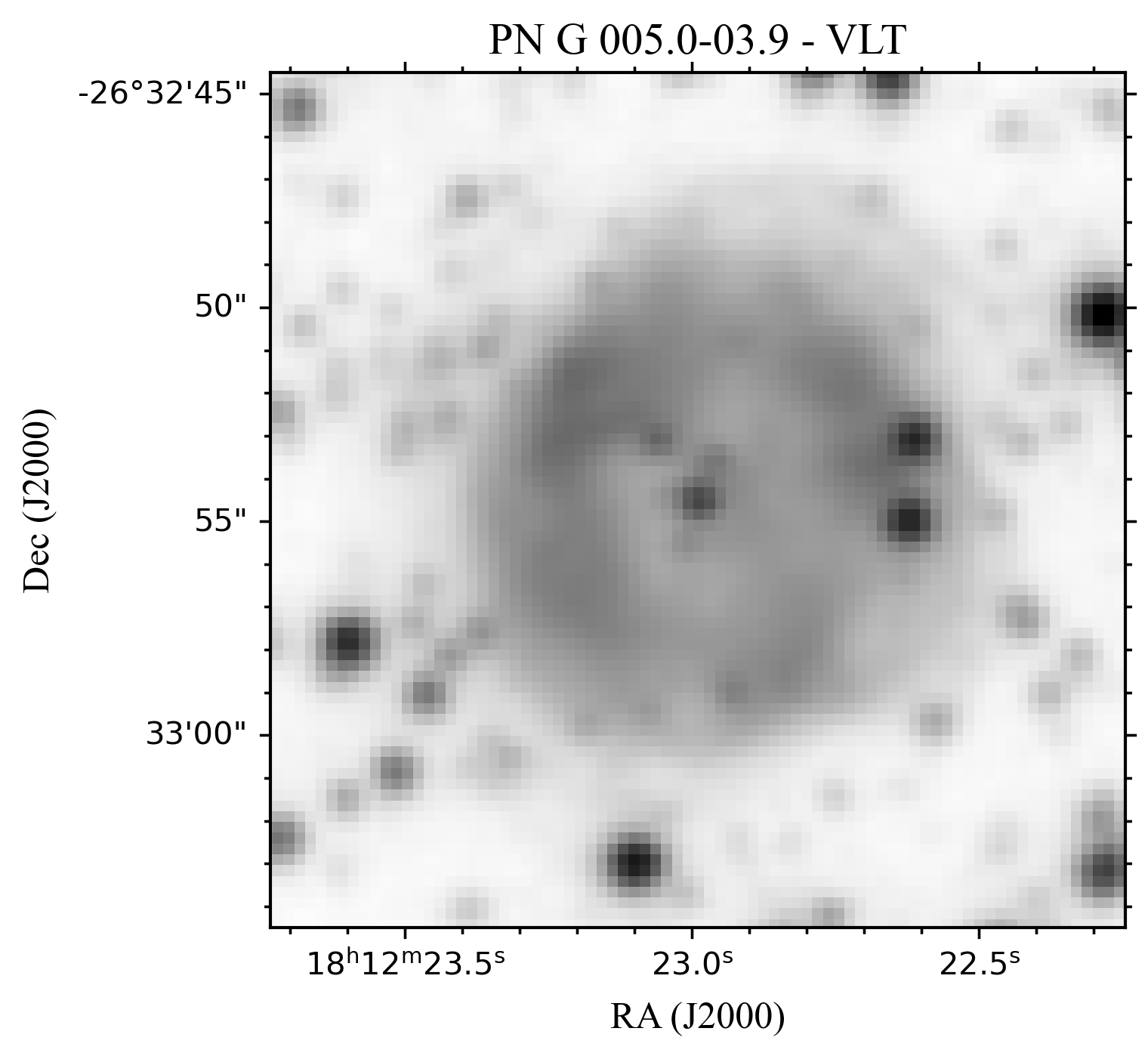}\hfill 
  \includegraphics[width=.32\linewidth]{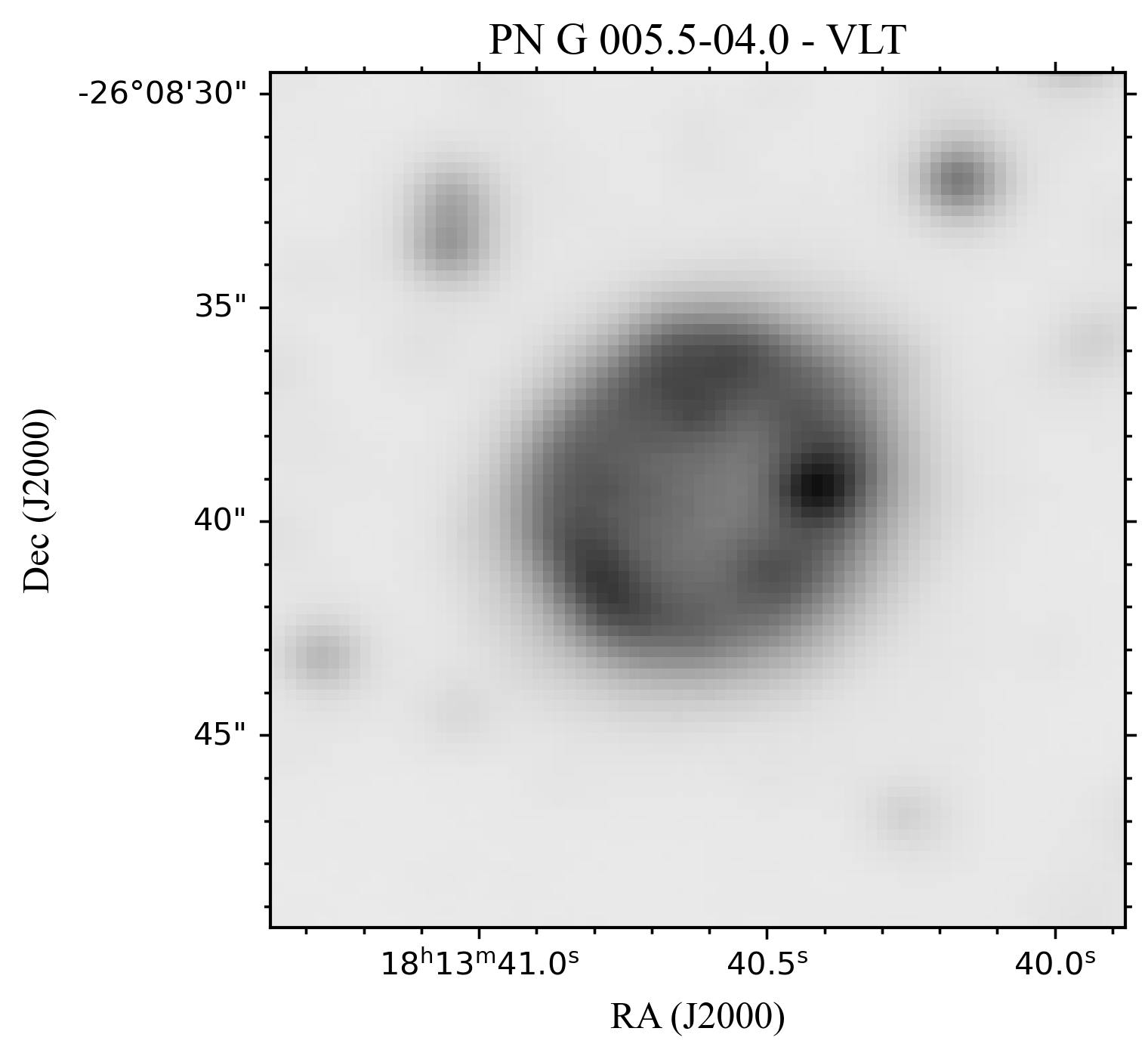}\hfill 
  \includegraphics[width=.32\linewidth]{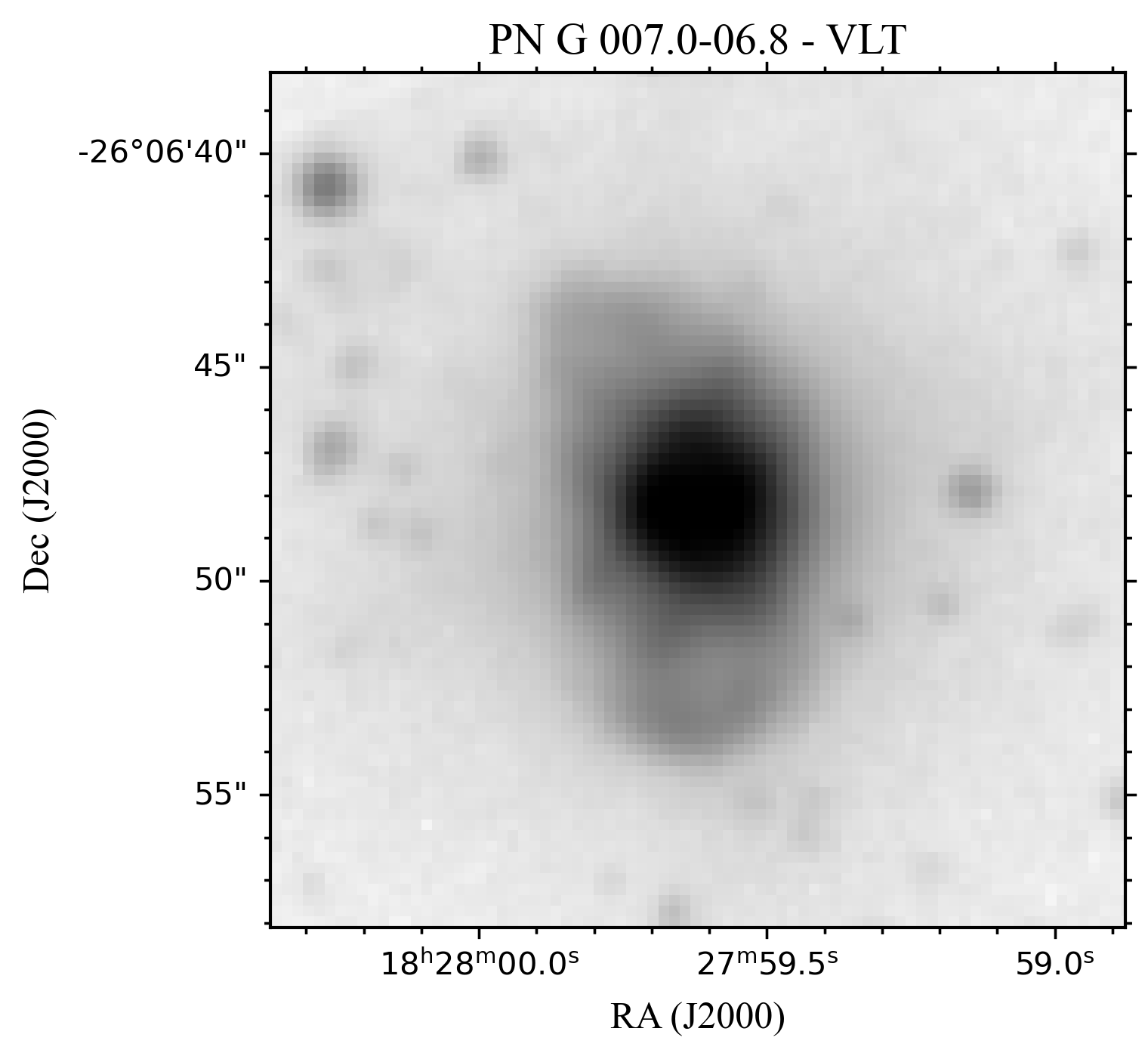}\hfill 
 \end{subfigure}\par\medskip 
\begin{subfigure}{\linewidth} 
  \includegraphics[width=.32\linewidth]{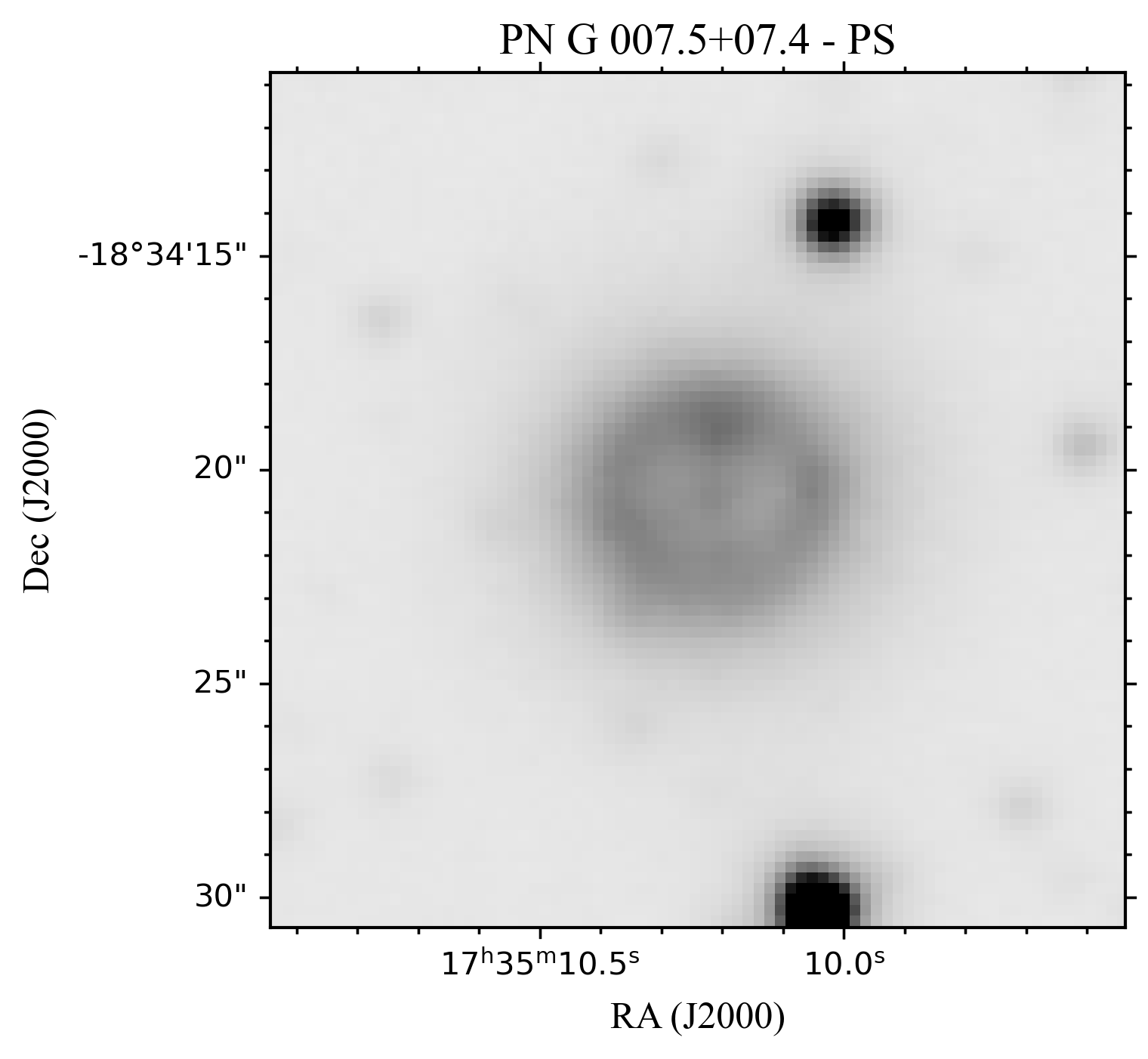}\hfill 
  \includegraphics[width=.32\linewidth]{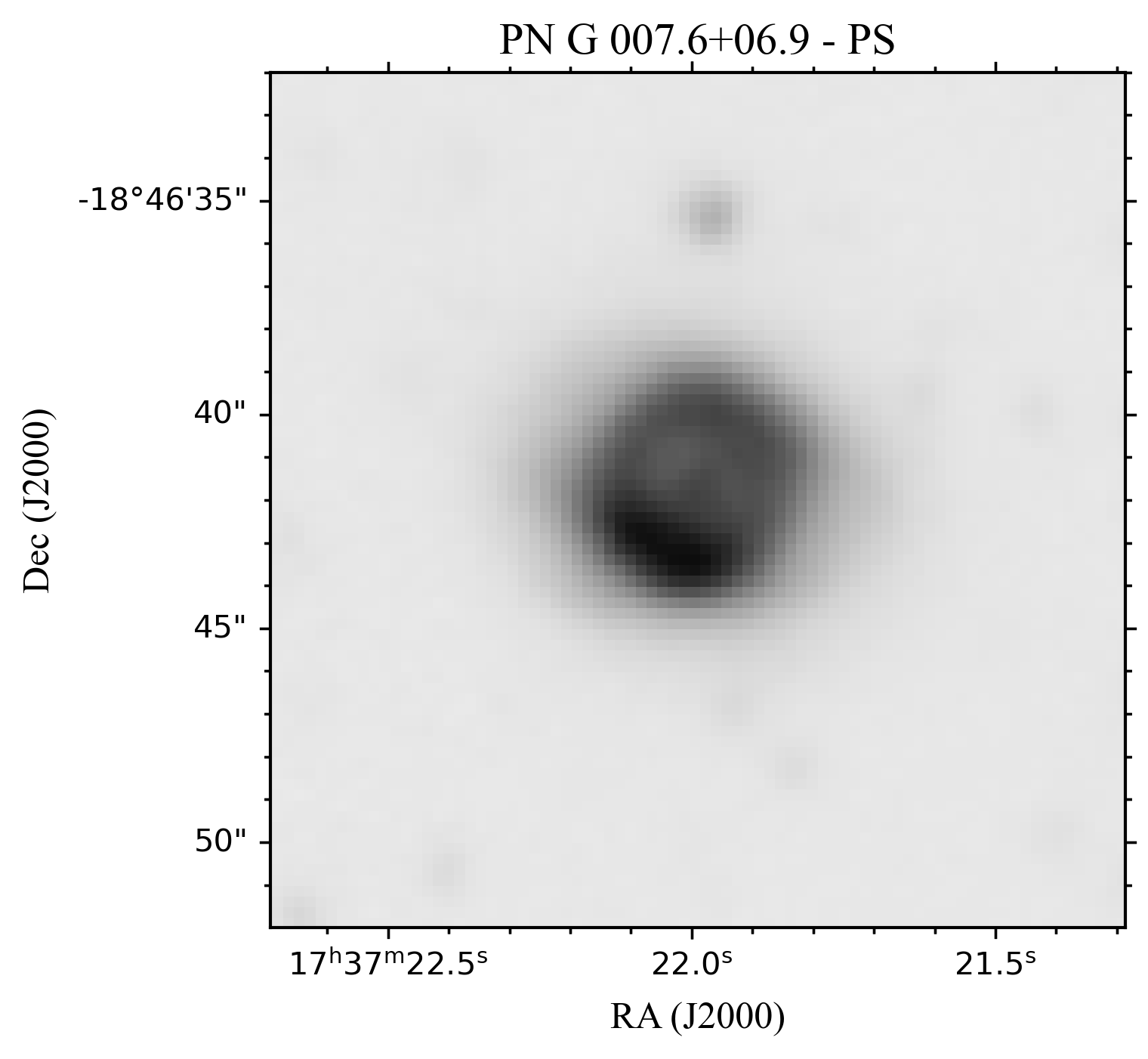}\hfill 
  \includegraphics[width=.32\linewidth]{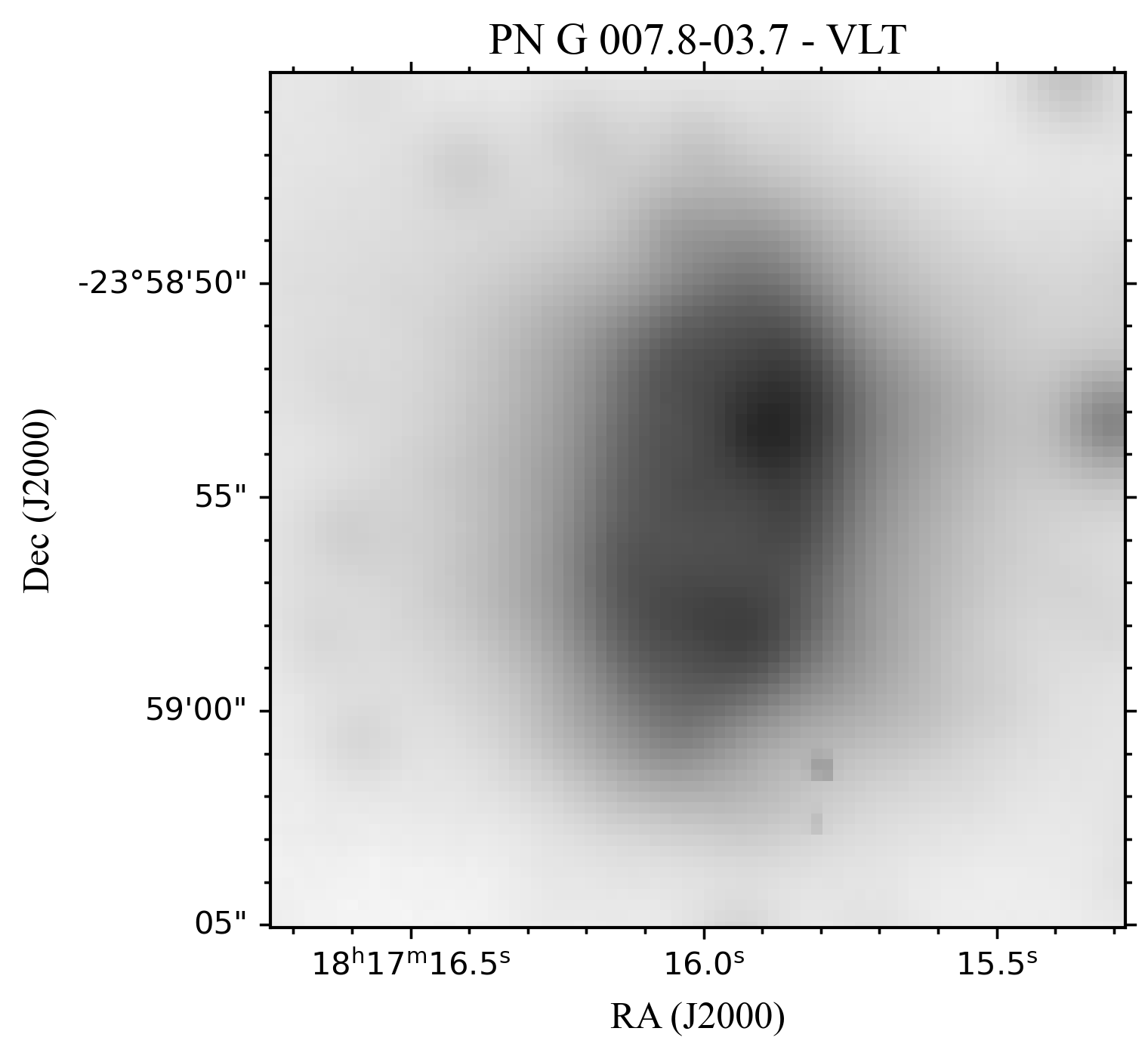}\hfill 
 \end{subfigure}\par\medskip 
\begin{subfigure}{\linewidth} 
  \includegraphics[width=.32\linewidth]{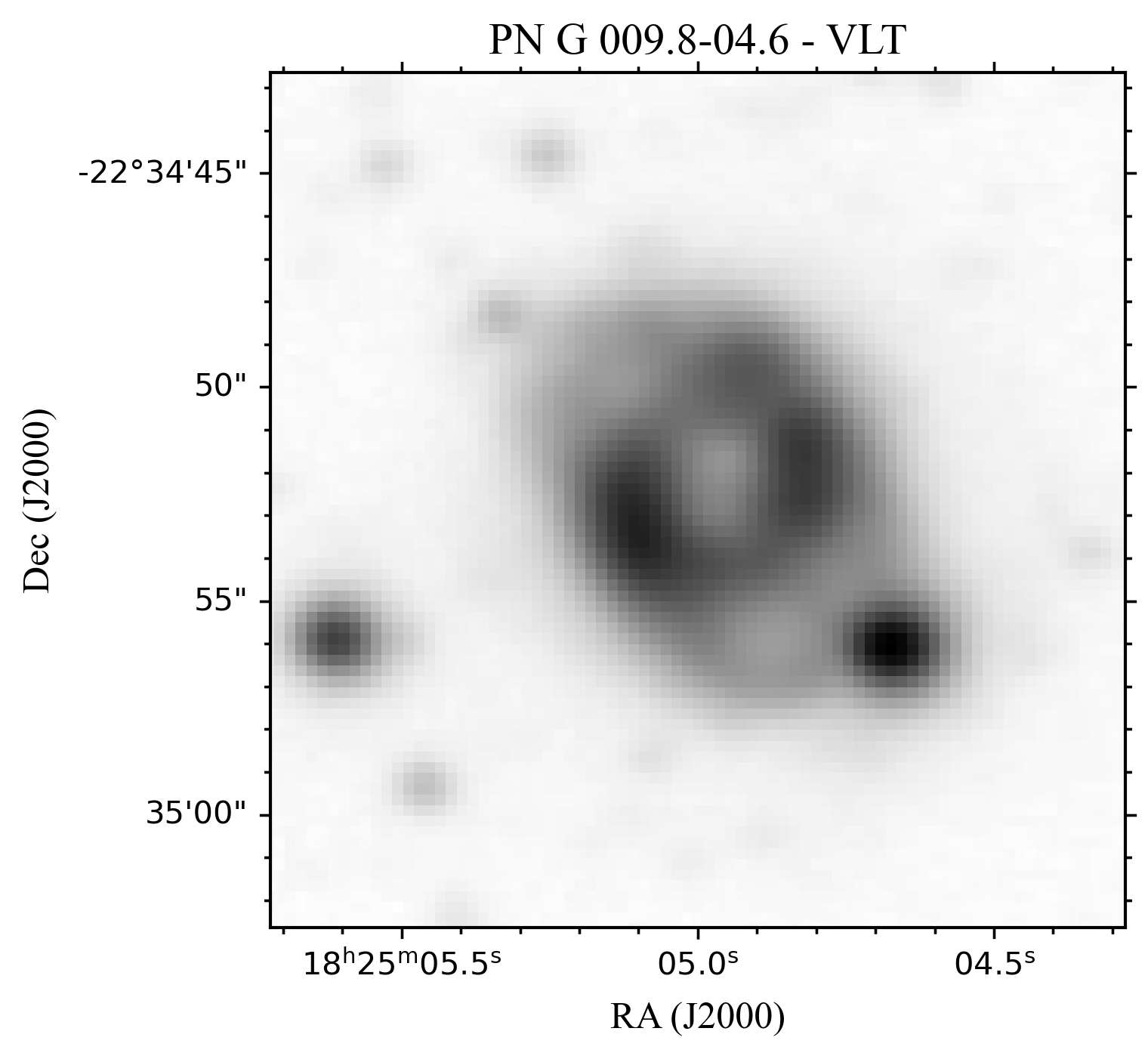}\hfill 
  \includegraphics[width=.32\linewidth]{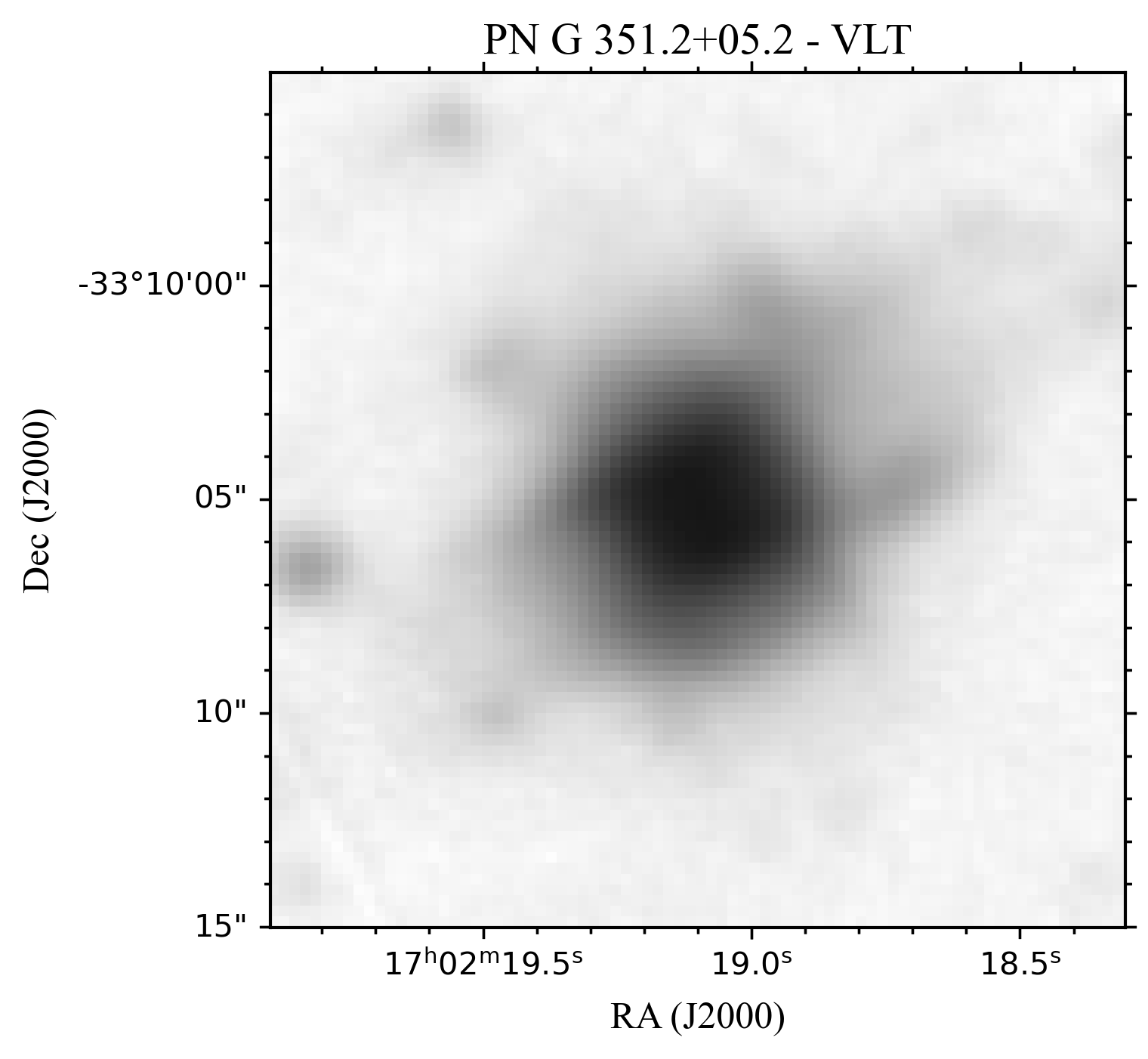}\hfill 
  \includegraphics[width=.32\linewidth]{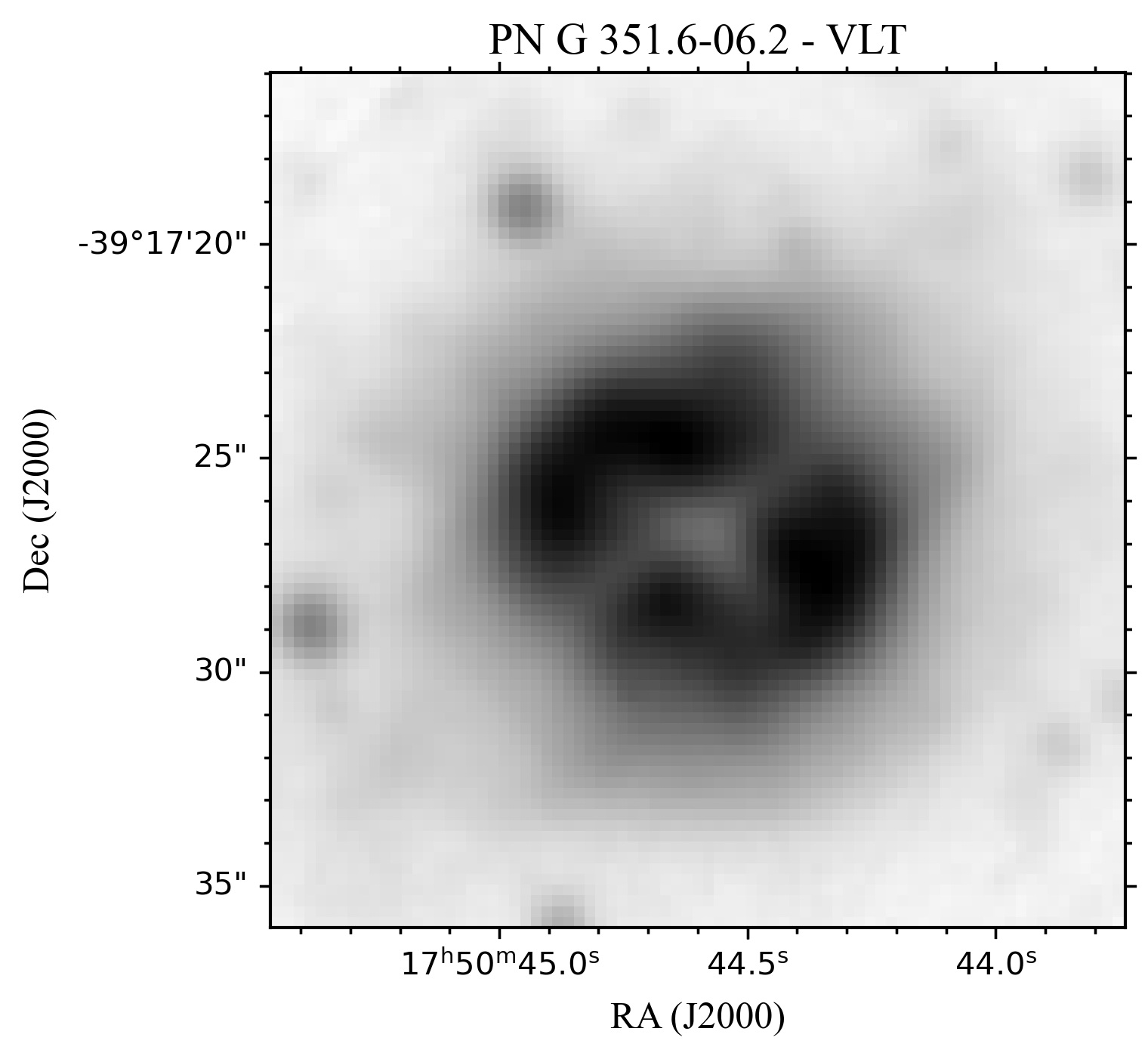}\hfill 
 \end{subfigure}\par\medskip 
  \end{figure} 
 \begin{figure} 
 \ContinuedFloat 
 \caption[]{continued:} 
\begin{subfigure}{\linewidth} 
  \includegraphics[width=.32\linewidth]{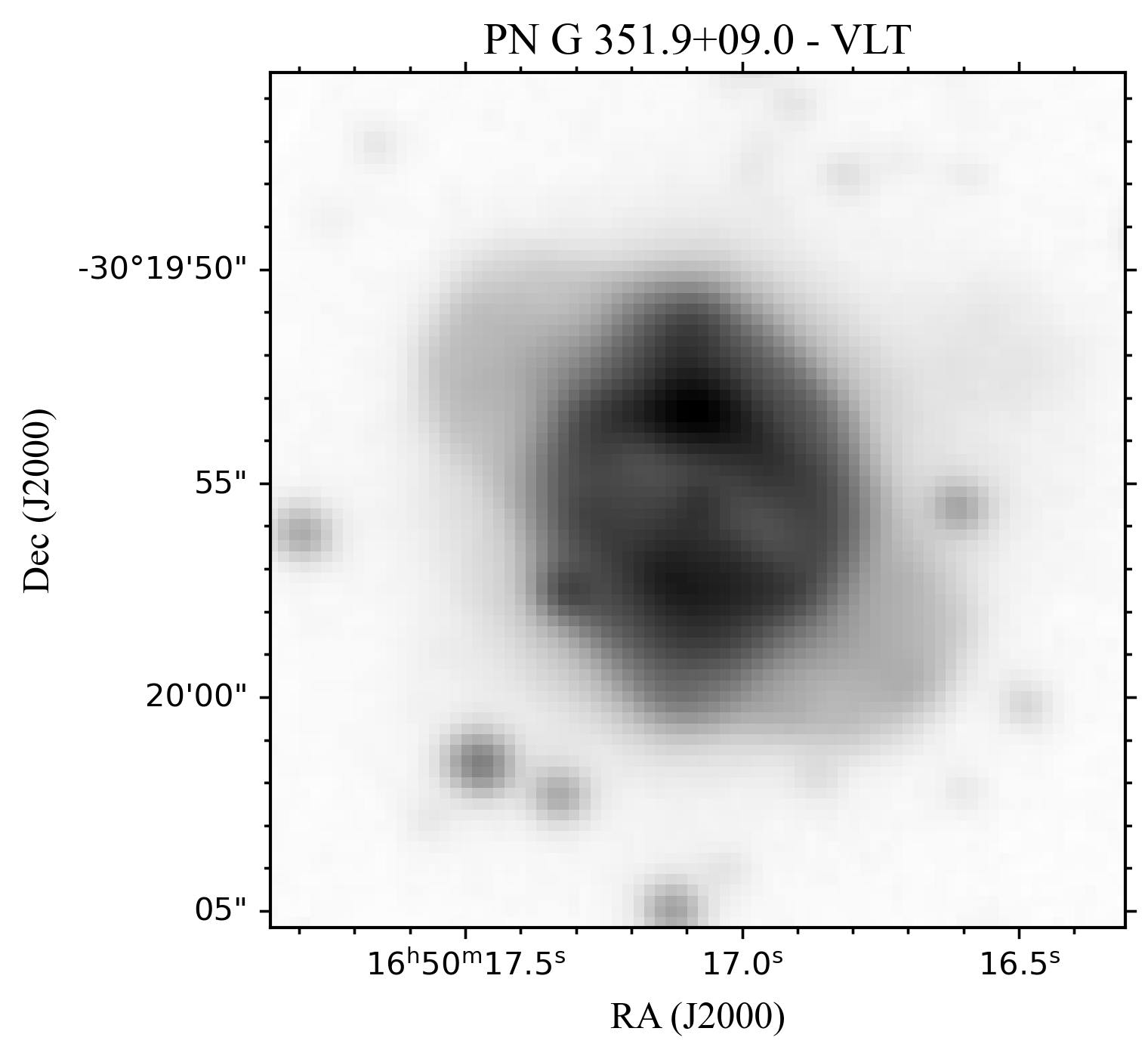}\hfill 
  \includegraphics[width=.32\linewidth]{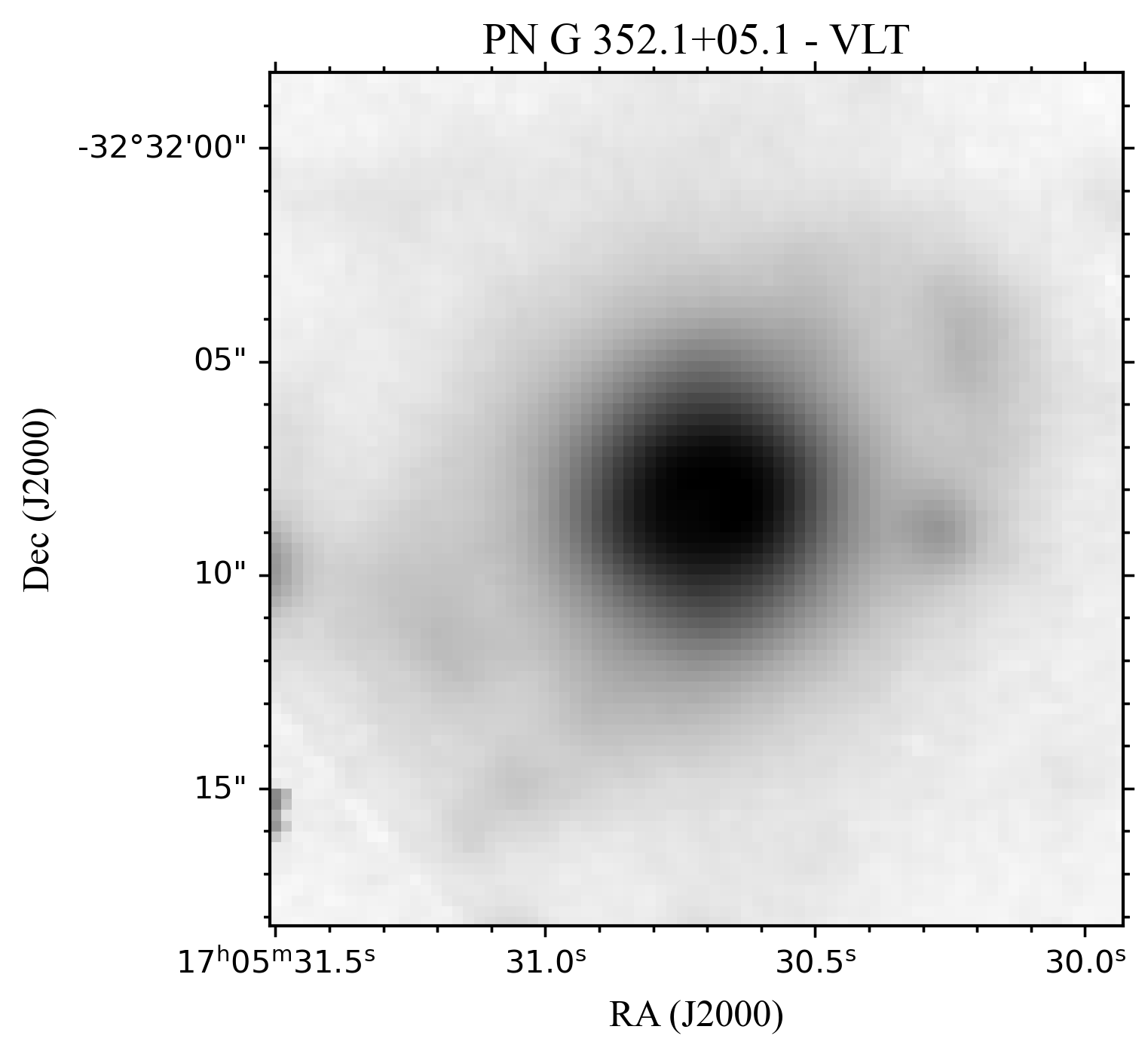}\hfill 
  \includegraphics[width=.32\linewidth]{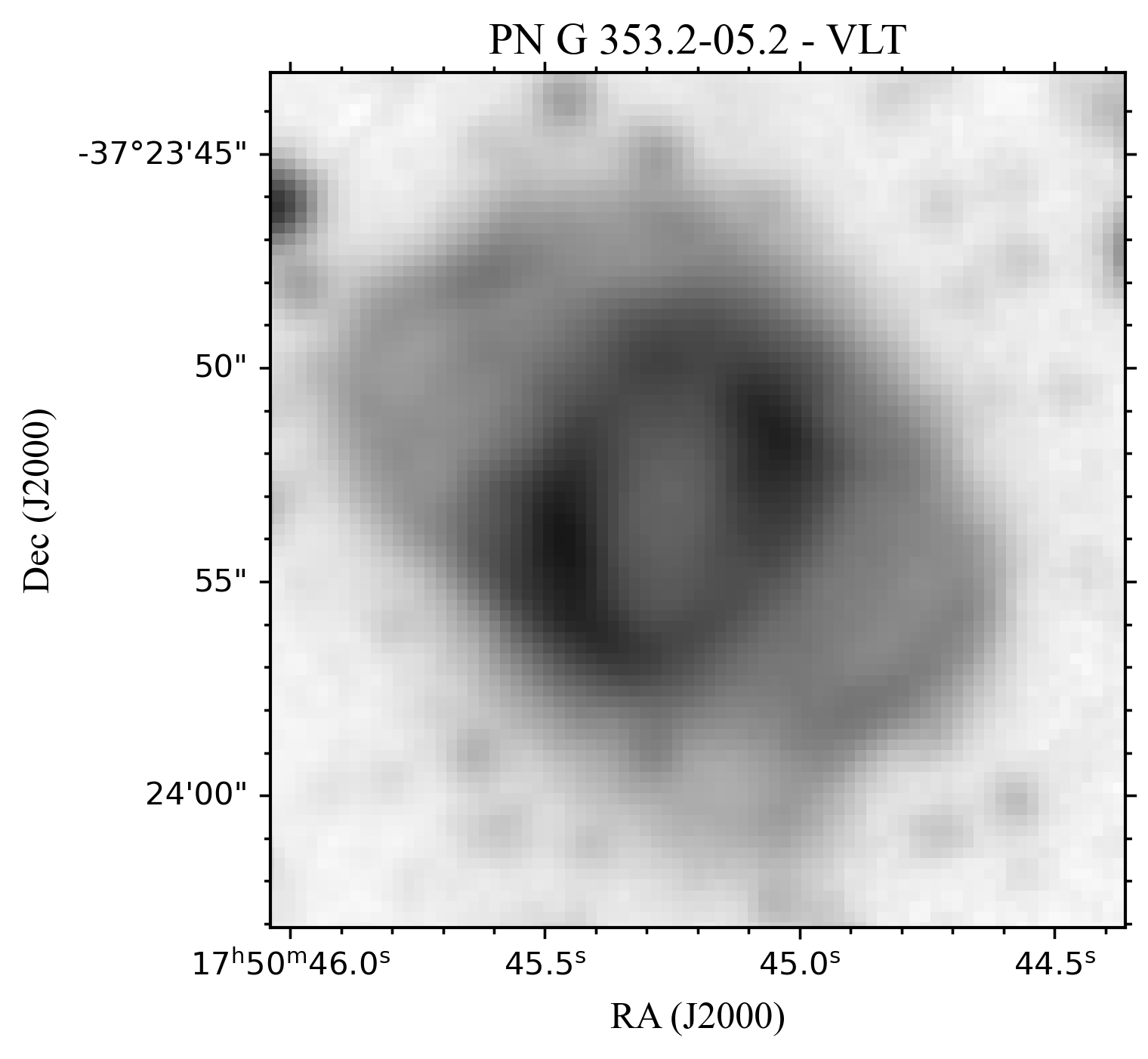}\hfill 
 \end{subfigure}\par\medskip 
\begin{subfigure}{\linewidth} 
  \includegraphics[width=.32\linewidth]{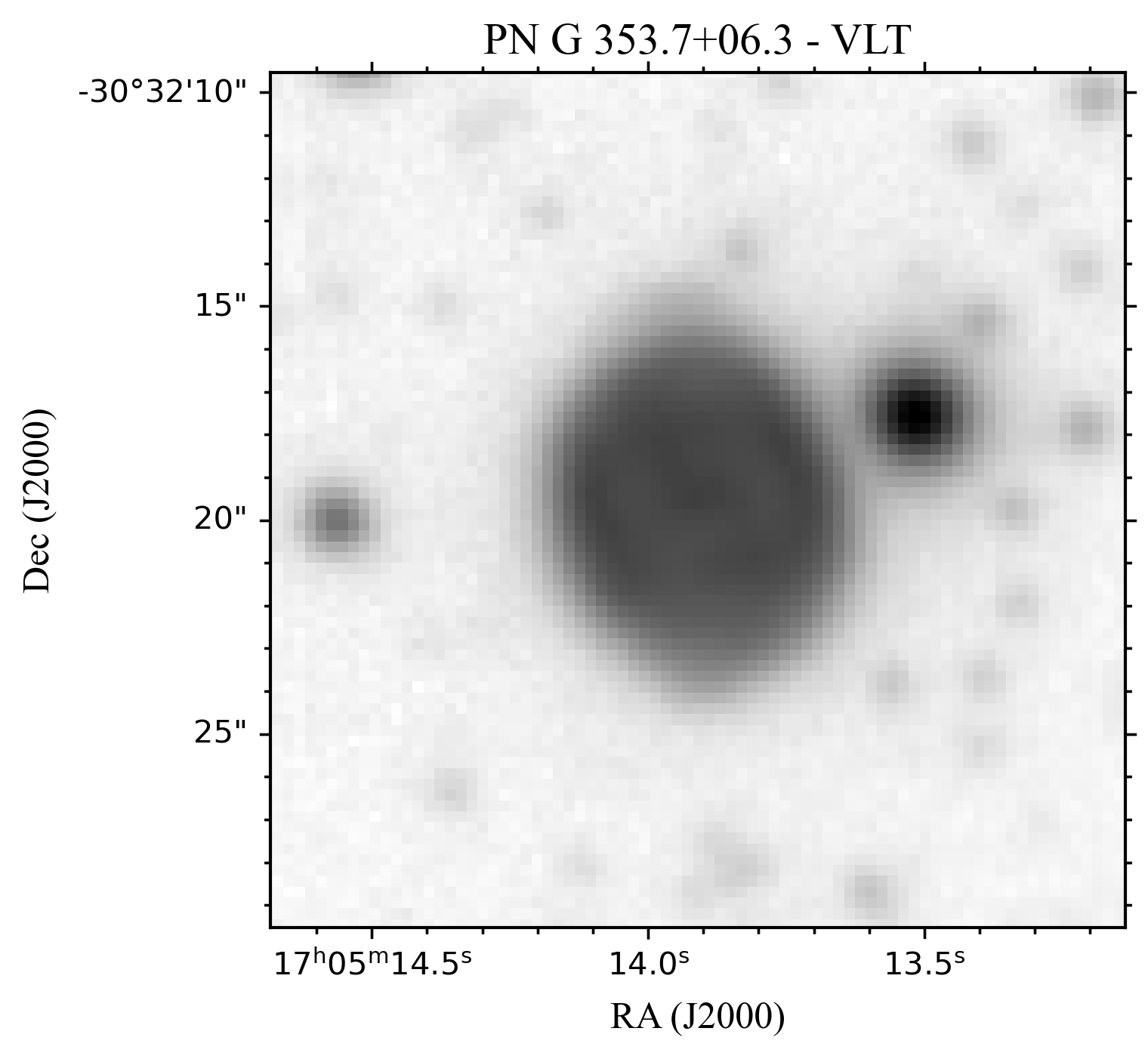}\hfill 
  \includegraphics[width=.32\linewidth]{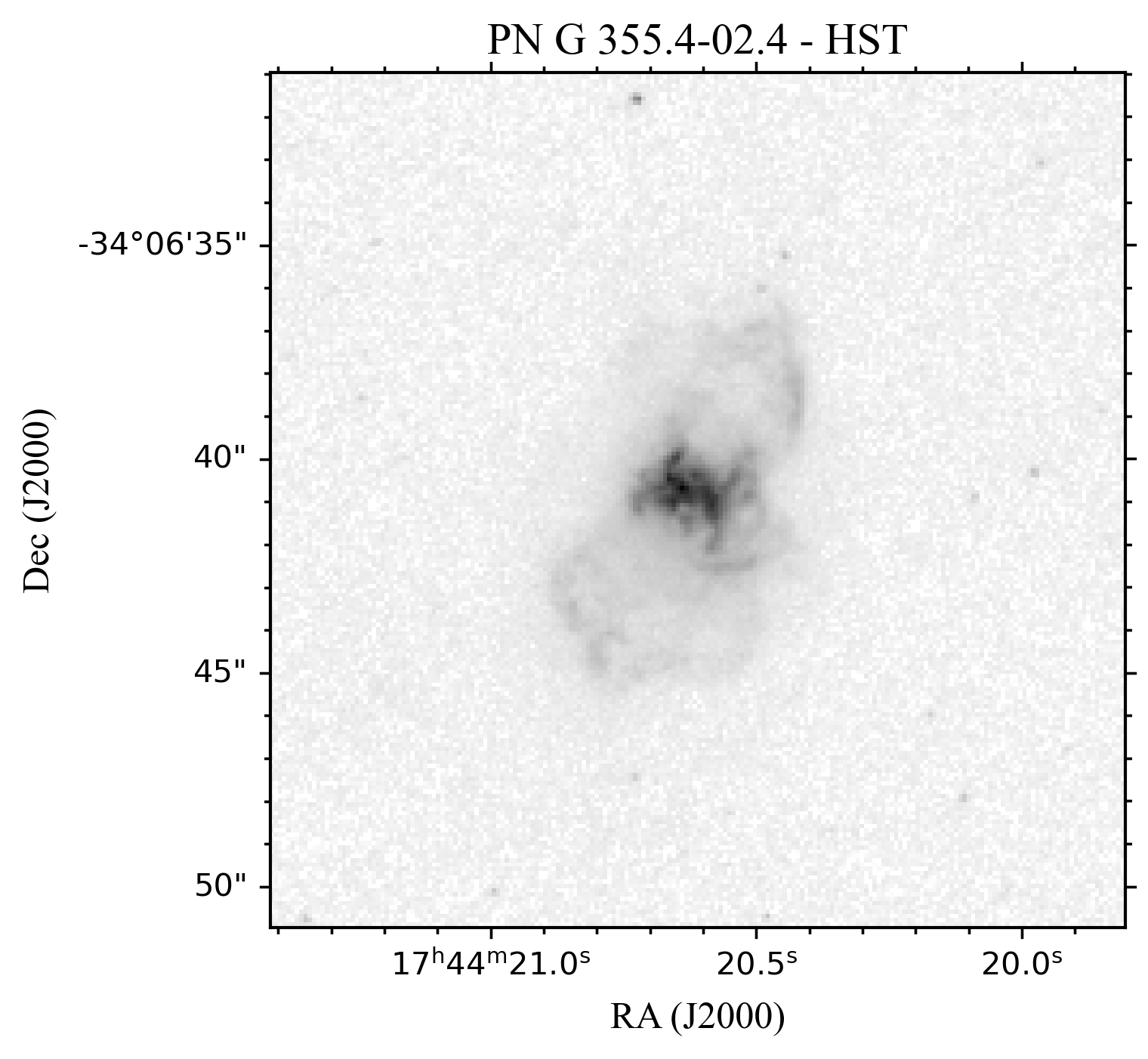}\hfill 
  \includegraphics[width=.32\linewidth]{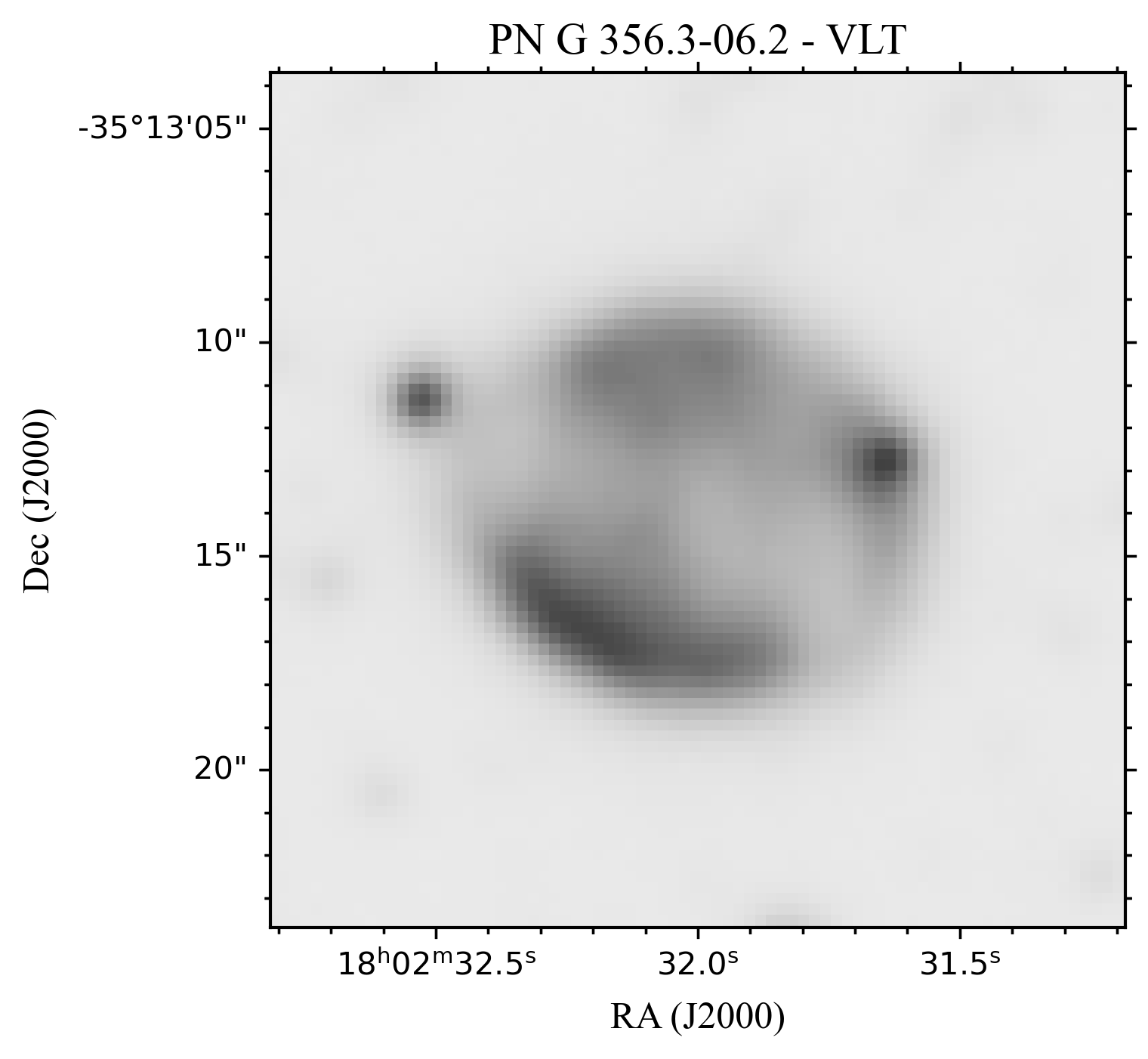}\hfill 
 \end{subfigure}\par\medskip 
\begin{subfigure}{\linewidth} 
  \includegraphics[width=.32\linewidth]{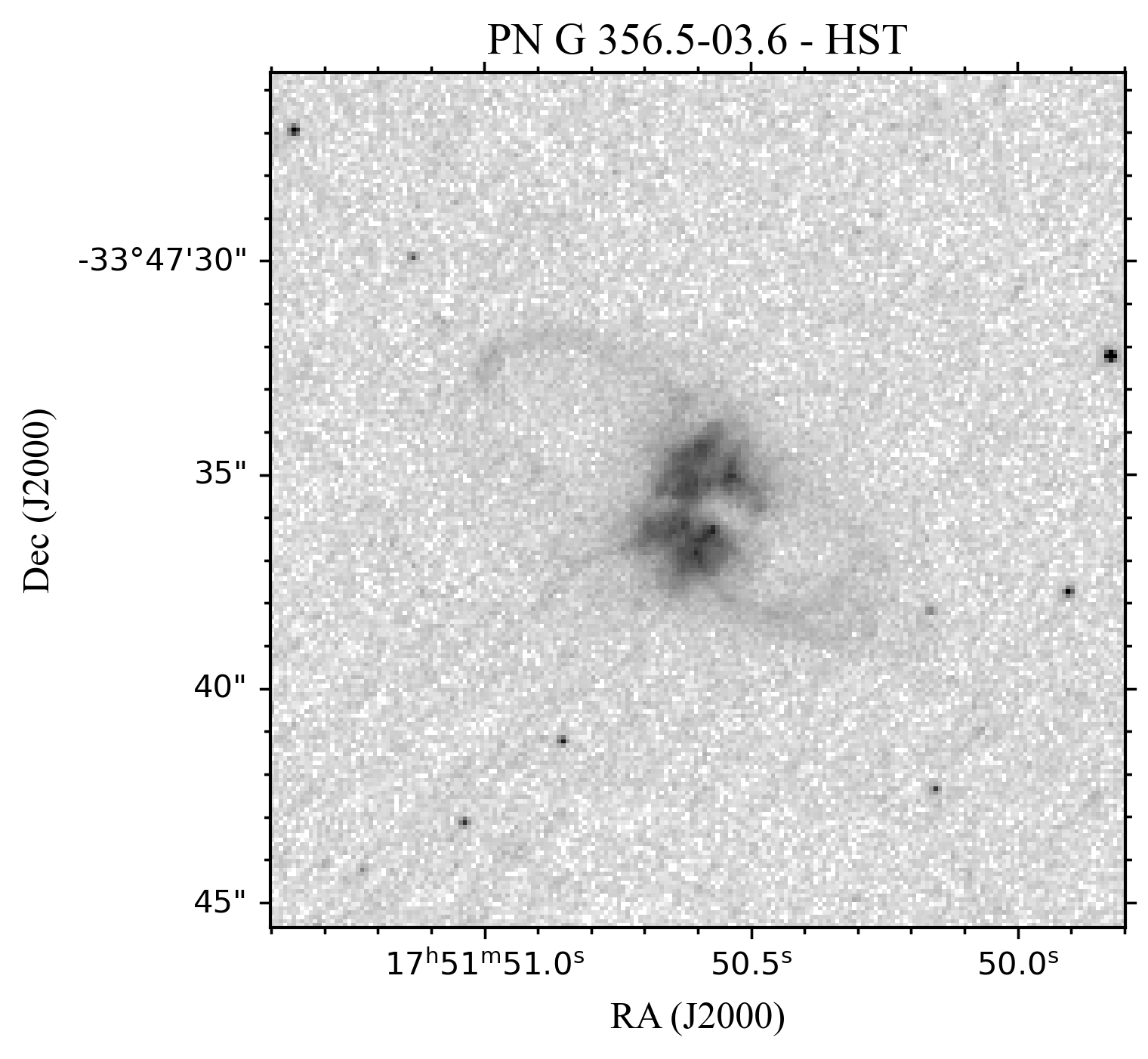}\hfill 
  \includegraphics[width=.32\linewidth]{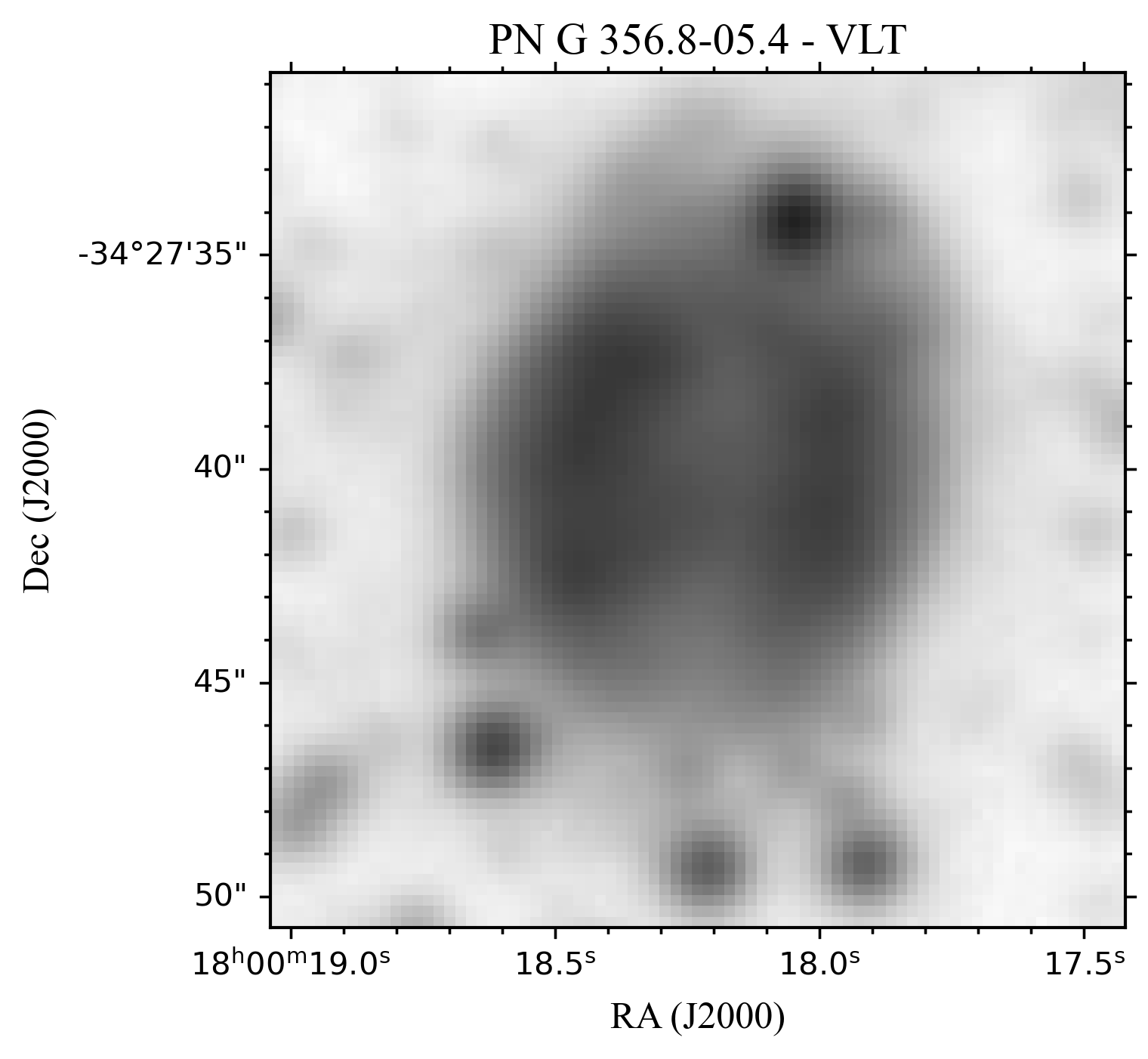}\hfill 
  \includegraphics[width=.32\linewidth]{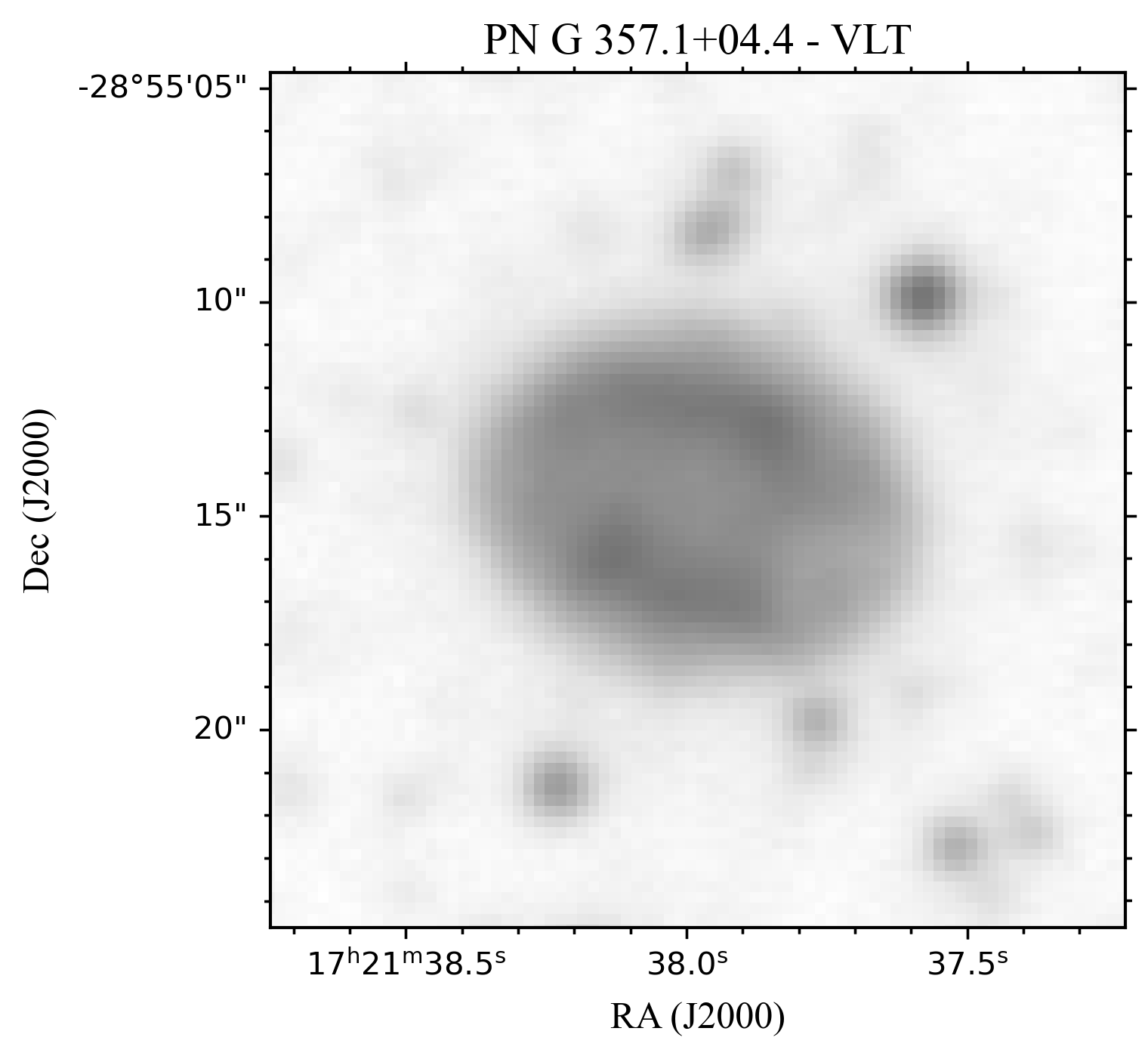}\hfill 
 \end{subfigure}\par\medskip 
\begin{subfigure}{\linewidth} 
  \includegraphics[width=.32\linewidth]{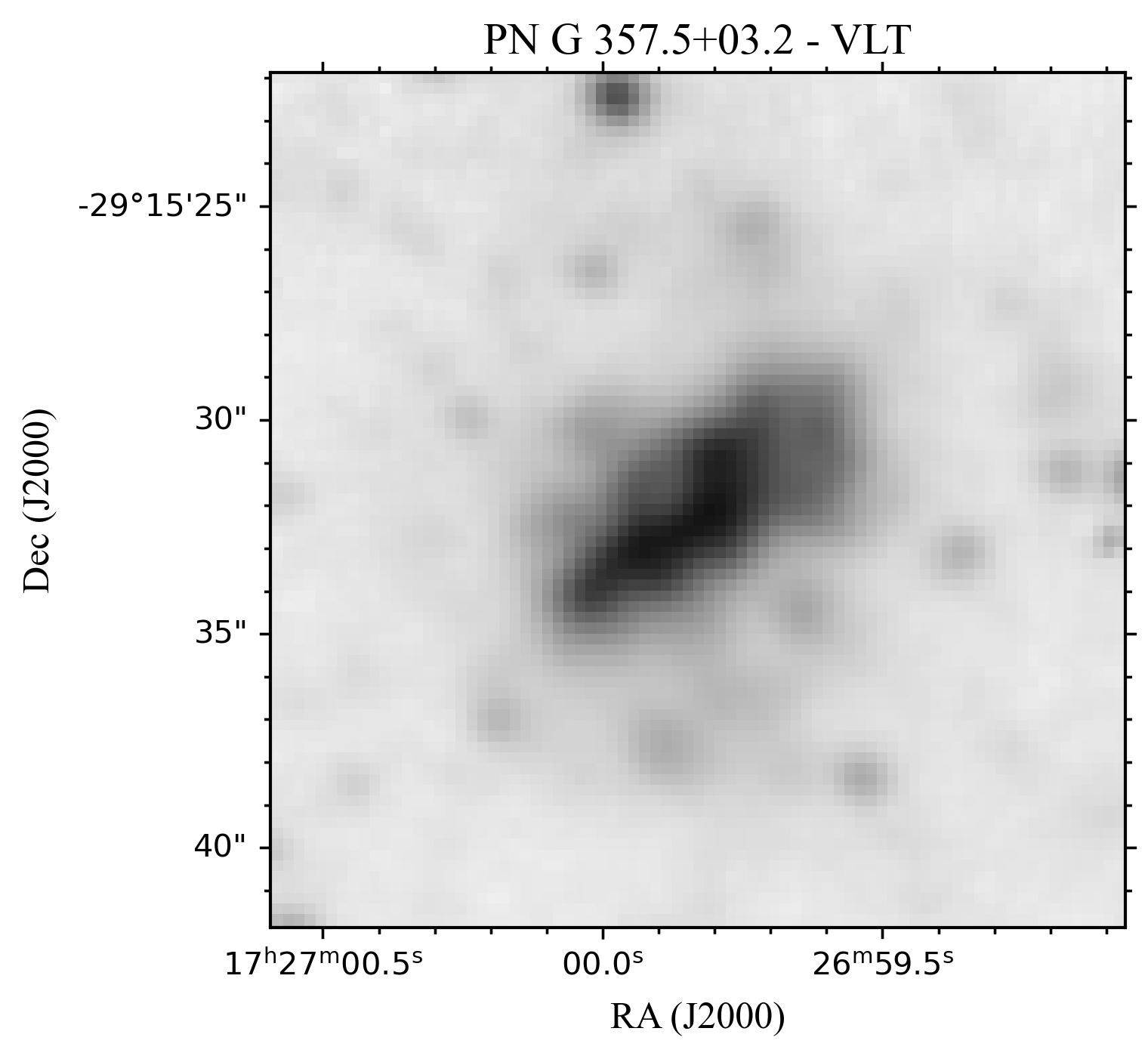}\hfill 
  \includegraphics[width=.32\linewidth]{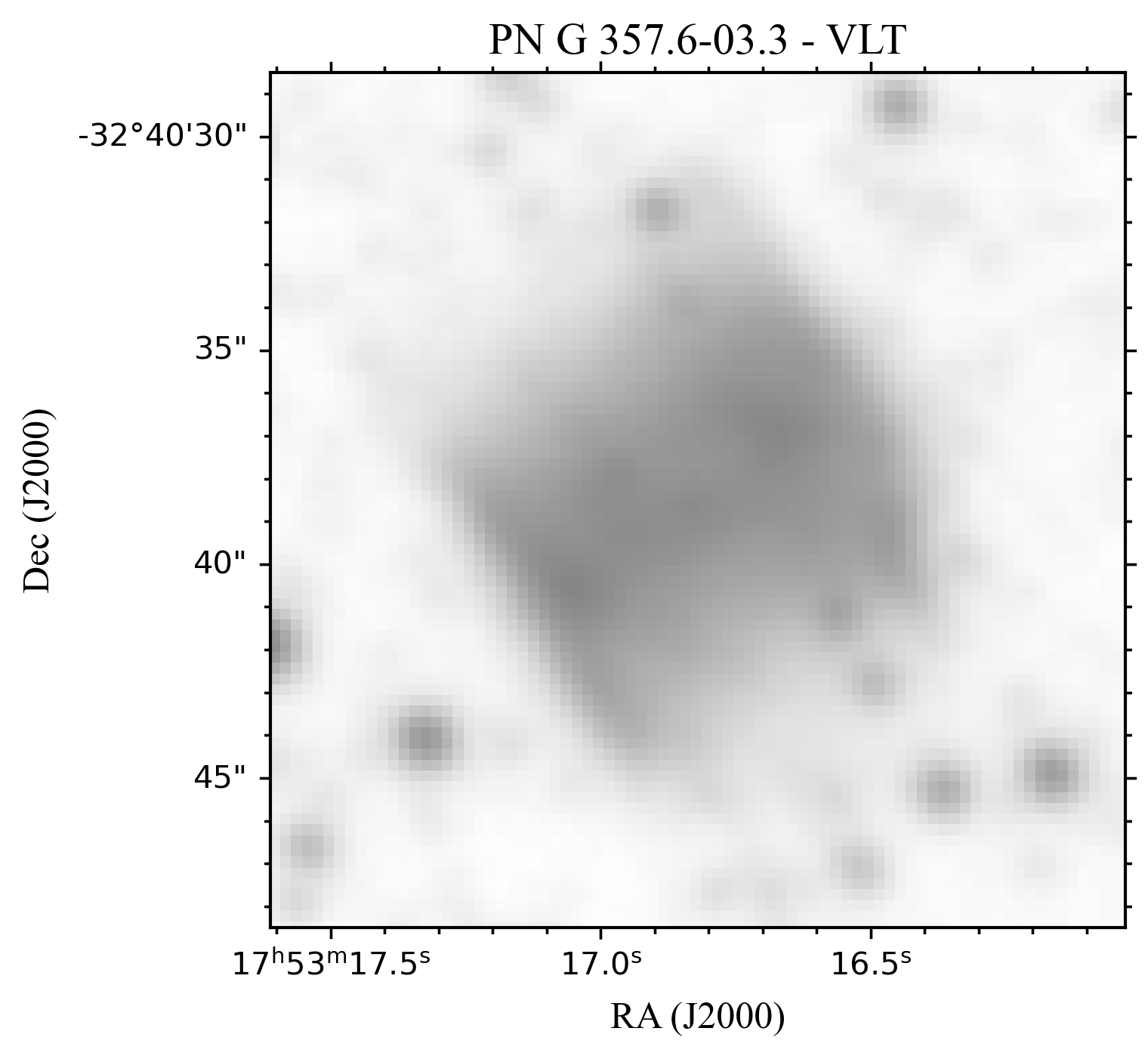}\hfill 
  \includegraphics[width=.32\linewidth]{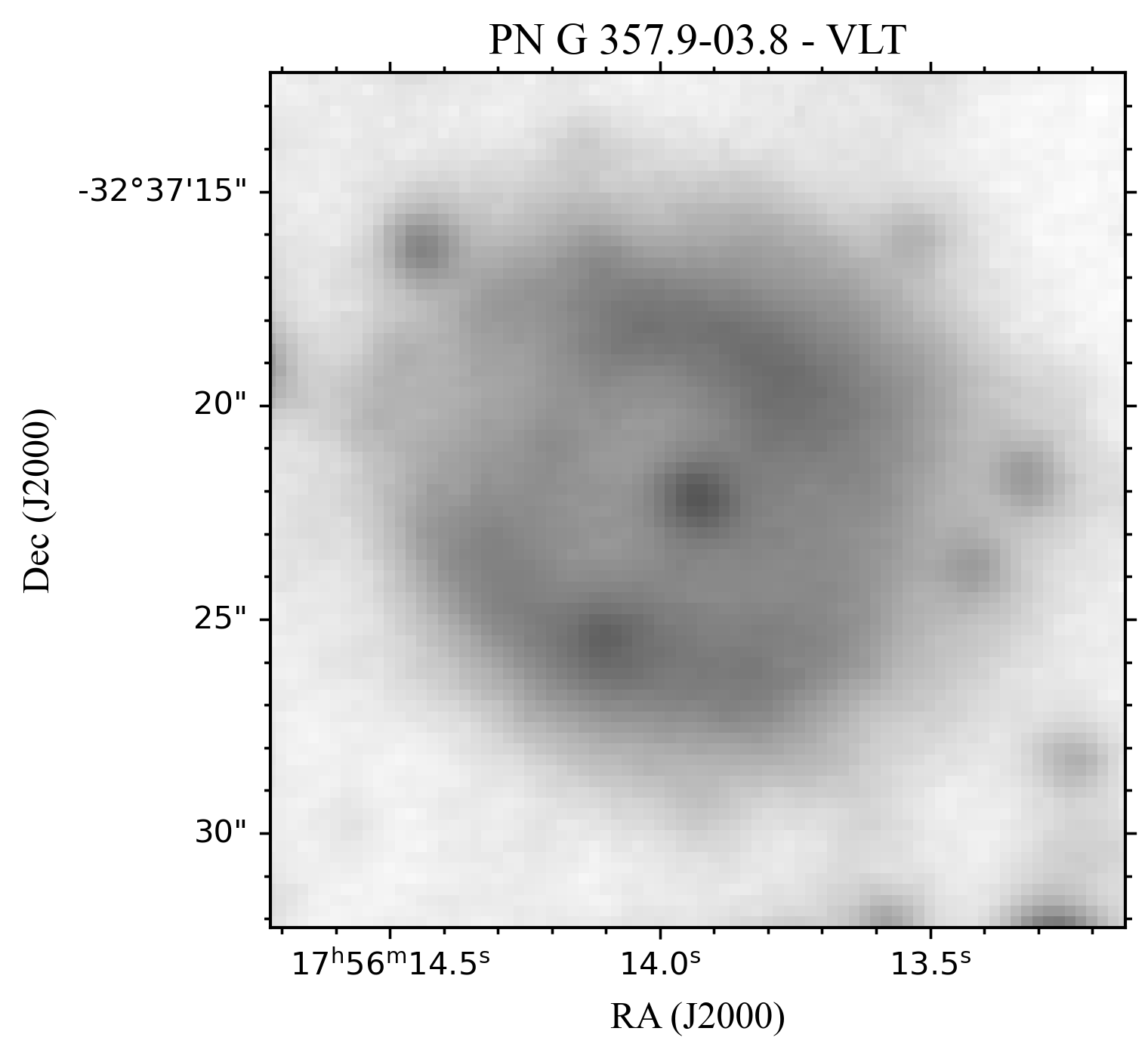}\hfill 
 \end{subfigure}\par\medskip 
  \end{figure} 
 \begin{figure} 
 \ContinuedFloat 
 \caption[]{continued:} 
\begin{subfigure}{\linewidth} 
  \includegraphics[width=.32\linewidth]{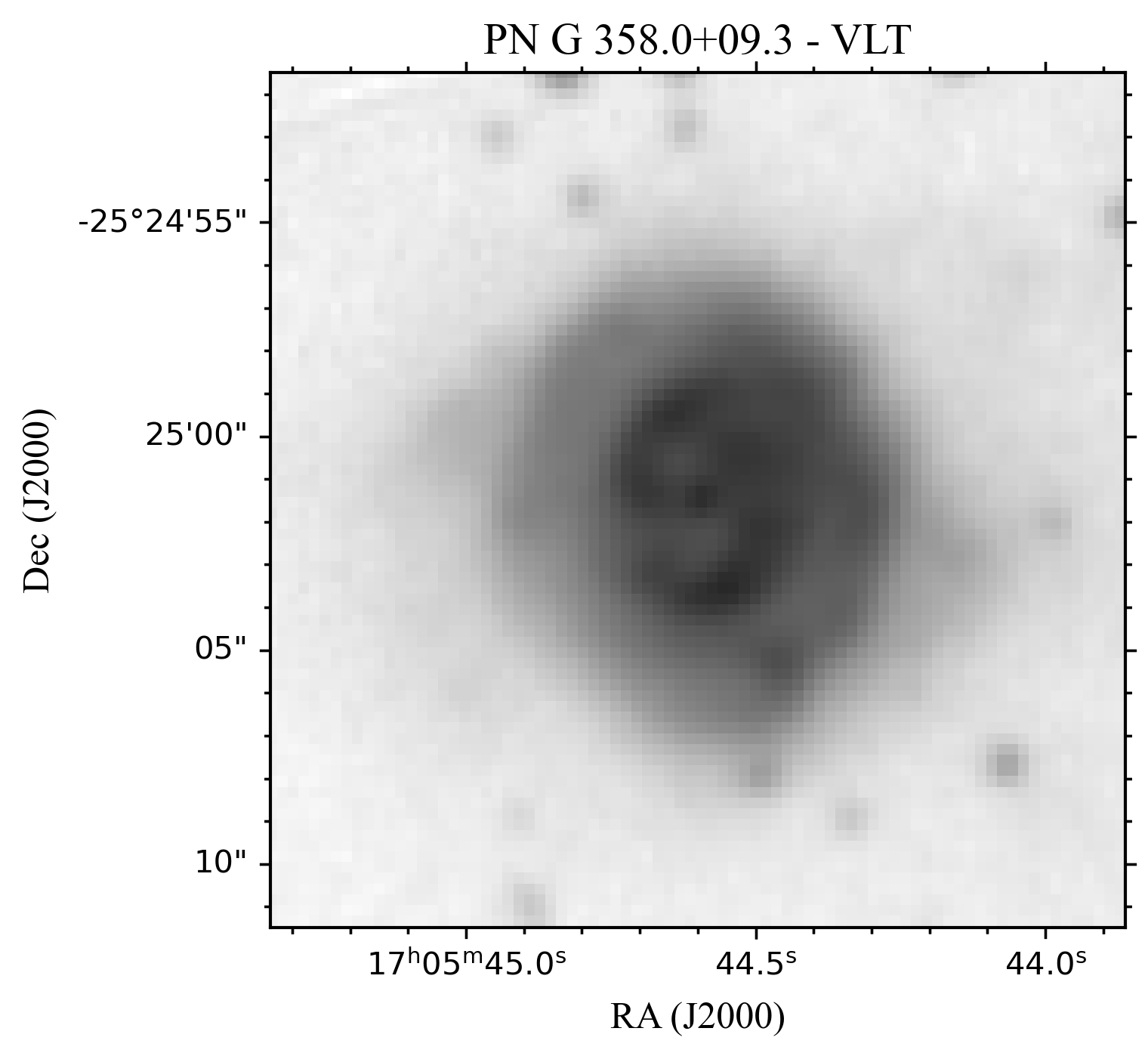}\hfill 
  \includegraphics[width=.32\linewidth]{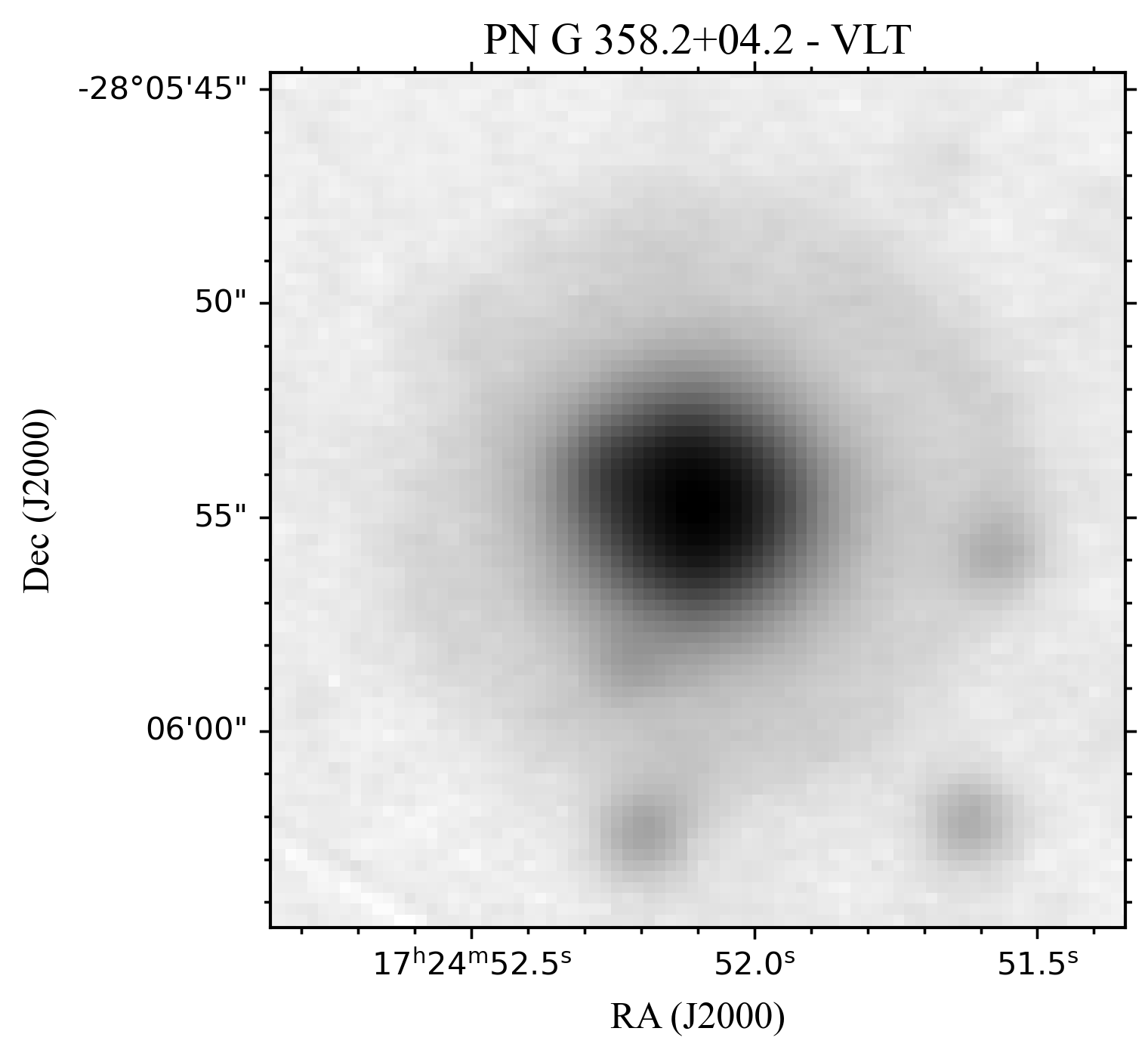}\hfill 
  \includegraphics[width=.32\linewidth]{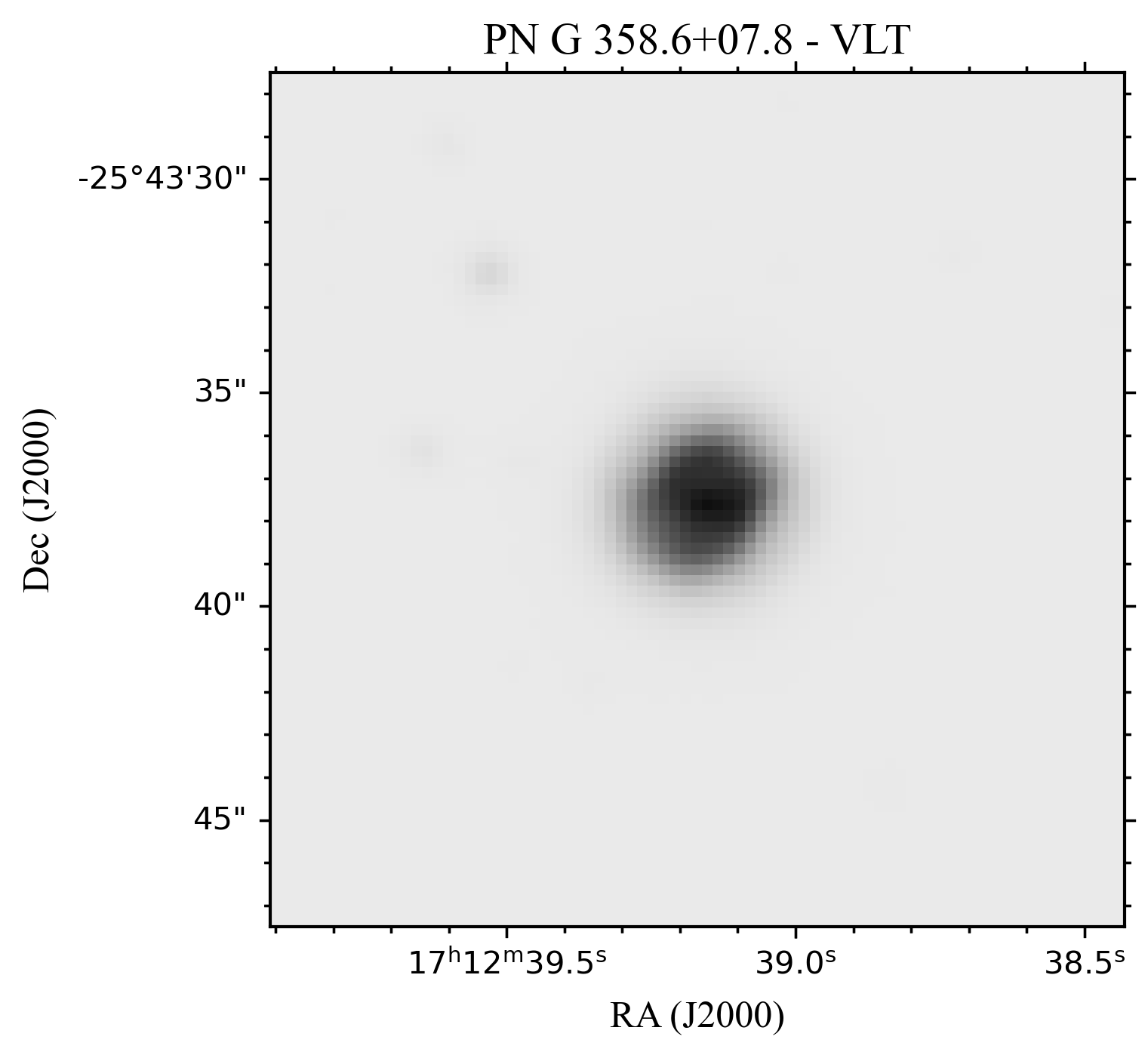}\hfill 
 \end{subfigure}\par\medskip 
\begin{subfigure}{\linewidth} 
  \includegraphics[width=.32\linewidth]{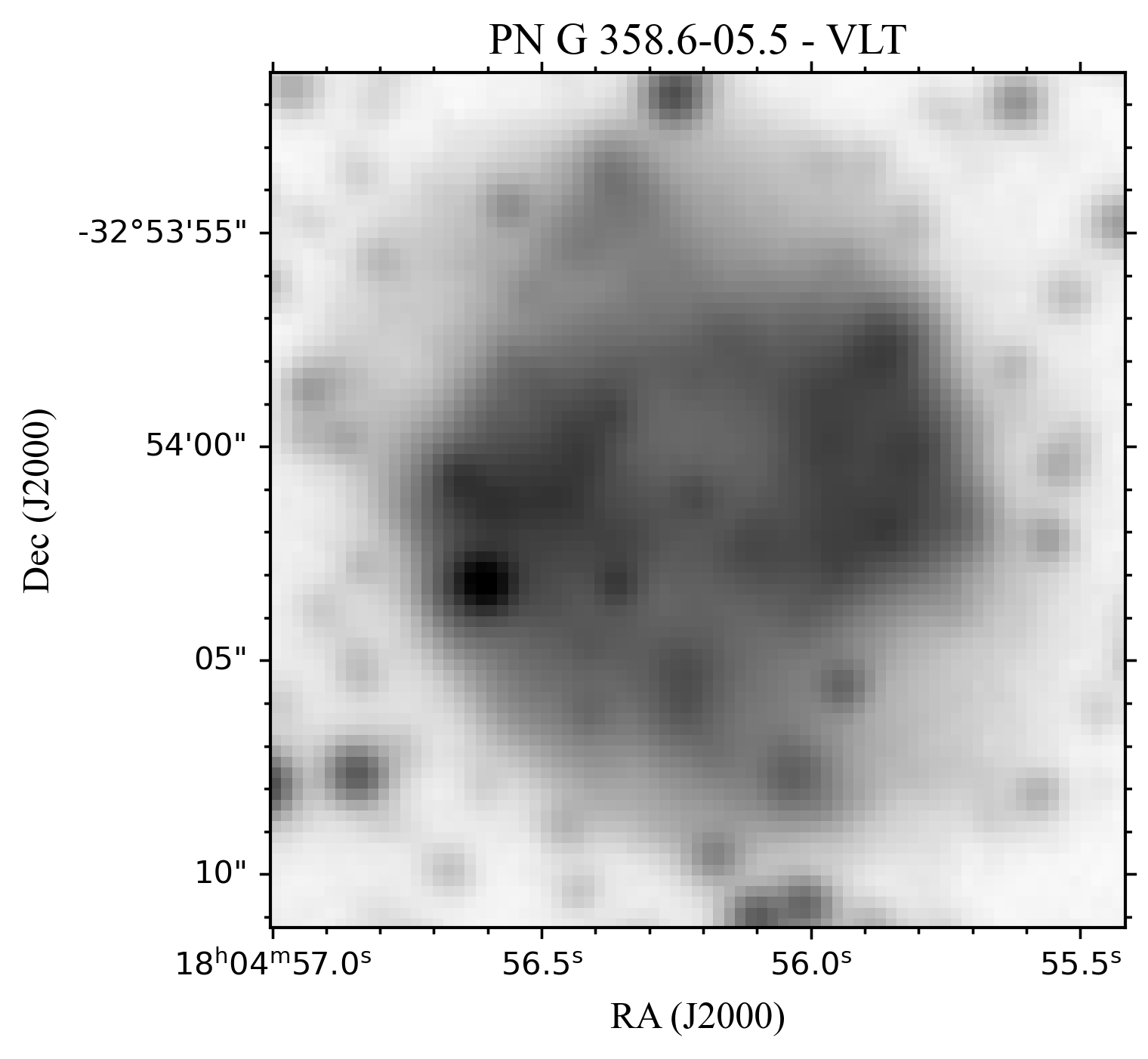}\hfill 
  \includegraphics[width=.32\linewidth]{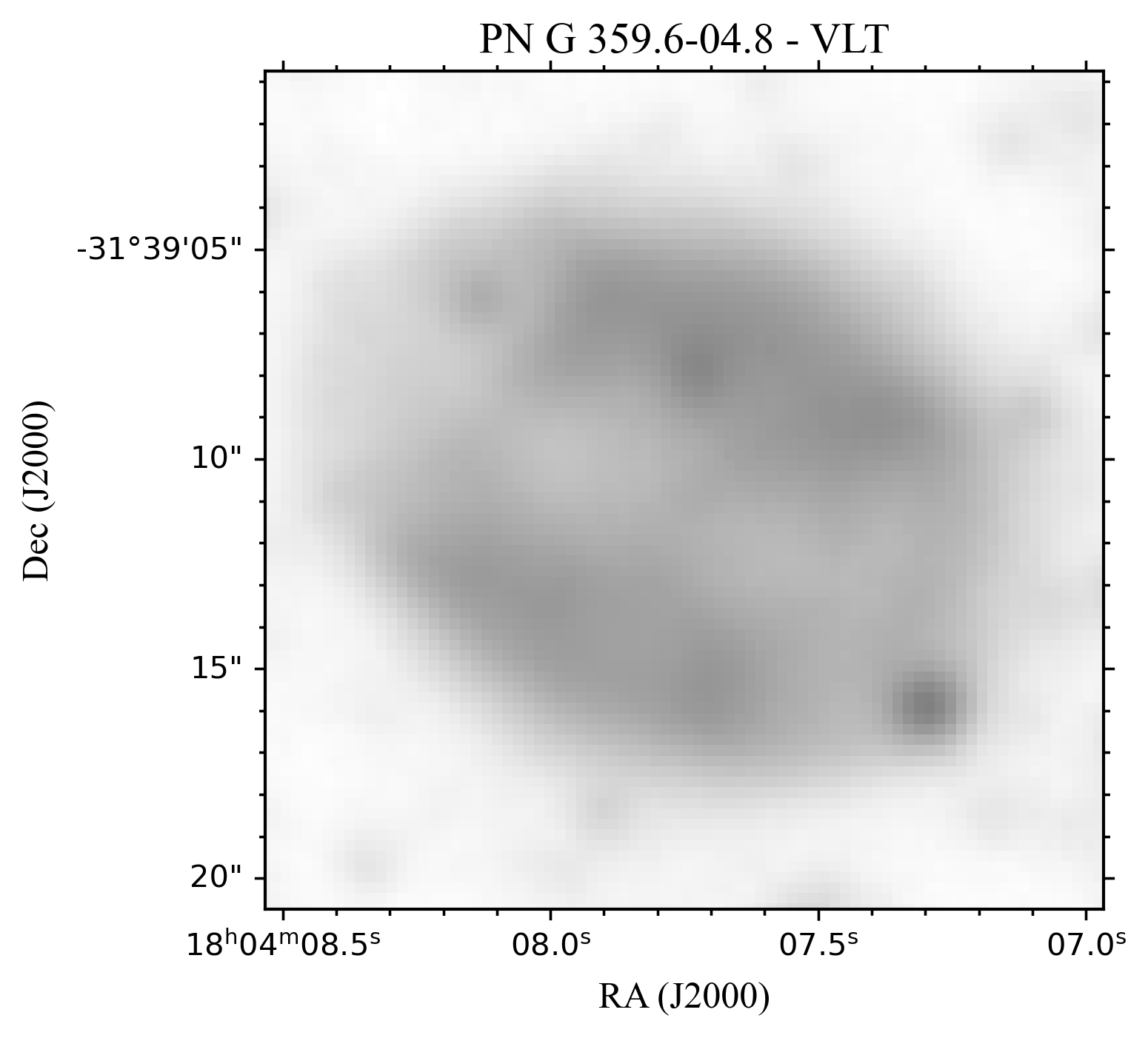}\hfill 
  \includegraphics[width=.32\linewidth]{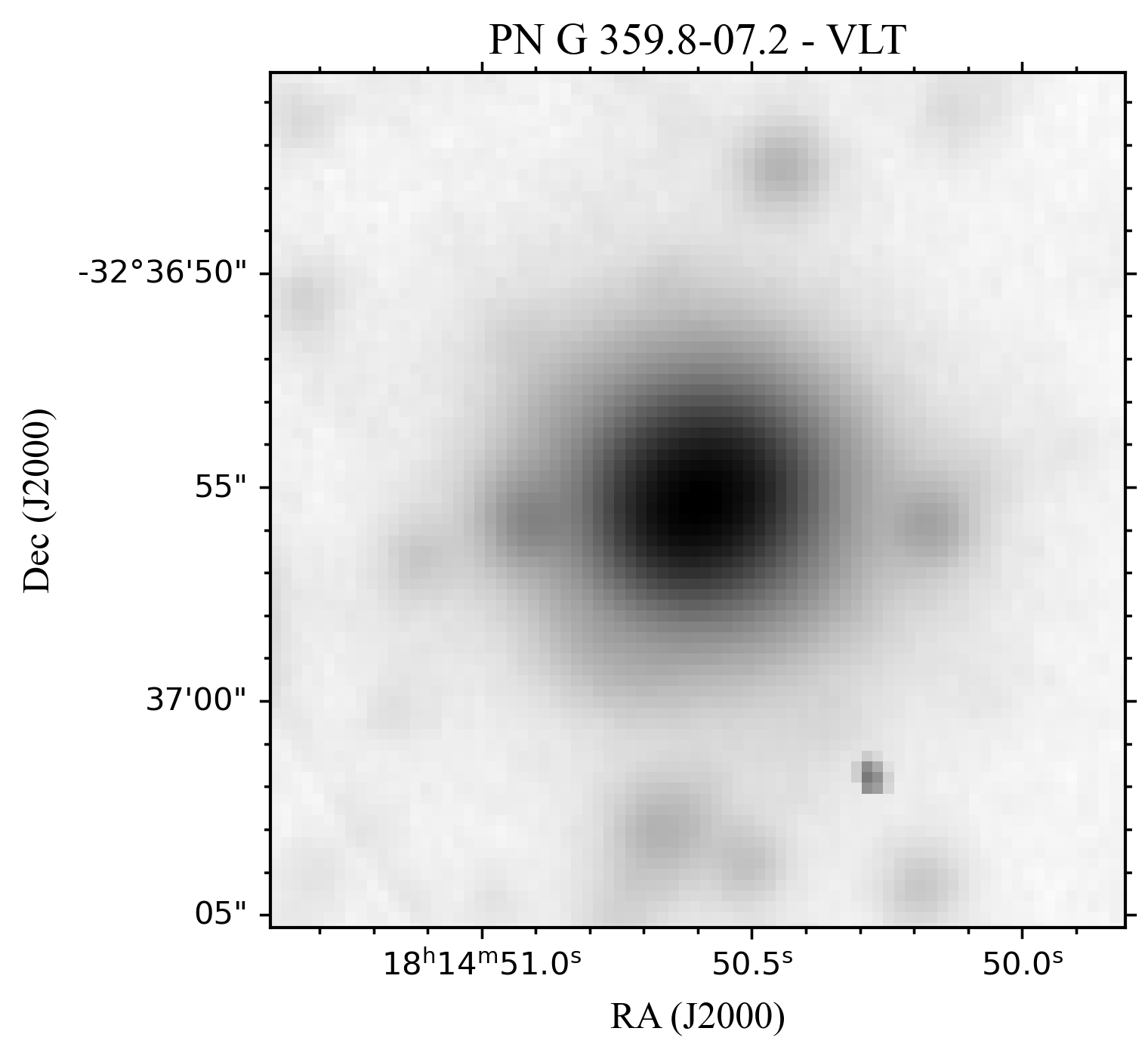}\hfill 
 \end{subfigure}\par\medskip 
\begin{subfigure}{\linewidth} 
\caption{PNe with angular sizes less than 10 arcsec. Each image has a field of view of 10" $\times$ 10".}
\includegraphics[width=.32\linewidth]{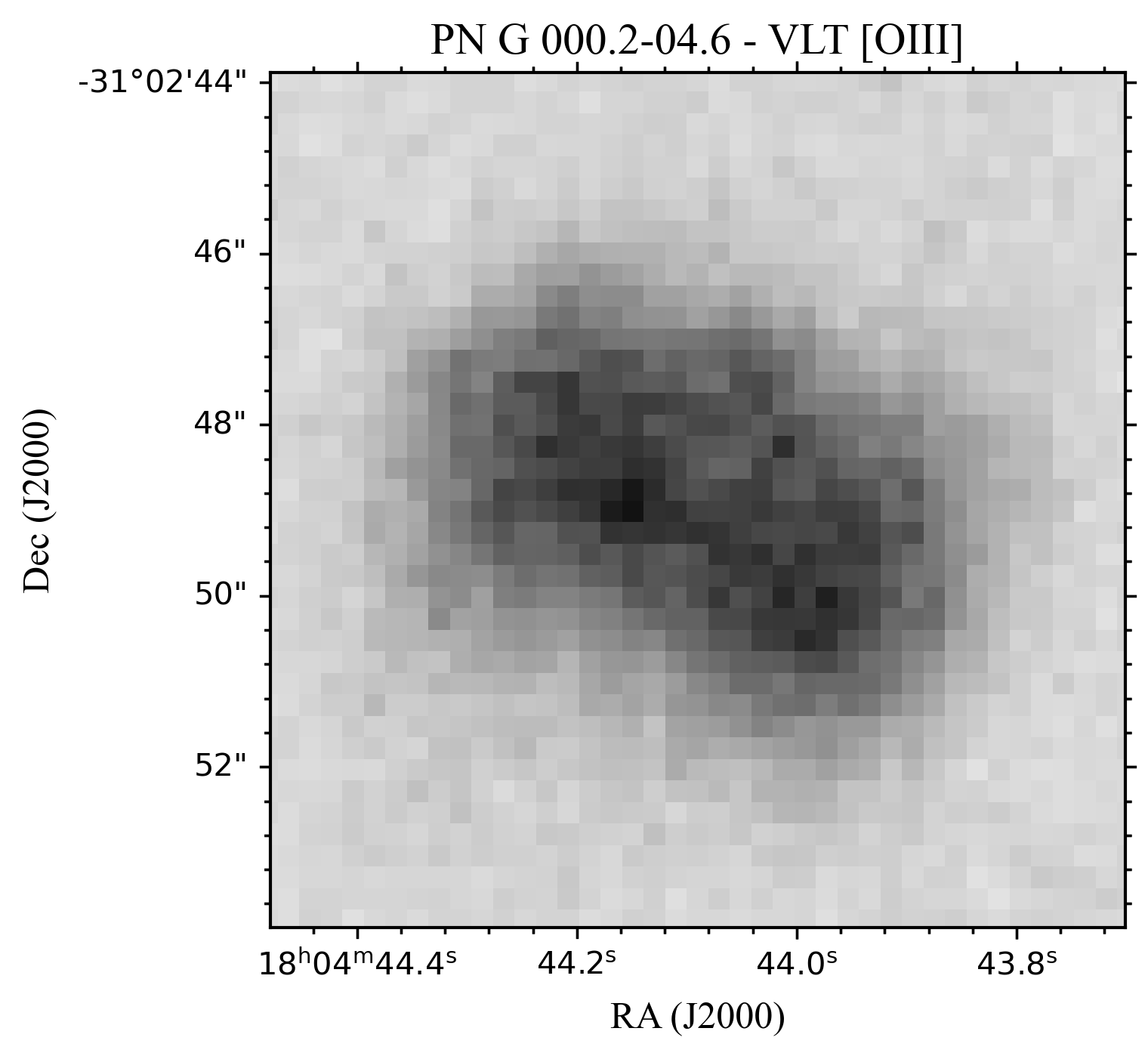}\hfill 
  \includegraphics[width=.32\linewidth]{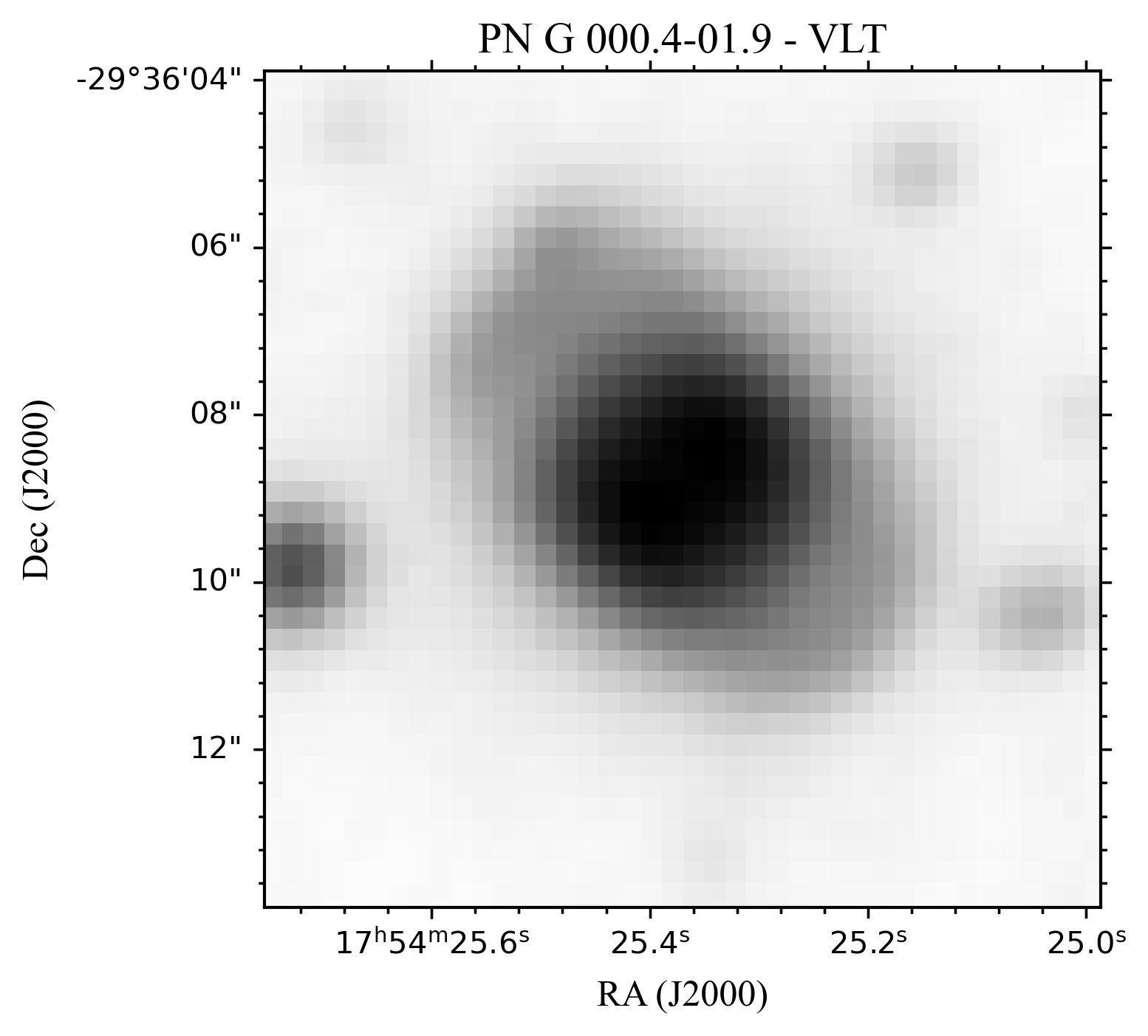}\hfill 
  \includegraphics[width=.32\linewidth]{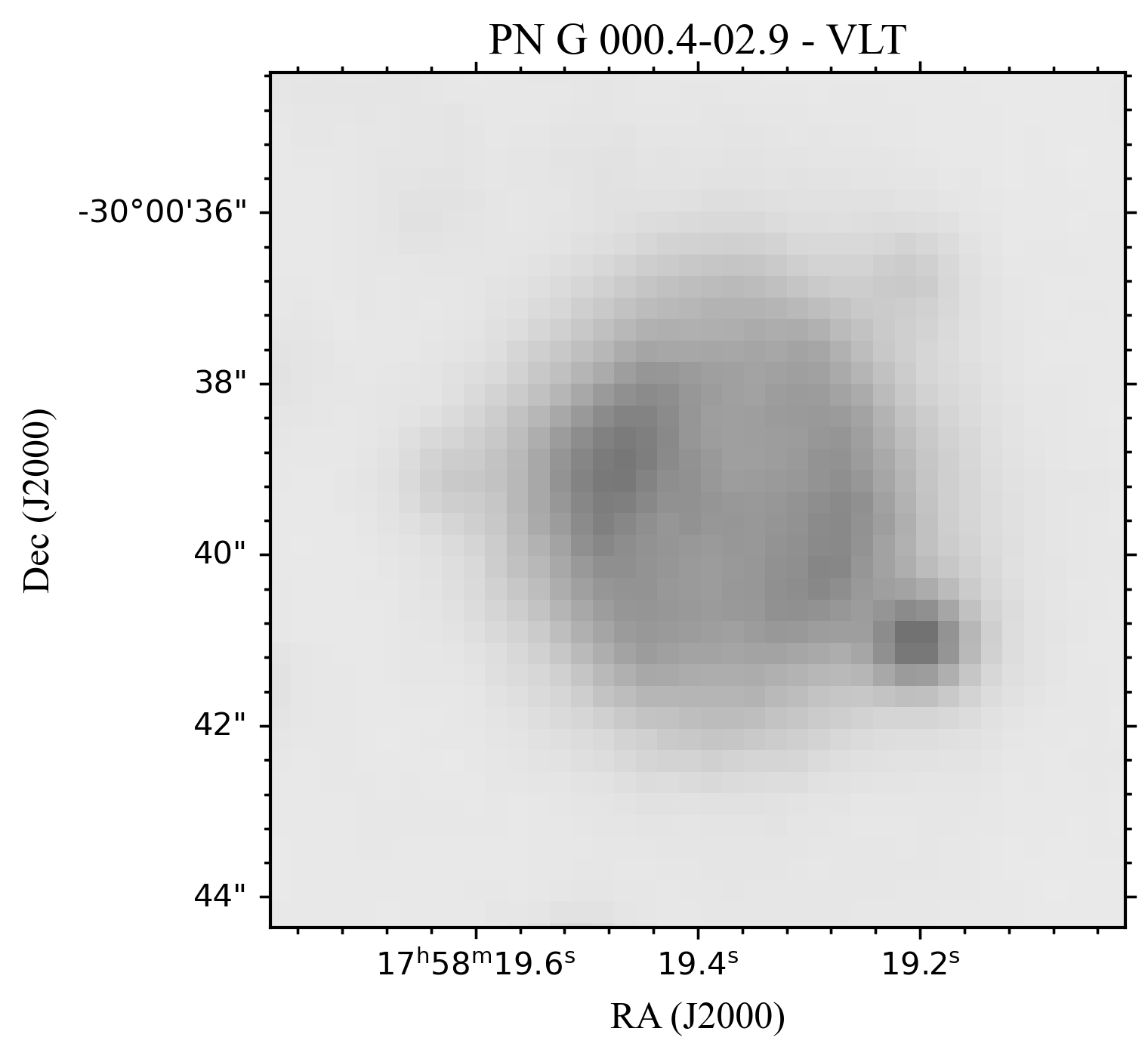}\hfill 
 \end{subfigure}\par\medskip 
\begin{subfigure}{\linewidth} 
  \includegraphics[width=.32\linewidth]{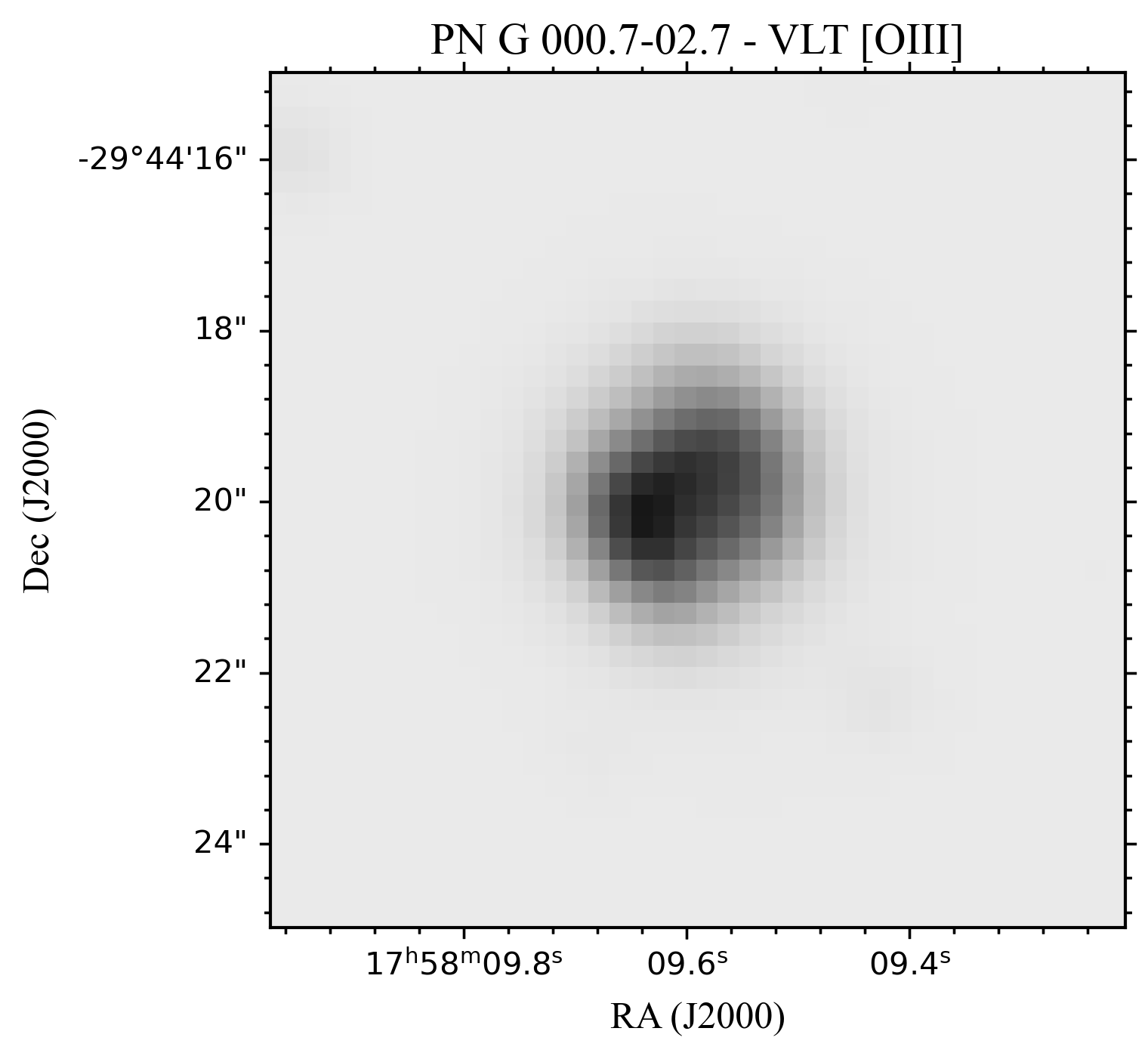}\hfill 
  \includegraphics[width=.32\linewidth]{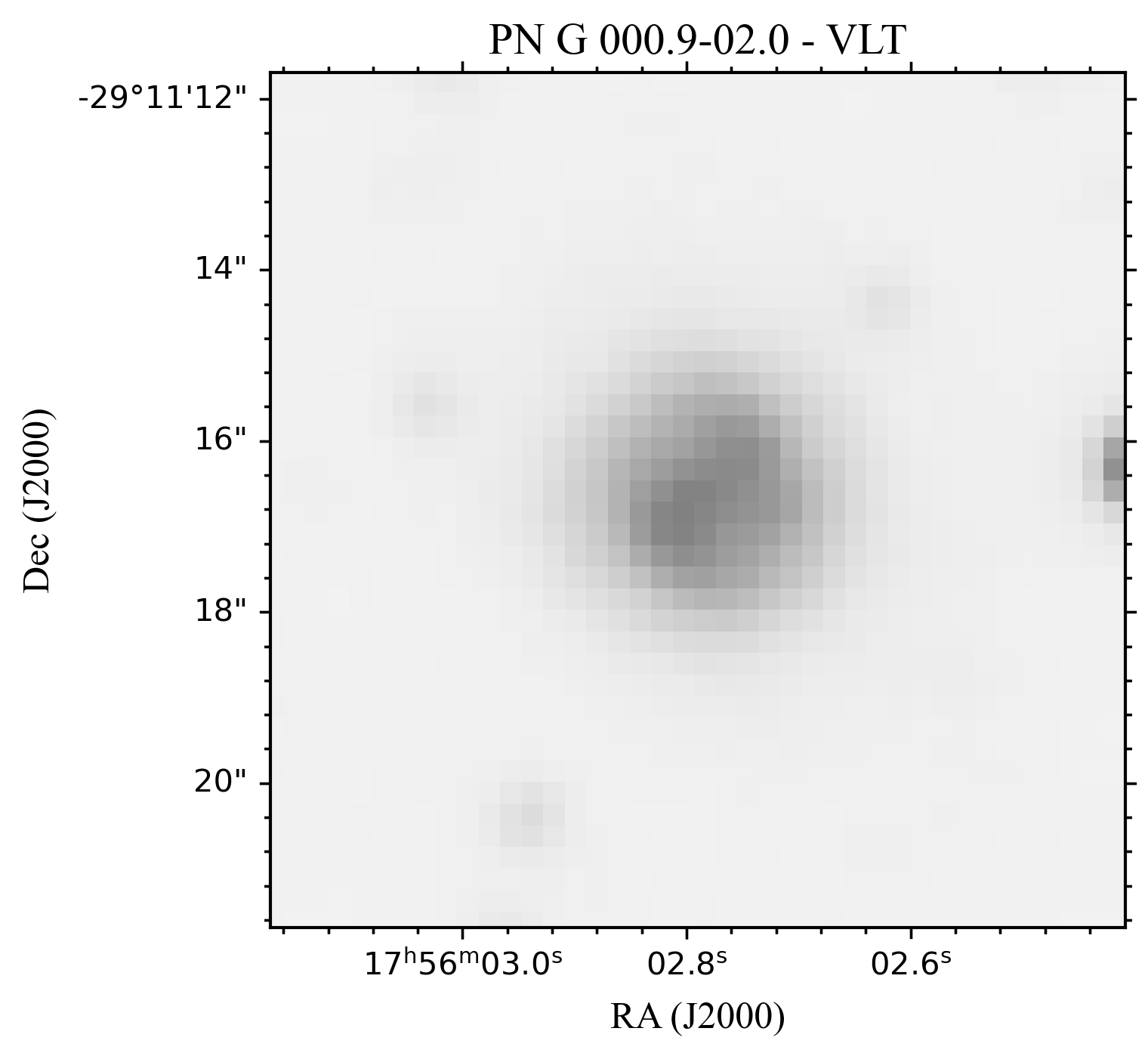}\hfill 
  \includegraphics[width=.32\linewidth]{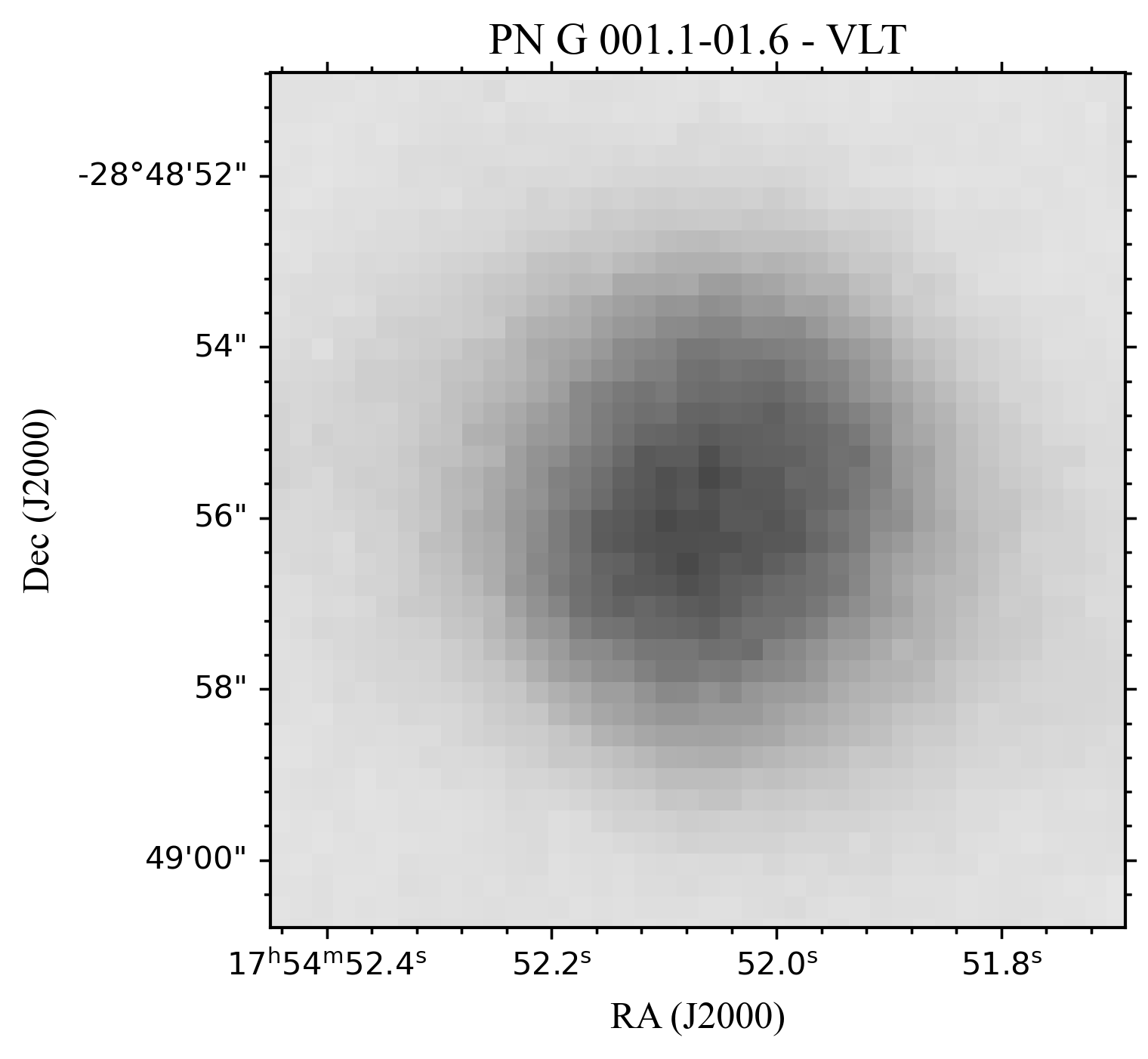}\hfill 
 \end{subfigure}\par\medskip 
  \end{figure} 
 \begin{figure} 
 \ContinuedFloat 
 \caption[]{continued:} 
\begin{subfigure}{\linewidth} 
  \includegraphics[width=.32\linewidth]{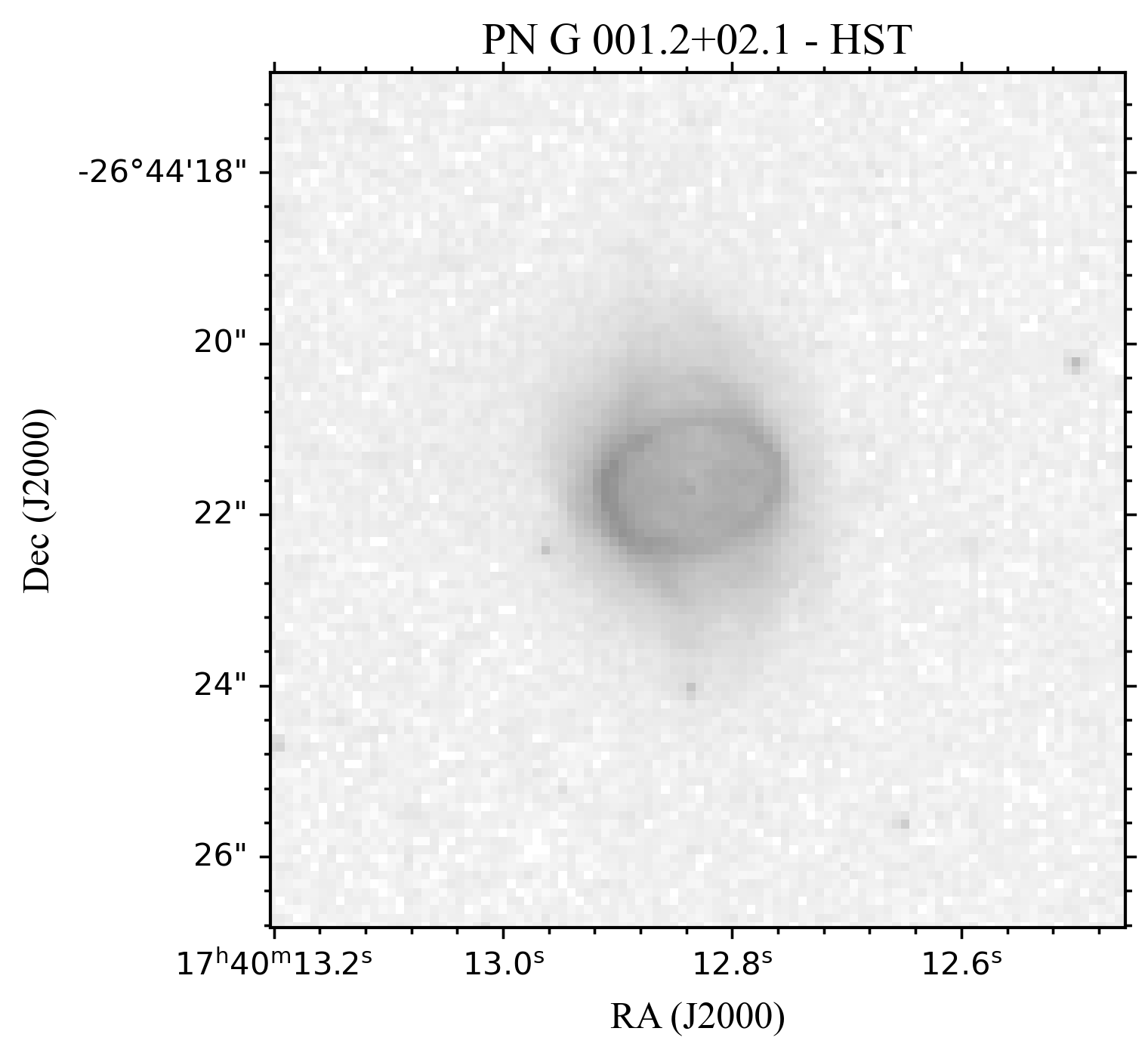}\hfill 
  \includegraphics[width=.32\linewidth]{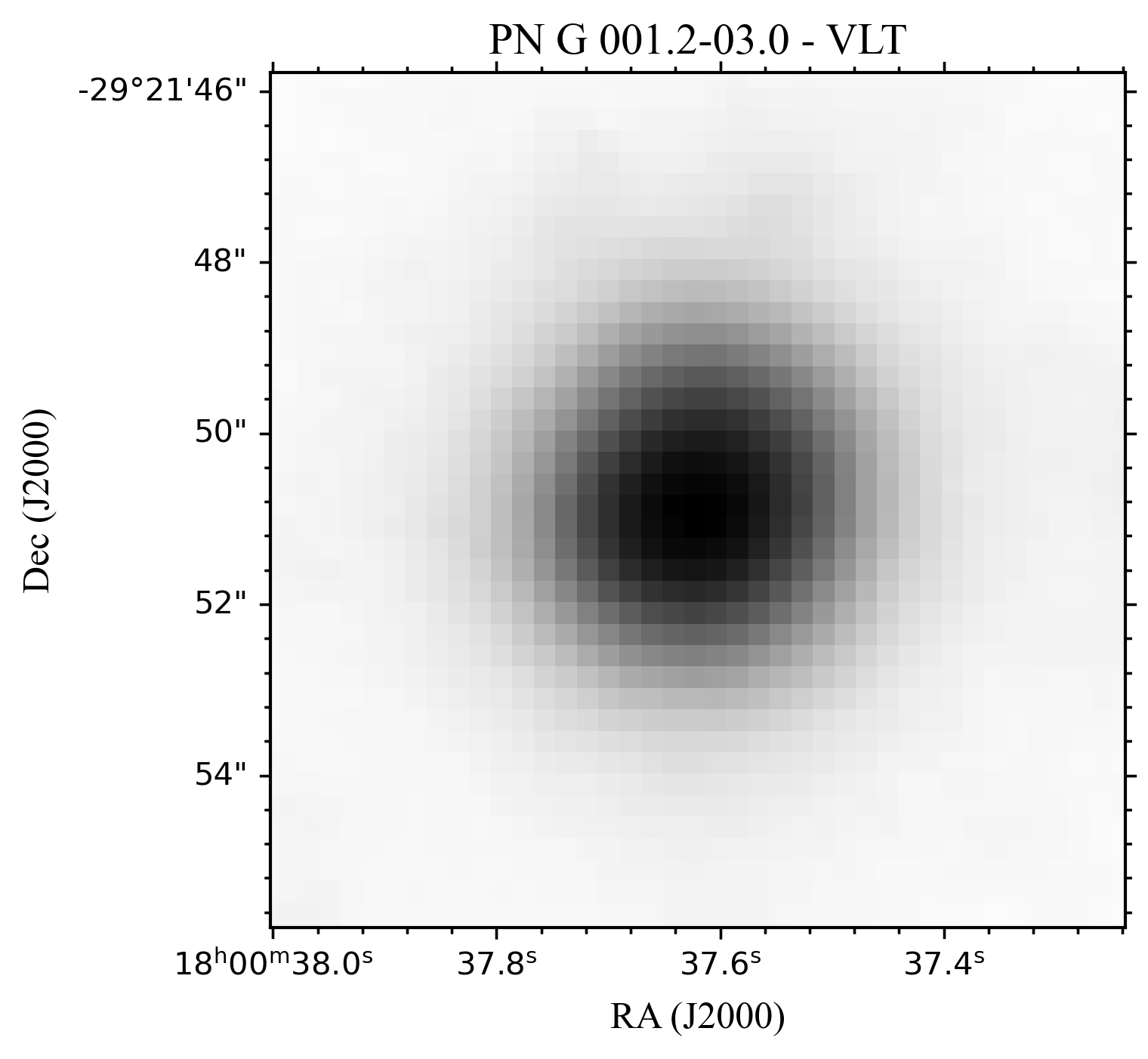}\hfill 
  \includegraphics[width=.32\linewidth]{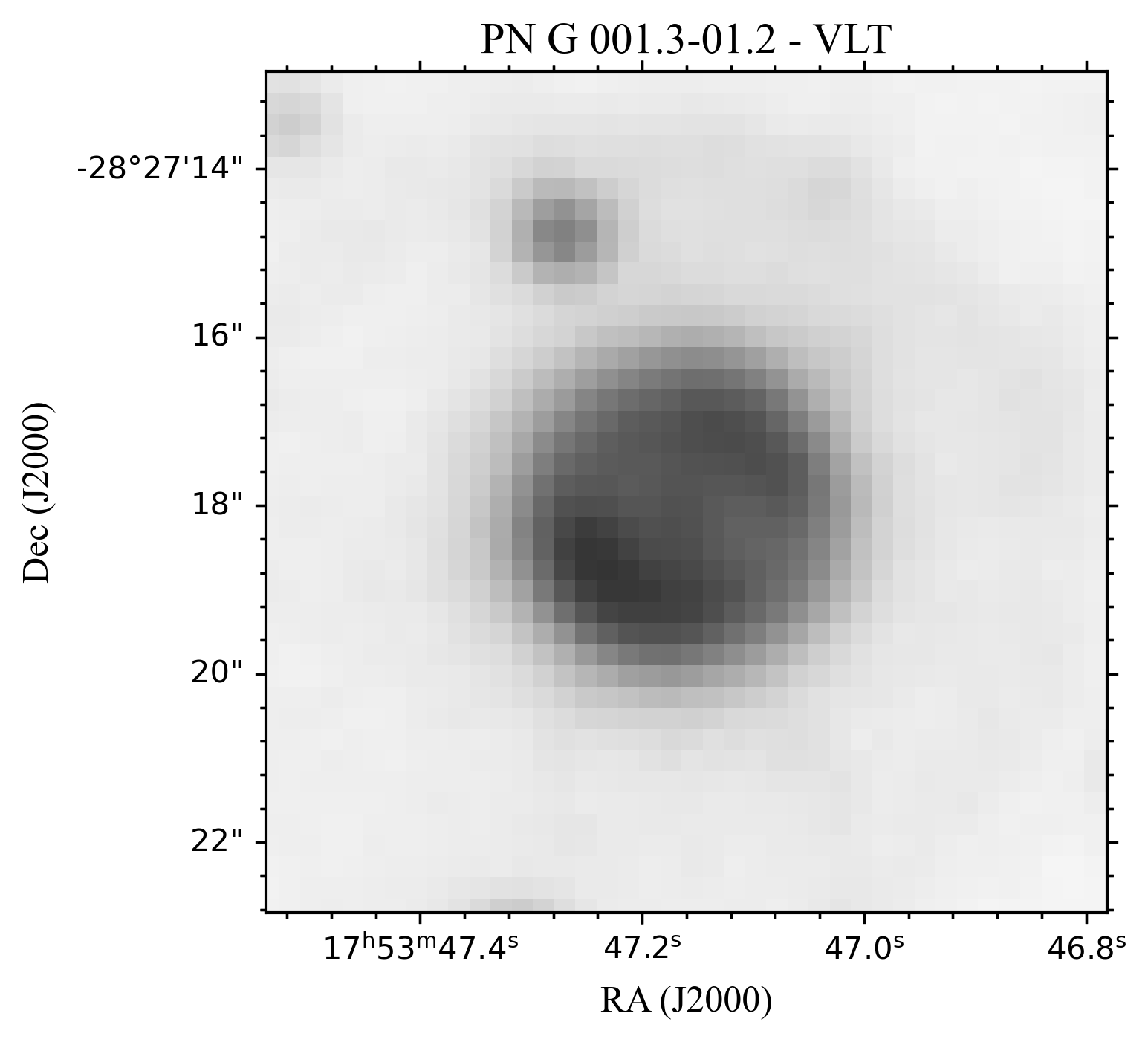}\hfill 
 \end{subfigure}\par\medskip 
\begin{subfigure}{\linewidth} 
  \includegraphics[width=.32\linewidth]{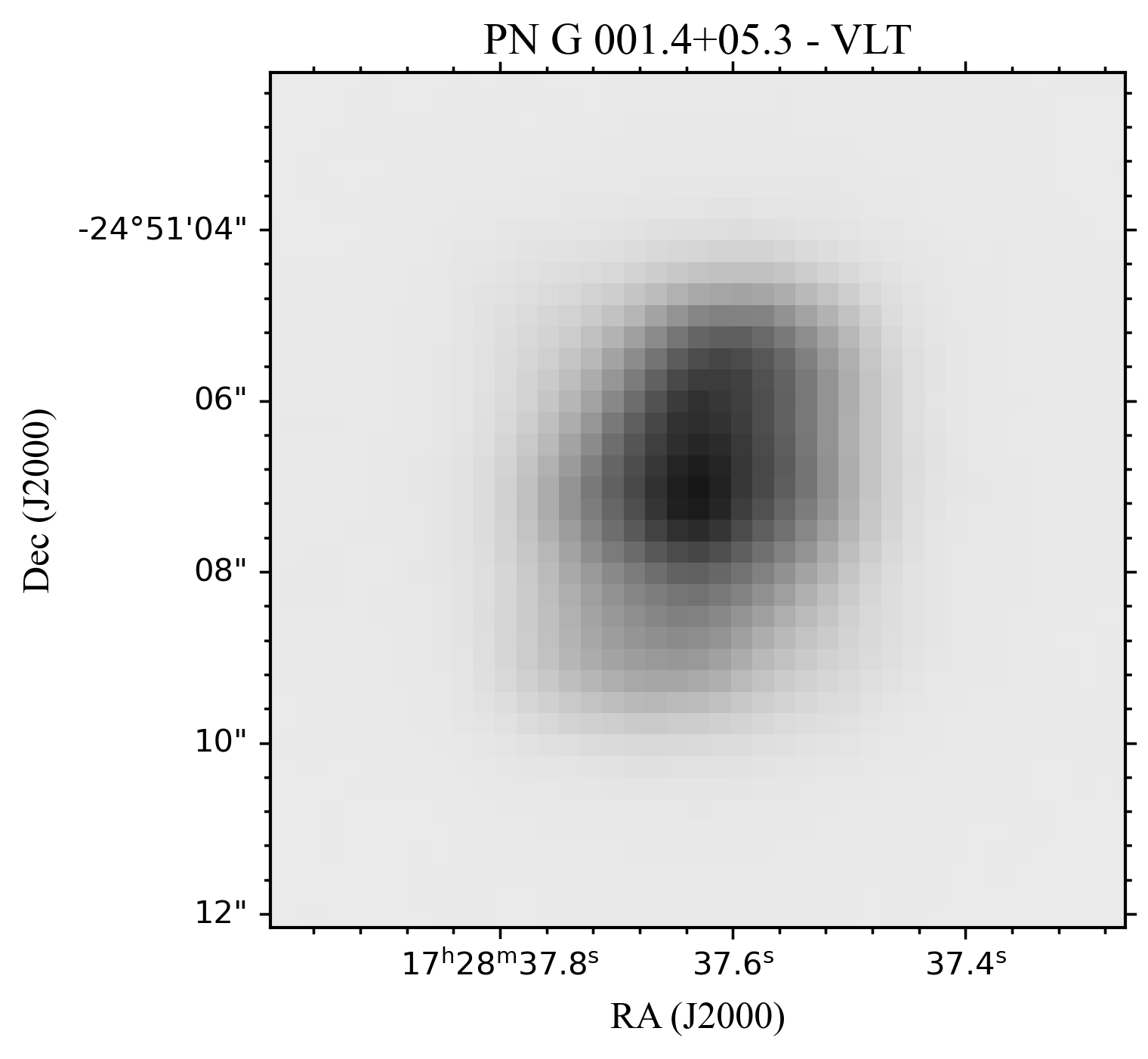}\hfill 
  \includegraphics[width=.32\linewidth]{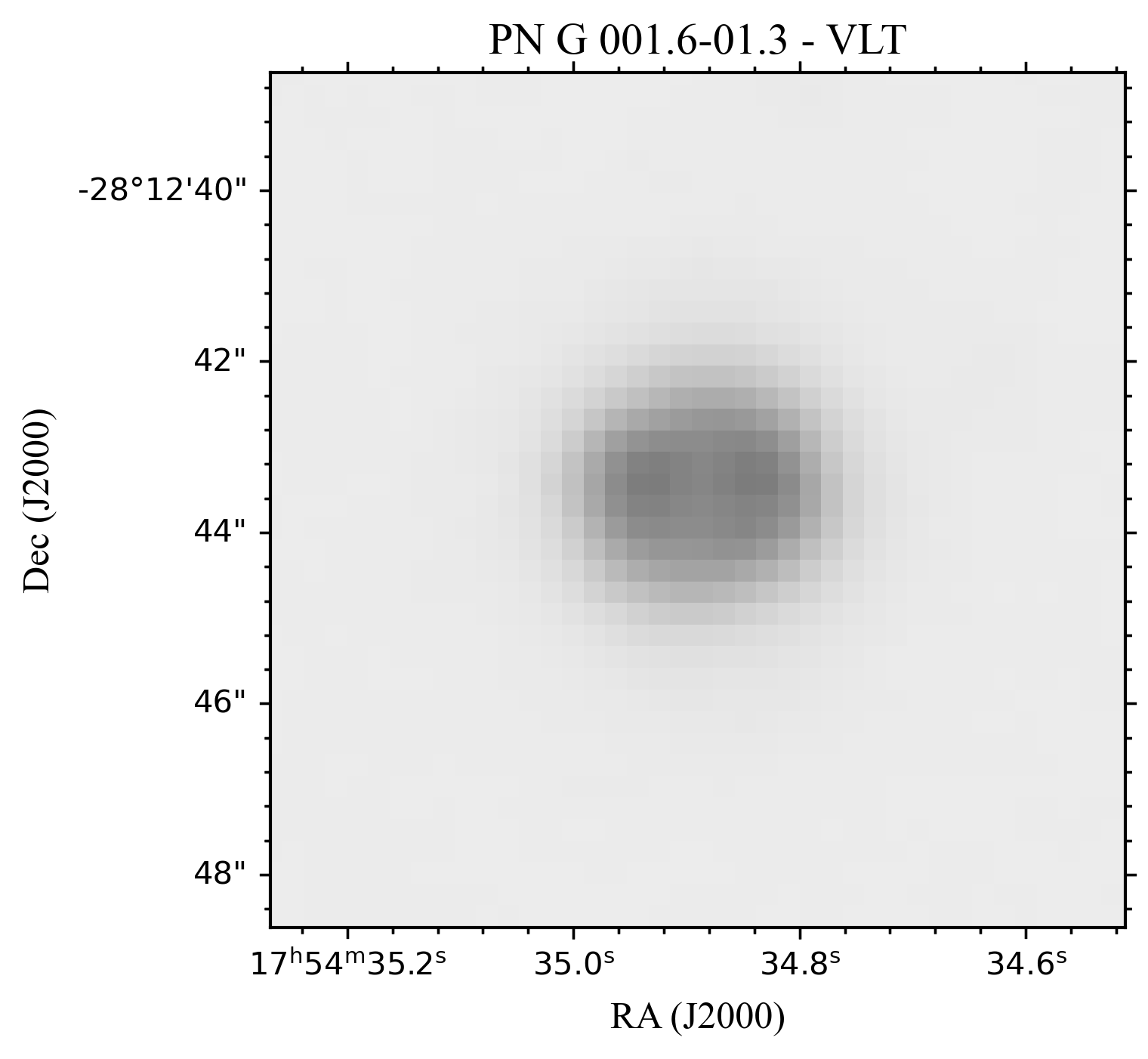}\hfill 
  \includegraphics[width=.32\linewidth]{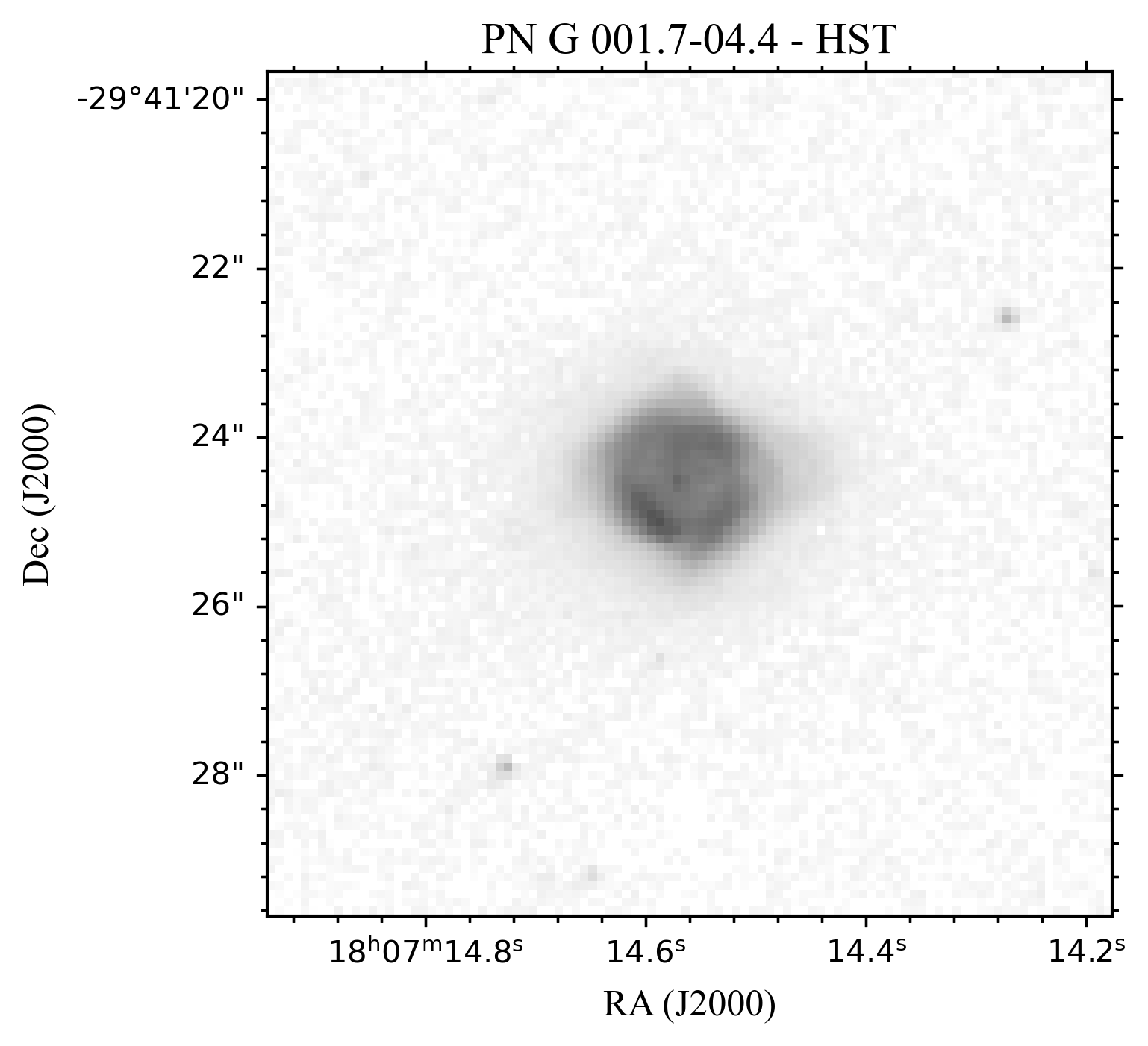}\hfill 
 \end{subfigure}\par\medskip 
\begin{subfigure}{\linewidth} 
  \includegraphics[width=.32\linewidth]{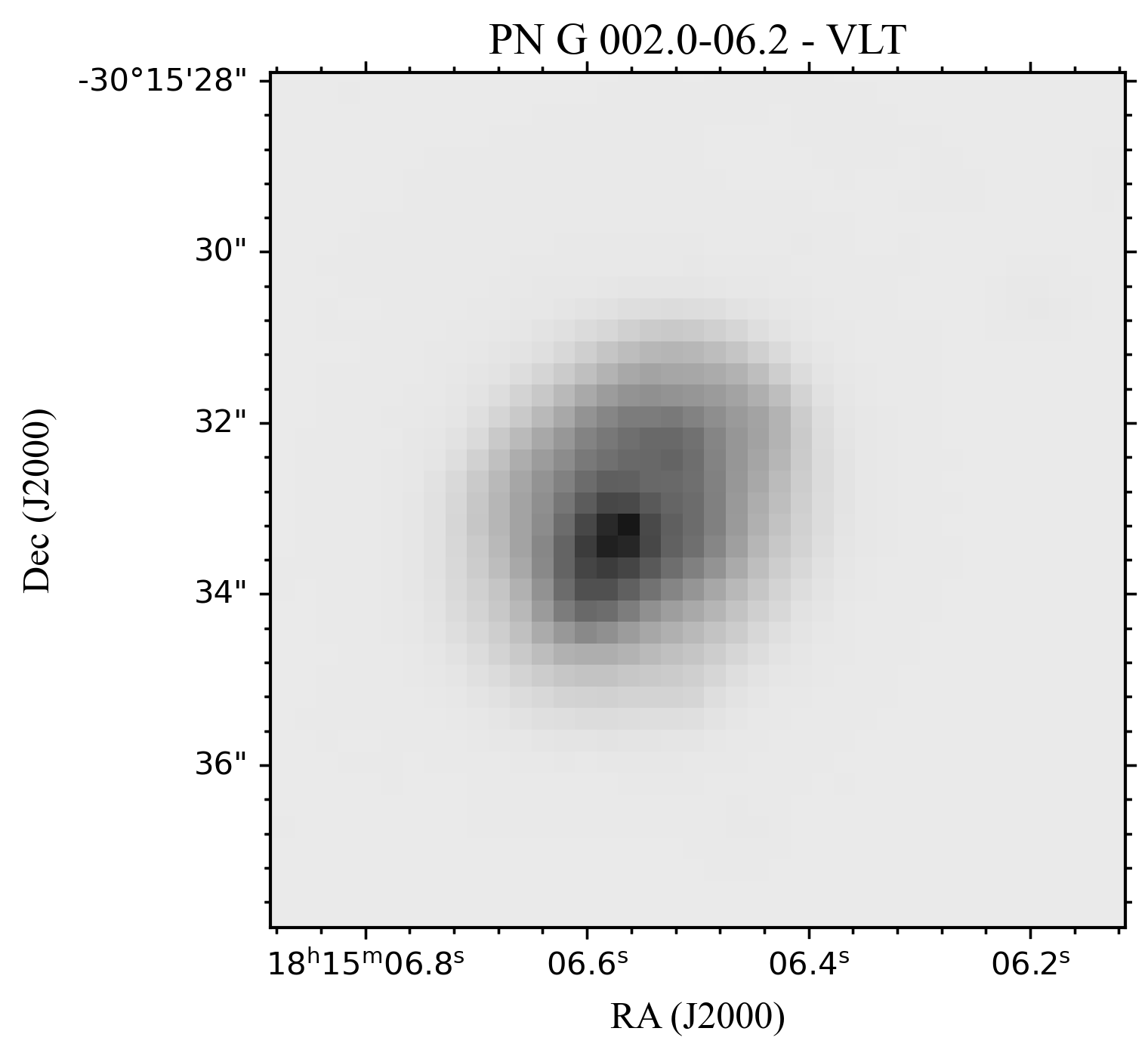}\hfill 
  \includegraphics[width=.32\linewidth]{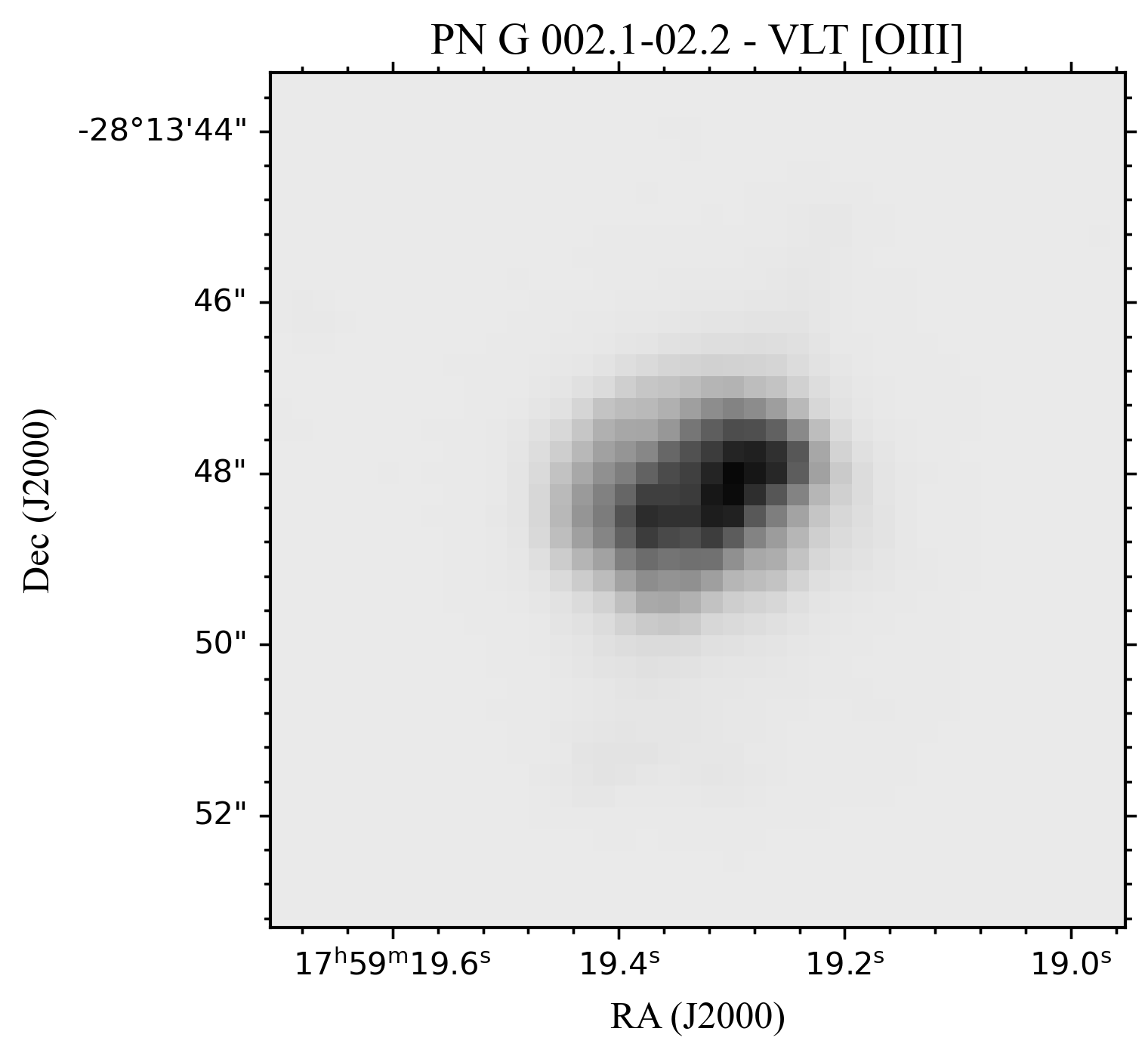}\hfill 
  \includegraphics[width=.32\linewidth]{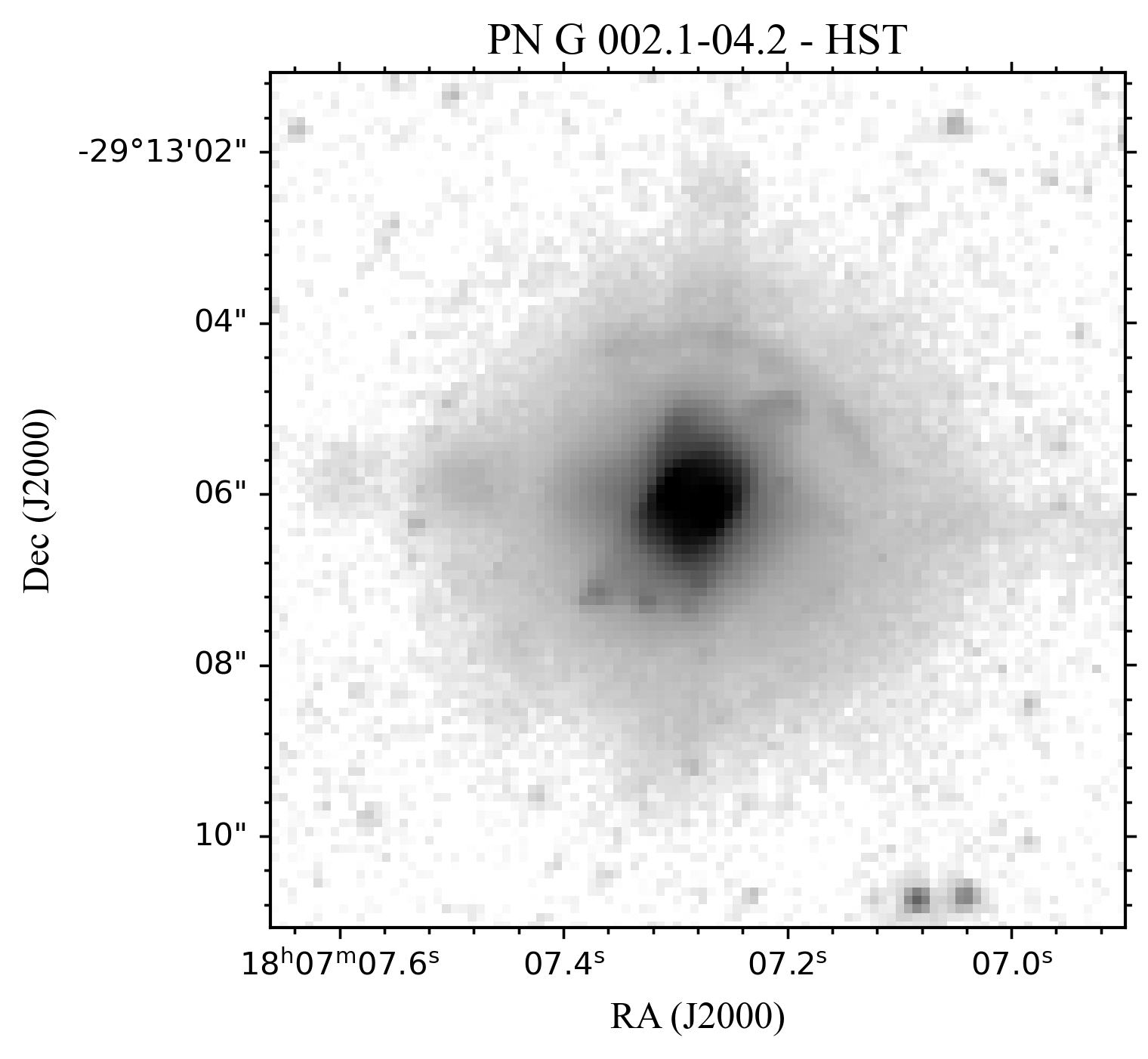}\hfill 
 \end{subfigure}\par\medskip 
\begin{subfigure}{\linewidth} 
  \includegraphics[width=.32\linewidth]{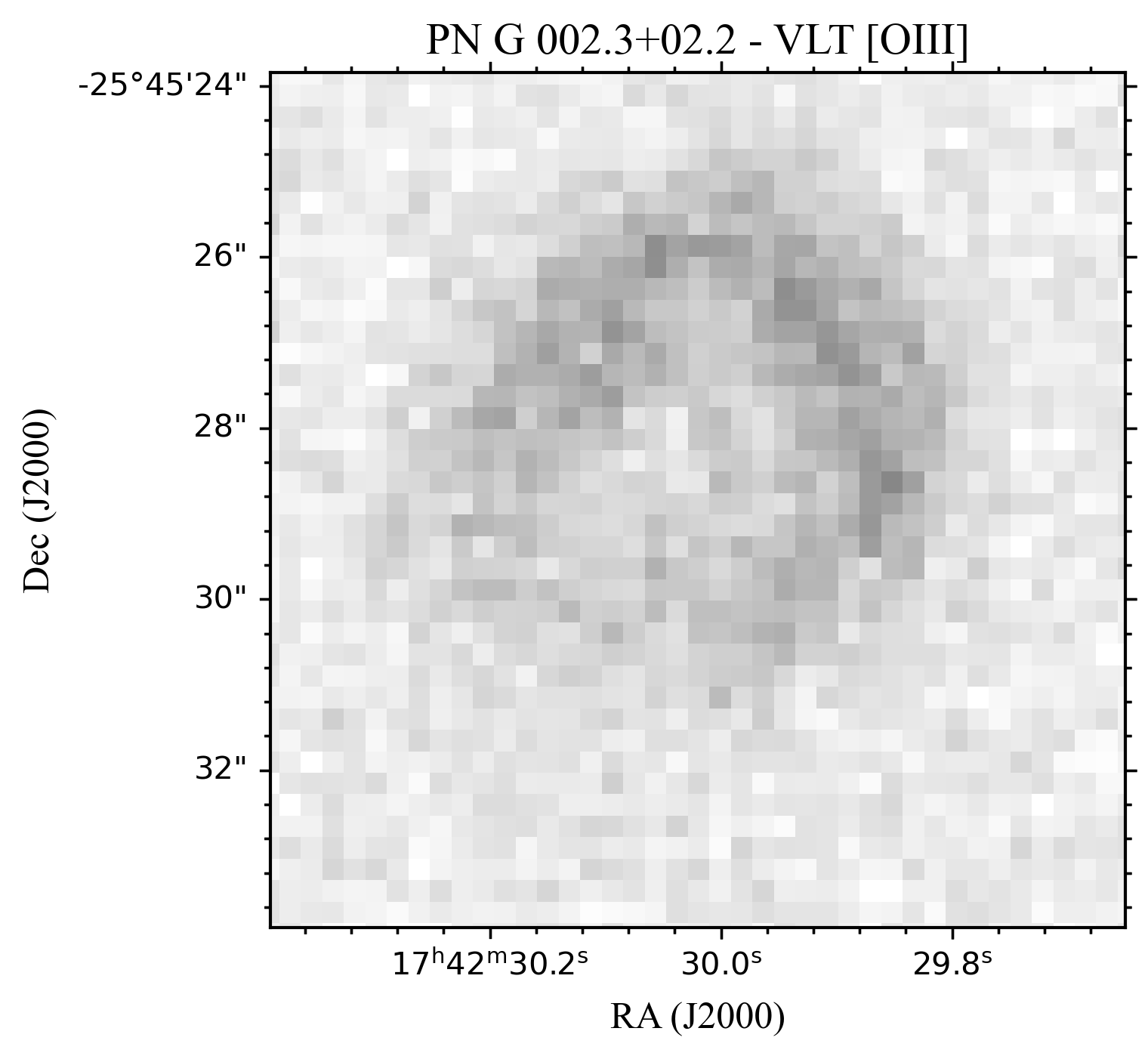}\hfill 
  \includegraphics[width=.32\linewidth]{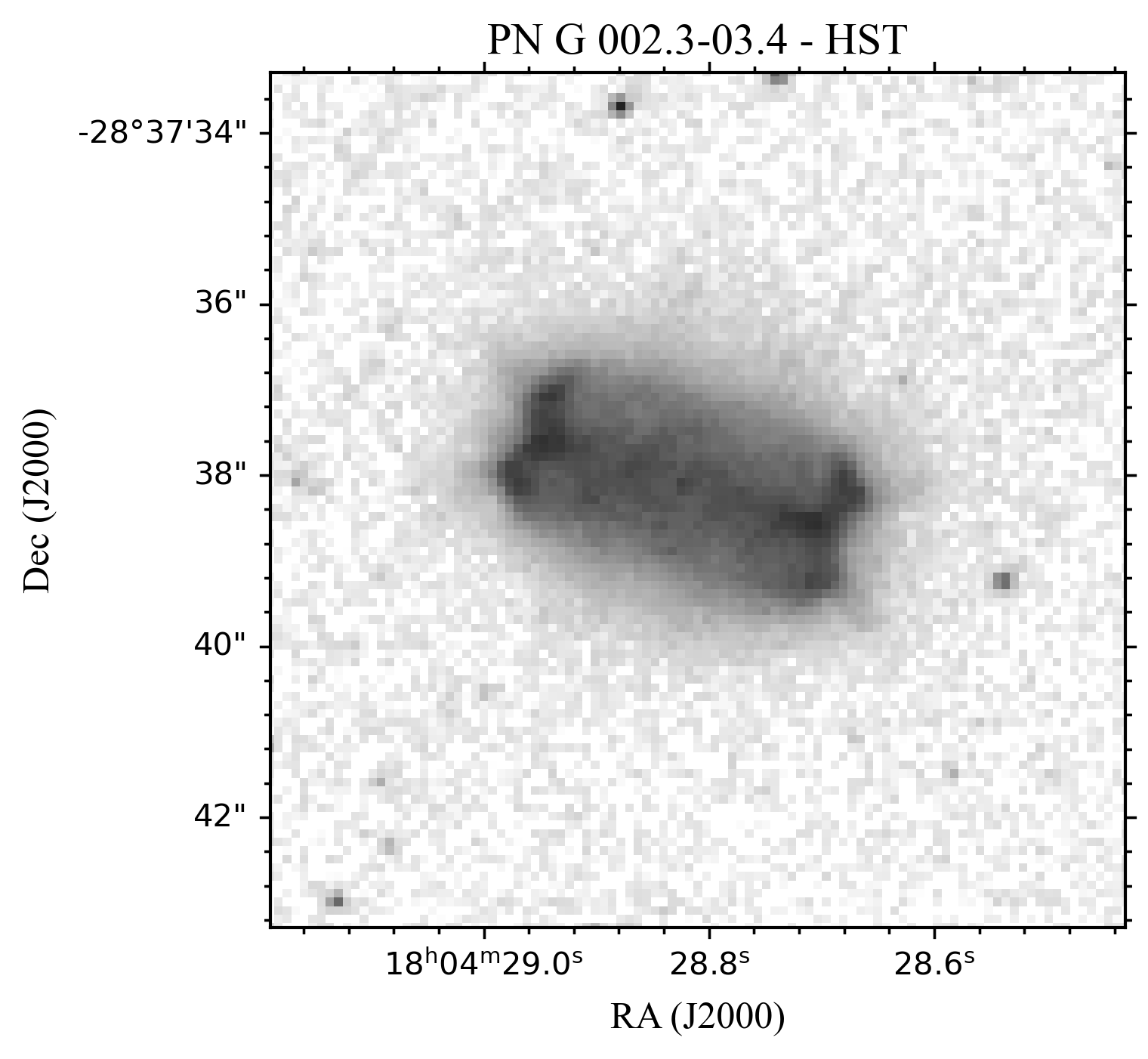}\hfill 
  \includegraphics[width=.32\linewidth]{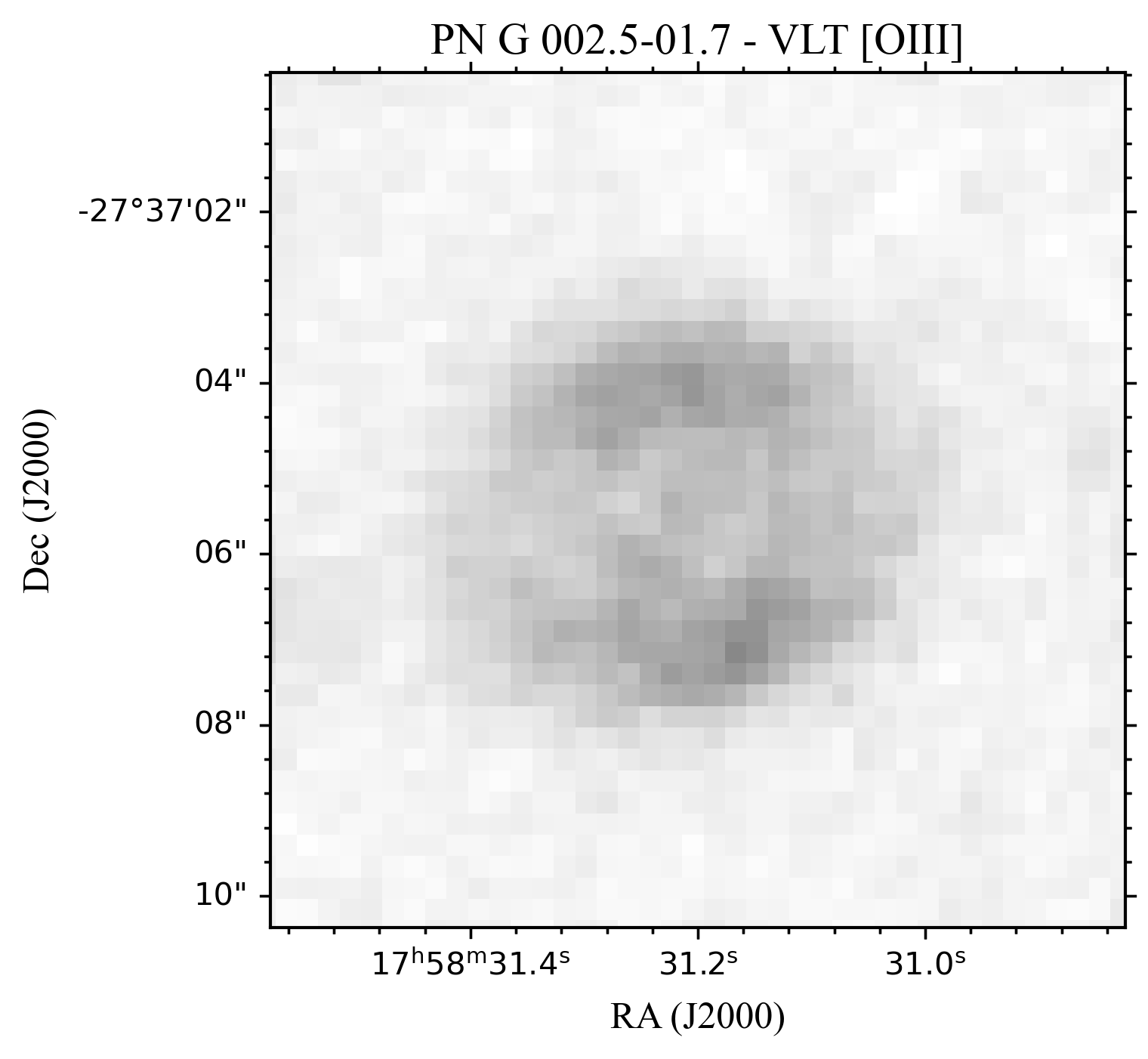}\hfill 
 \end{subfigure}\par\medskip 
  \end{figure} 
 \begin{figure} 
 \ContinuedFloat 
 \caption[]{continued:} 
\begin{subfigure}{\linewidth} 
  \includegraphics[width=.32\linewidth]{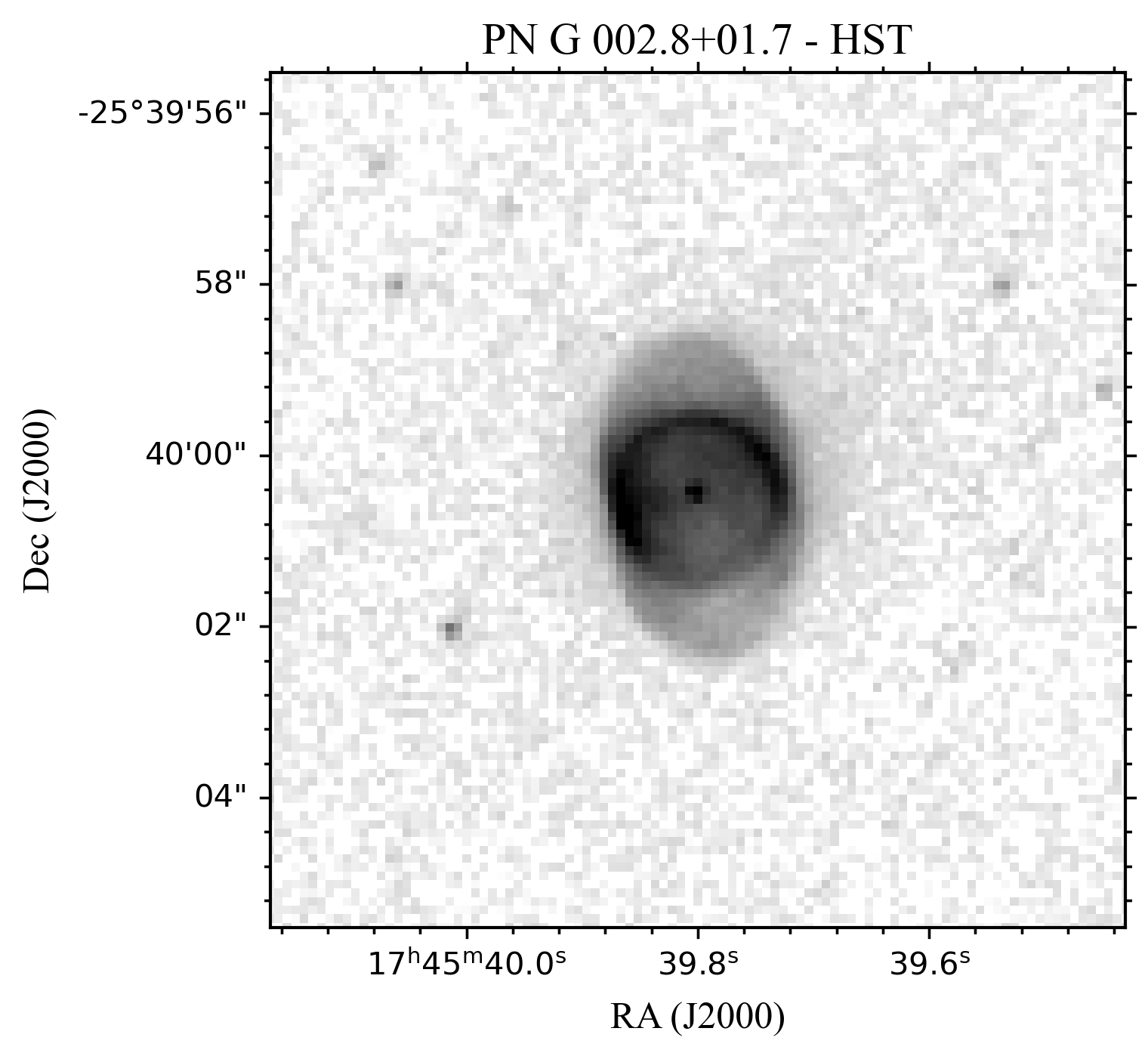}\hfill 
  \includegraphics[width=.32\linewidth]{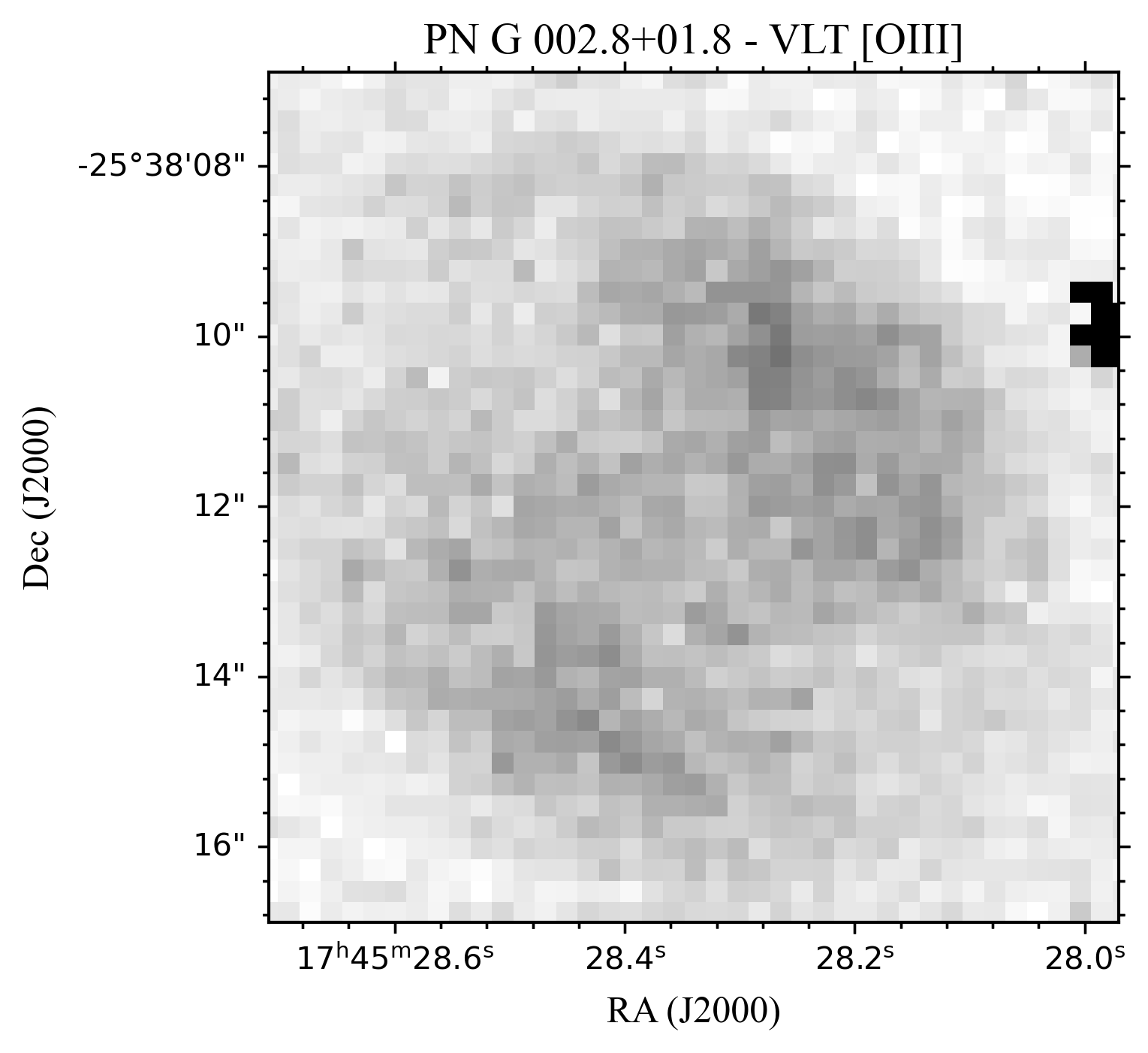}\hfill 
  \includegraphics[width=.32\linewidth]{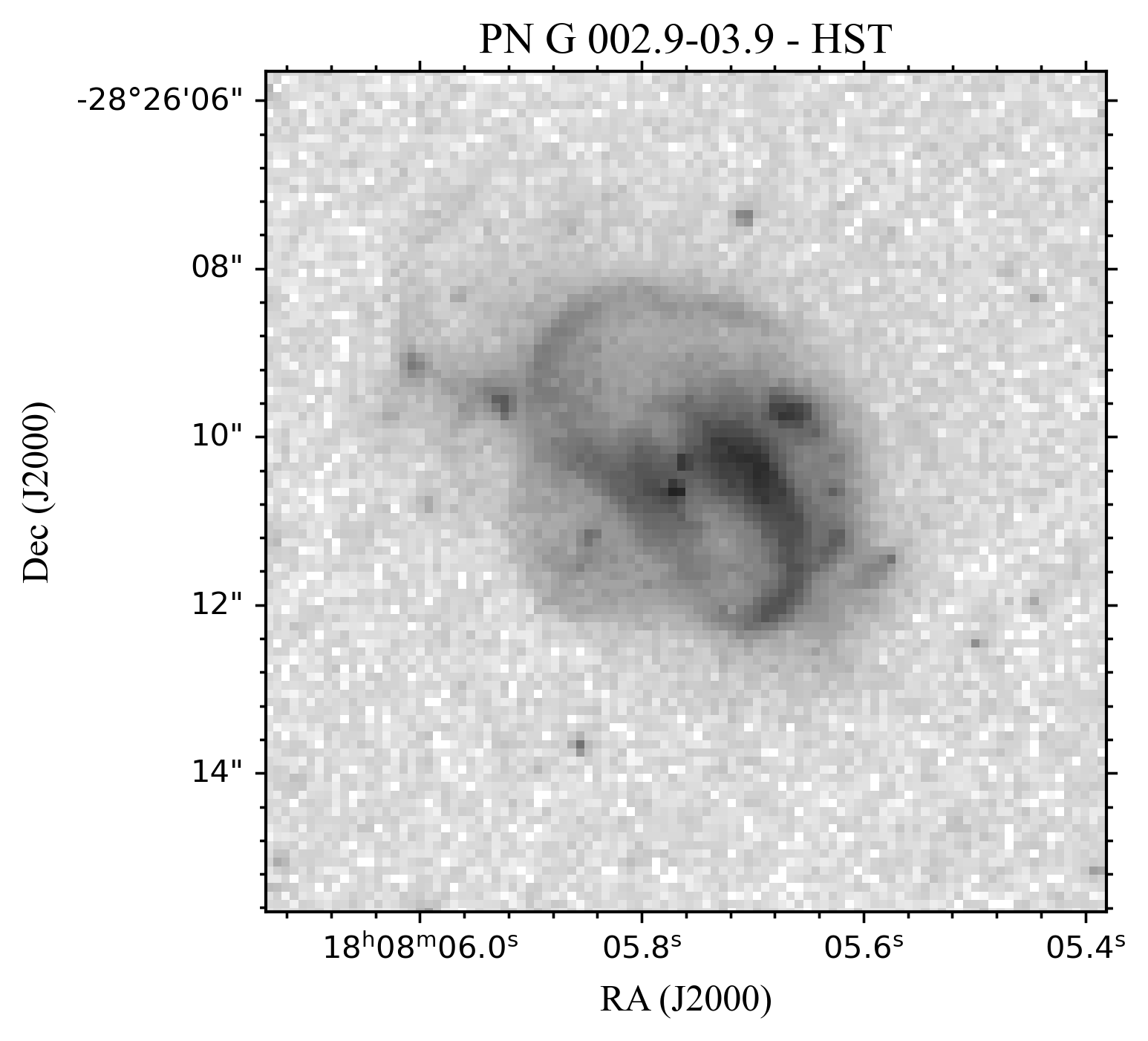}\hfill 
 \end{subfigure}\par\medskip 
\begin{subfigure}{\linewidth} 
  \includegraphics[width=.32\linewidth]{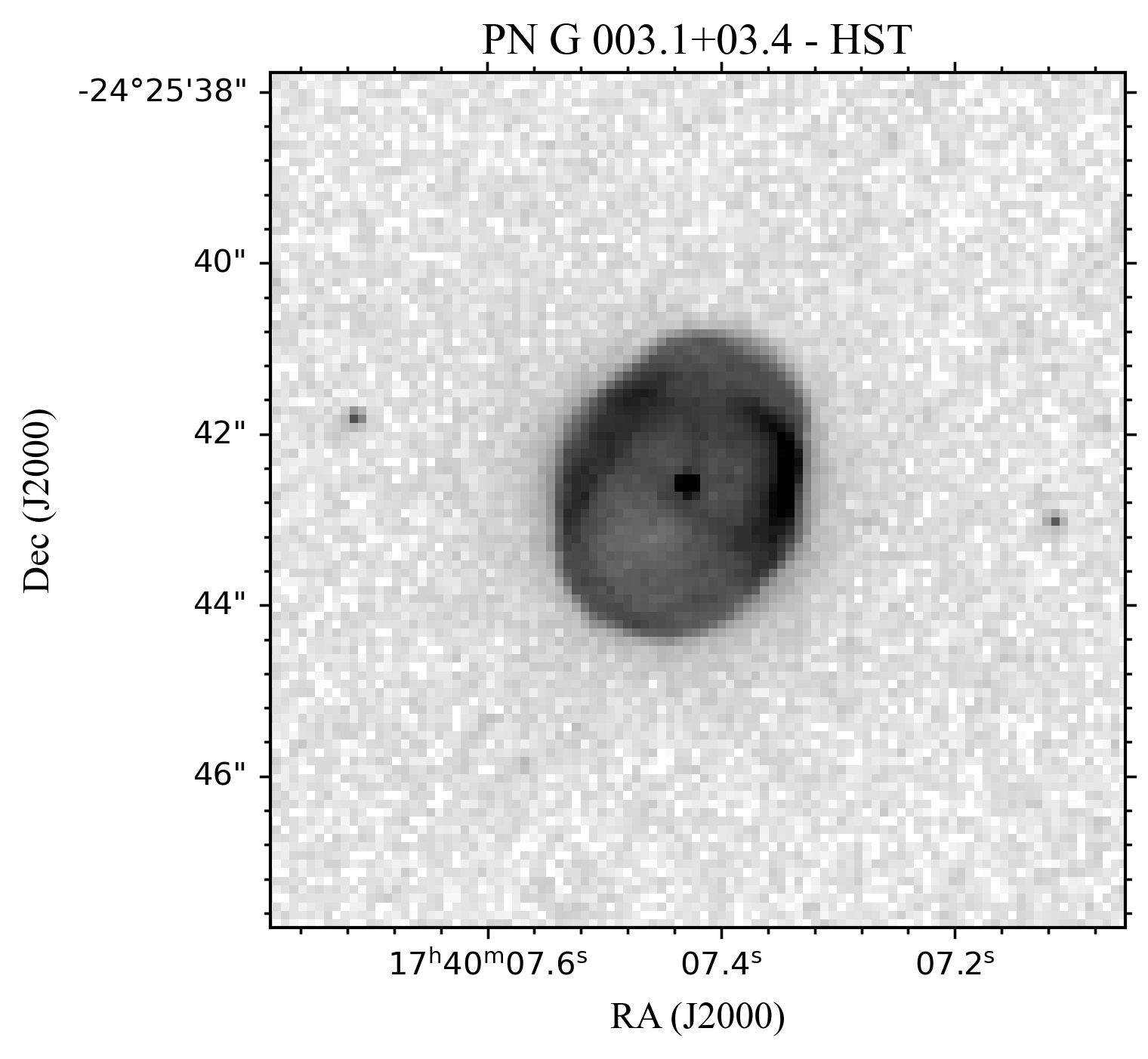}\hfill 
  \includegraphics[width=.32\linewidth]{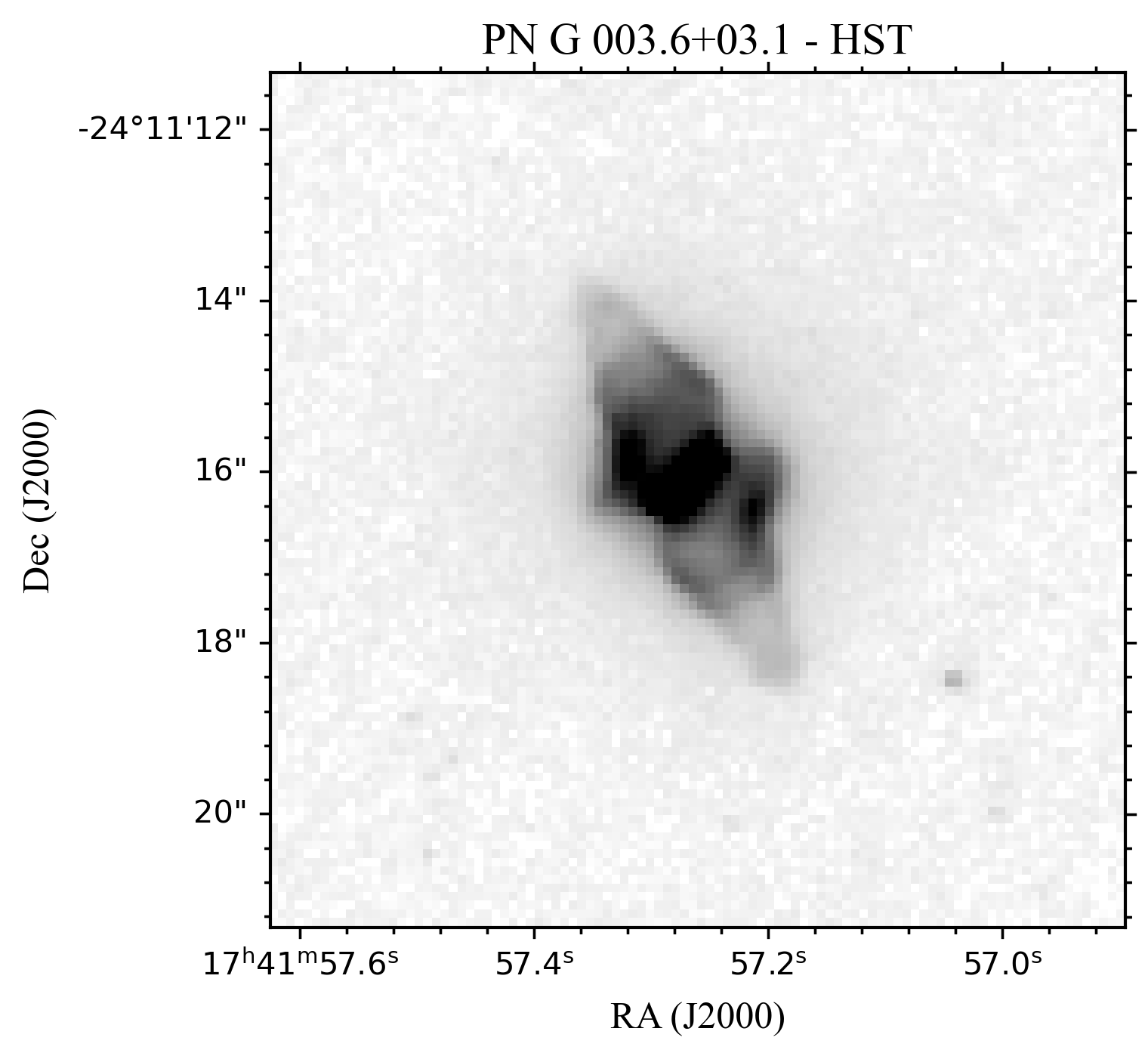}\hfill 
  \includegraphics[width=.32\linewidth]{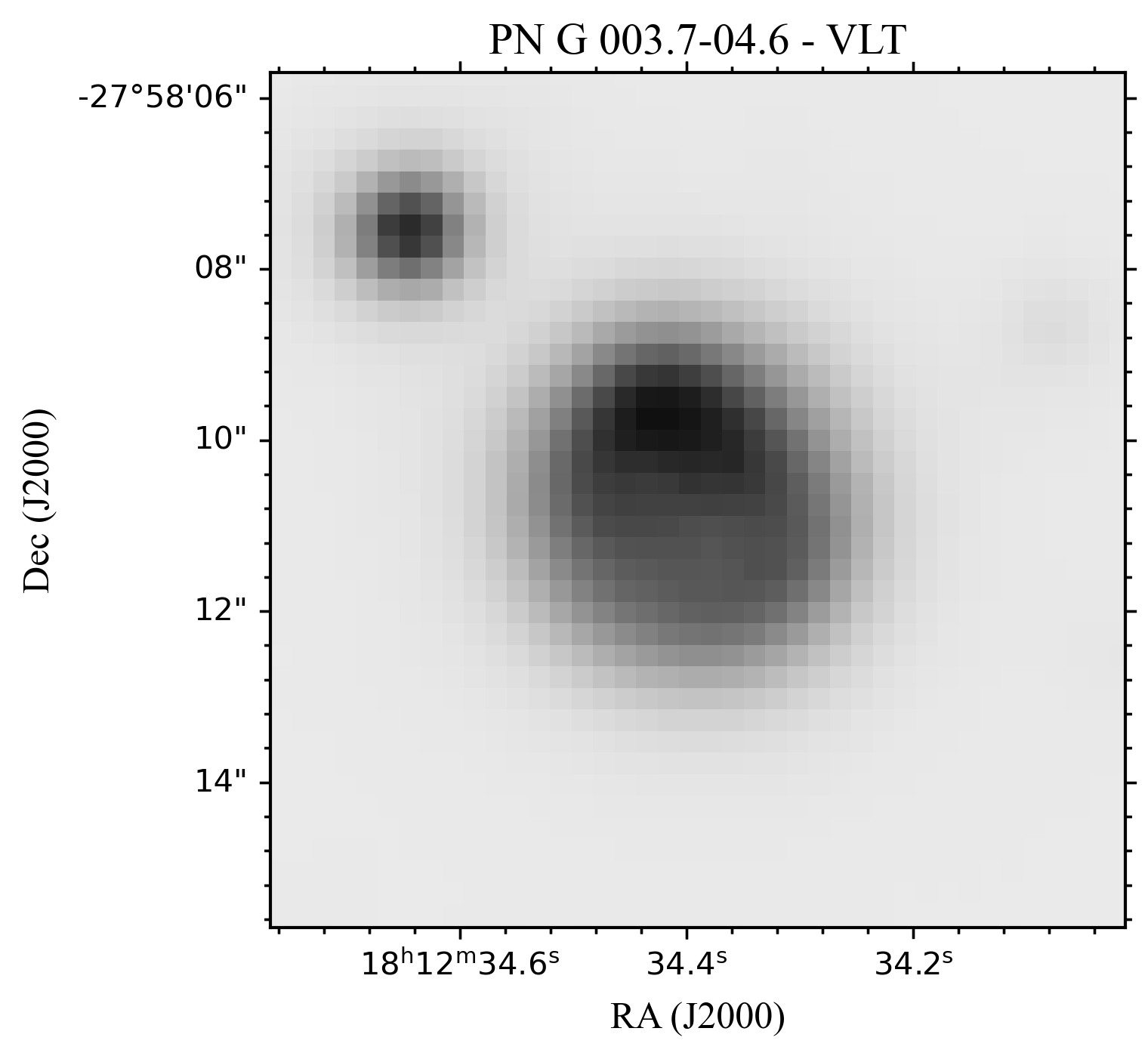}\hfill 
 \end{subfigure}\par\medskip 
\begin{subfigure}{\linewidth} 
  \includegraphics[width=.32\linewidth]{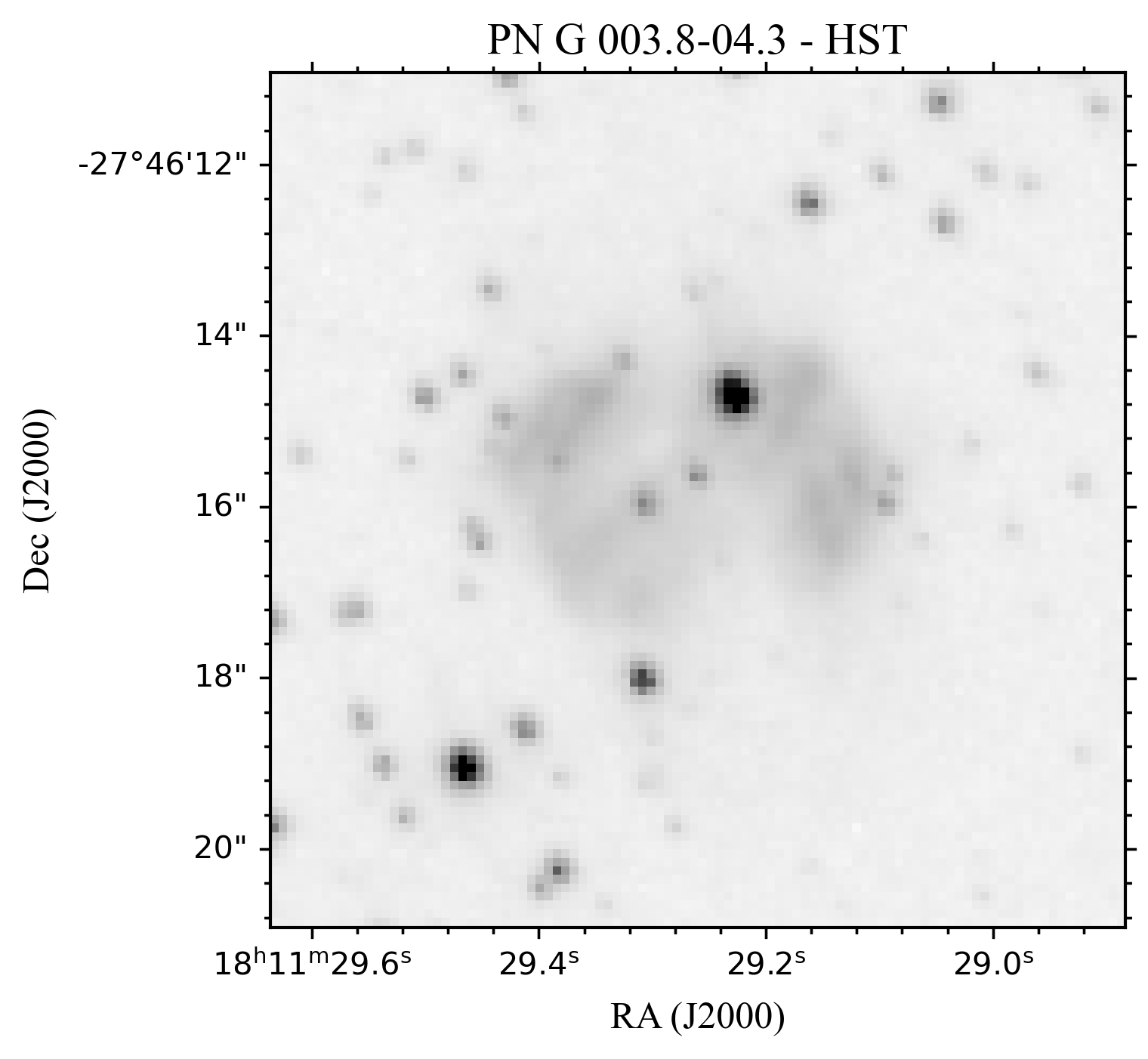}\hfill 
  \includegraphics[width=.32\linewidth]{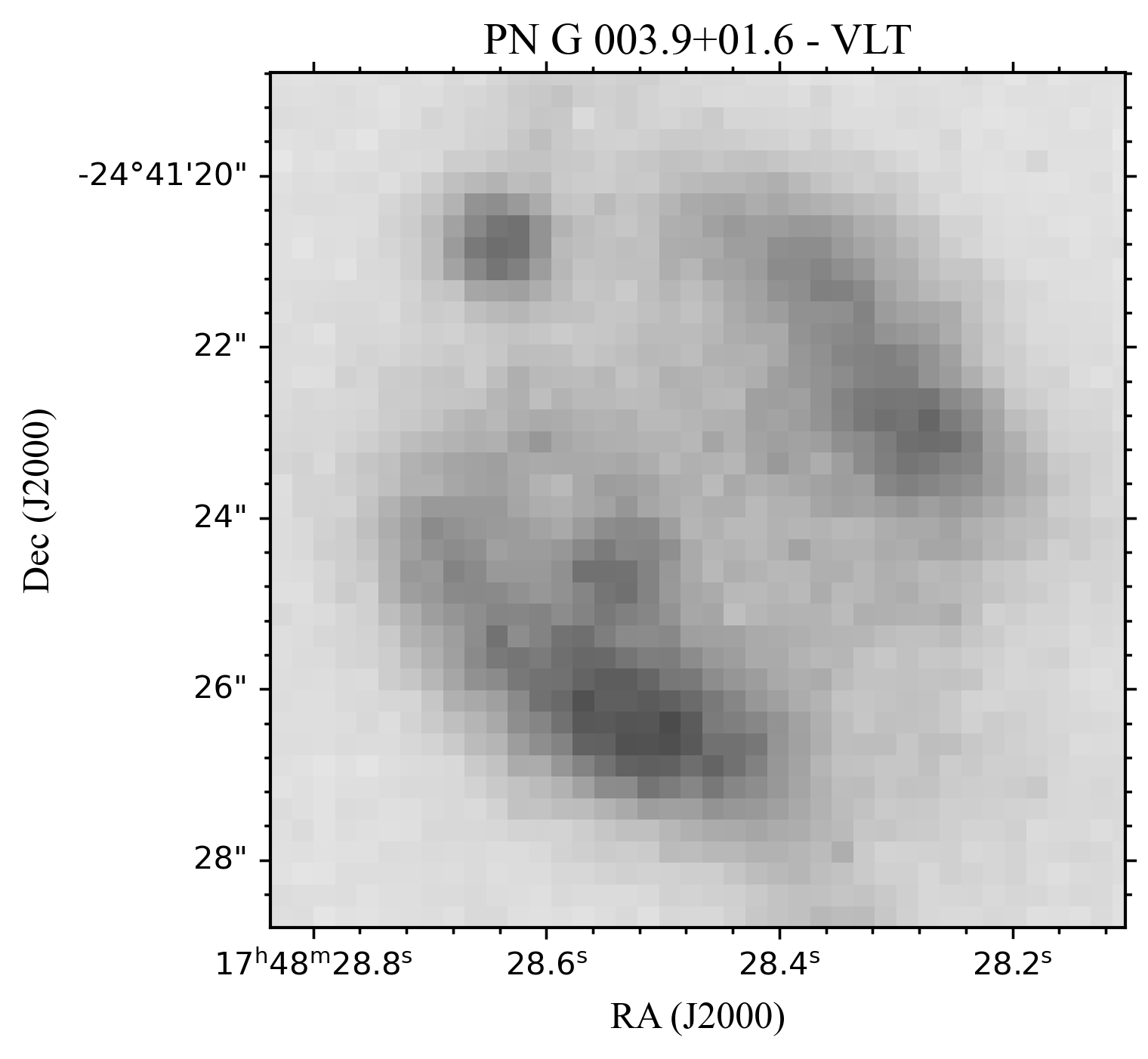}\hfill 
  \includegraphics[width=.32\linewidth]{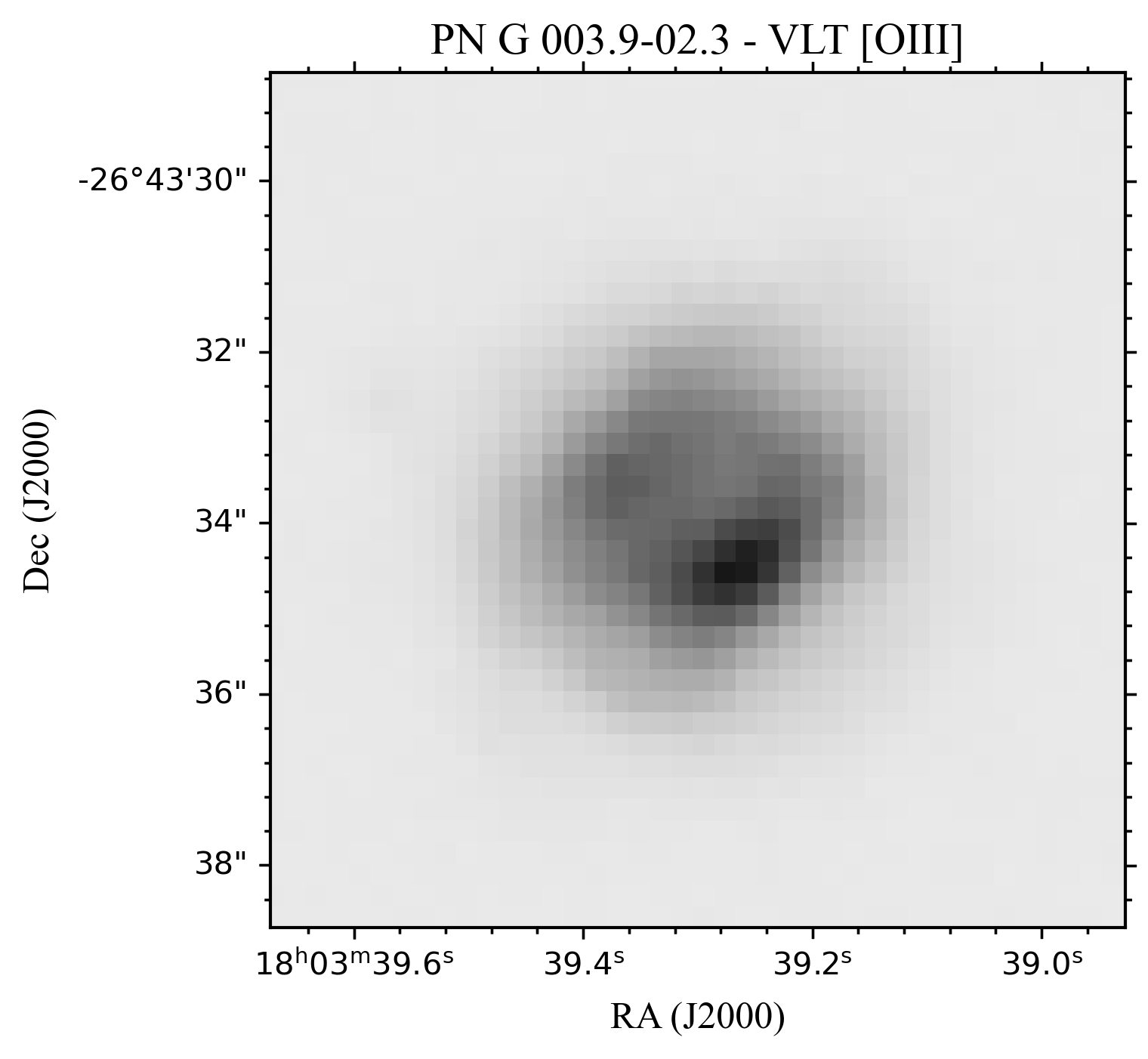}\hfill 
 \end{subfigure}\par\medskip 
\begin{subfigure}{\linewidth} 
  \includegraphics[width=.32\linewidth]{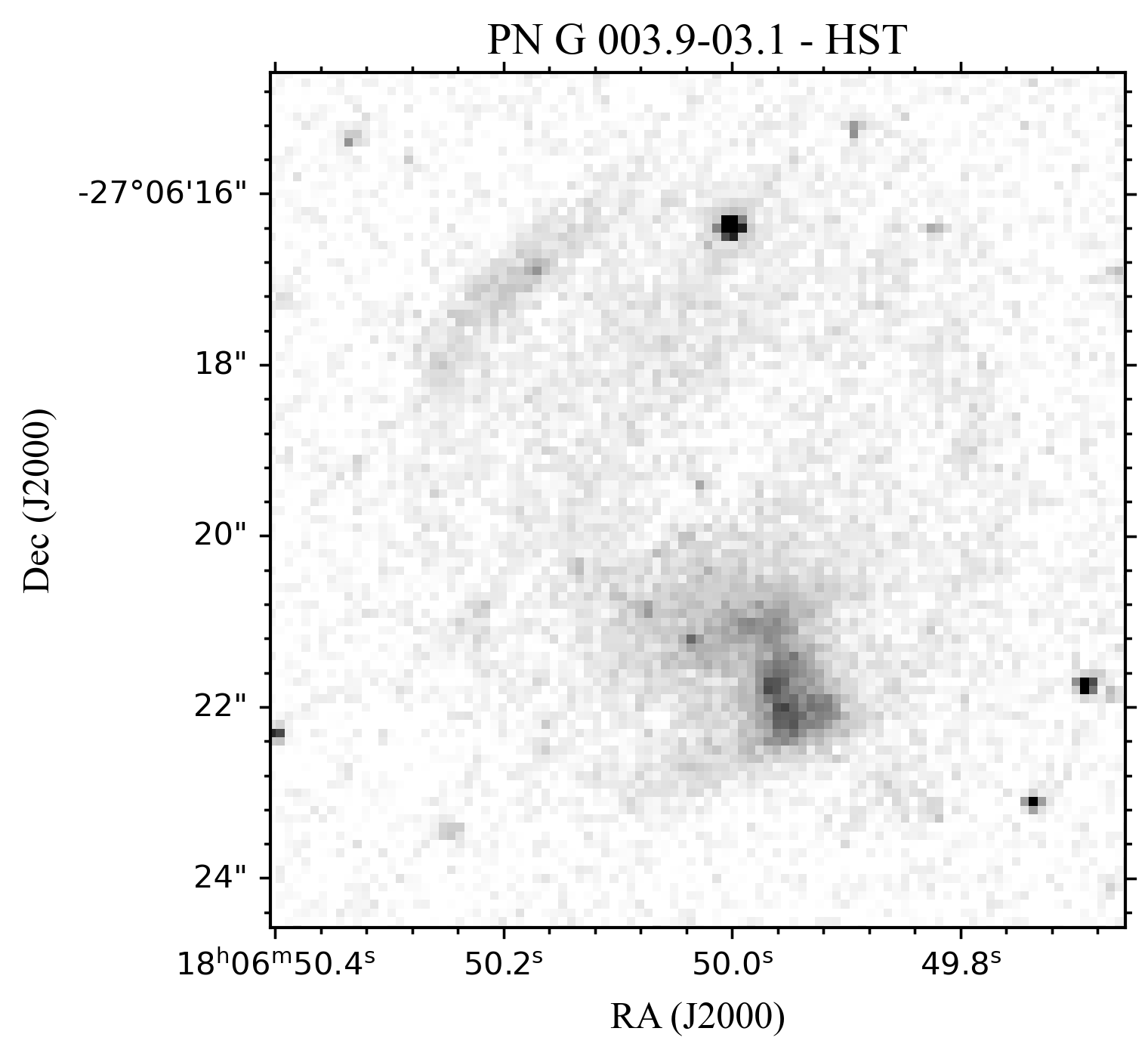}\hfill 
  \includegraphics[width=.32\linewidth]{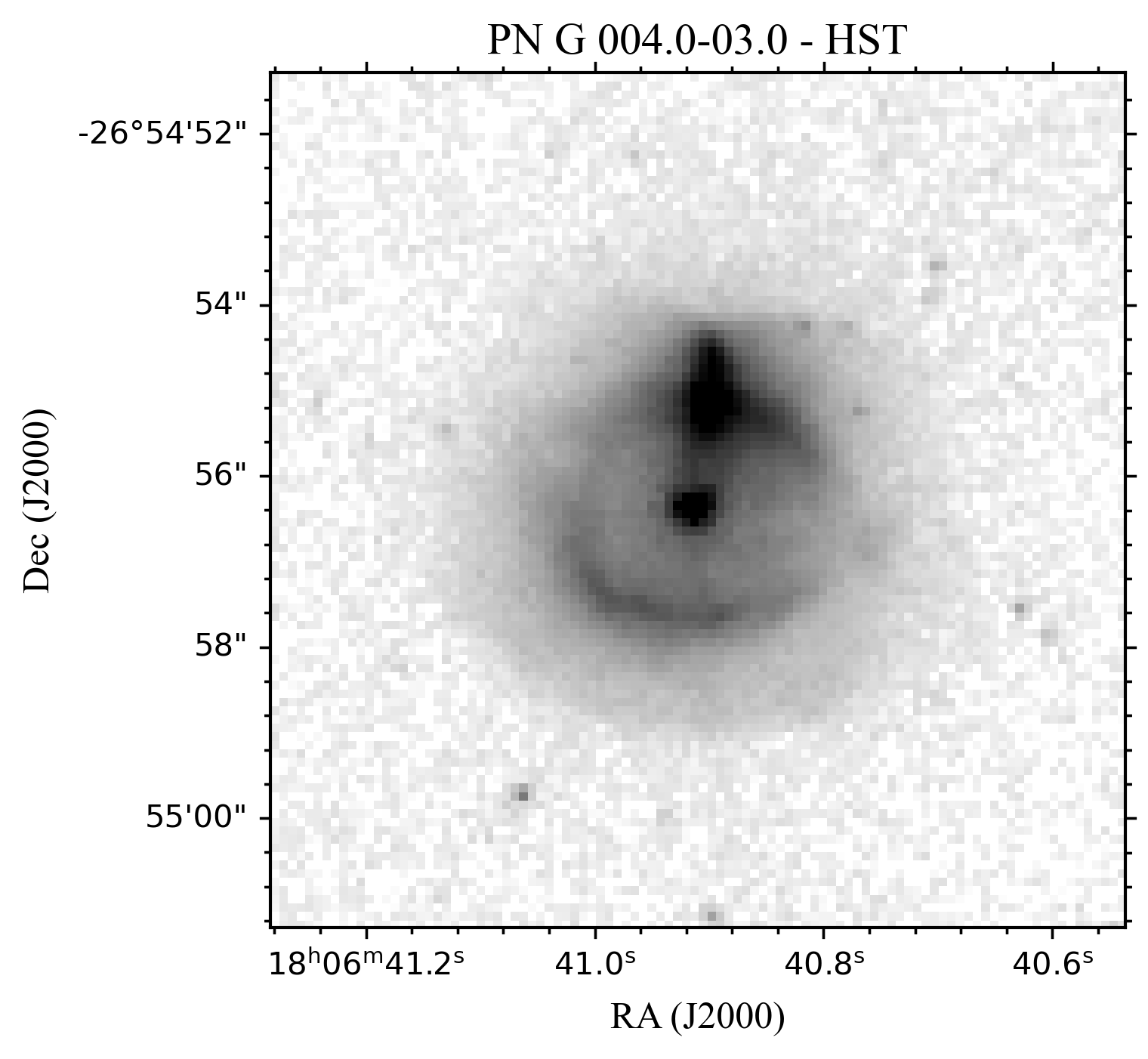}\hfill 
  \includegraphics[width=.32\linewidth]{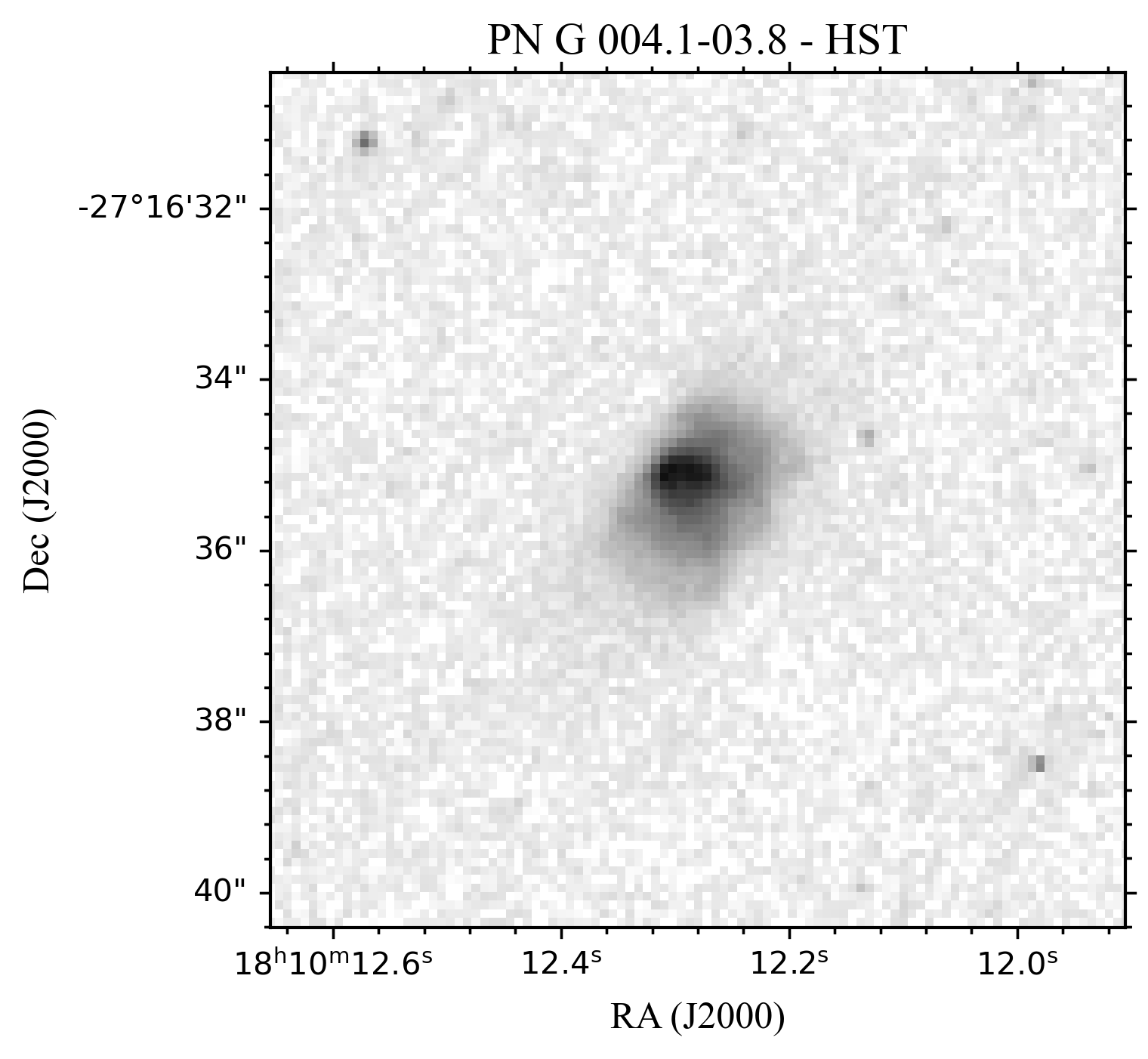}\hfill 
 \end{subfigure}\par\medskip 
  \end{figure} 
 \begin{figure} 
 \ContinuedFloat 
 \caption[]{continued:} 
\begin{subfigure}{\linewidth} 
  \includegraphics[width=.32\linewidth]{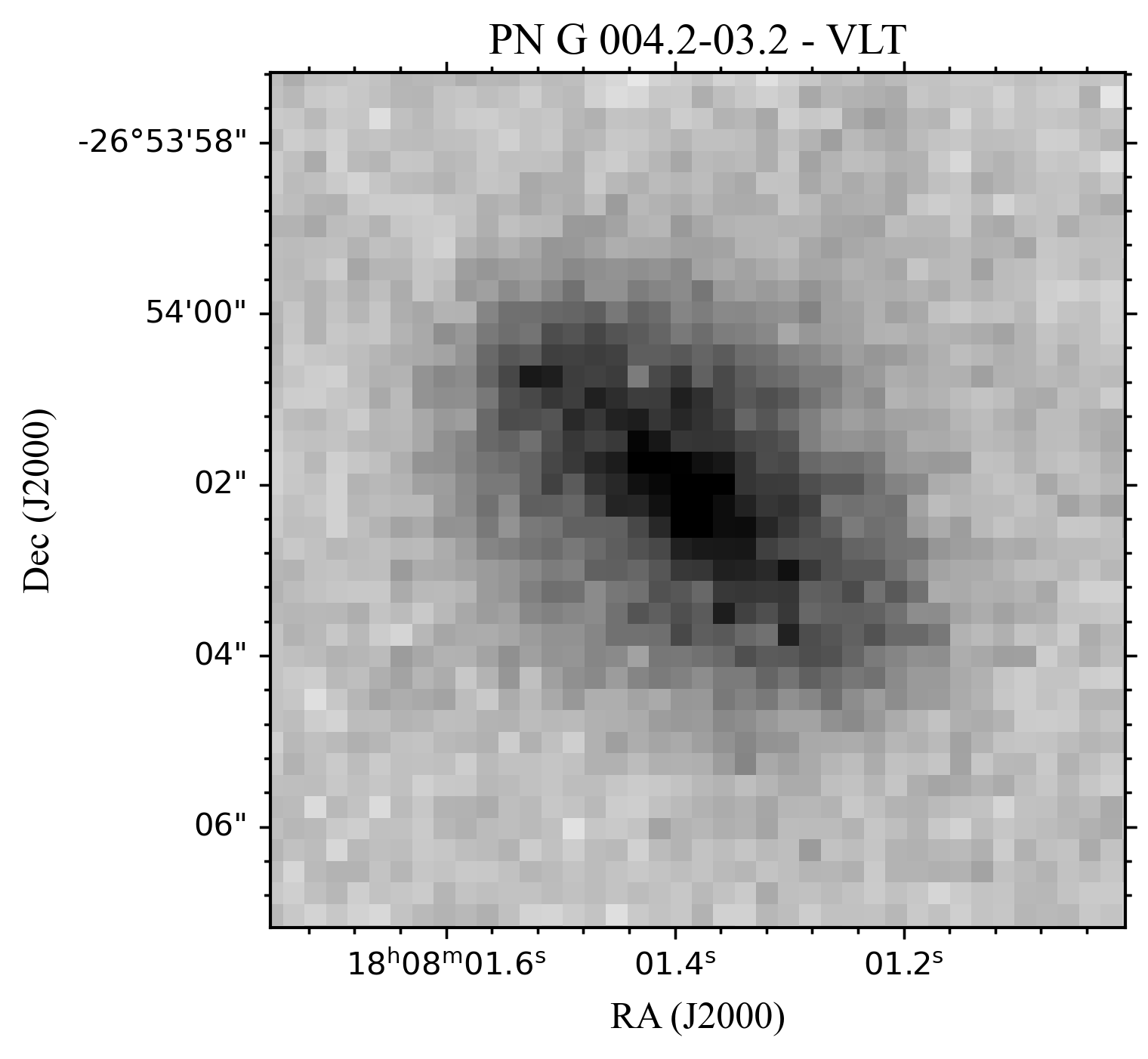}\hfill 
  \includegraphics[width=.32\linewidth]{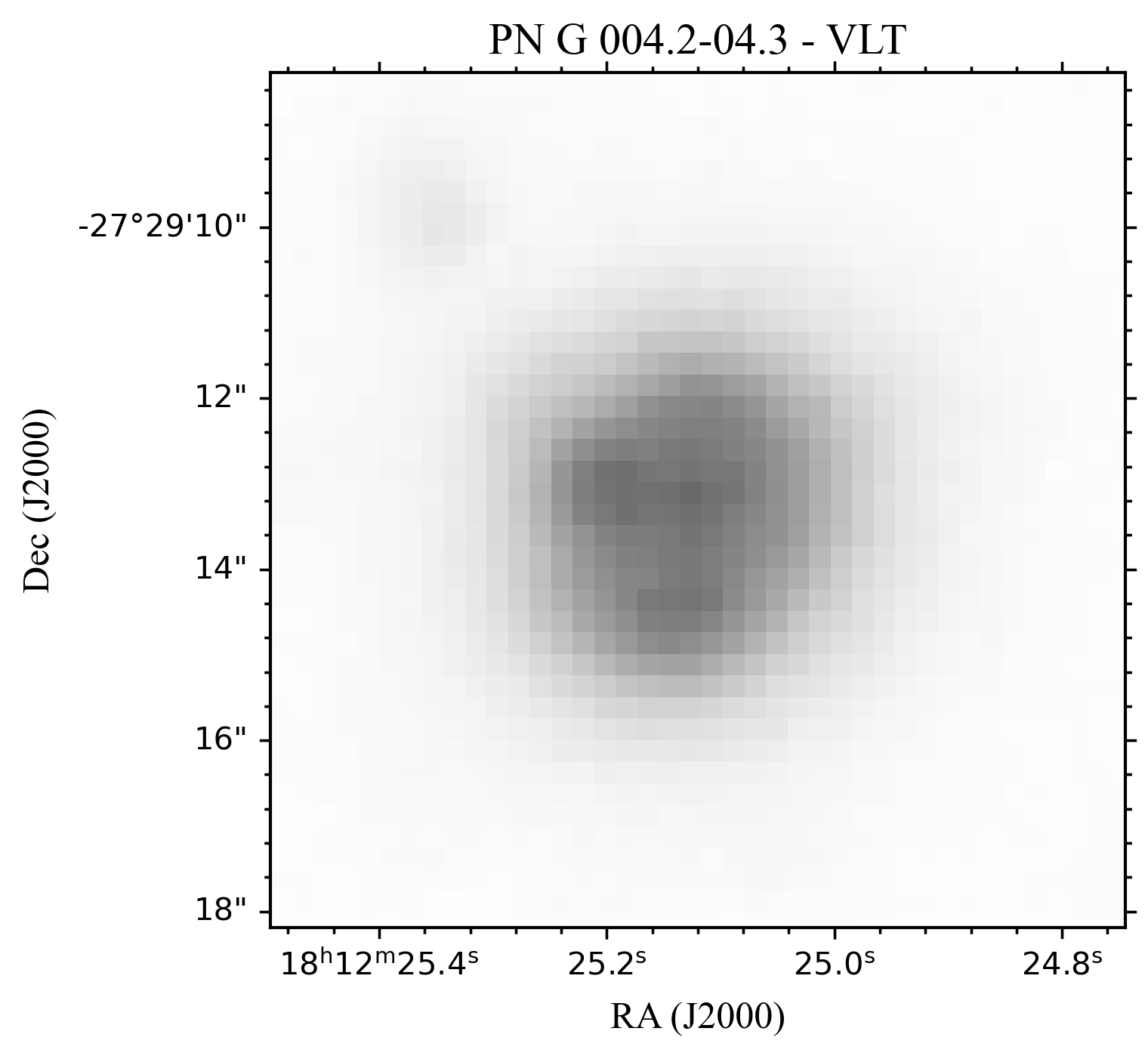}\hfill 
  \includegraphics[width=.32\linewidth]{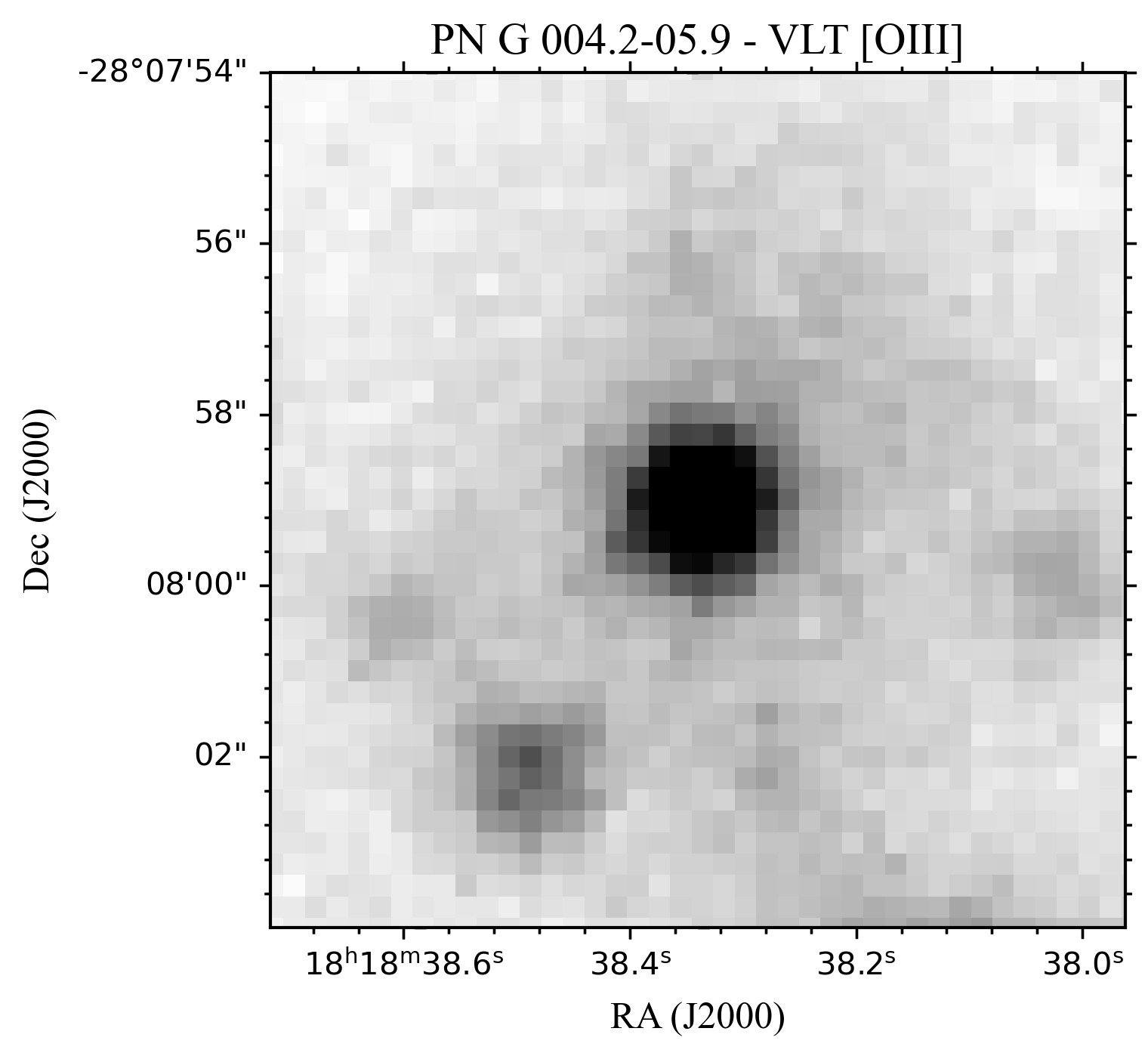}\hfill 
 \end{subfigure}\par\medskip 
\begin{subfigure}{\linewidth} 
  \includegraphics[width=.32\linewidth]{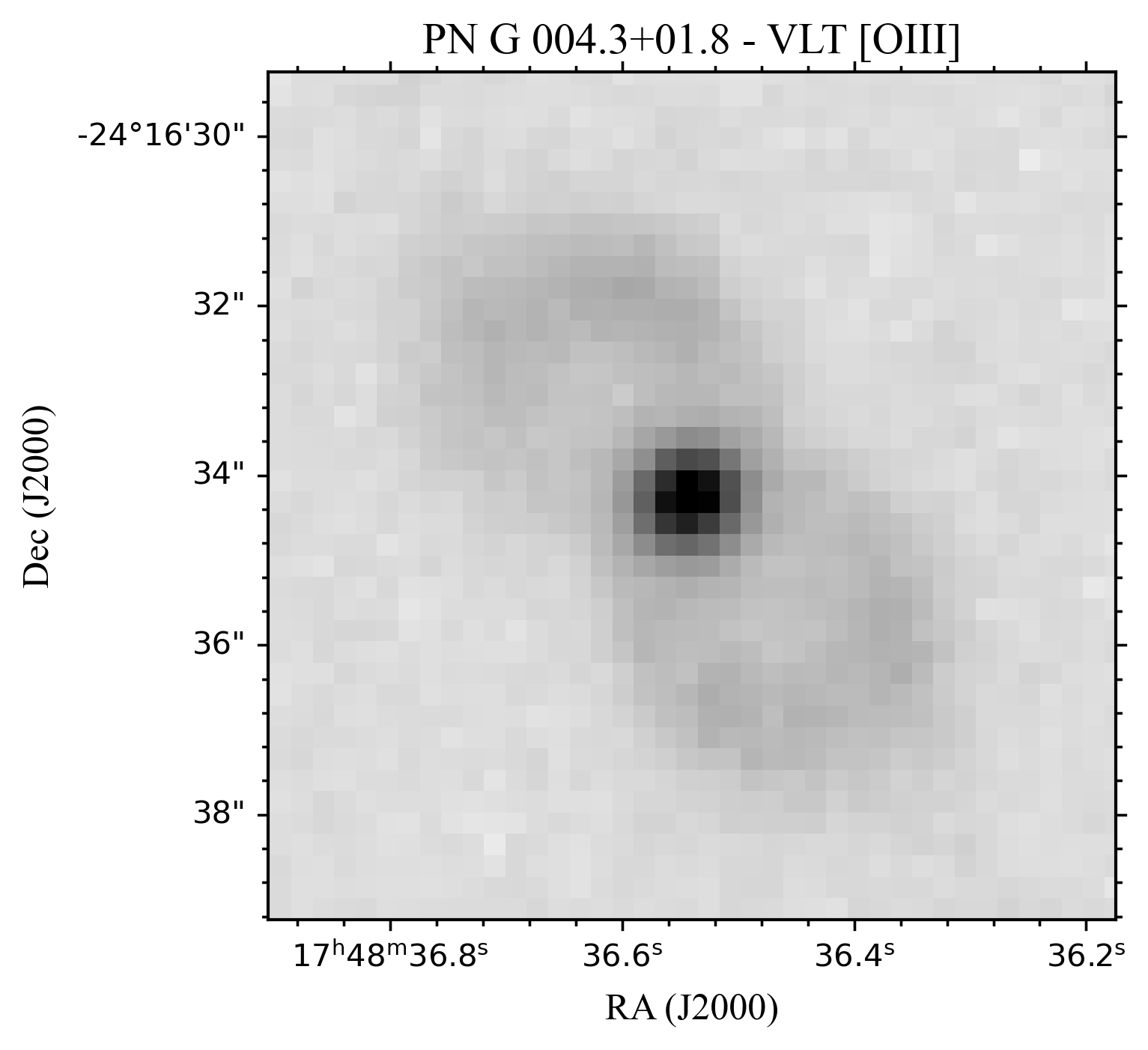}\hfill 
  \includegraphics[width=.32\linewidth]{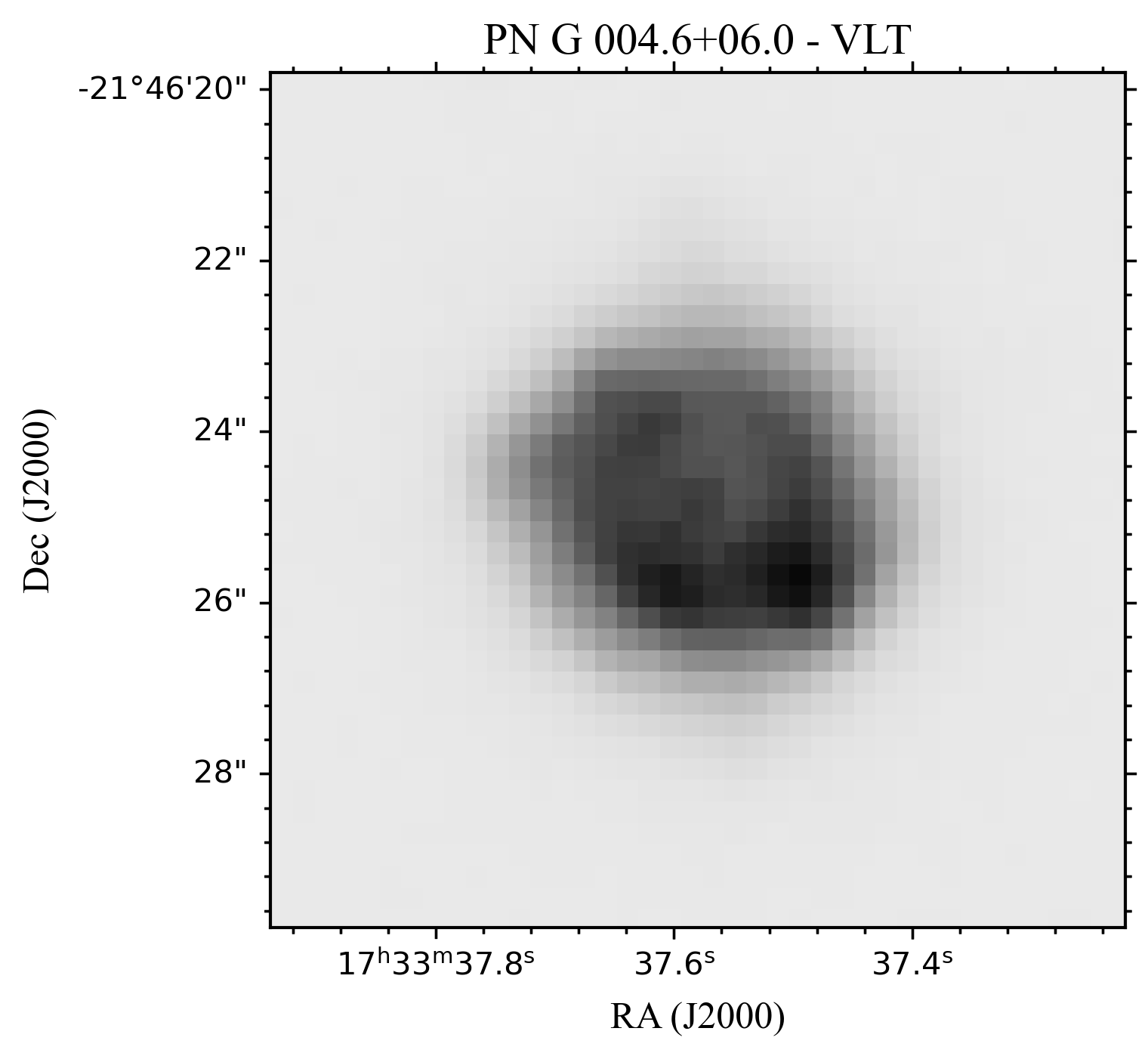}\hfill 
  \includegraphics[width=.32\linewidth]{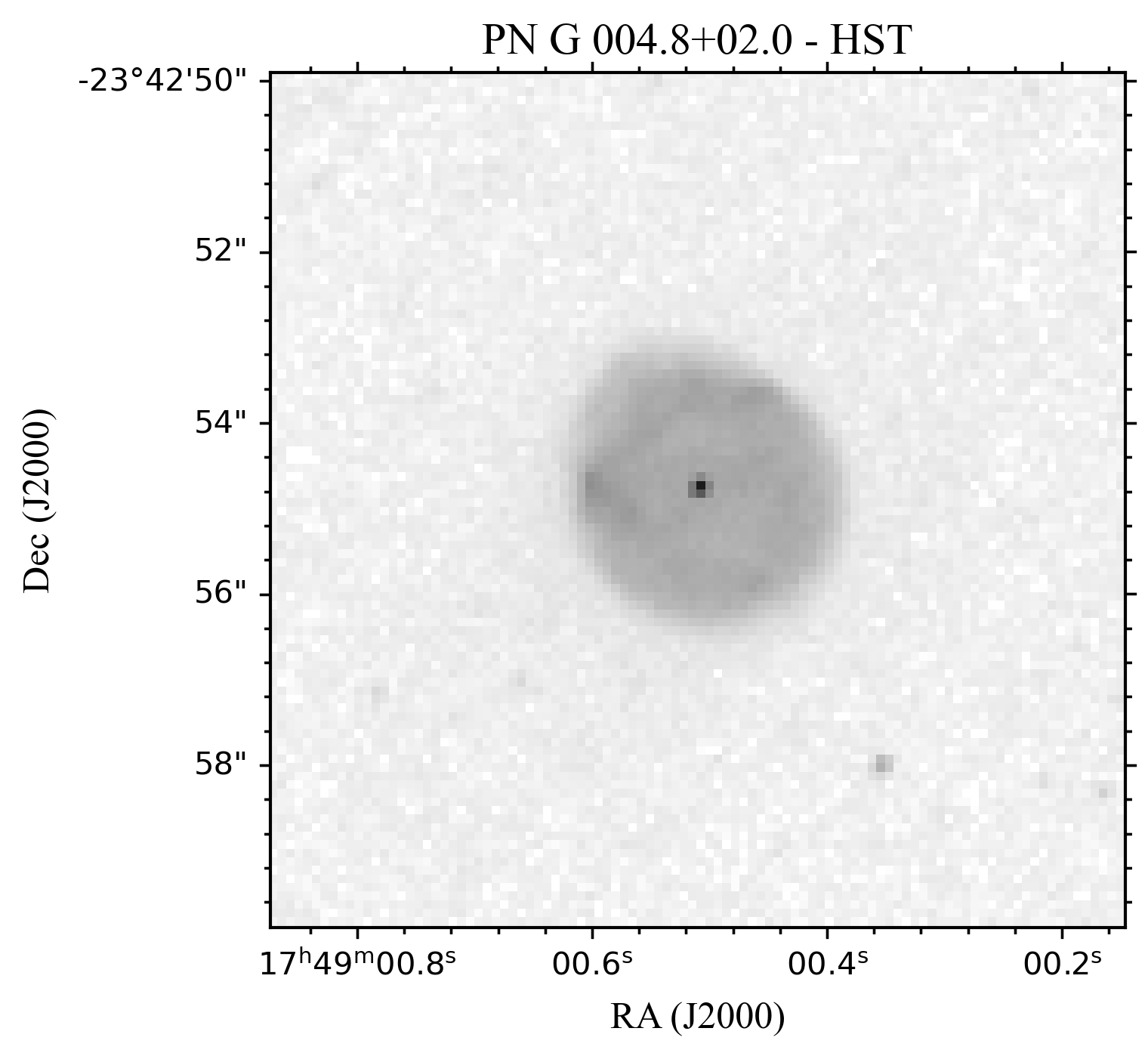}\hfill 
 \end{subfigure}\par\medskip 
\begin{subfigure}{\linewidth} 
  \includegraphics[width=.32\linewidth]{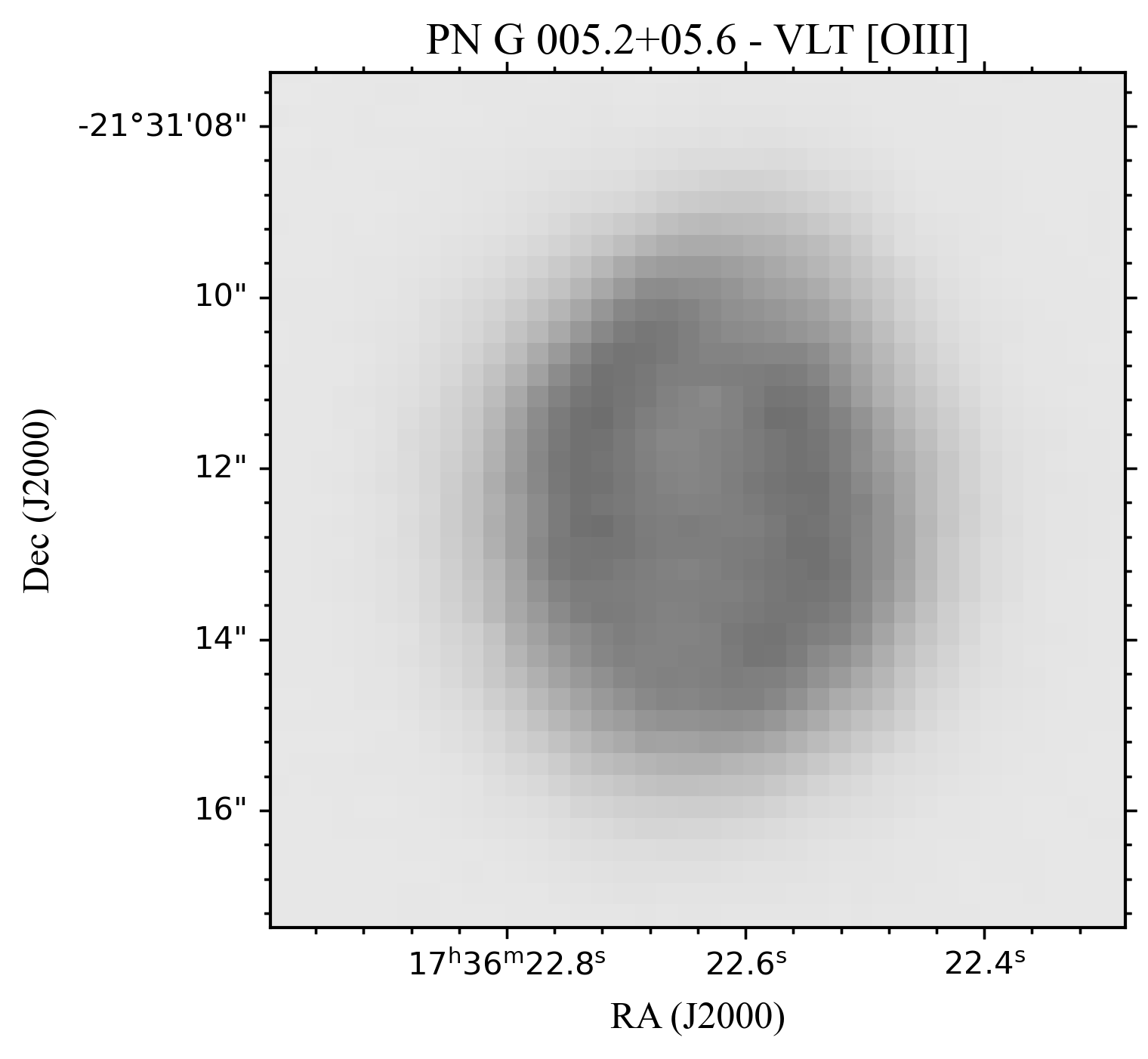}\hfill 
  \includegraphics[width=.32\linewidth]{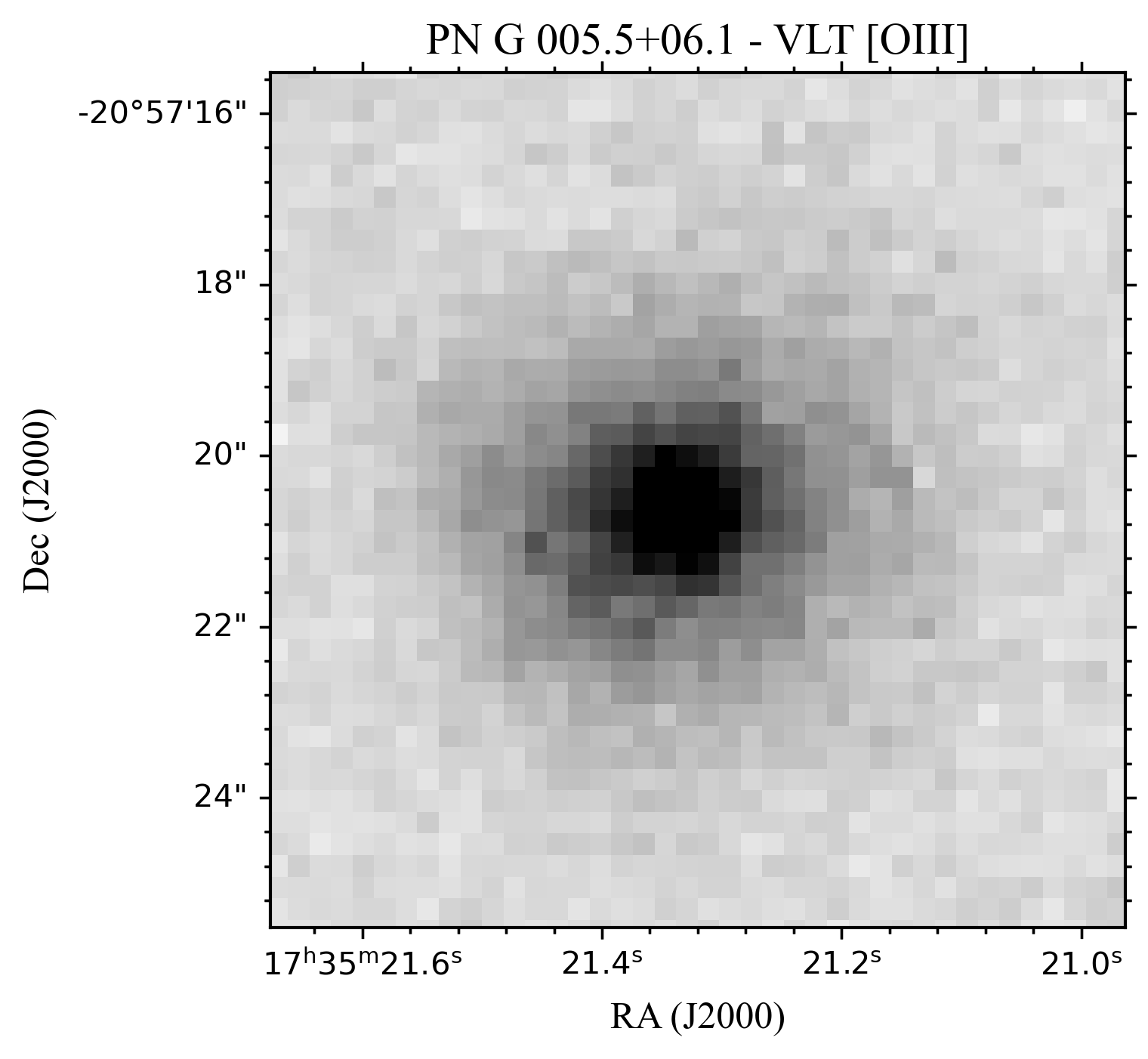}\hfill 
  \includegraphics[width=.32\linewidth]{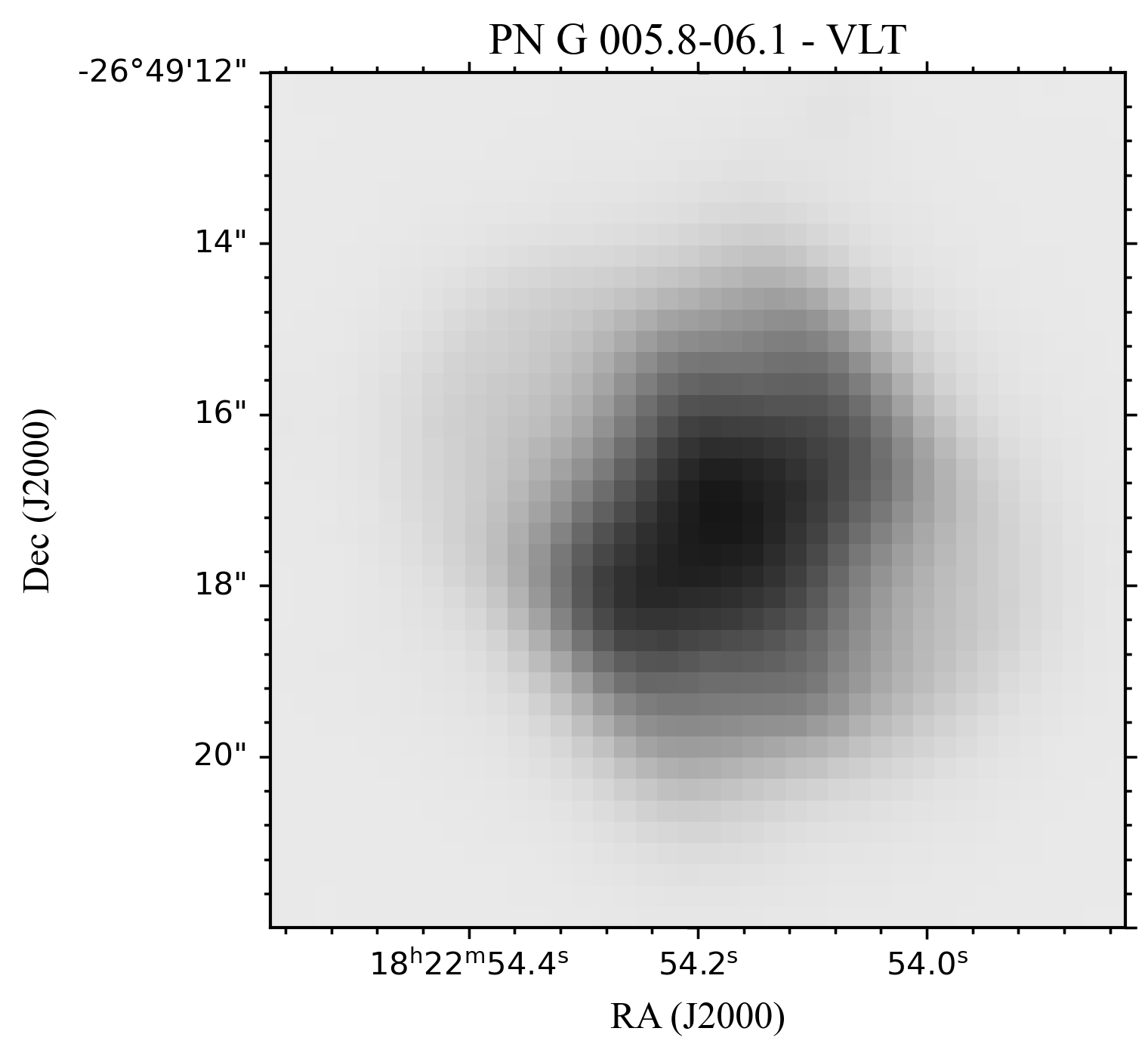}\hfill 
 \end{subfigure}\par\medskip 
\begin{subfigure}{\linewidth} 
  \includegraphics[width=.32\linewidth]{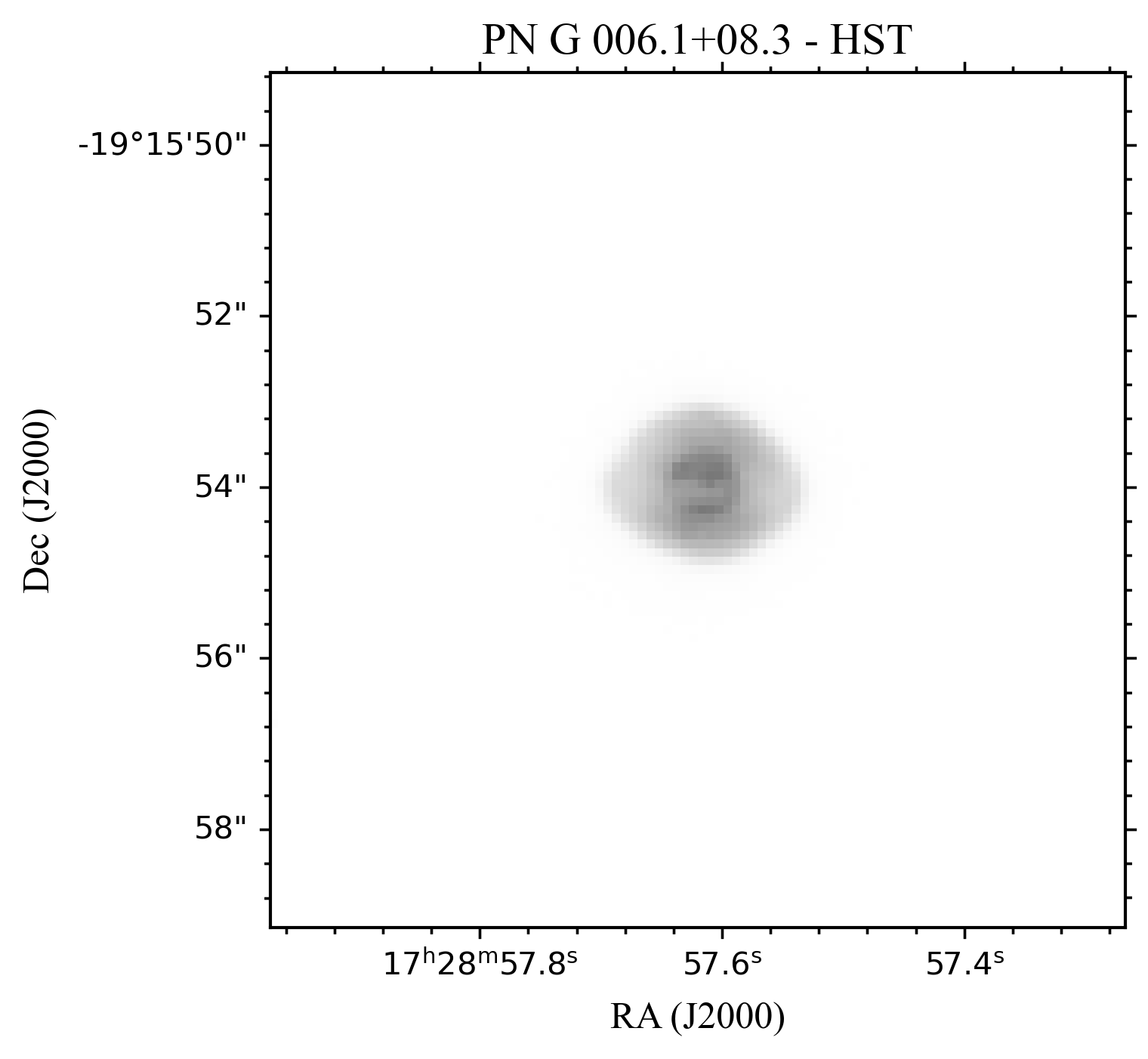}\hfill 
  \includegraphics[width=.32\linewidth]{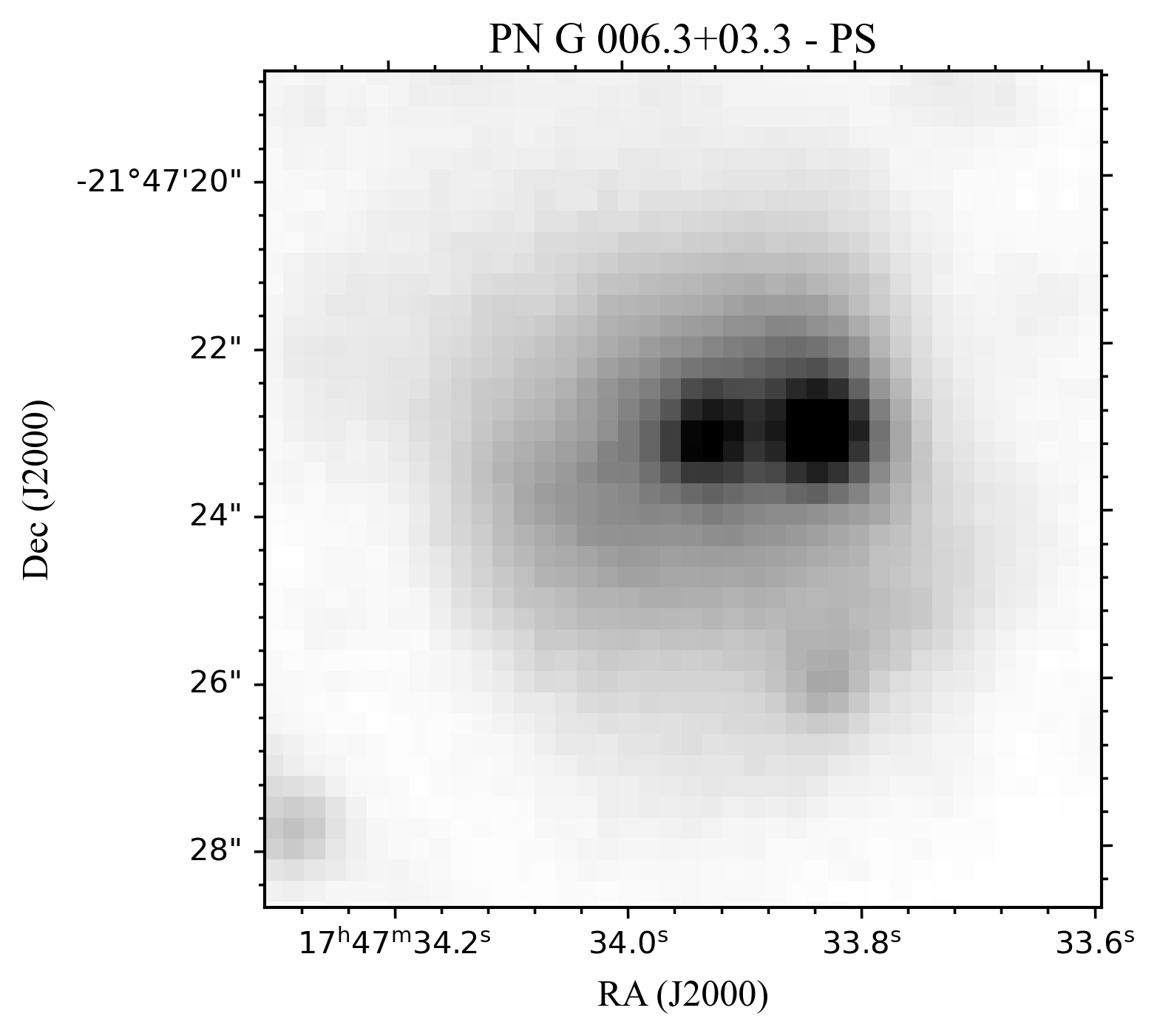}\hfill 
  \includegraphics[width=.32\linewidth]{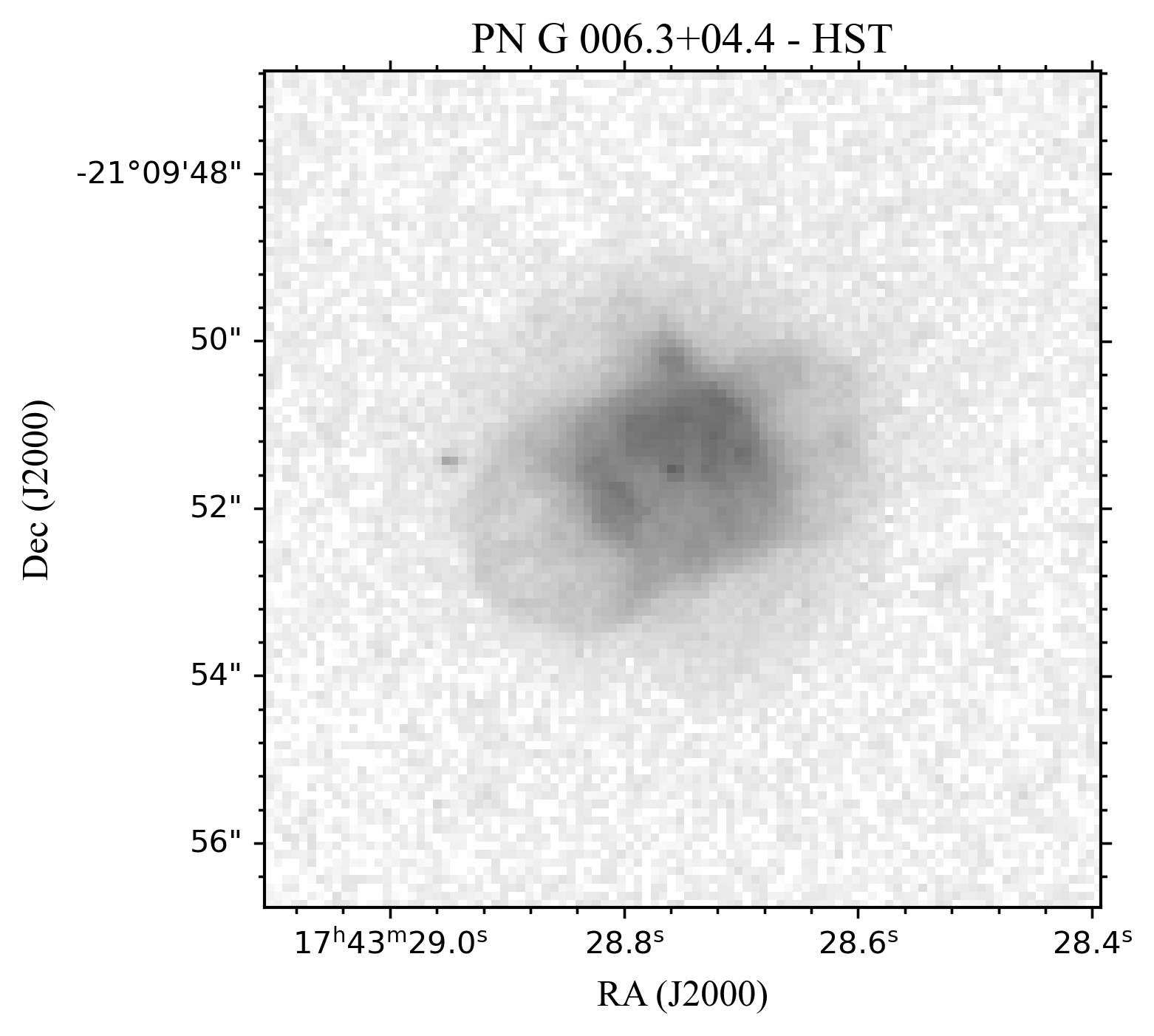}\hfill 
 \end{subfigure}\par\medskip 
  \end{figure} 
 \begin{figure} 
 \ContinuedFloat 
 \caption[]{continued:} 
\begin{subfigure}{\linewidth} 
  \includegraphics[width=.32\linewidth]{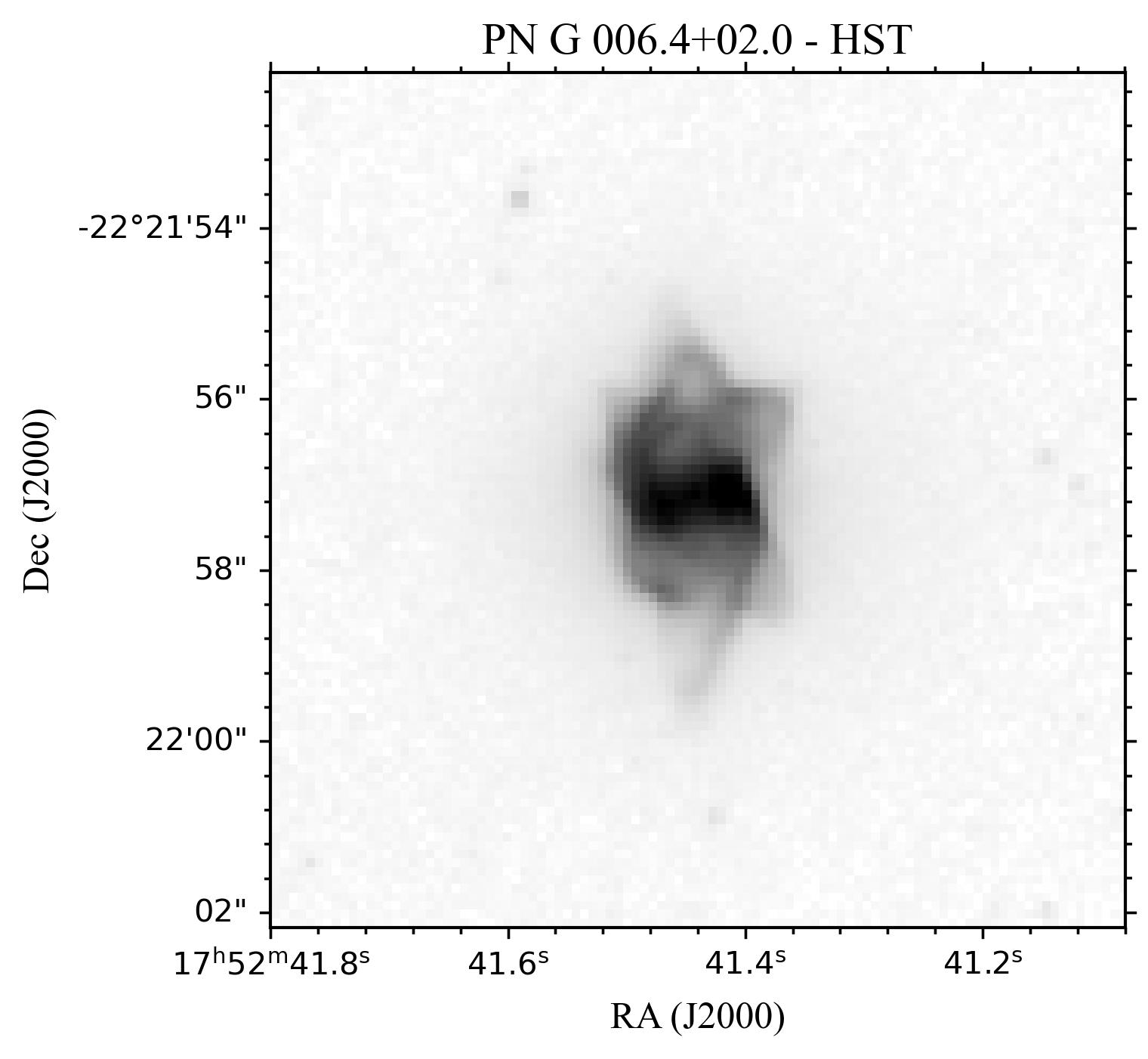}\hfill 
  \includegraphics[width=.32\linewidth]{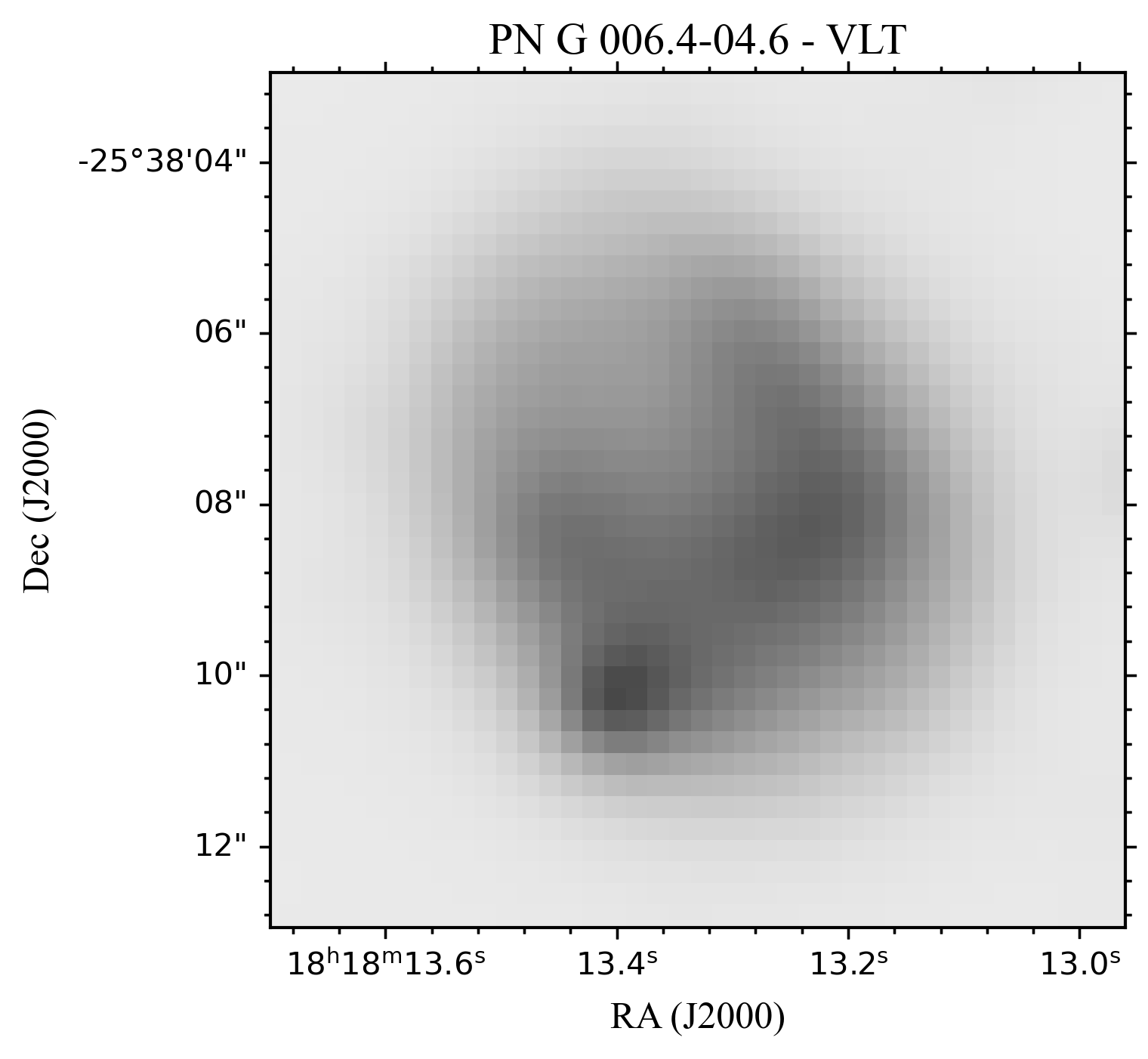}\hfill 
  \includegraphics[width=.32\linewidth]{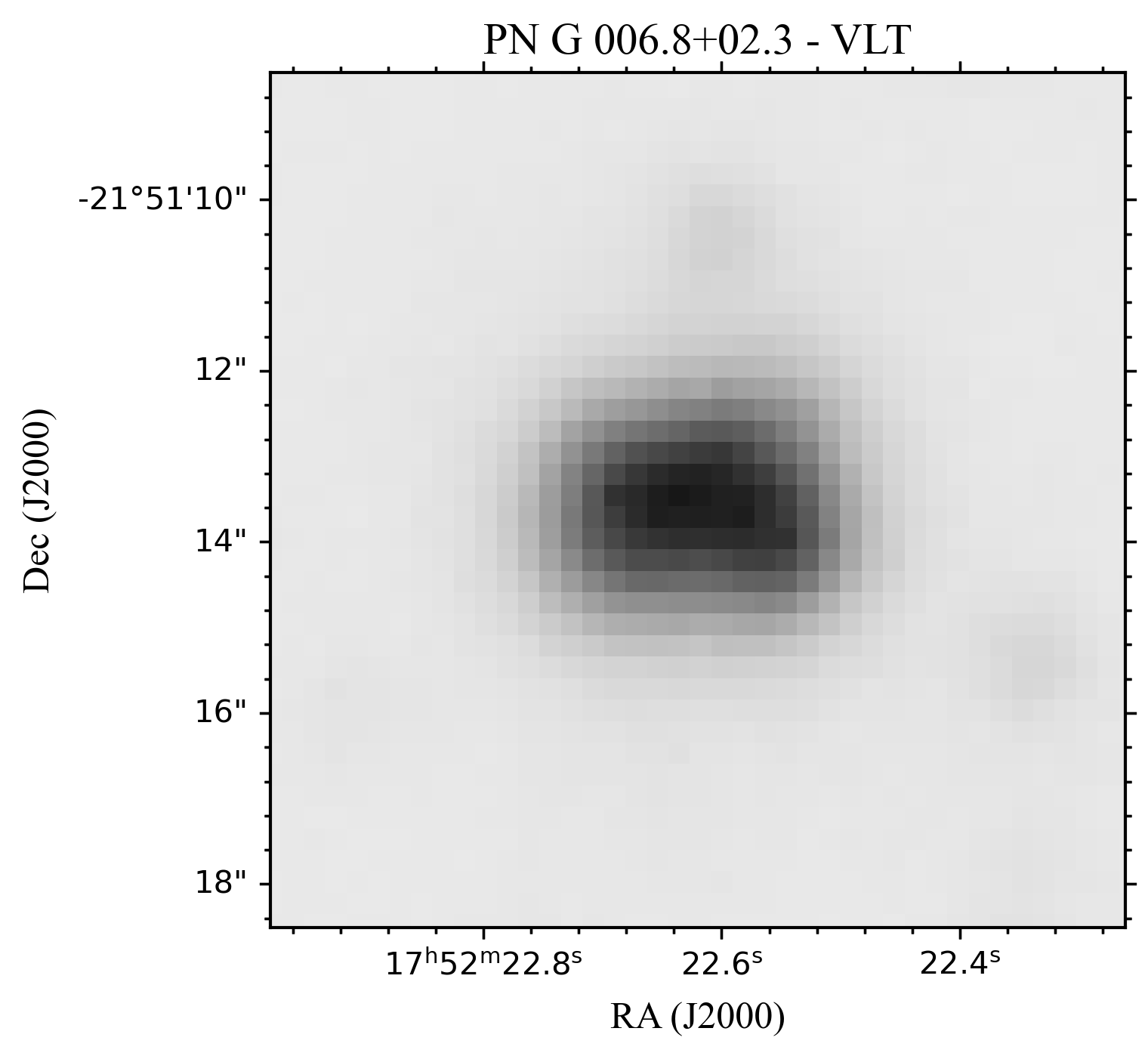}\hfill 
 \end{subfigure}\par\medskip 
\begin{subfigure}{\linewidth} 
  \includegraphics[width=.32\linewidth]{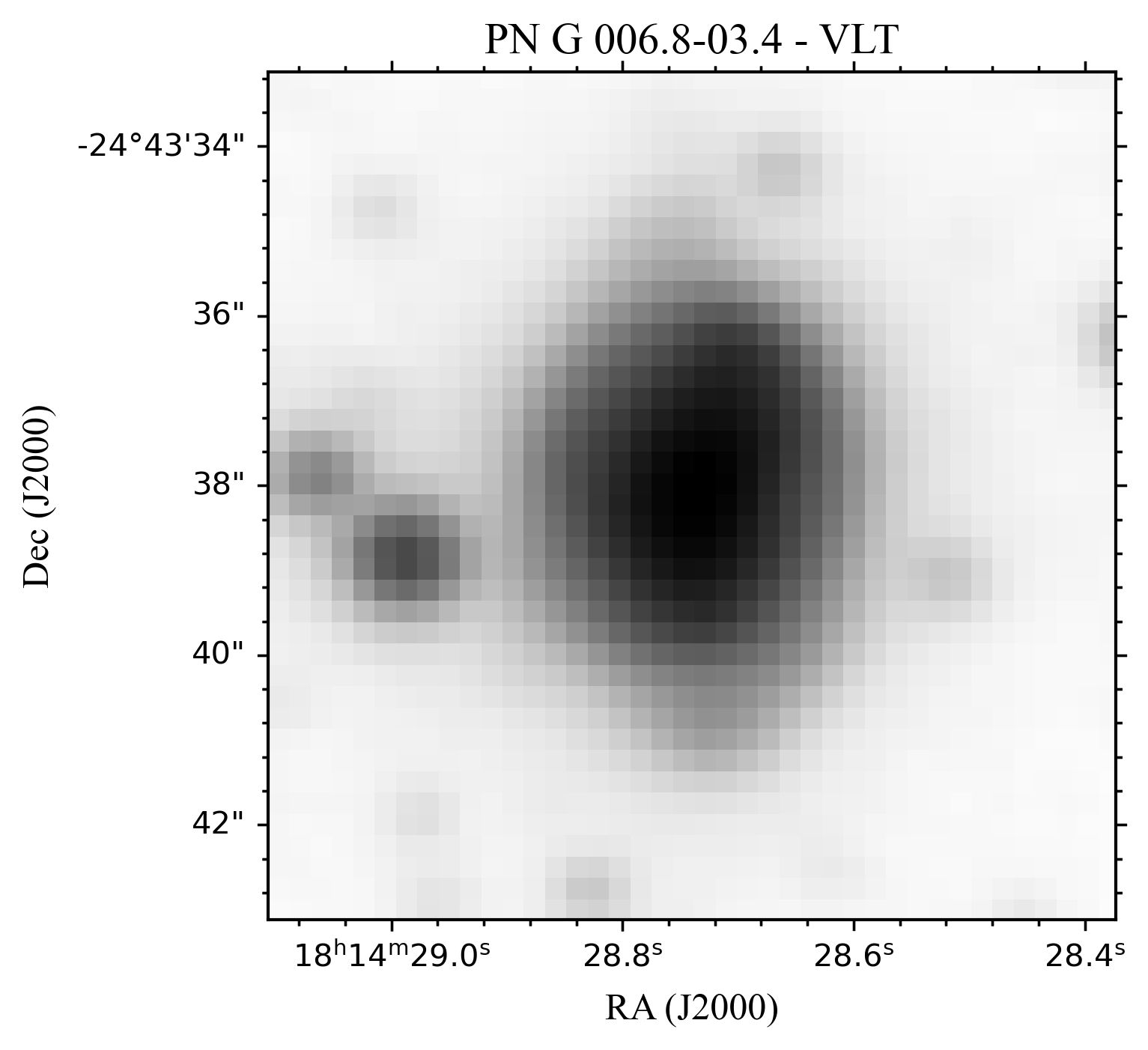}\hfill 
  \includegraphics[width=.32\linewidth]{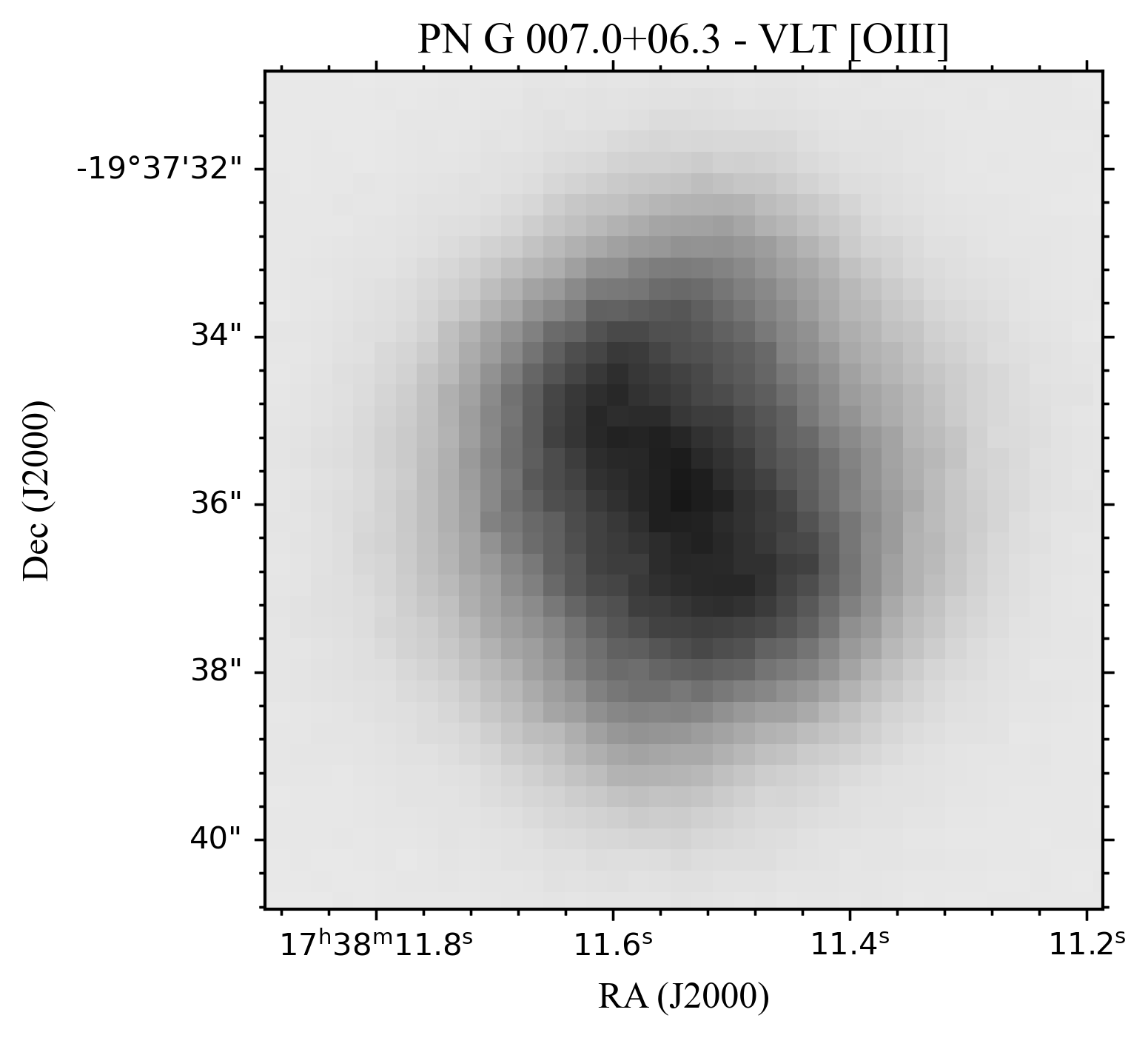}\hfill 
  \includegraphics[width=.32\linewidth]{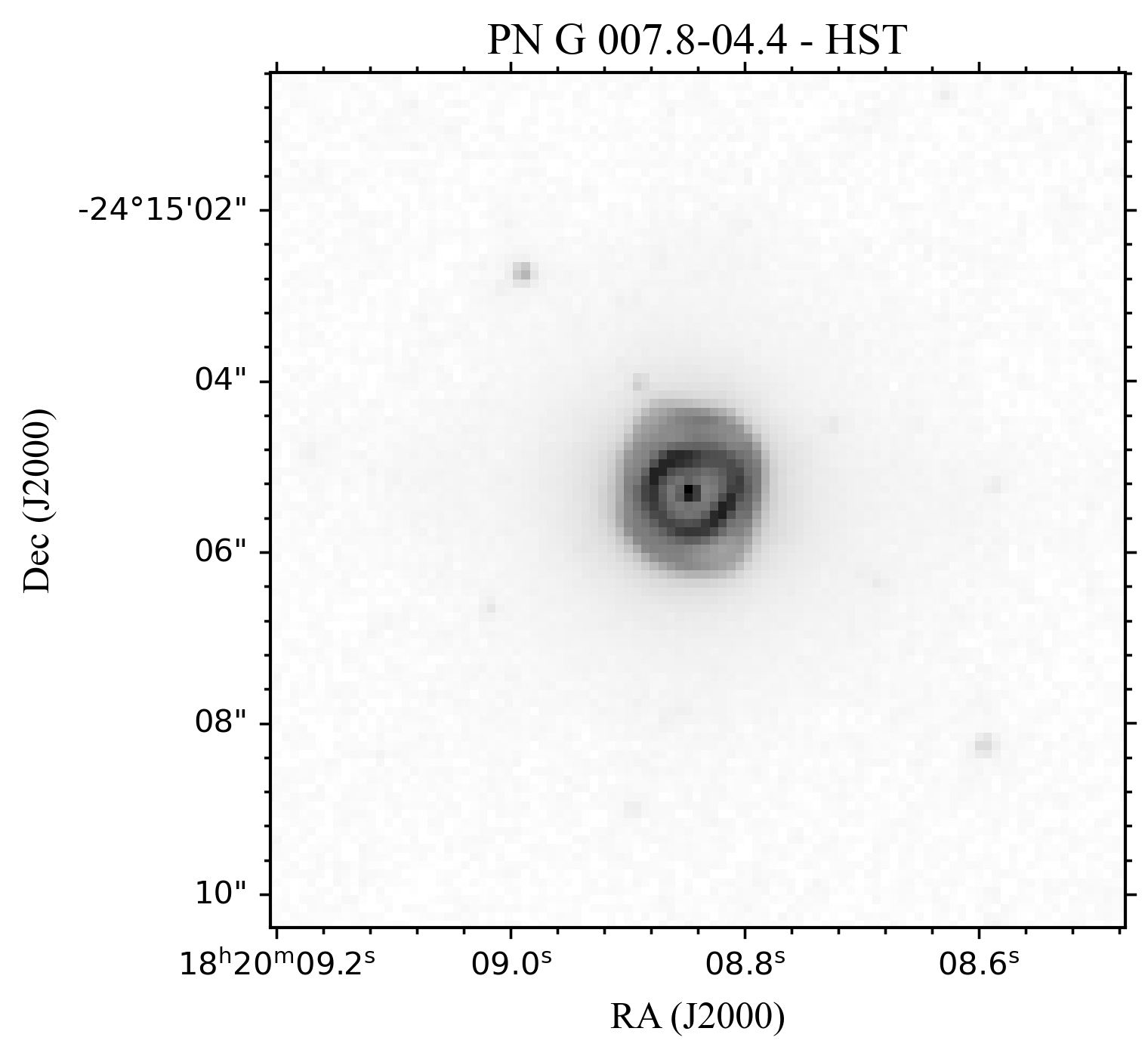}\hfill 
 \end{subfigure}\par\medskip 
\begin{subfigure}{\linewidth} 
  \includegraphics[width=.32\linewidth]{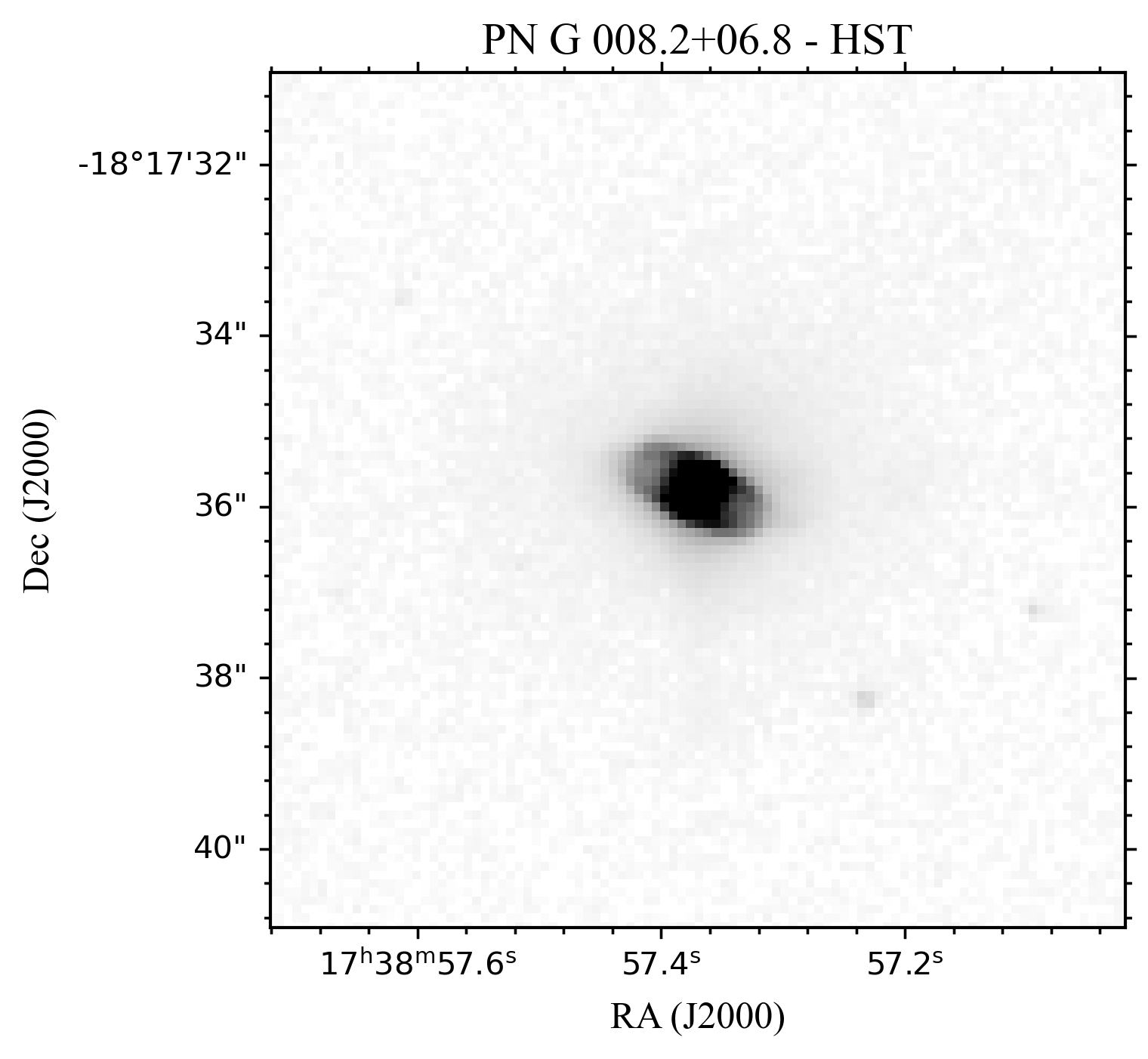}\hfill 
  \includegraphics[width=.32\linewidth]{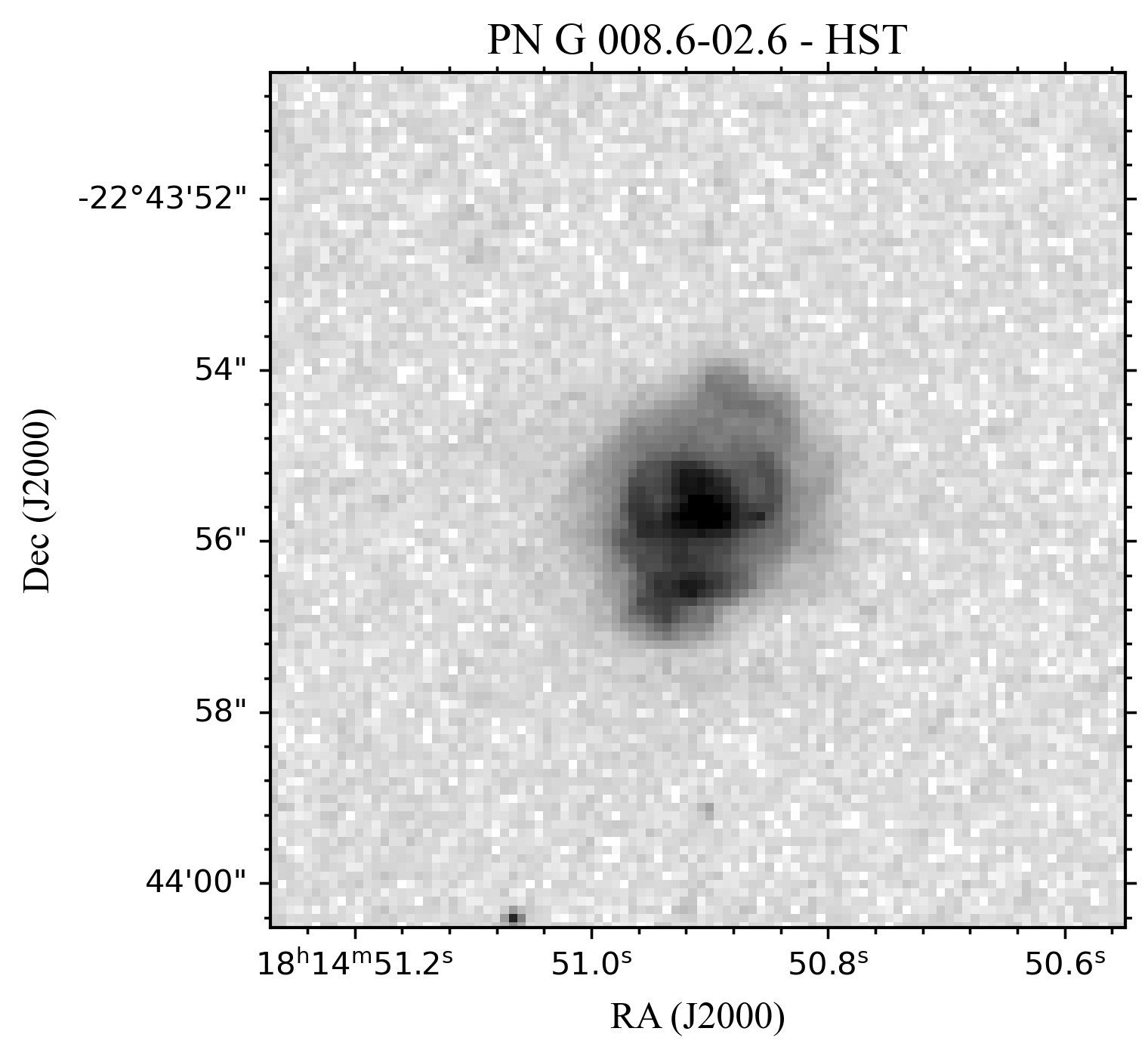}\hfill 
  \includegraphics[width=.32\linewidth]{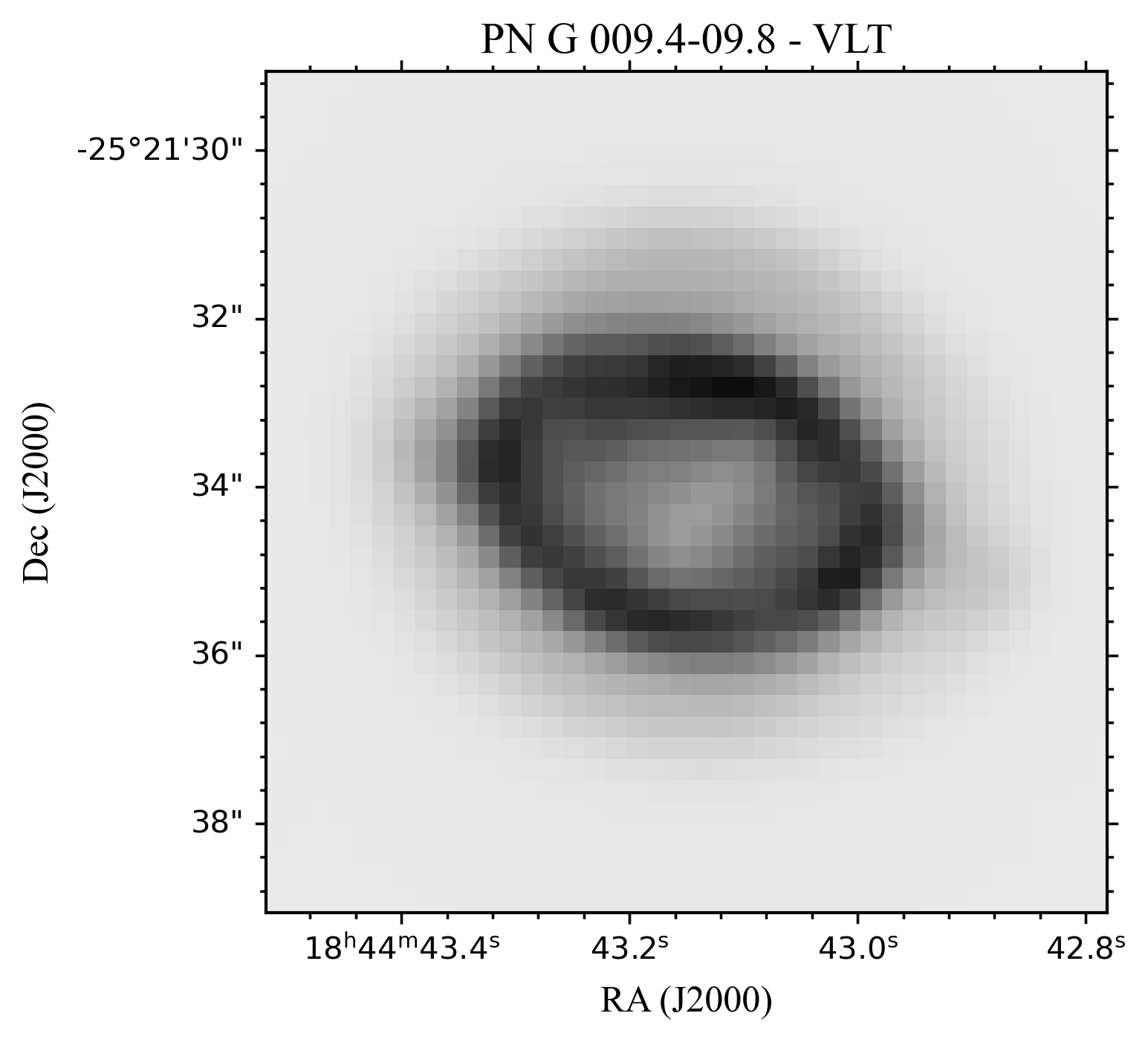}\hfill 
 \end{subfigure}\par\medskip 
\begin{subfigure}{\linewidth} 
  \includegraphics[width=.32\linewidth]{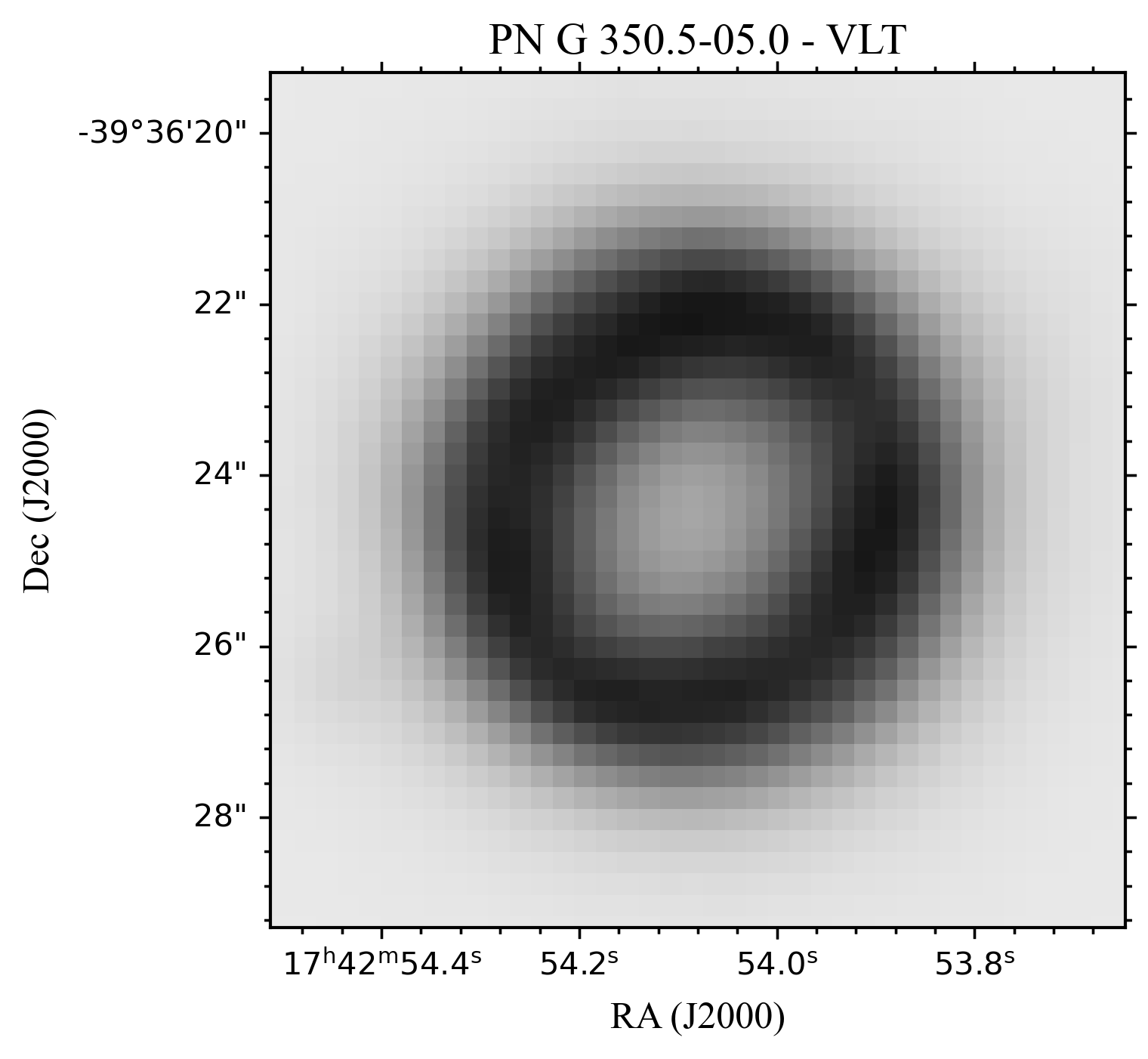}\hfill 
  \includegraphics[width=.32\linewidth]{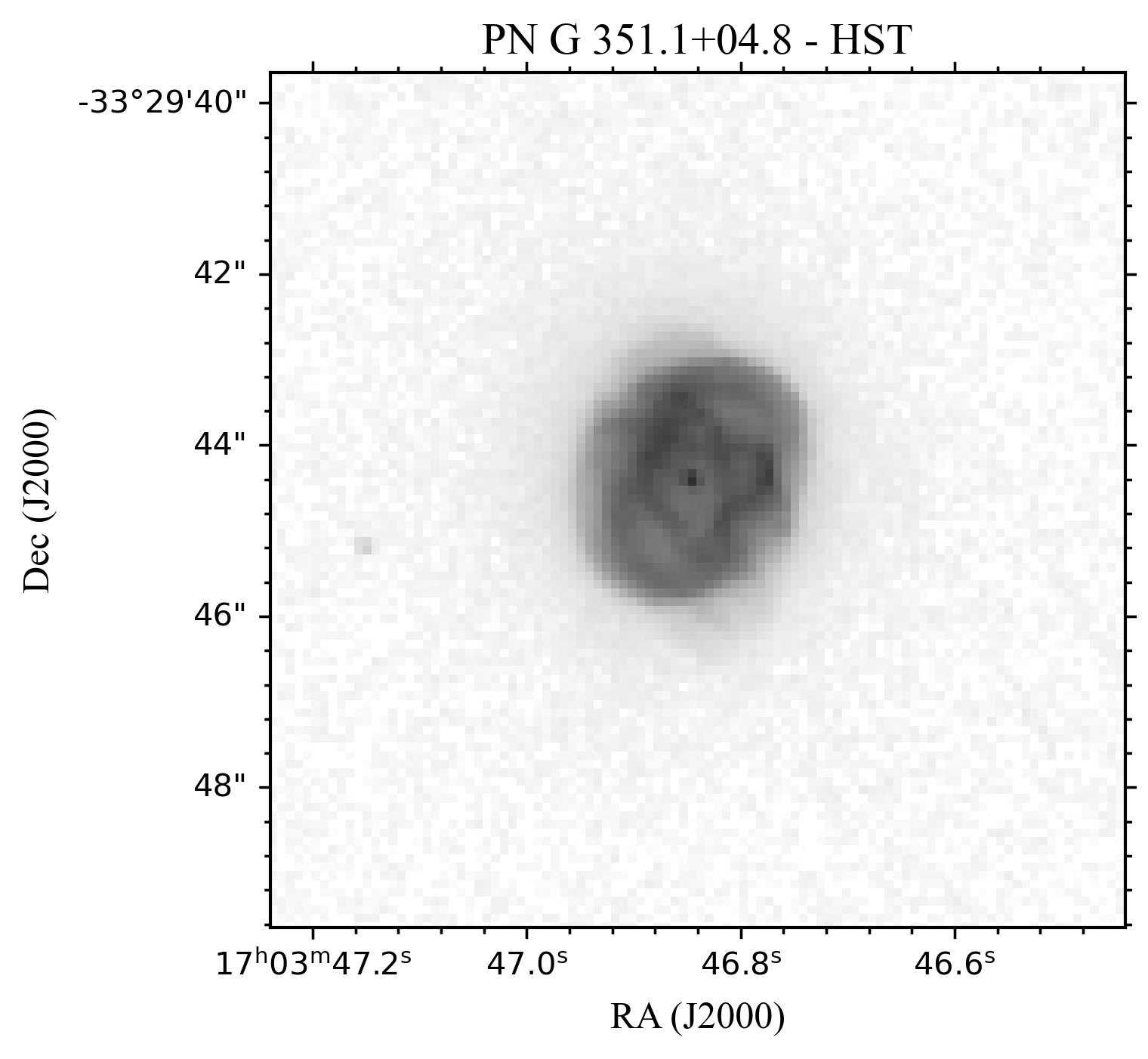}\hfill 
  \includegraphics[width=.32\linewidth]{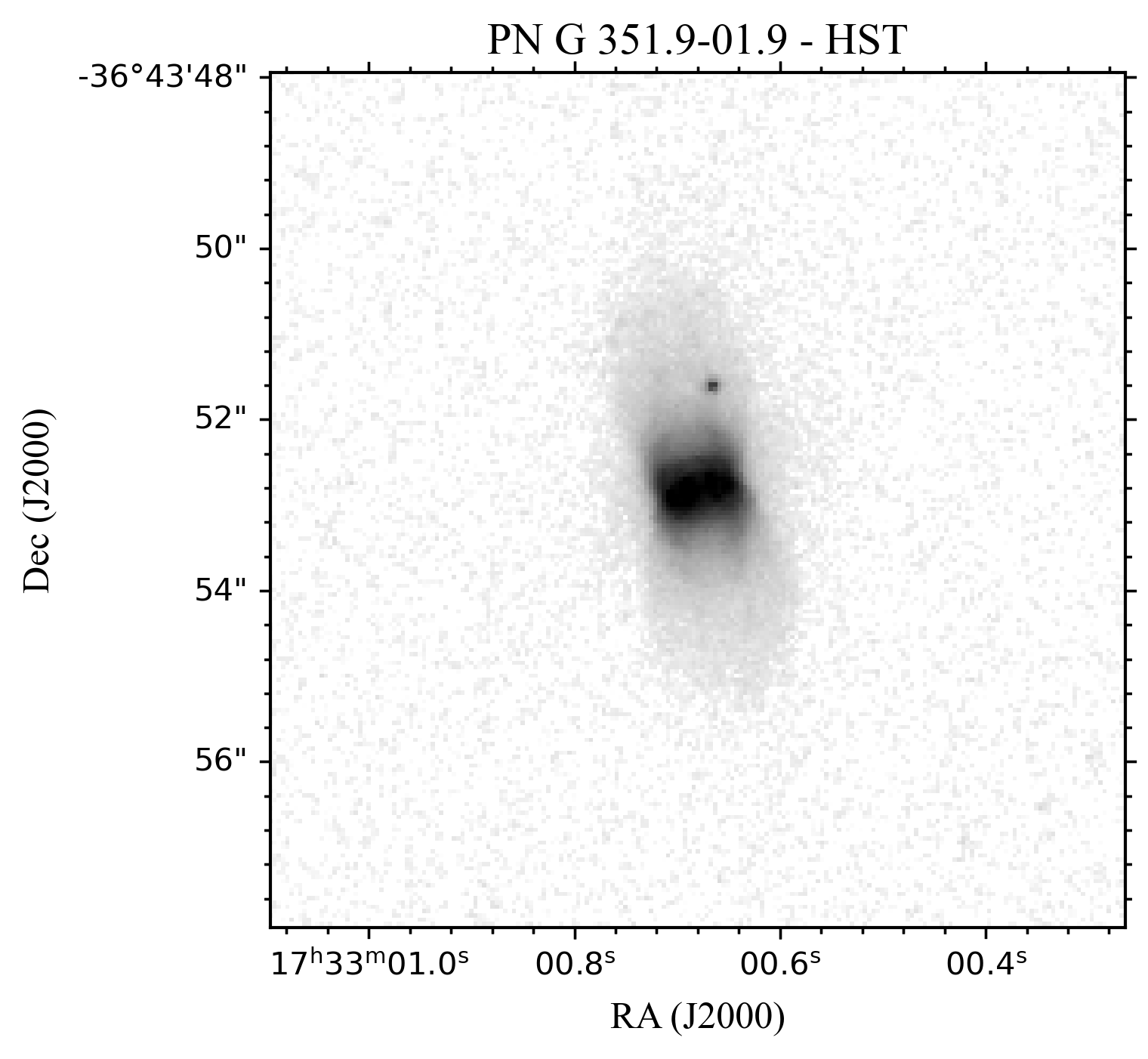}\hfill 
 \end{subfigure}\par\medskip 
  \end{figure} 
 \begin{figure} 
 \ContinuedFloat 
 \caption[]{continued:} 
\begin{subfigure}{\linewidth} 
  \includegraphics[width=.32\linewidth]{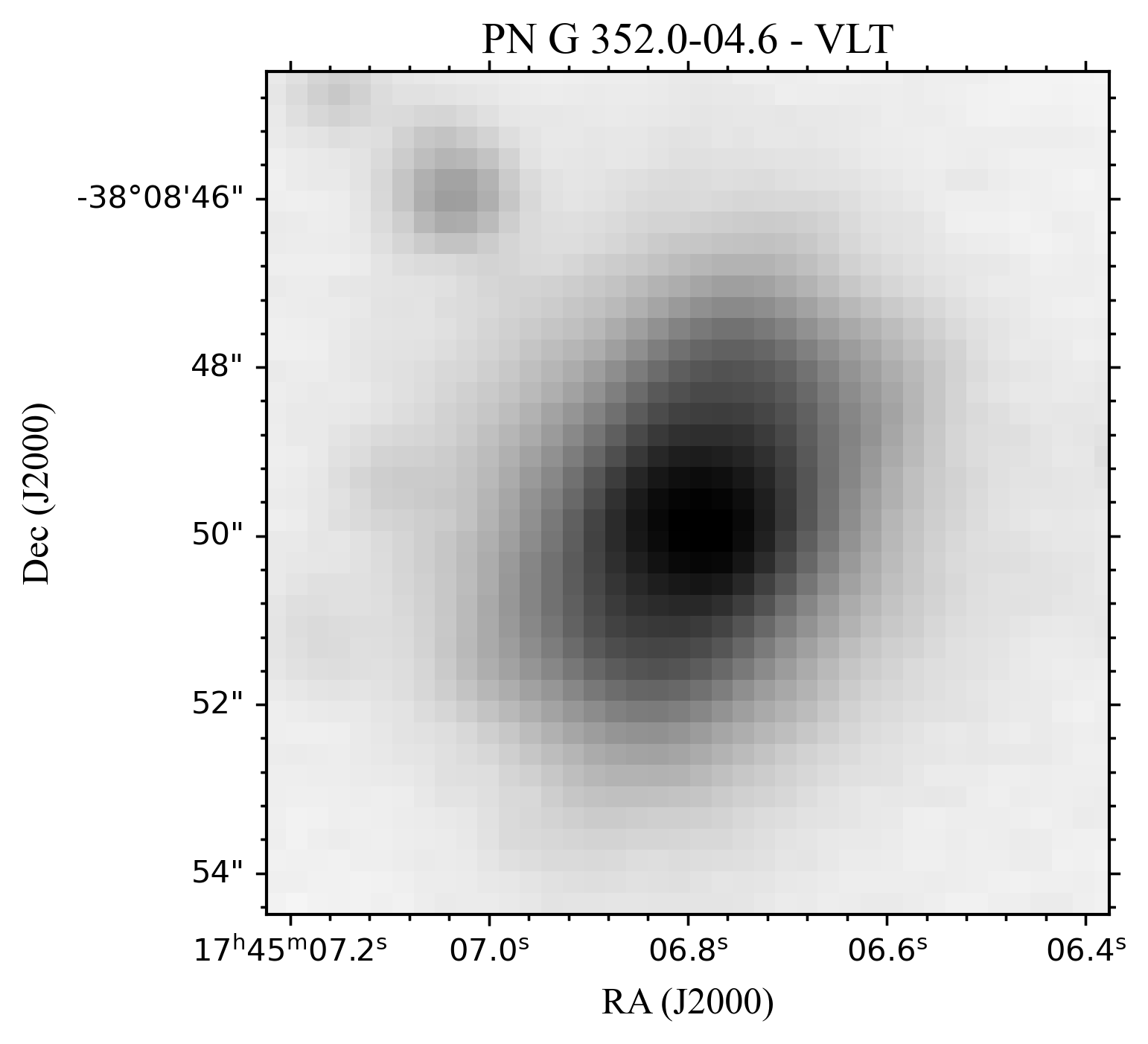}\hfill 
  \includegraphics[width=.32\linewidth]{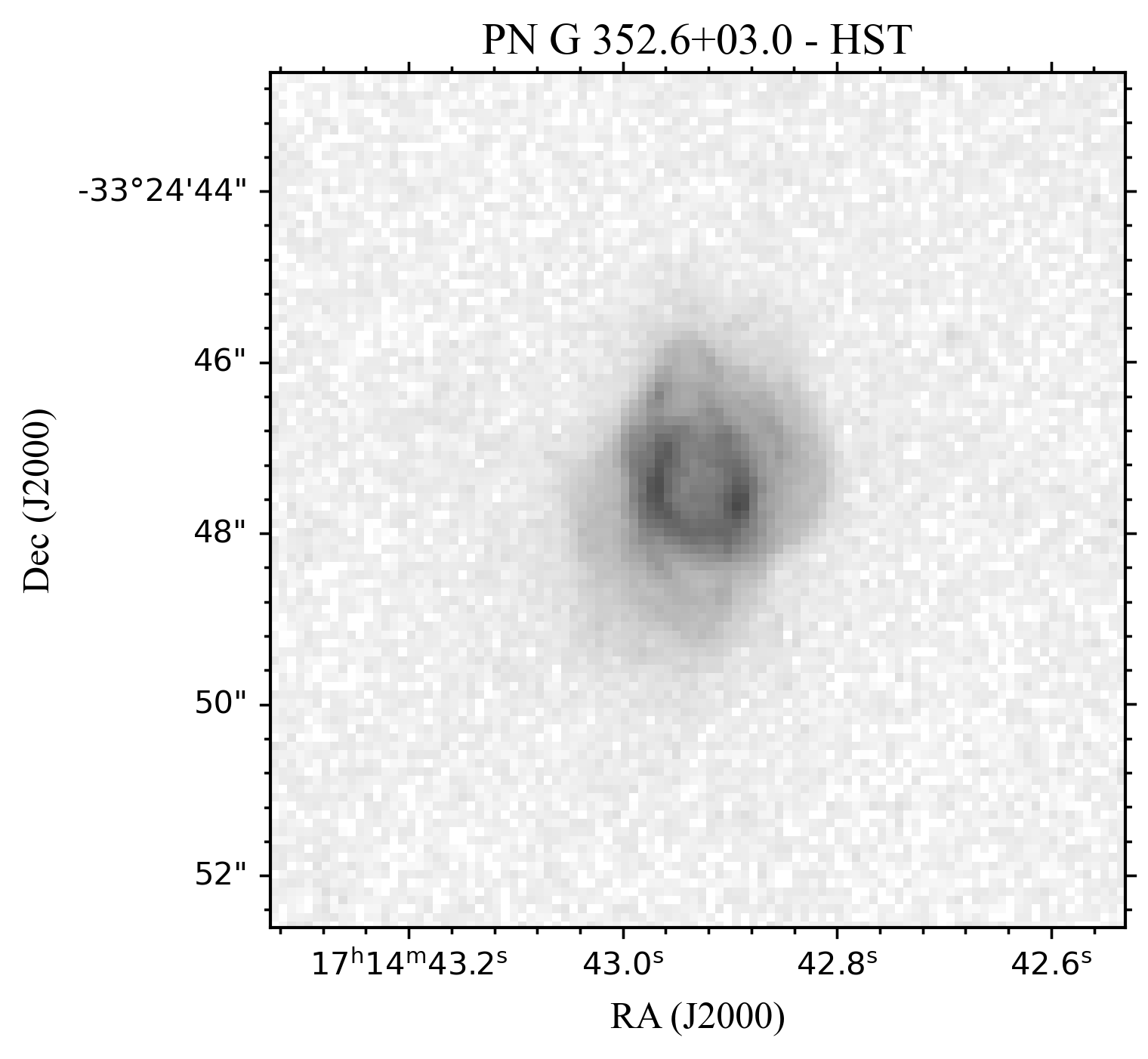}\hfill 
  \includegraphics[width=.32\linewidth]{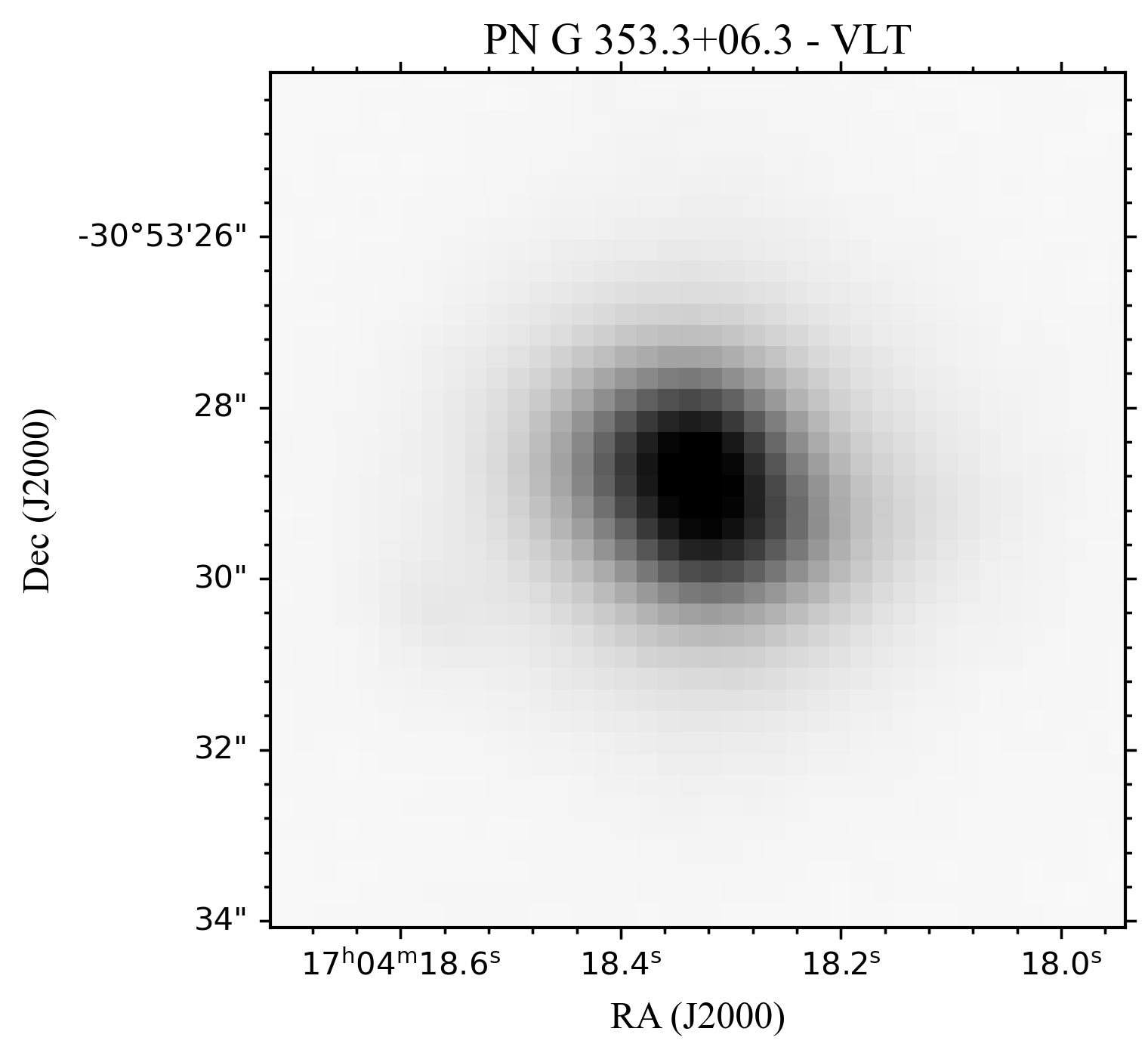}\hfill 
 \end{subfigure}\par\medskip 
\begin{subfigure}{\linewidth} 
  \includegraphics[width=.32\linewidth]{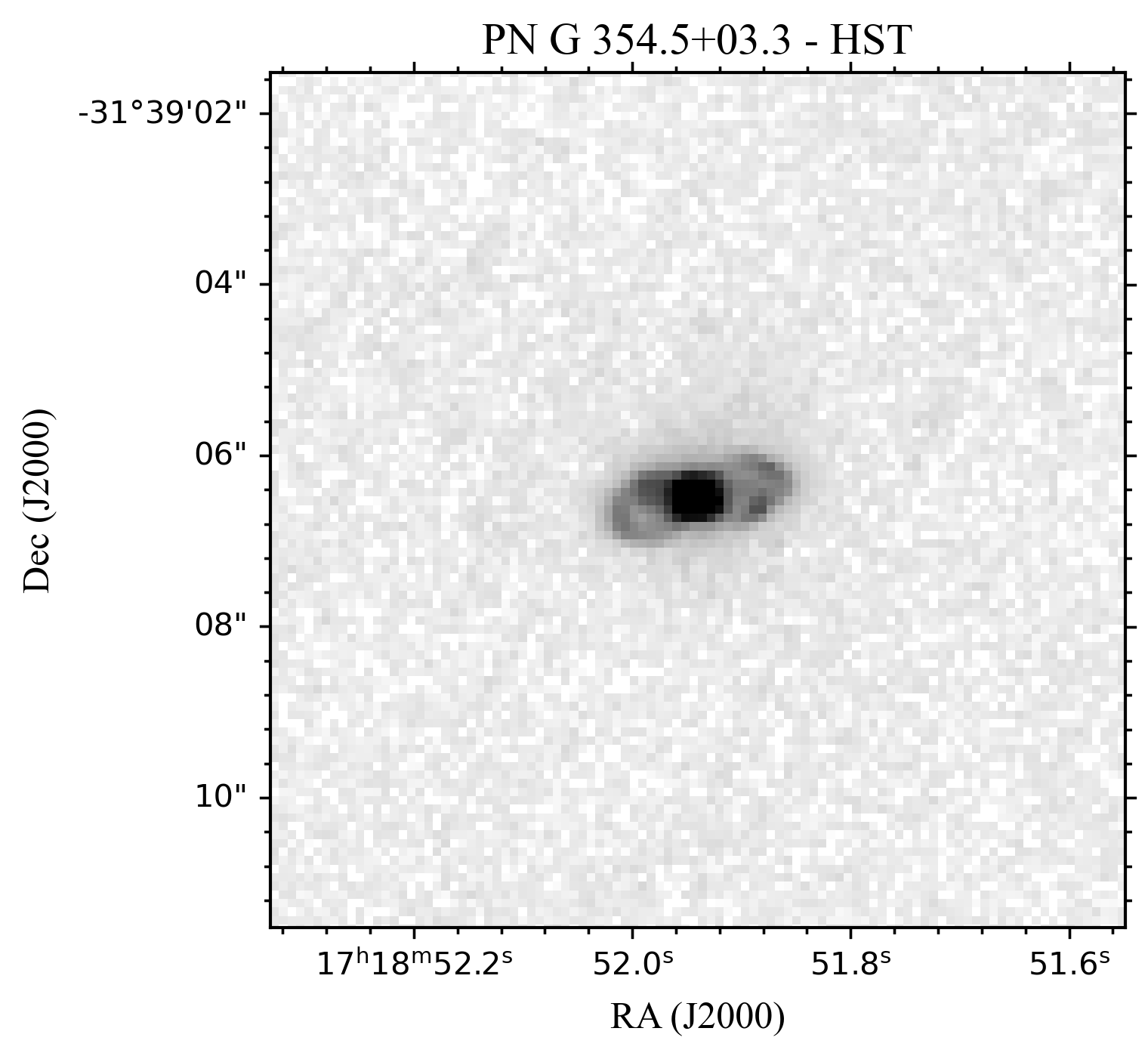}\hfill 
  \includegraphics[width=.32\linewidth]{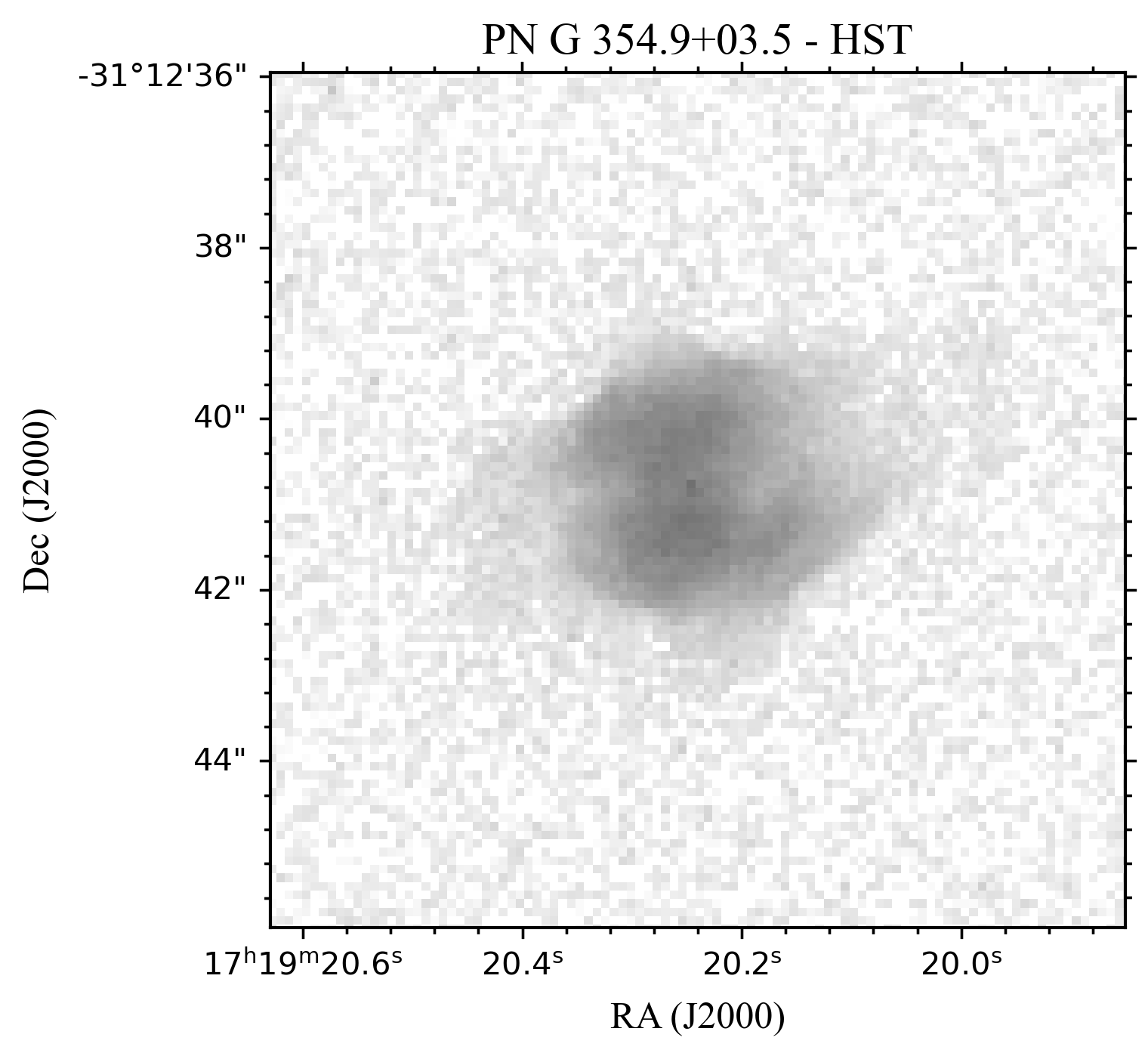}\hfill 
  \includegraphics[width=.32\linewidth]{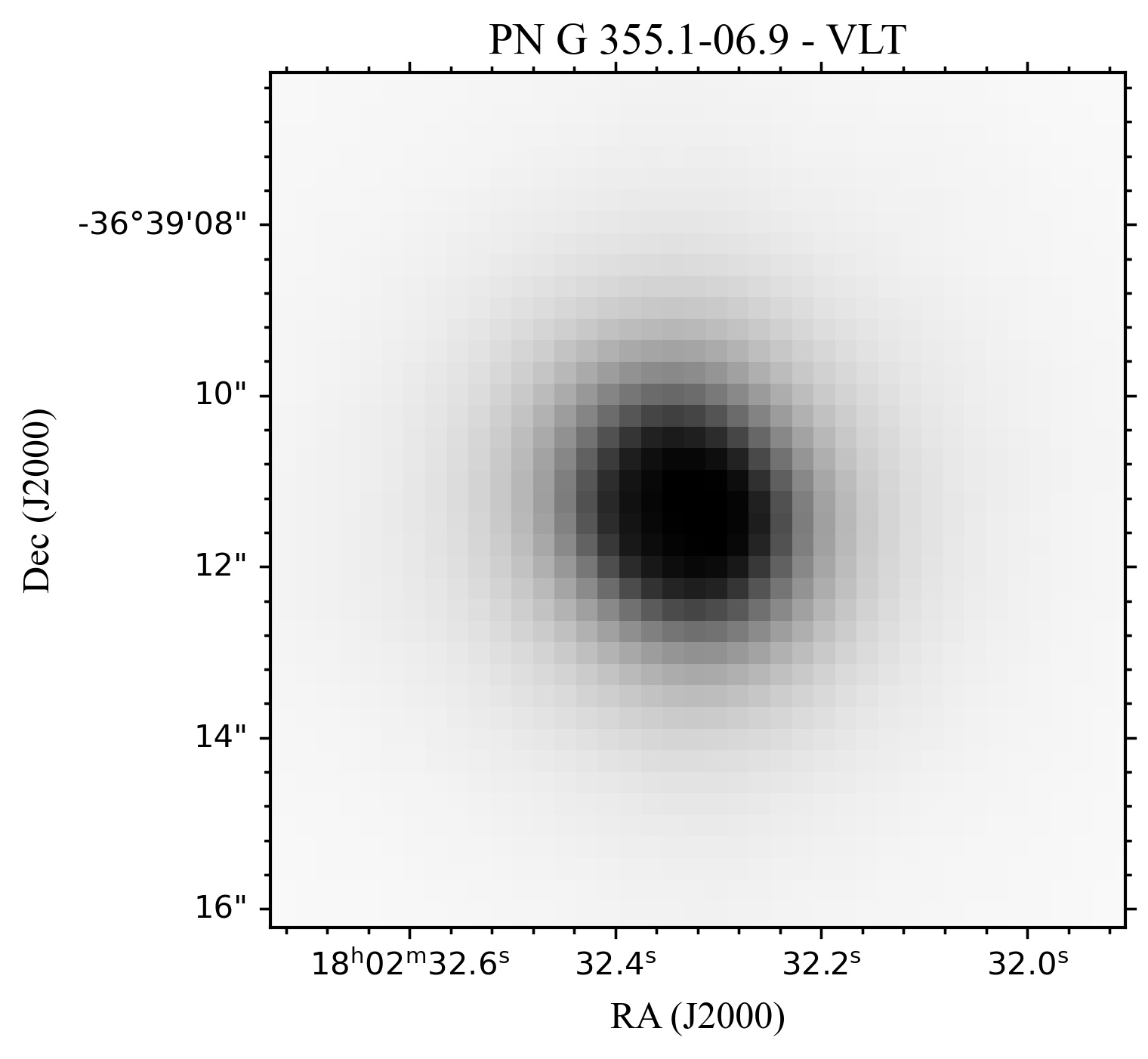}\hfill 
 \end{subfigure}\par\medskip 
\begin{subfigure}{\linewidth} 
  \includegraphics[width=.32\linewidth]{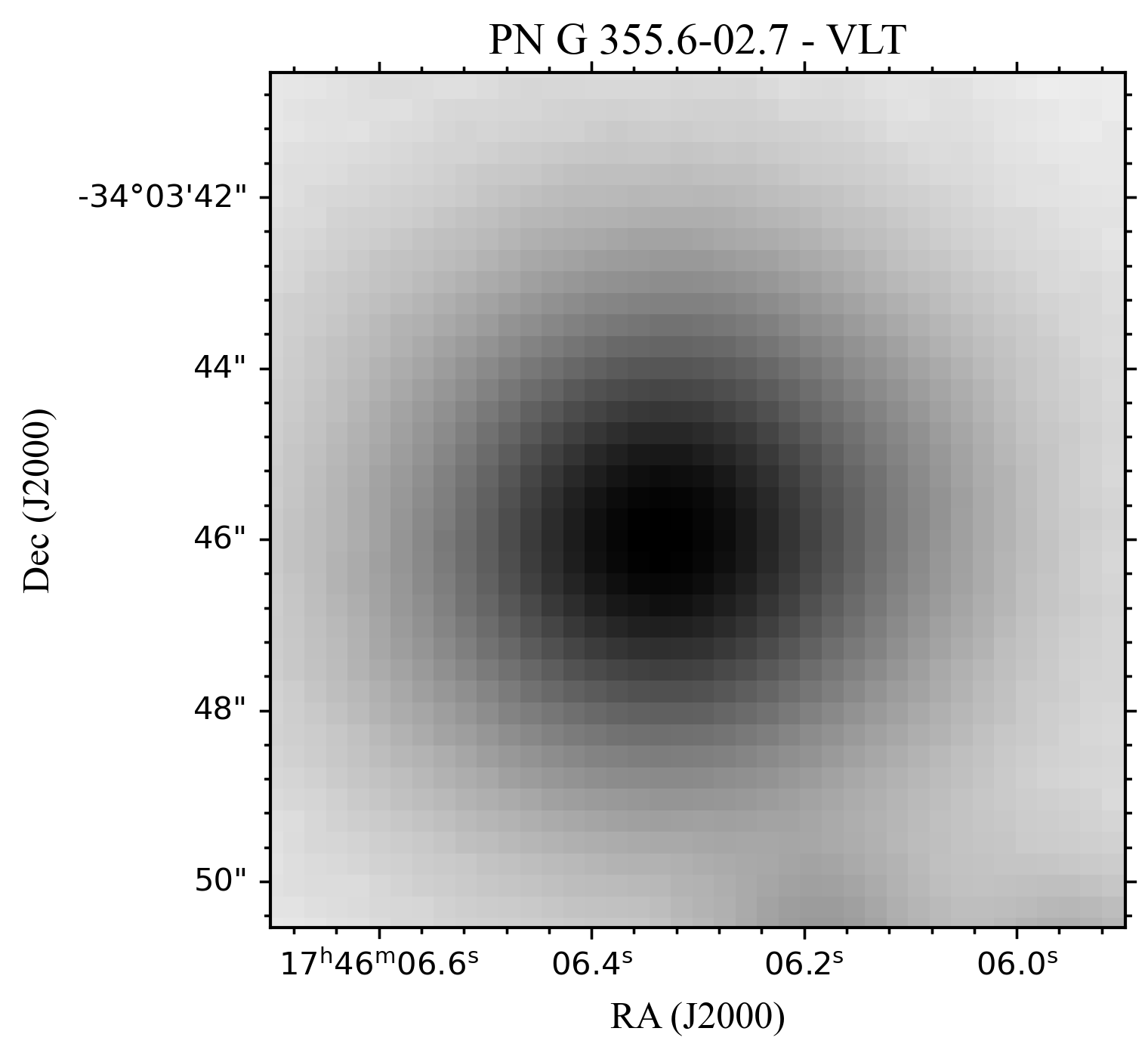}\hfill 
  \includegraphics[width=.32\linewidth]{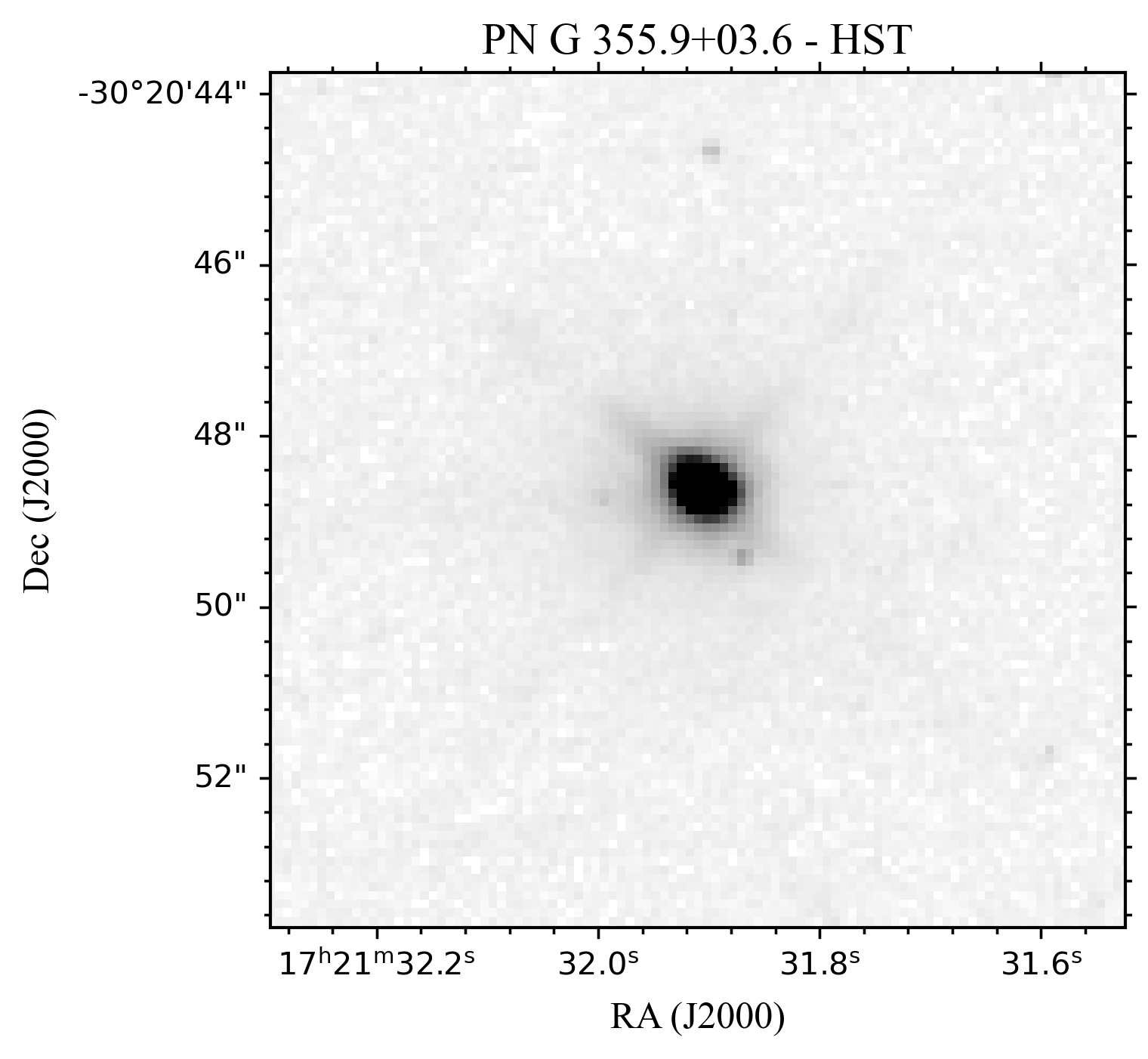}\hfill 
  \includegraphics[width=.32\linewidth]{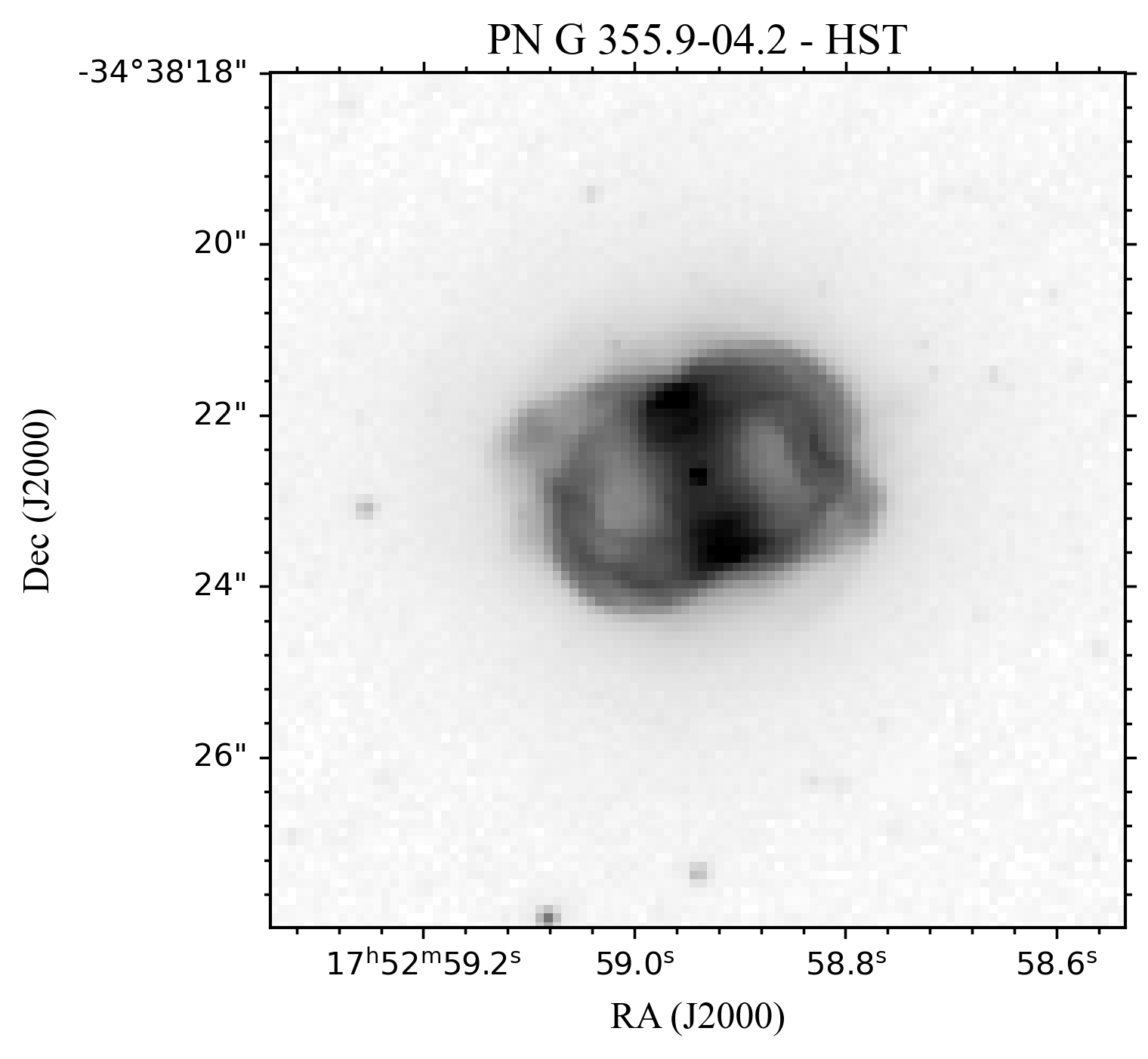}\hfill 
 \end{subfigure}\par\medskip 
\begin{subfigure}{\linewidth} 
  \includegraphics[width=.32\linewidth]{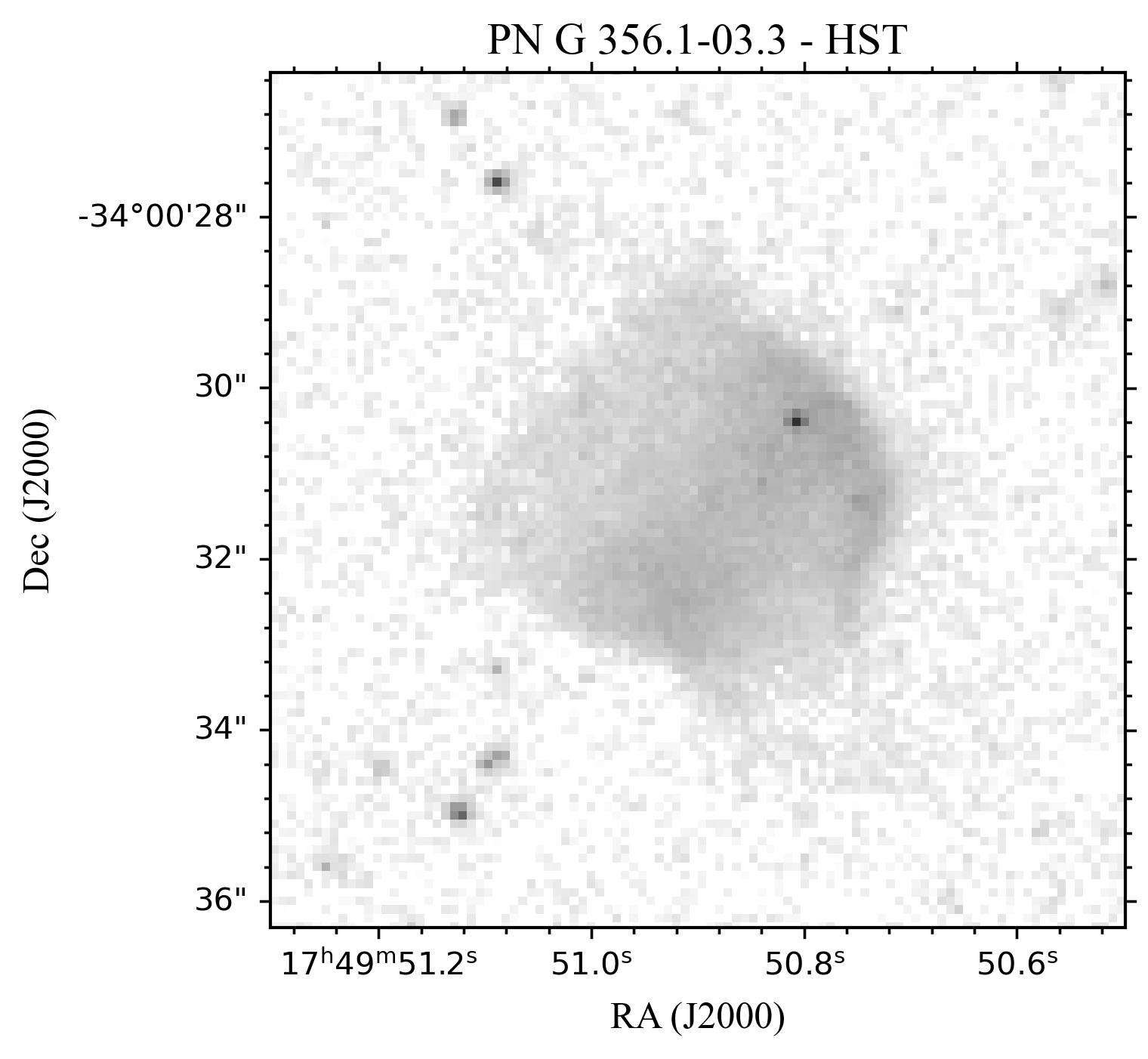}\hfill 
  \includegraphics[width=.32\linewidth]{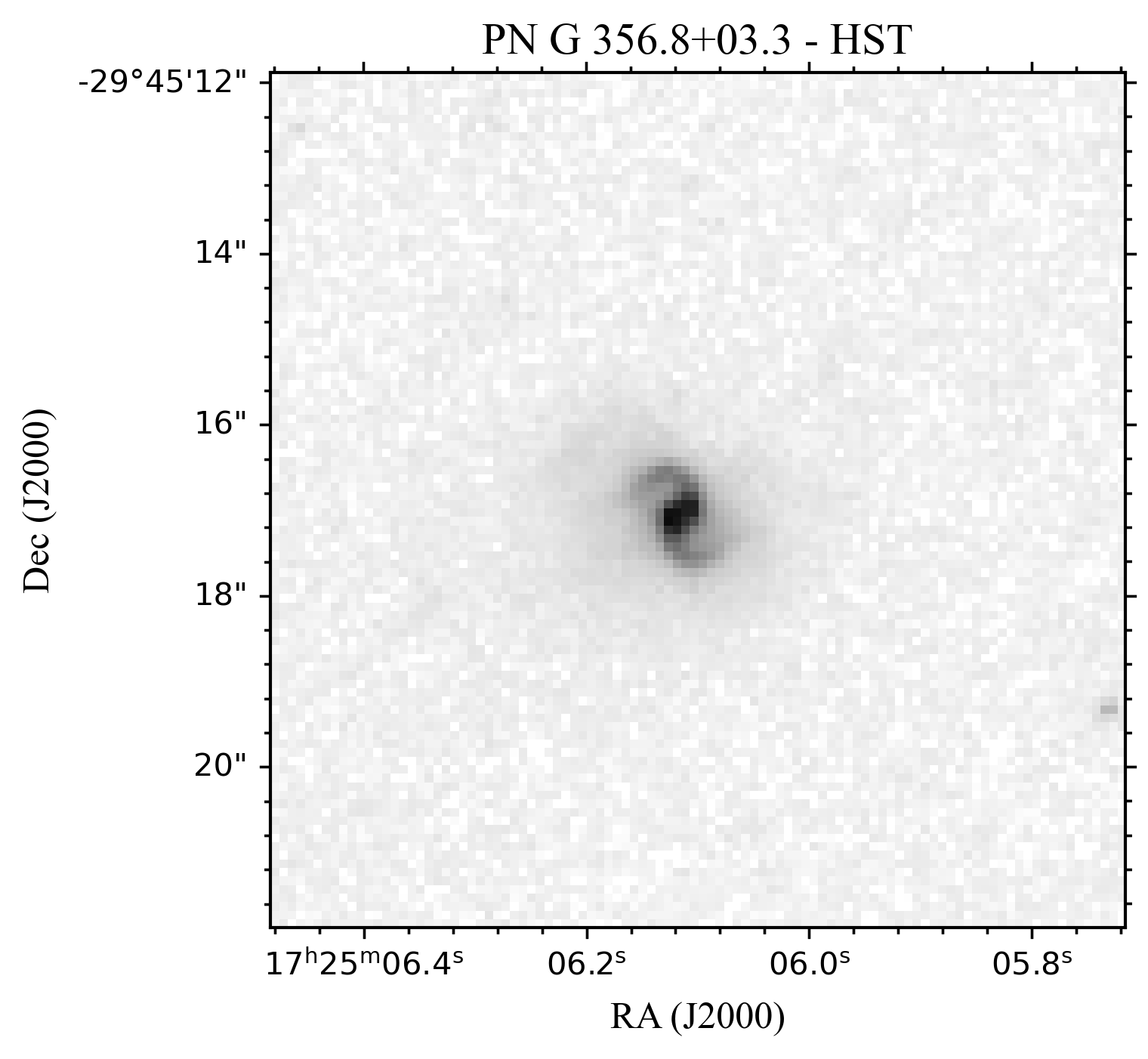}\hfill 
  \includegraphics[width=.32\linewidth]{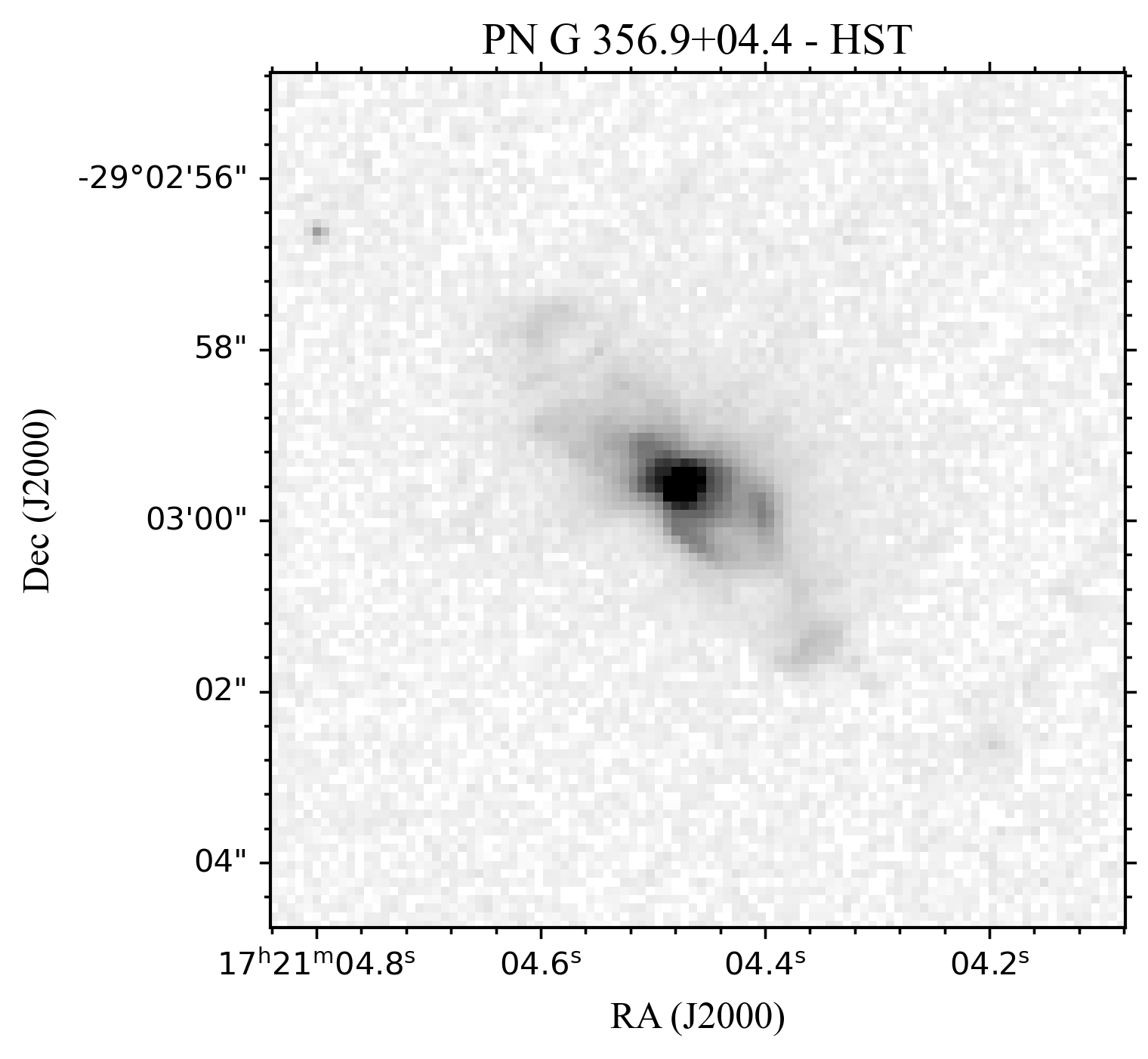}\hfill 
 \end{subfigure}\par\medskip 
  \end{figure} 
 \begin{figure} 
 \ContinuedFloat 
 \caption[]{continued:} 
\begin{subfigure}{\linewidth} 
  \includegraphics[width=.32\linewidth]{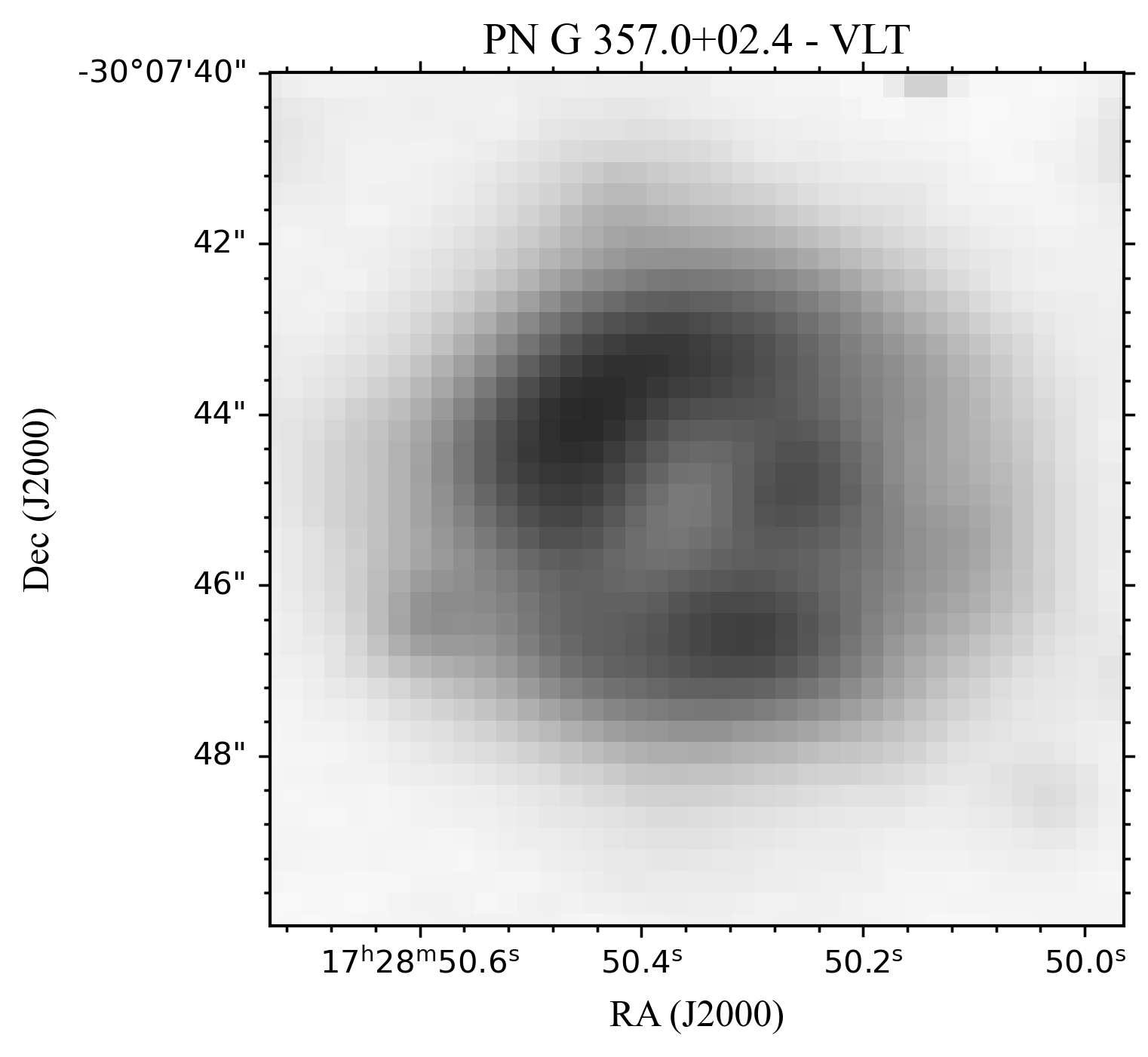}\hfill 
  \includegraphics[width=.32\linewidth]{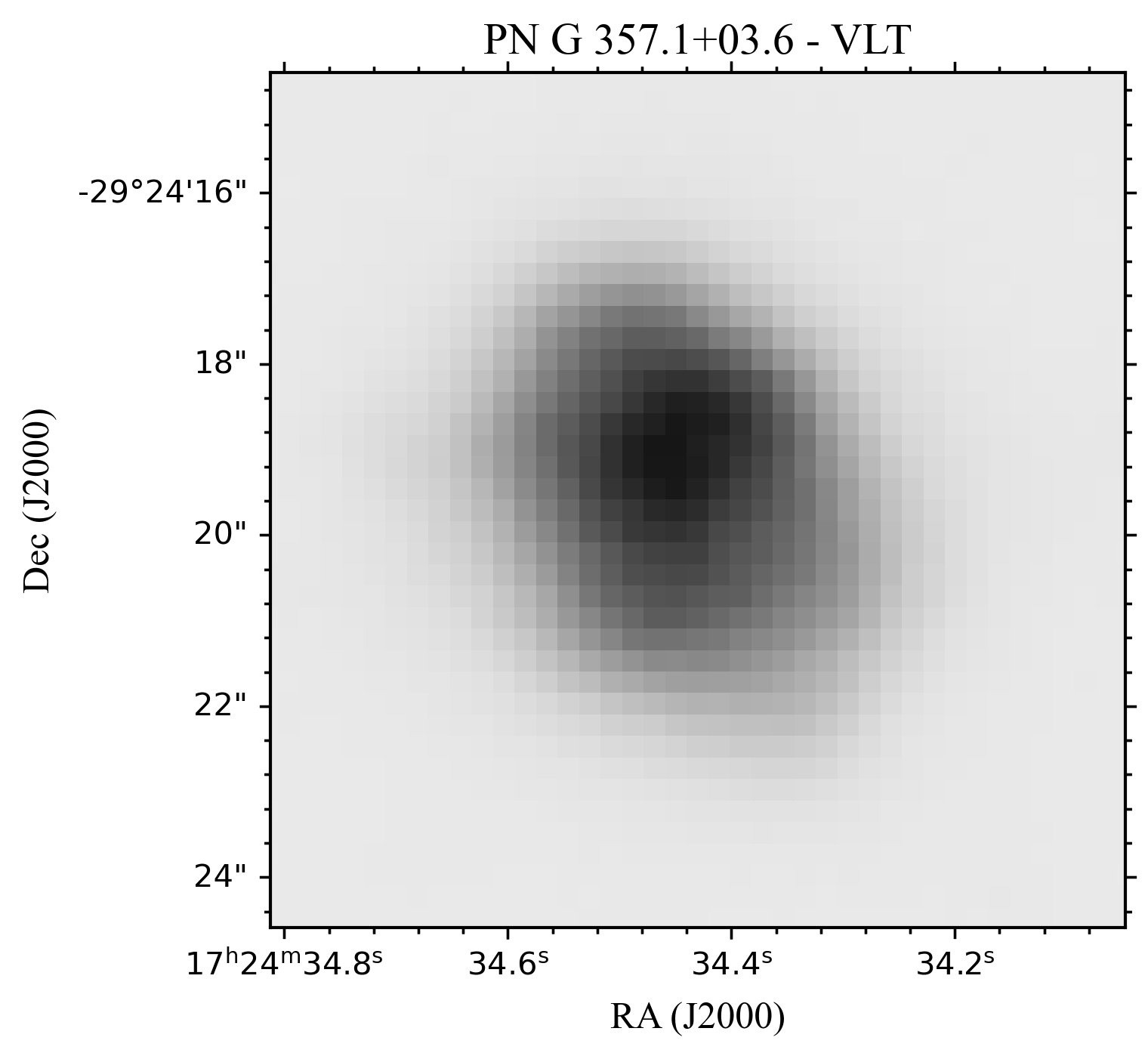}\hfill 
  \includegraphics[width=.32\linewidth]{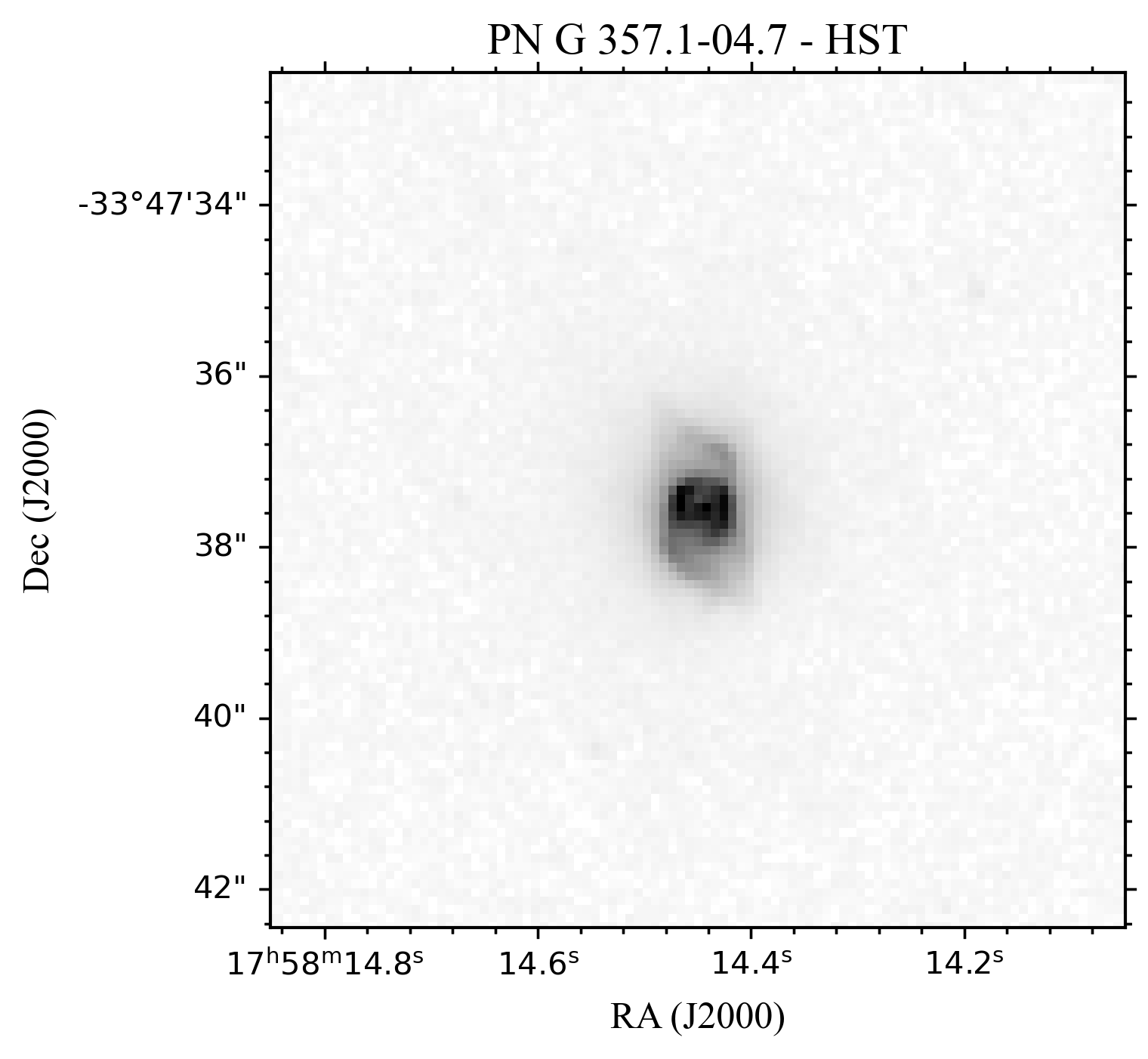}\hfill 
 \end{subfigure}\par\medskip 
\begin{subfigure}{\linewidth} 
  \includegraphics[width=.32\linewidth]{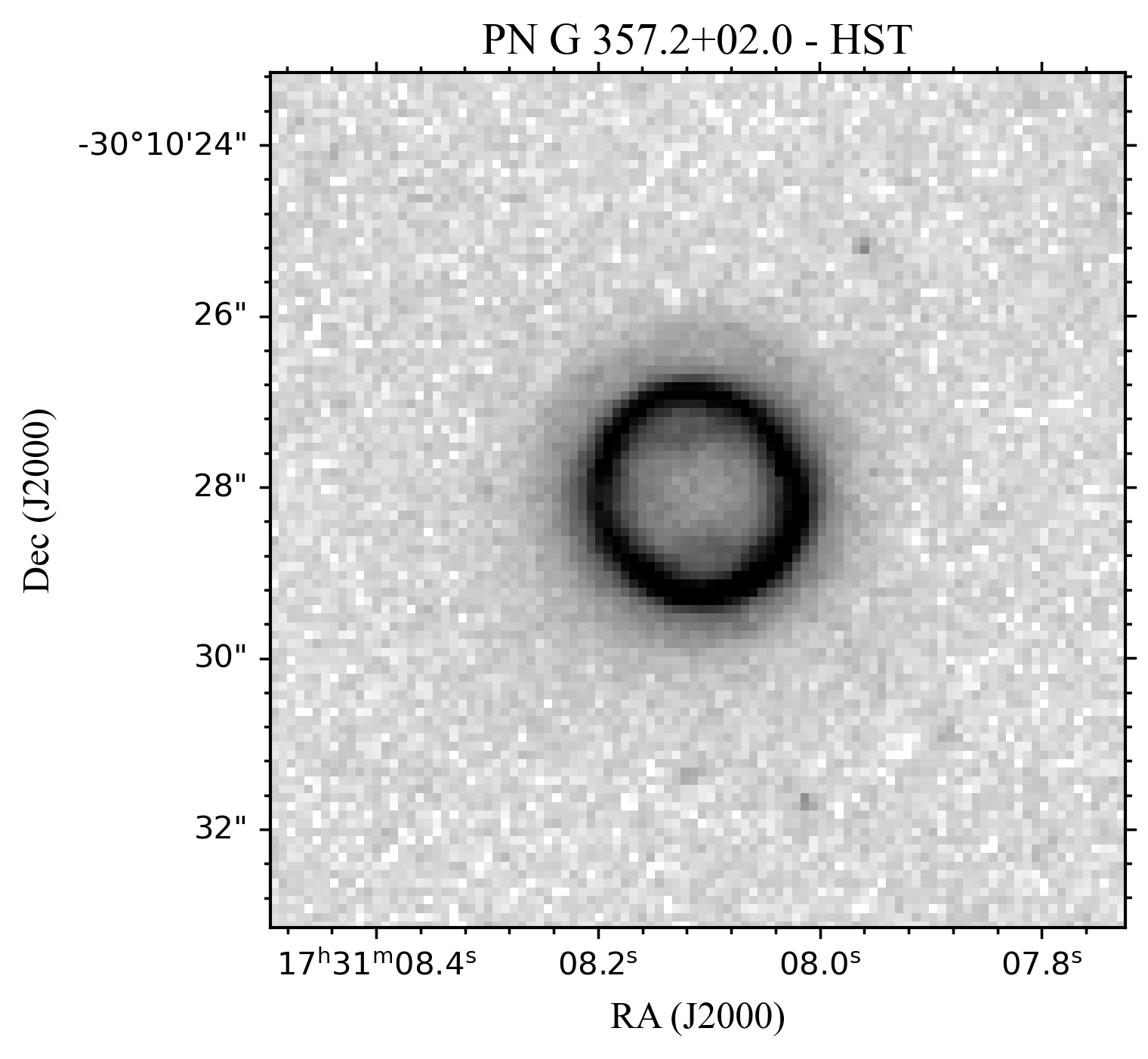}\hfill 
  \includegraphics[width=.32\linewidth]{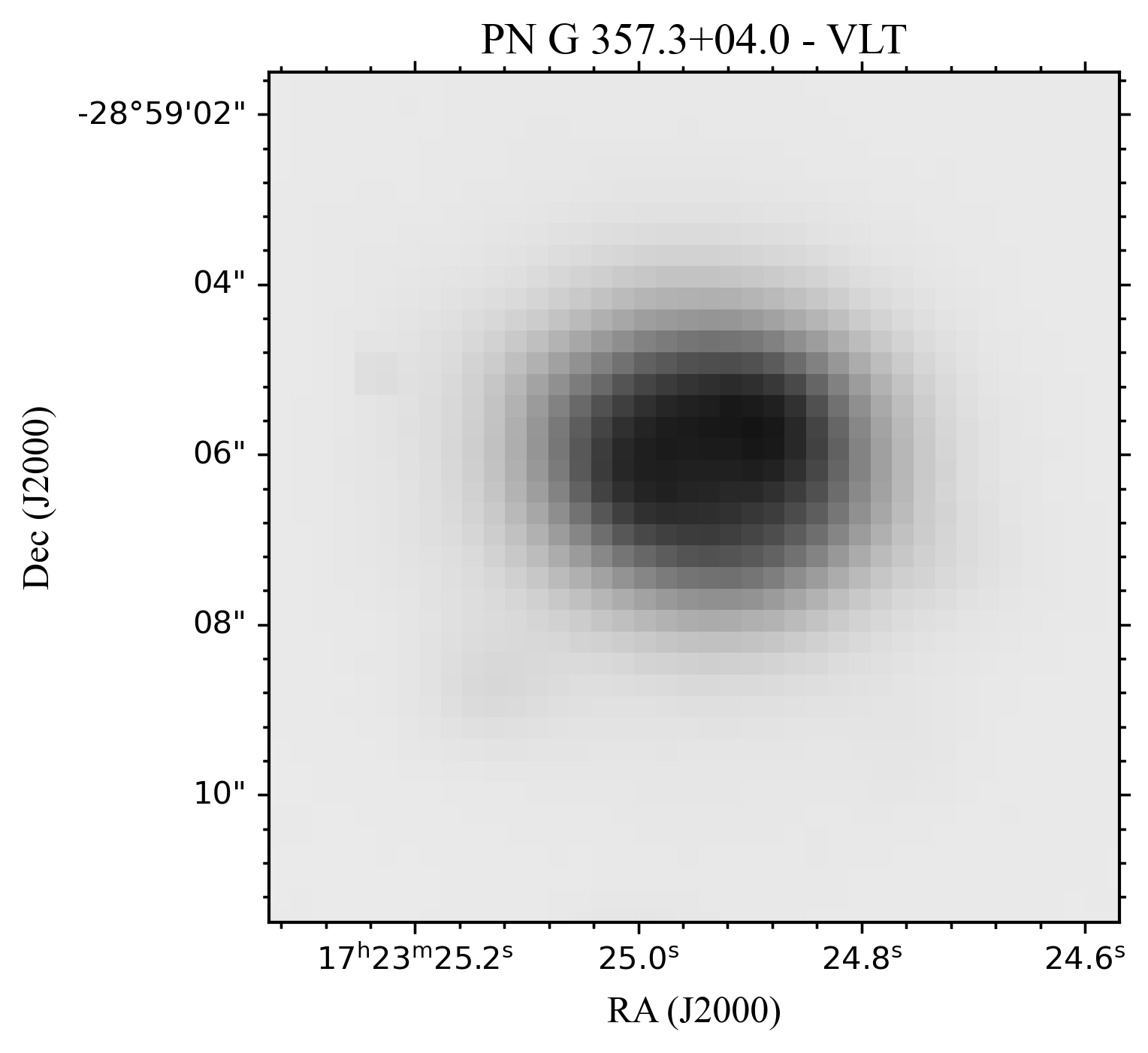}\hfill 
  \includegraphics[width=.32\linewidth]{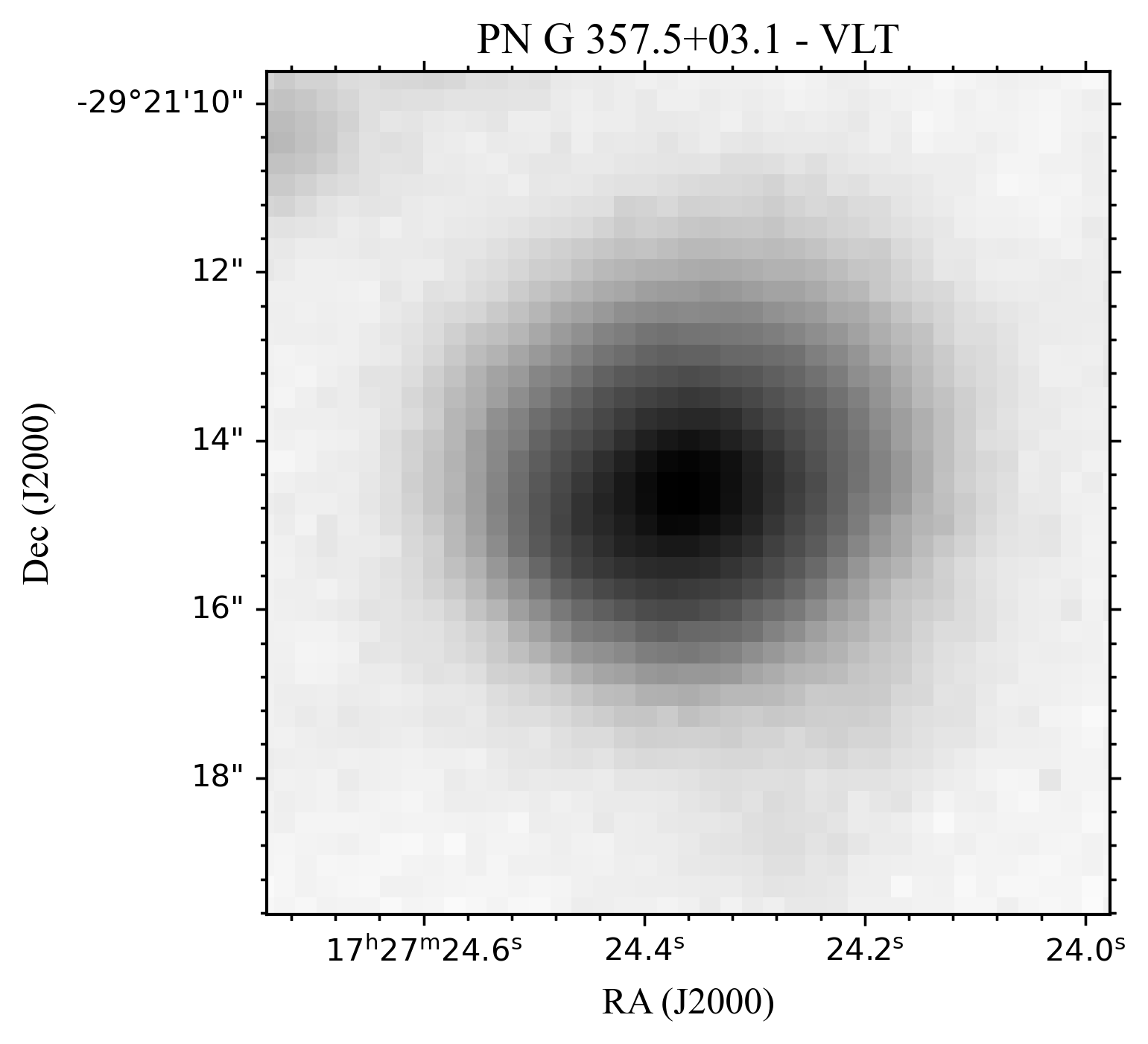}\hfill 
 \end{subfigure}\par\medskip 
\begin{subfigure}{\linewidth} 
  \includegraphics[width=.32\linewidth]{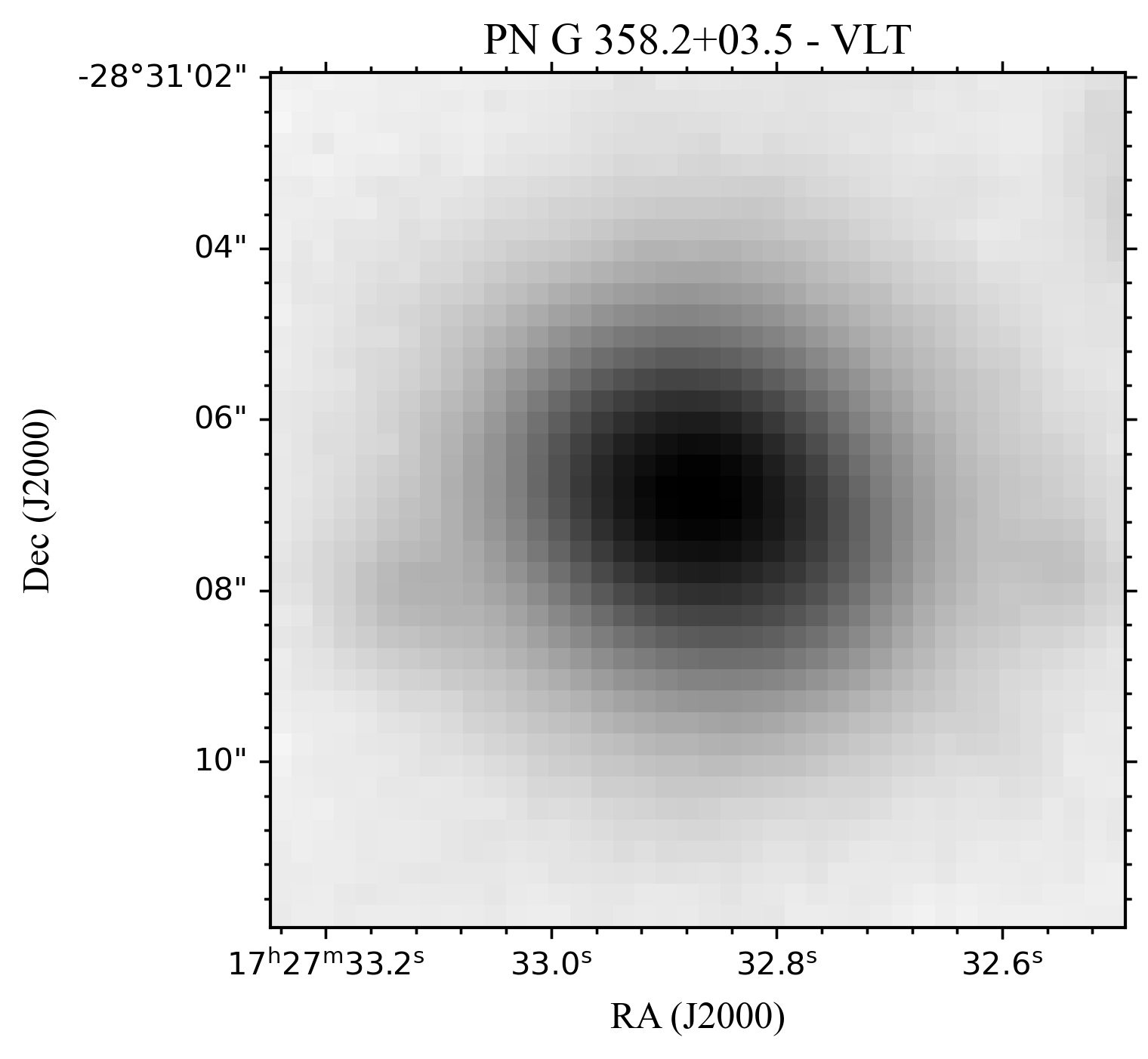}\hfill 
  \includegraphics[width=.32\linewidth]{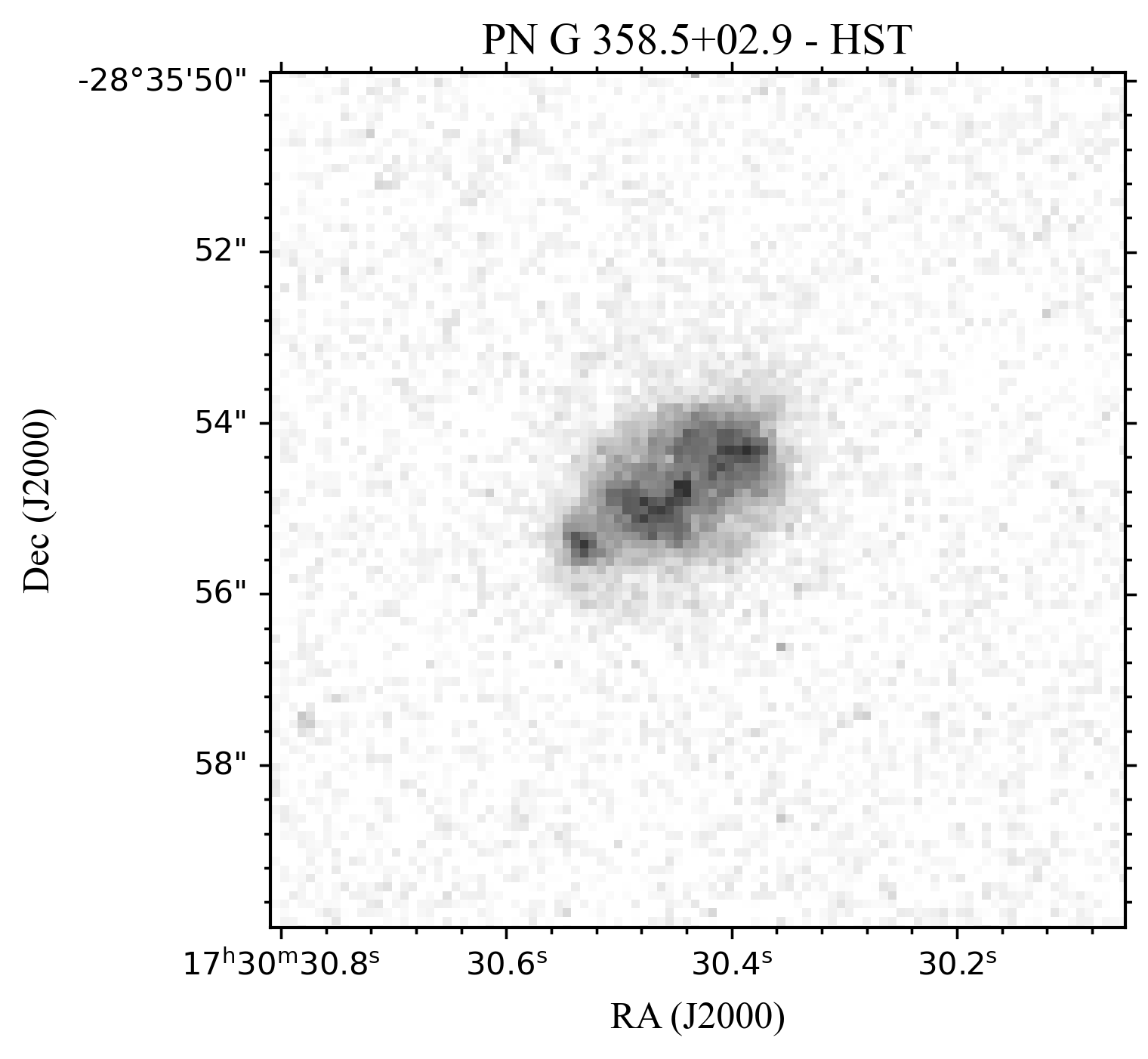}\hfill 
  \includegraphics[width=.32\linewidth]{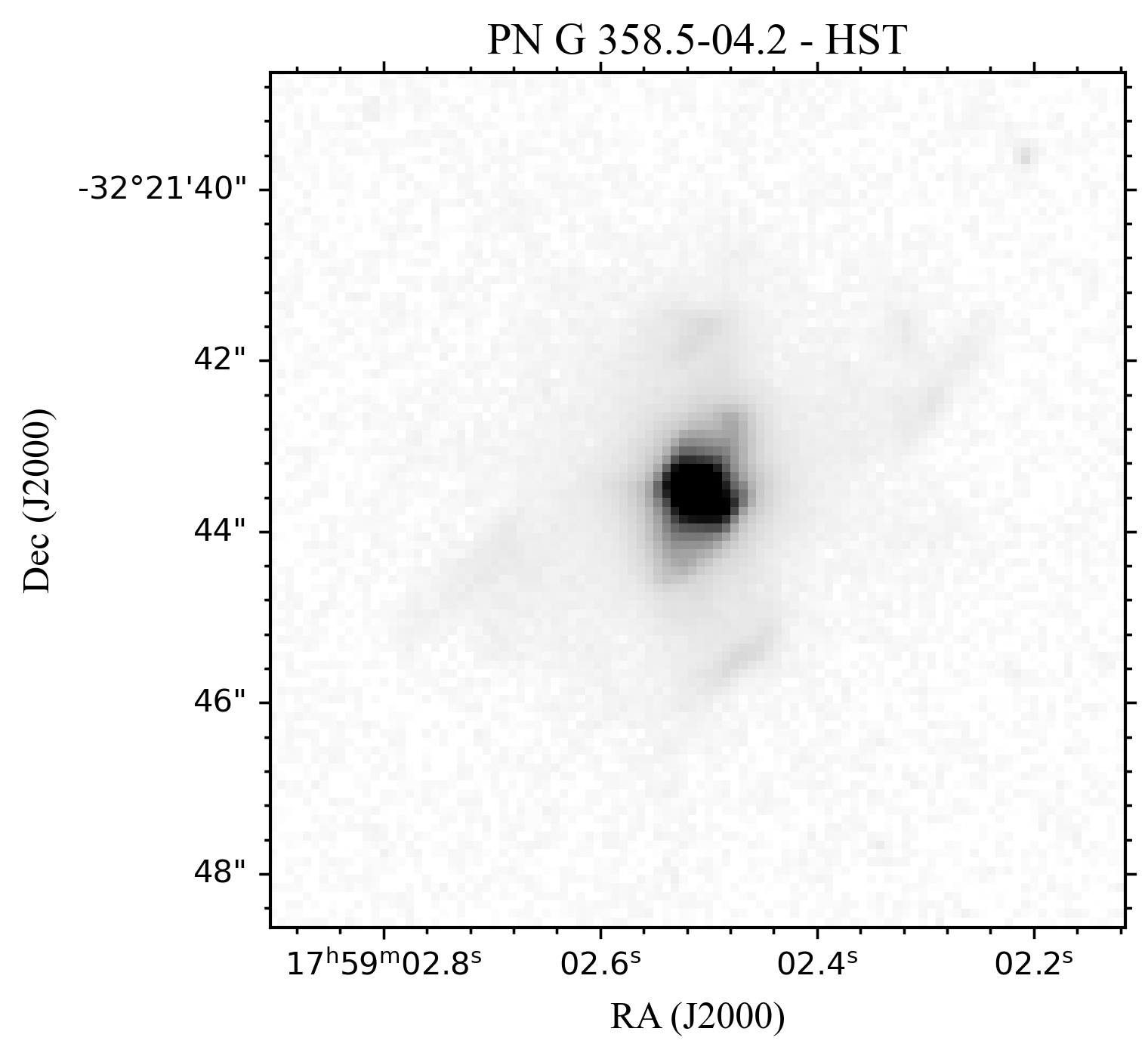}\hfill 
 \end{subfigure}\par\medskip 
\begin{subfigure}{\linewidth} 
  \includegraphics[width=.32\linewidth]{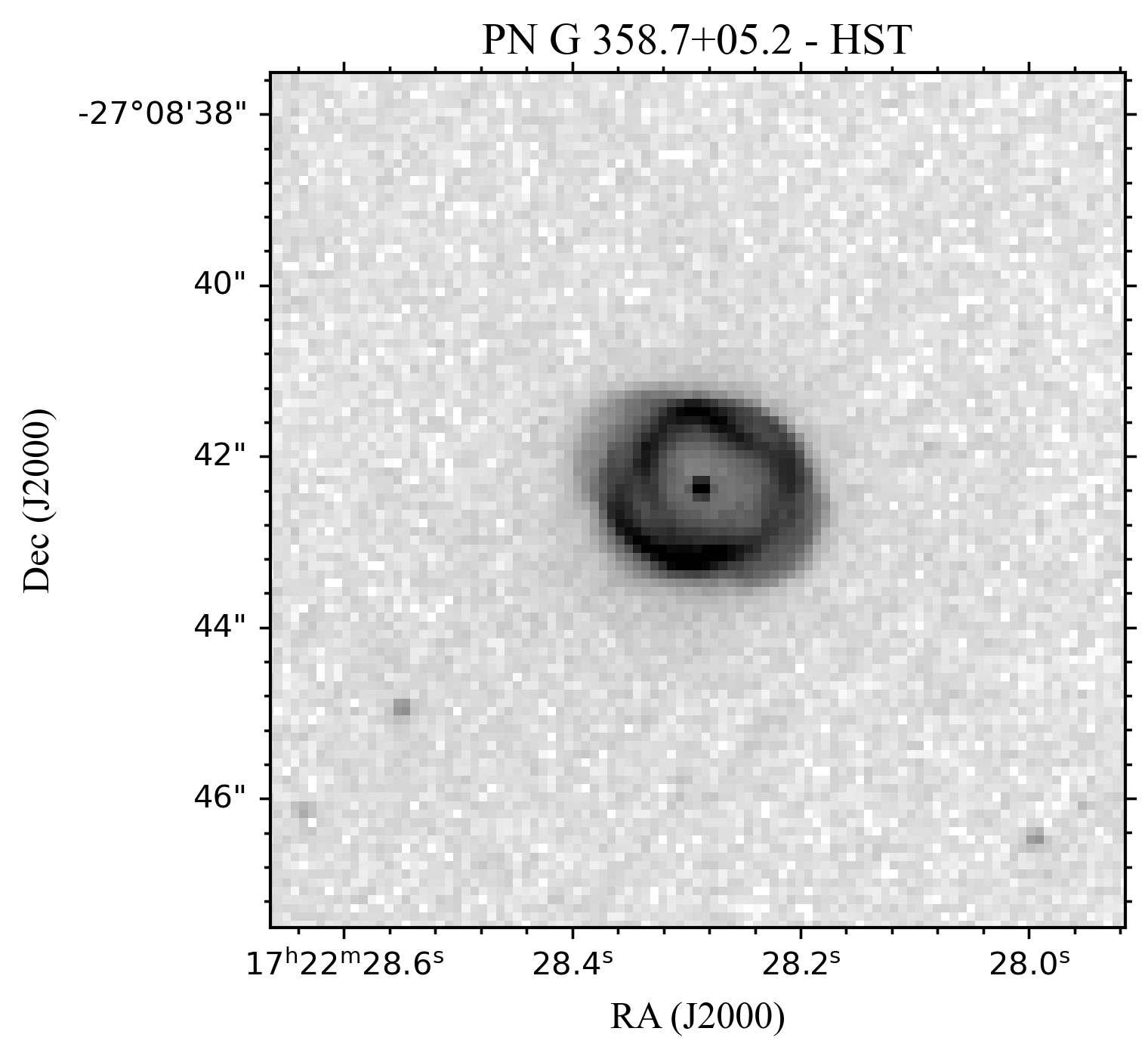}\hfill 
  \includegraphics[width=.32\linewidth]{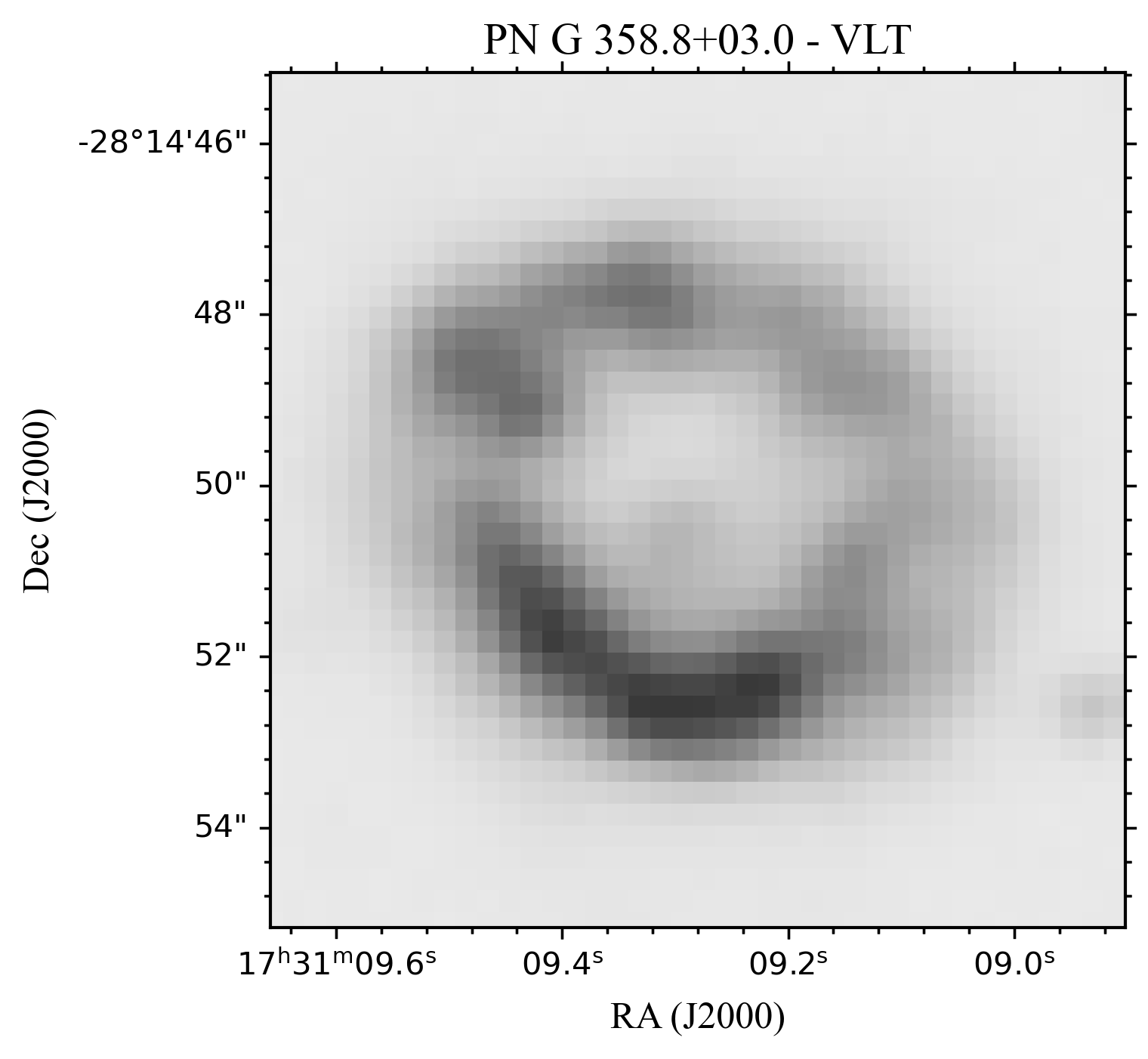}\hfill 
  \includegraphics[width=.32\linewidth]{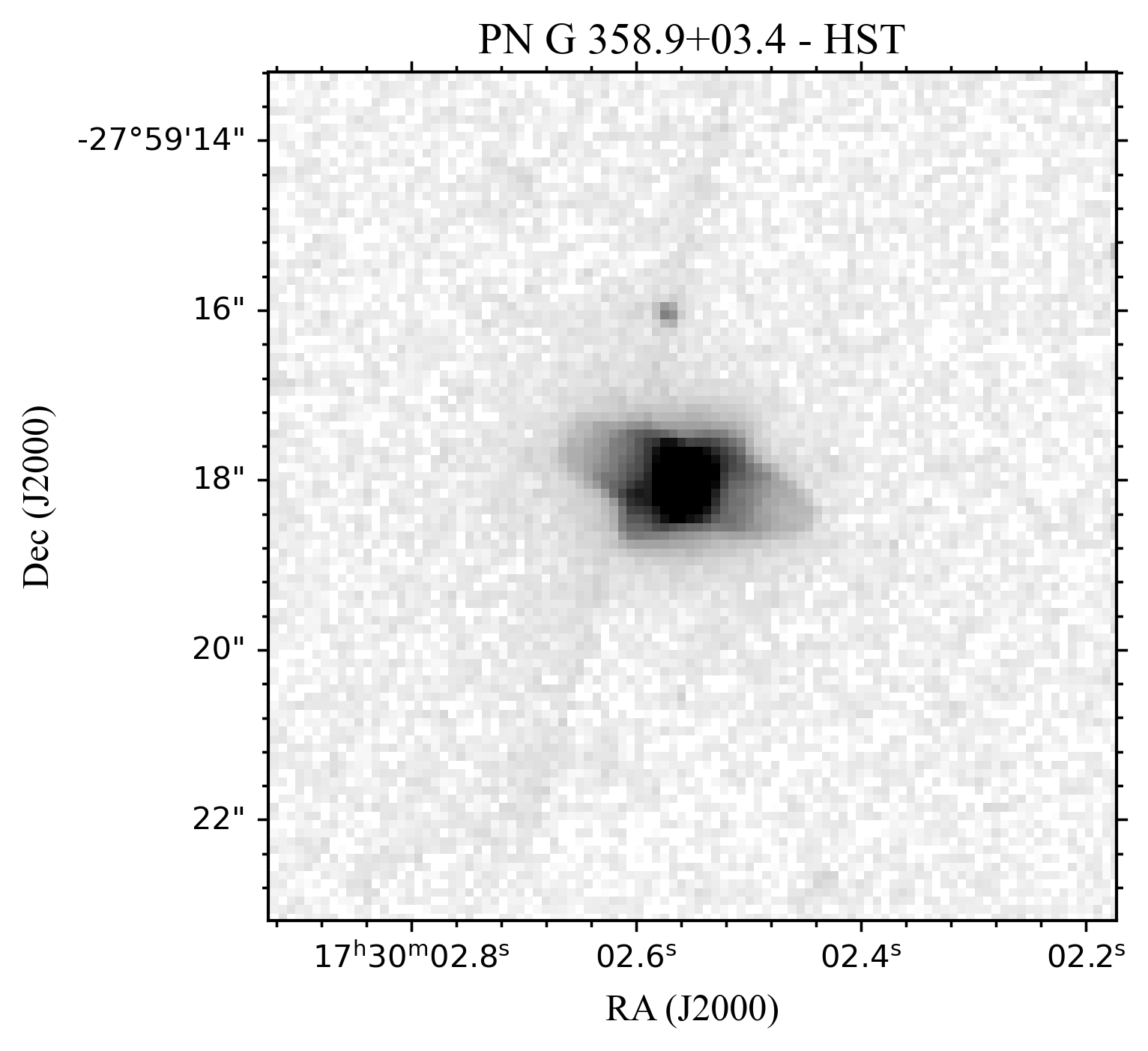}\hfill 
 \end{subfigure}\par\medskip 
  \end{figure} 
 \begin{figure} 
 \ContinuedFloat 
 \caption[]{continued:} 
\begin{subfigure}{\linewidth} 
  \includegraphics[width=.32\linewidth]{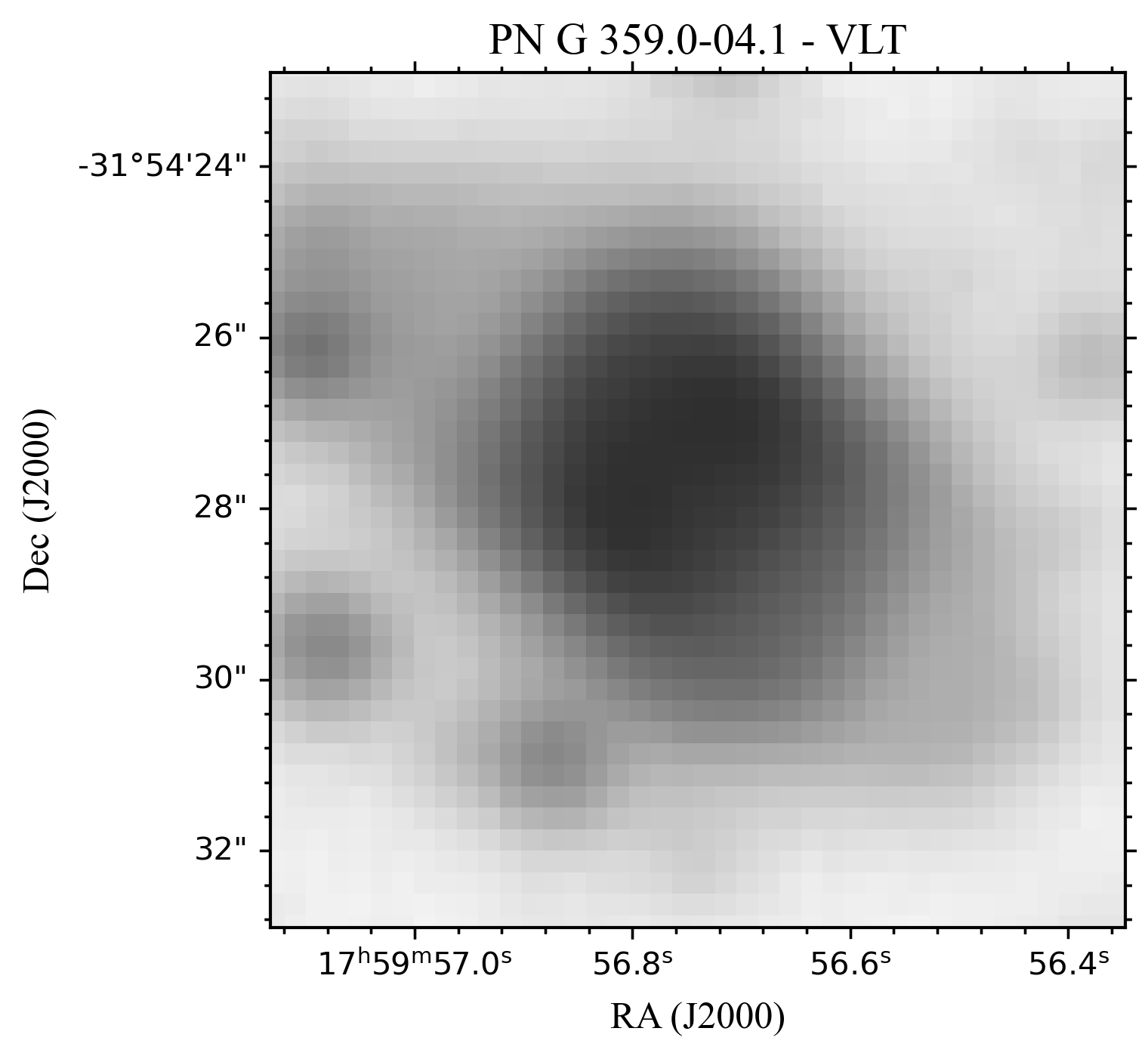}\hfill 
  \includegraphics[width=.32\linewidth]{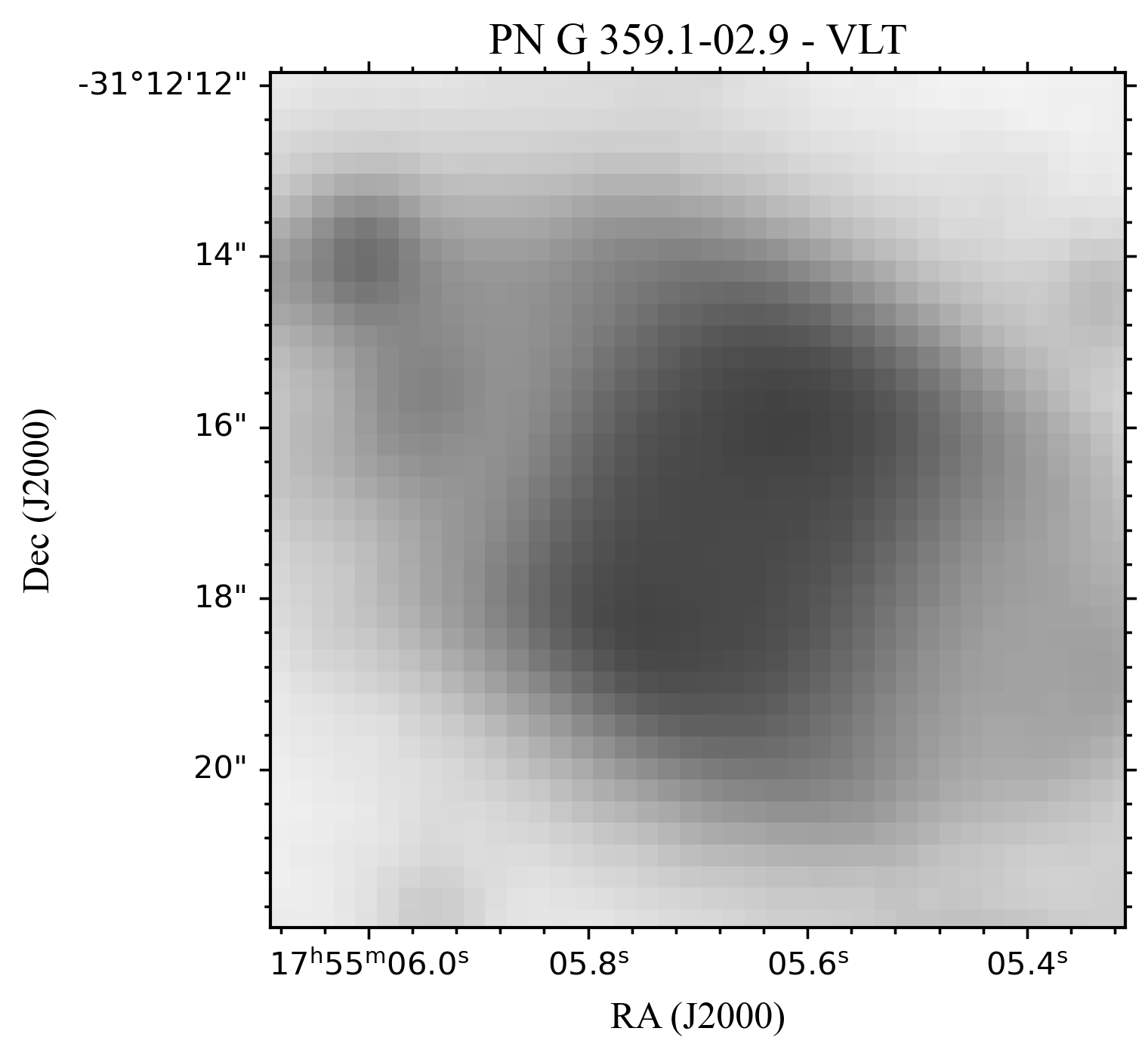}\hfill 
  \includegraphics[width=.32\linewidth]{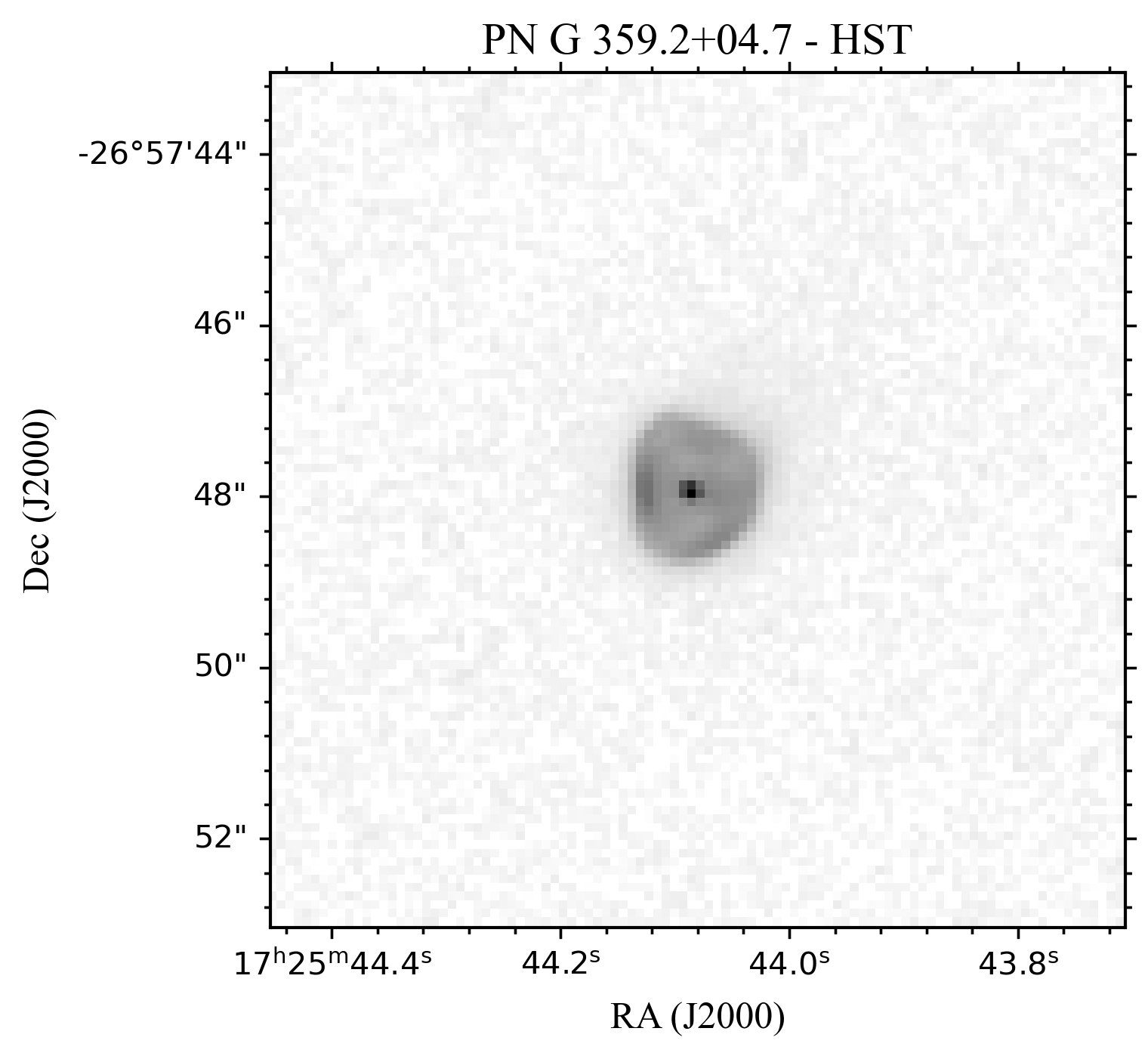}\hfill 
 \end{subfigure}\par\medskip 
\begin{subfigure}{\linewidth} 
  \includegraphics[width=.32\linewidth]{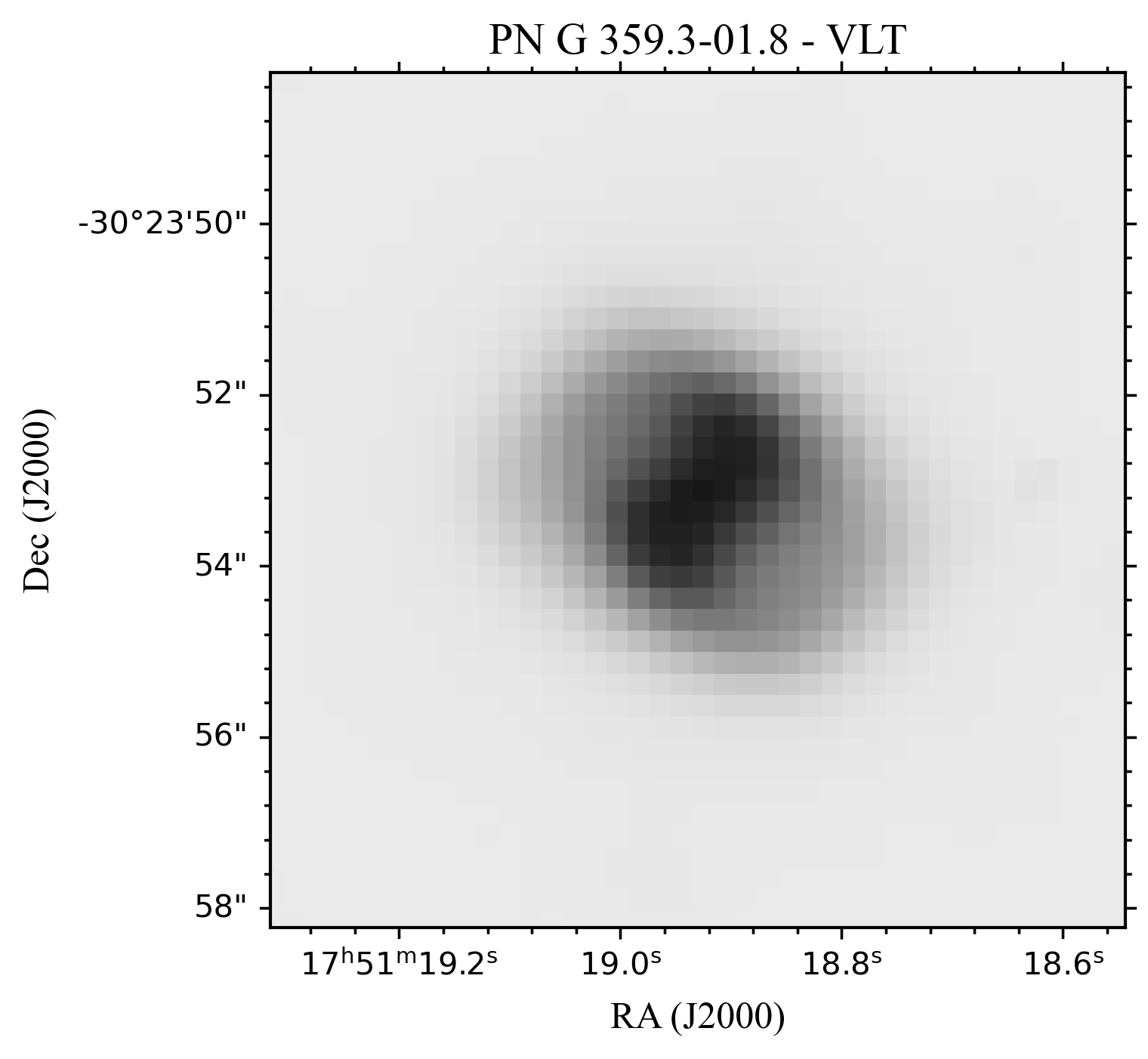}\hfill 
  \includegraphics[width=.32\linewidth]{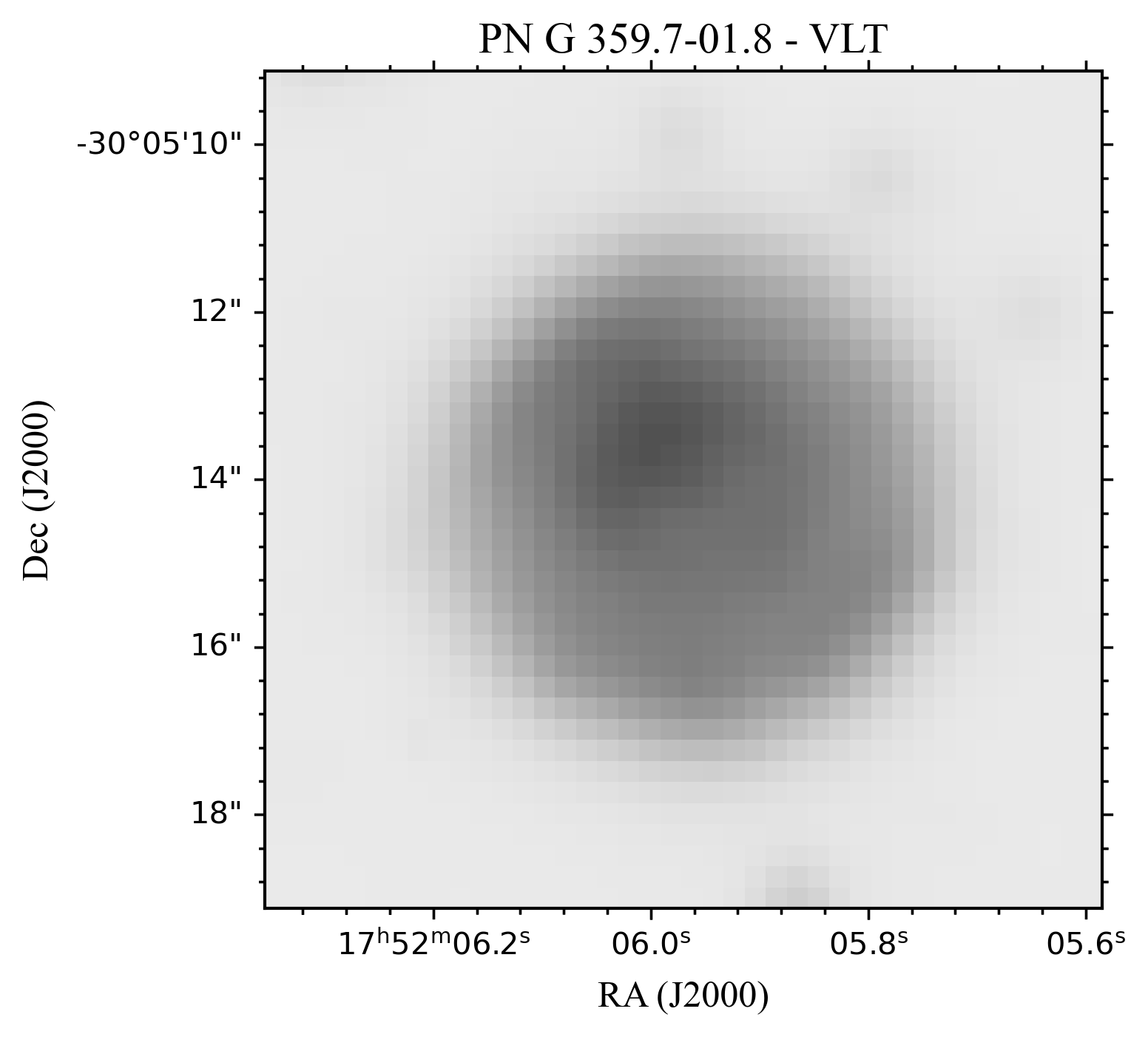}\hfill 
  \includegraphics[width=.32\linewidth]{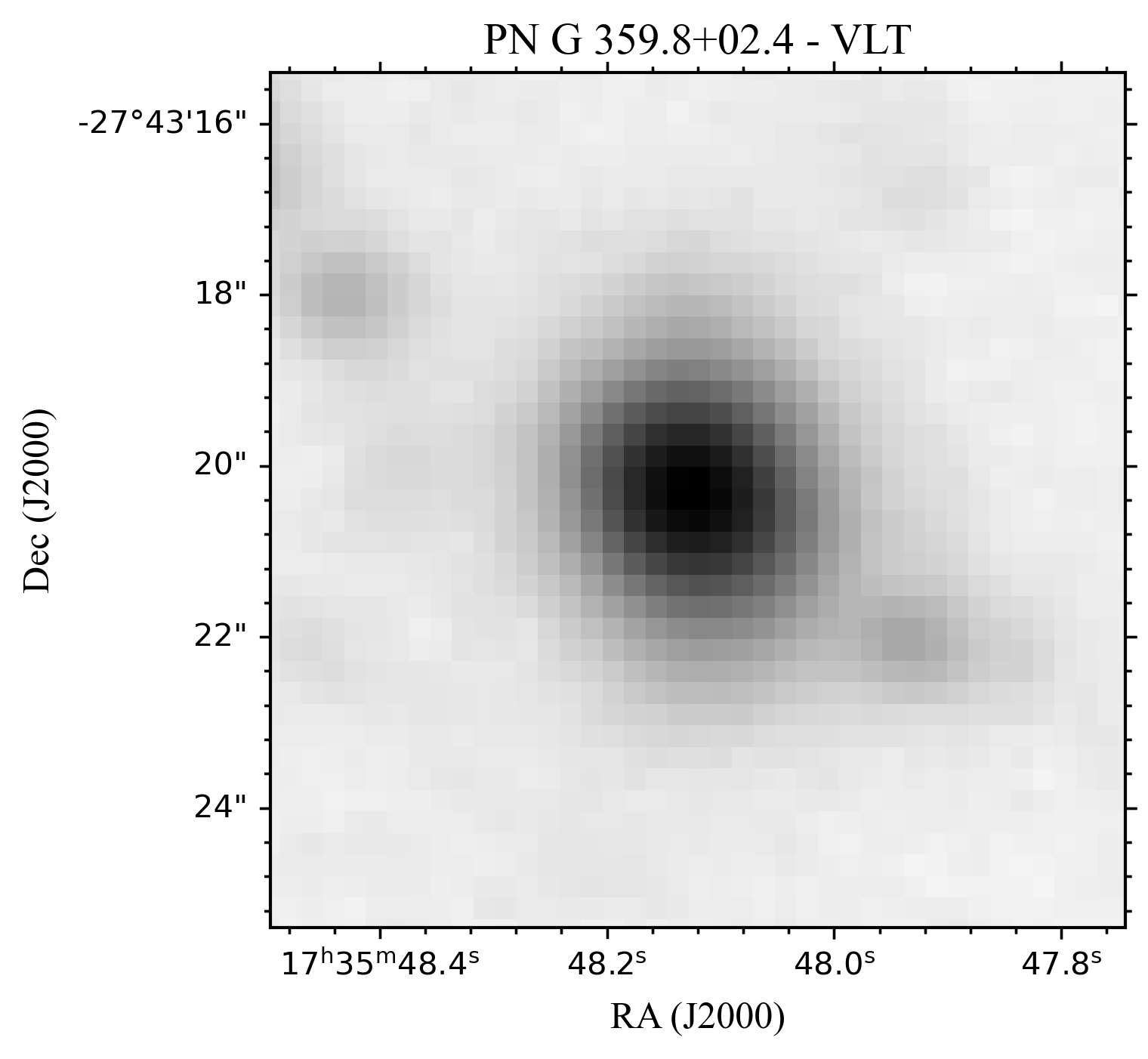}\hfill 
 \end{subfigure}\par\medskip 
\begin{subfigure}{\linewidth} 
  \includegraphics[width=.32\linewidth]{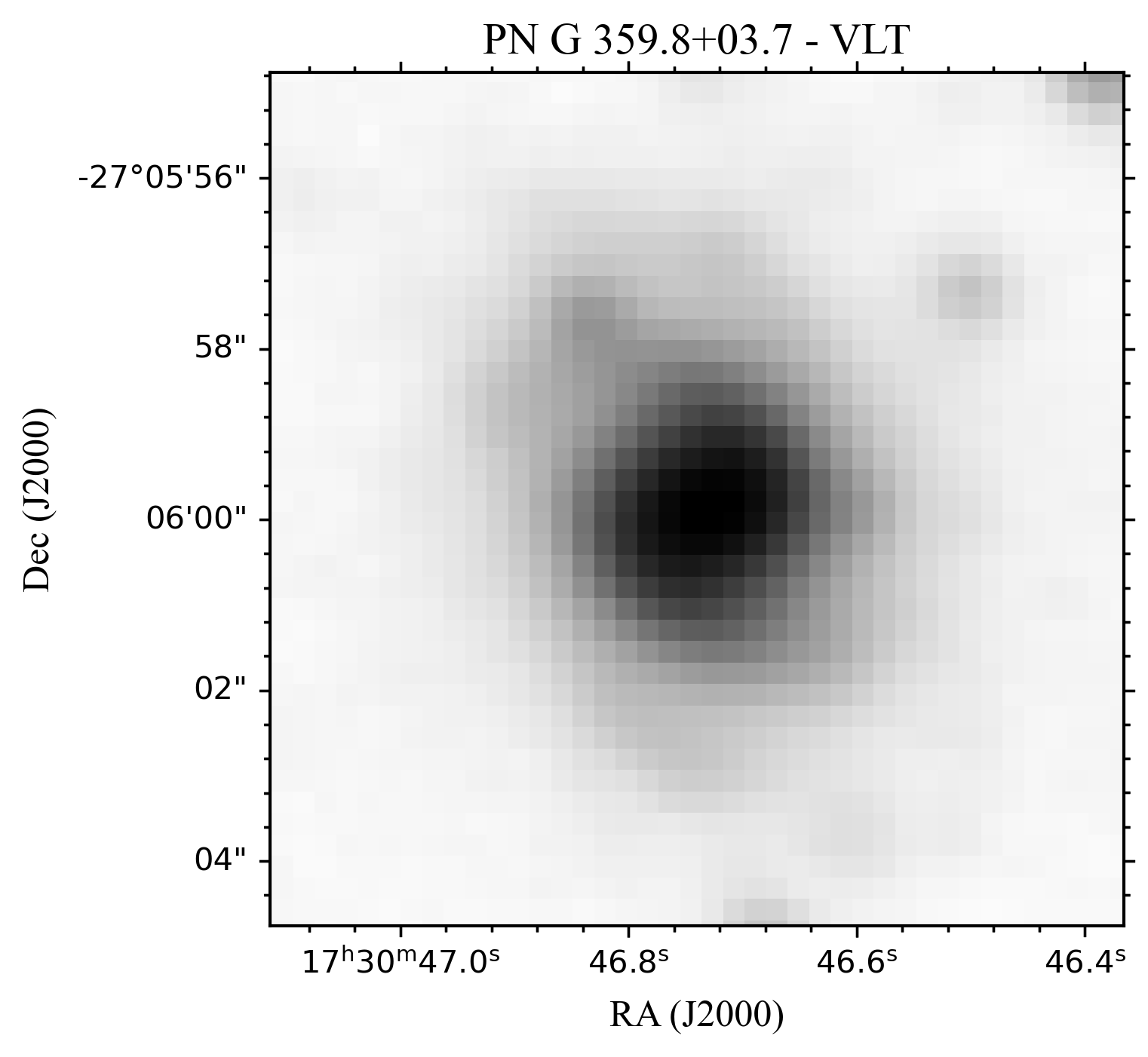}\hfill 
  \includegraphics[width=.32\linewidth]{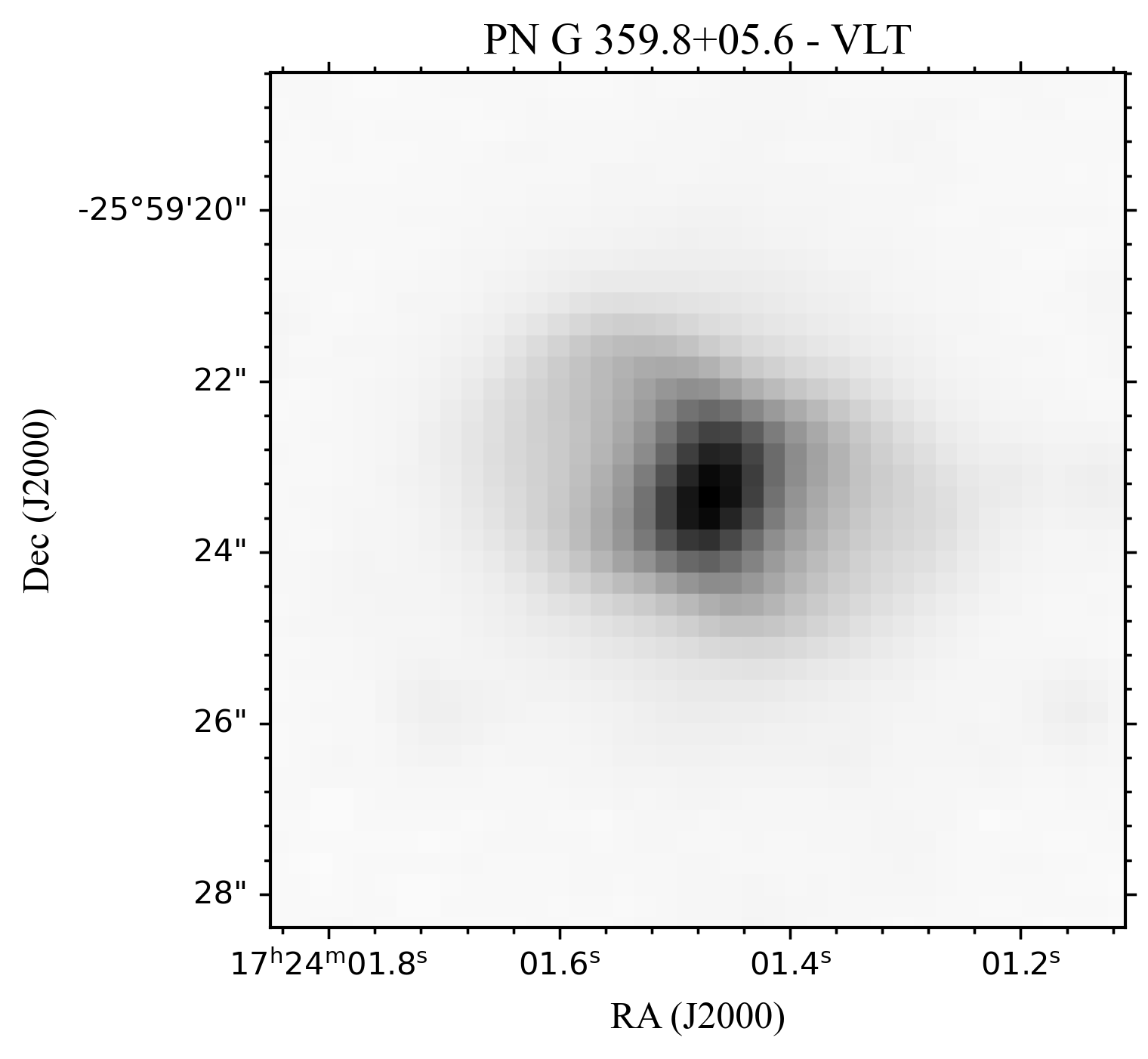}\hfill 
  \includegraphics[width=.32\linewidth]{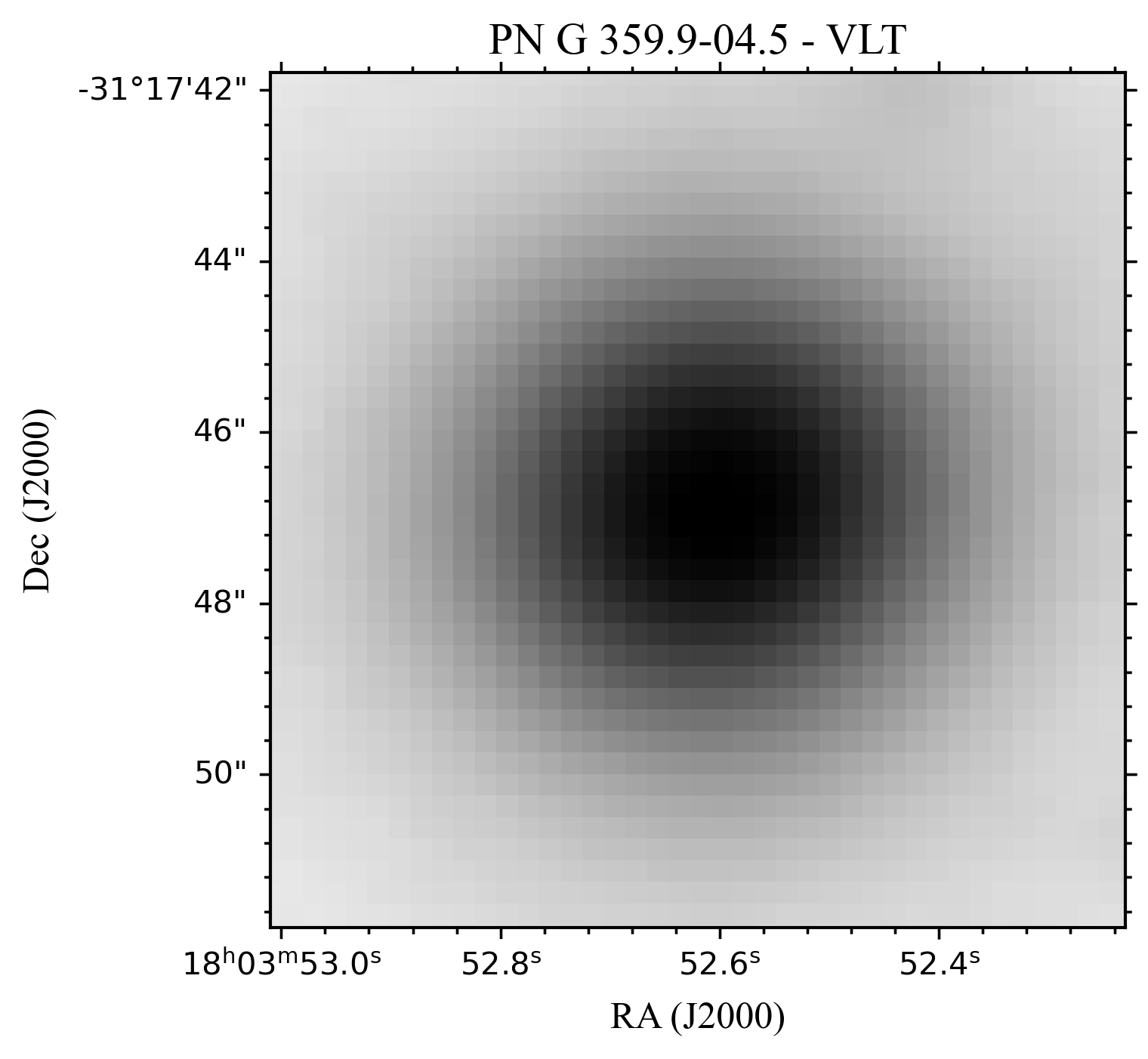}\hfill 
 \end{subfigure}\par\medskip 
 \end{figure}